\documentclass[12pt,twoside]{book}
\usepackage{amssymb}

\usepackage{graphicx}
\usepackage{amsmath}

\begin{document}

\date{}
\title{Phenomenological analysis of the Two Higgs Doublet Model.\\
Ph.D. Thesis}
\author{Rodolfo Alexander Diaz Sanchez \\
Universidad Nacional de Colombia.\\
Bogot\'{a}, Colombia.}
\maketitle
\listoffigures
\tableofcontents

\newpage \pagestyle{plain}

\begin{center}
{\Large \textbf{Acknowledgments} }
\end{center}

{\small I am strongly indebted to my advisor Dr. Roberto Martinez M. for his
academical, personal and financial support throughout my Ph. D. program. I
am also especially indebted to Drs. Marcela Carena for giving me the
opportunity to interact and work with her and some other colleagues at the
Fermi National Accelerator Laboratory.}

{\small I thank to Alexis Rodriguez for his important contributions in many
topics of this work. I also acknowledge to my thesis examiners: Marcela
Carena, and Marek Nowakowski for their useful comments and suggestions to
improve this work. I also acknowledge to Gabriel L\'{o}pez Castro and Enrico
Nardi for their comments about the thesis proposal.}

{\small Further, I am grateful to all members of the group of ``F\'{\i}sica
Te\'{o}rica de Altas Energ\'{\i}as'' of the Universidad Nacional de
Colombia, especially to Diego Torres for helping me with the edition of the
manuscript, and John Id\'{a}rraga for his useful support in some
computational issues. Finally, I am very grateful to Debajyoti Choudhury,
Heather Logan and Carlos Wagner for helping me to understand several topics
developed at Fermilab, especial acknowledgment to Heather for her patience
and unconditional collaboration.}

{\small On the other hand, I thank to Universidad Nacional de Colombia for
accepting me in its Ph.D. program, and for providing me with many of the
necessary resources. I also thank to Colciencias, DIB, DINAIN, and Banco de
la Rep\'{u}blica for the financial support to carry out the Ph.D. program. I
am indebted to the CLAF for the financial support to assist at the\ Third
Latin American Symposium of High Energy Physics (SILAFAE III) held at
Cartagena (Colombia). I also acknowledge as well to Sociedad Mexicana de
F\'{i}sica for its invitation and financial support to participate at the IX
Mexican School of Particles and Fields, and at the\ Workshop on Ultrahigh
Energy Cosmic Rays, realized at Puebla (Mexico). I am grateful to CERN and
CLAF for the financial support to participate at the First CERN-CLAF School
of High Energy Physics held in itacuruca (Brazil), and to the ICTP for the
financial support to participate in the Summer School on Particle Physics
2001, held at Trieste (Italy). I am besides indebted to Fermilab (USA), for
its kind hospitality and financial support.}

\begin{center}
{\Large \textbf{Abstract} }
\end{center}

{\small The Two Higgs Doublet model and its phenomenological implications
are discussed. A brief survey on the present status of this model is given.
In particular, we concentrate on the Two Higgs Doublet Model with Flavor
Changing Neutral Currents. First, we develop some new parametrizations of
the 2HDM type III, in such a way that the relation of it with models type I
and type II become apparent, based on two of these parametrizations we get
some bounds on the mixing vertex involving the second and third lepton
generations, as well as some lower bounds on the mass of the pseudoscalar
Higgs boson; such bounds are obtained from the }$g-2\;${\small muon factor.}

{\small Further, by using a parametrization in which one of the vacuum
expectation values vanishes, we constrain some lepton flavor violating
vertices, assuming that the lightest scalar Higgs mass }$m_{h^{0}}\;${\small %
is about }$115\;${\small GeV\ and that the pseudoscalar Higgs is heavier
than }$h^{0}${\small . Specifically, based on the }$g-2\;${\small muon
factor and the decay width of}$\;\mu \rightarrow e\gamma ${\small ,}$\;$%
{\small the following quite general bounds are obtained:}$\;7.62\times
10^{-4}\lesssim \xi _{\mu \tau }^{2}\lesssim 4.44\times 10^{-2},\;${\small \ 
}$\xi _{e\tau }^{2}\lesssim 2.77\times 10^{-14}${\small . Additionally,
based on the processes }$\tau \rightarrow \mu \gamma ${\small , and }$\tau
\rightarrow \mu \mu \mu ${\small , bounds on }$\xi _{\tau \tau }\;${\small %
and }$\xi _{\mu \mu }\;${\small are also gotten, such constraints on these
parameters}$\;${\small still give enough room for either a strong suppression%
}$\;${\small or strong enhancement on the coupling of any Higgs boson to a
pair of tau leptons or a pair of muons.\ Moreover, upper limits on the decay
widths of the leptonic decays }$\tau \rightarrow e\gamma ,\;${\small and }$%
\tau \rightarrow eee\;${\small are calculated, finding them to stay far from
the reach of near future experiments.}

{\small Finally, the Flavor Changing Charged Current decay }$\mu \rightarrow
\nu _{e}e\overline{\nu }_{\mu }\;${\small is considered in the framework of
the 2HDM type III as well. Since FCNC generates in turn flavor changing
charged currents in the lepton sector, this process appears at tree level
mediated by a charged Higgs boson exchange. From the experimental upper
limit for this decay, we obtain the bound }$\left| \xi _{\mu e}/m_{H^{\pm
}}\right| \leq 3.8\times 10^{-3}${\small \ where}$\;m_{H^{\pm }}\;${\small %
denotes the mass of the charged Higgs boson. This bound is independent on
the other free parameters of the model.}

{\small On the other hand, as an addendum to this work, we study the
production and decays of top squarks (stops) at the Tevatron collider in
models of low-energy supersymmetry breaking. We consider the case where the
lightest Standard Model (SM) superpartner is a light neutralino that
predominantly decays into a photon and a light gravitino. Considering the
lighter stop to be the next-to-lightest Standard Model superpartner, we
analyze stop signatures associated with jets, photons and missing energy,
which lead to signals naturally larger than the associated SM backgrounds.
We consider both 2-body and 3-body decays of the top squarks and show that
the reach of the Tevatron can be significantly larger than that expected
within either the standard supergravity models or models of low-energy
supersymmetry breaking in which the stop is the lightest SM superpartner.
For a modest projection of the final Tevatron luminosity, }$L\simeq 4$%
{\small \ fb}$^{-1}${\small , stop masses of order 300 GeV are accessible at
the Tevatron collider in both 2-body and 3-body decay modes. We also
consider the production and decay of ten degenerate squarks that are the
supersymmetric partners of the five light quarks. In this case we find that
common squark masses up to 360 GeV are easily accessible at the Tevatron
collider, and that the reach increases further if the gluino is light.}

\newpage

\pagenumbering{arabic}

\chapter{Introduction\label{introduction}}

The Standard Model (SM) of particle physics has been very succesful in
describing most of the smallest scale phenomenology known so far. However,
it possesses some problems whose solution could imply physics beyond its
scope. In order to motivate the study of the two Higgs doublet model (2HDM)
which is one of the simplest extensions of the SM, it is necessary to
discuss two important issues. Perhaps the two most fundamental ideas from
which the success of the SM comes are, (1) the extension of the gauge
invariance principle as a local concept (inspired in classical
electrodynamics) and (2) the implementation of the Spontaneous Symmetry
Breaking (SSB) phenomenon.

The introduction of local gauge invariance generates the so called \emph{%
gauge bosons} as well as the interactions of these gauge bosons with matter
(fermions), and also among the gauge bosons themselves (the latter only when
the gauge group is non abelian). On the other hand, the combination of local
gauge invariance with SSB leads naturally to the \emph{Higgs mechanism}
which in turn generates the masses of weak vector bosons and fermions. I
shall give a brief survey of both ideas emphasizing in the Higgs mechanism,
since the Two Higgs Doublet Model (2HDM) is an extension on the symmetry
breaking sector.

\section{Local gauge invariance\label{local gauge}}

It is well known from classical electrodynamics that Maxwell's equations are
invariant under a \emph{``local'' gauge transformation }of the form\emph{:\ }%
$A_{\mu }\rightarrow A_{\mu }+$ $\partial _{\mu }\lambda (x)\;$where$%
\;A_{\mu }\;$is the four-vector potential. On the other hand, taking the
free Dirac Lagrangian

\begin{equation}
\pounds _{free}=\overline{\Psi }(i\gamma ^{\mu }\partial _{\mu }-m)\Psi
\label{Dirac}
\end{equation}
We can see that such Lagrangian is invariant under the global phase shift $%
\Psi \rightarrow e^{i\theta }\Psi .\;$Nevertheless, inspired in the local
gauge symmetry in electrodynamics explained above, we could ask the
following question, is it possible to extend the global symmetry and demand
it to be local? if yes, what are the physical consequences?. It is
straightforward to check that such locality could be accomplished by
replacing the ``normal'' derivative $\partial _{\mu }\;$by the ``covariant
derivative'' $D_{\mu }\equiv \partial _{\mu }+iqA_{\mu },\;$where $A_{\mu
}\; $is a four vector field that transforms as $A_{\mu }\rightarrow A_{\mu
}+\partial _{\mu }\lambda \left( x\right) \;$when the local transformation
of $\Psi \rightarrow \exp \left( -iq\lambda \left( x\right) \right) \Psi \;$%
is realized.$\;$The Lagrangian (\ref{Dirac}) becomes

\begin{equation*}
\pounds =\overline{\Psi }(i\gamma ^{\mu }D_{\mu }-m)\Psi =\overline{\Psi }%
(i\gamma ^{\mu }\partial _{\mu }-m)\Psi -qA_{\mu }\overline{\Psi }\gamma
^{\mu }\Psi =\pounds _{free}-J^{\mu }A_{\mu }
\end{equation*}
it is easy to see that this new Lagrangian is invariant under the combined
transformations $\Psi \rightarrow e^{-iq\lambda \left( x\right) }\Psi
,\;A_{\mu }\rightarrow A_{\mu }+\partial _{\mu }\lambda \left( x\right) .\;$%
Now, if we interpret $A_{\mu }\;$as the four vector electromagnetic
potential, then $J^{\mu }$ is the four-vector electromagnetic current. To
complete the Lagrangian of Quantum Electrodynamics (QED) we just add the
kinetic term that describe the propagation of free photons

\begin{eqnarray*}
\pounds _{QED} &=&\pounds _{free}-J^{\mu }A_{\mu }-\frac{1}{4}F^{\mu \nu
}F_{\mu \nu } \\
F_{\mu \nu } &\equiv &\partial _{\mu }A_{\nu }-\partial _{\nu }A_{\mu }
\end{eqnarray*}
And the kinetic term for the free photons (which leads to the Maxwell's
equations) is also local gauge invariant. Therefore, the matter-radiation
coupling has been generated from the imposition of ``locality'' to the gauge
principle. Additionally, to preserve locality we have introduced into the
covariant derivative a four-vector field (the four-vector potential $A_{\mu
} $), which is called a \emph{gauge field,\ }and also a parameter $q\;$which
acts as the generator for the local group transformations $\widehat{U}\left(
x\right) =\exp \left( -iq\lambda \left( x\right) \right) $.\emph{\ }In this
case, to analize the symmetries we have used the unidimensional rotation
group in the complex space, this group is called $U\left( 1\right) ,$\ the
group of unitary complex matrices $1\times 1$. In the electroweak SM we use
besides, the group of unitary matrices $2\times 2$ of determinant one (known
as $SU\left( 2\right) $), the latter is a non abelian group whose generators
obey the Lie algebra of the rotation group in three dimensions $SO\left(
3\right) \;$(i.e. they are isomorphic). After applying local gauge
invariance to the whole electroweak group $SU\left( 2\right) \times U\left(
1\right) ,\;$four gauge fields appear and they in turn generate after some
additional transformations the three weak vector bosons and the photon.

\section{Spontaneous symmetry breaking and Higgs mechanism\label{SSB}}

Using local gauge invariance as a dynamical principle is not enough to
predict particle physics phenomenology since it leads to massless gauge
bosons that do not correspond to physical reality. SM predicts that these
vector bosons acquire their masses from a Spontaneous Symmetry Breaking
(SSB) phenomenon explained below.

Sometimes when the vacuum (minimum of the potential) of a system is
degenerate, after choosing a particular one\footnote{%
Quantum field theory demands the vacuum to be unique such that perturbation
expansions are calculated around that point}, this minimum is not invariant
under the symmetry of the Lagrangian, when the vacuum does not have the
symmetry of the Lagrangian we say that the symmetry has been spontaneously
broken. When this phenomenon occurs some other massless particles called
Goldstone bosons arise in the spectrum. However, if the Lagrangian posseses
a local gauge symmetry an interrelation among gauge and Goldstone bosons
endows the former with a physical mass, while the latter dissapear from the
spectrum, this phenomenon is called the \emph{Higgs mechanism\ }\cite{Higgs
mech}.\emph{\ }To explain the mechanism we shall use a toy model describing
a couple of self interacting complex scalar fields $\left( \phi \;\text{and }%
\phi ^{\ast }\right) $ whose Lagrangian is local gauge invariant

\begin{equation}
\pounds \;=\frac{1}{2}\left| D^{\mu }\phi \right| ^{2}-V\left( \phi ^{\ast
}\phi \right) -\frac{1}{4}F_{\mu \nu }F^{\mu \nu }\;;\;V\left( \phi \right)
\equiv -\frac{1}{2}\mu ^{2}\left| \phi \right| ^{2}+\frac{1}{4}\lambda
^{2}(\phi ^{\ast }\phi )^{2}  \label{toy}
\end{equation}
where

\begin{equation*}
\phi =\phi _{1}+i\phi _{2},
\end{equation*}
is a complex field and

\begin{equation*}
D_{\mu }\equiv \partial _{\mu }+iqA_{\mu }\;\;\;;\;\;\;\;F_{\mu \nu }\equiv
\partial _{\mu }A_{\nu }-\partial _{\nu }A_{\mu }
\end{equation*}
This Lagrangian has already the local gauge invariance described by the
simultaneous transformations 
\begin{equation}
\phi (x)\rightarrow e^{-iq\lambda (x)}\phi (x),\ A_{\mu }(x)\rightarrow
A_{\mu }(x)+\partial _{\mu }\lambda (x)  \label{local inv}
\end{equation}
Observe that the imposition of locality generates the interaction of the
complex scalar fields with a four vector field. If $\mu ^{2}<0,$ The
potential $V\left( \phi \right) $ posseses a unique minimum at $\phi =0$
which preserves the symmetry of the Lagrangian. However, if $\mu ^{2}>0$,
the Lagrangian has a continuum degenerate set of vacua (minima) lying on a
circle of radius $\mu /\lambda $ 
\begin{equation*}
\langle \left| \phi \right| ^{2}\rangle =\langle \phi _{1}\rangle
^{2}+\langle \phi _{2}\rangle ^{2}=\frac{\mu ^{2}}{\lambda ^{2}}\equiv \nu
^{2}
\end{equation*}
any of them might be chosen as the fundamental state, but no one of them is
invariant under a local phase rotation\footnote{%
The set of all ground states is invariant under the symmetry but the
obligation to choose (in order to set up a perturbation formalism) only one
of the vacua, leads us to the breaking of the symmetry.}. According to the
definition made above, the symmetry of the Lagrangian has been spontaneously
broken. Choosing a particular minimum: 
\begin{equation*}
\langle \phi _{1}\rangle =\frac{\mu }{\lambda }\equiv \nu \;\;;\;\;\langle
\phi _{2}\rangle =0
\end{equation*}
we say that the field $\phi _{1}\;$has acquired a Vacuum Expectation Value
(VEV) $\langle \phi _{1}\rangle $. It is convenient to introduce new fields 
\begin{equation*}
\eta \equiv \phi _{1}-v\;\;;\;\;\xi \equiv \phi _{2}
\end{equation*}
and expanding the Lagrangian in terms of these new fields we obtain:

\begin{center}
\begin{eqnarray*}
\pounds &=&\left[ \frac{1}{2}(\partial _{\mu }\eta )(\partial ^{\mu }\eta
)-\mu ^{2}\eta ^{2}\right] +\frac{1}{2}\left[ (\partial _{\mu }\xi
)(\partial ^{\mu }\xi )\right] +\left[ -\frac{1}{4}F_{\mu \nu }F^{\mu \nu }+%
\frac{q^{2}v^{2}}{2}A_{\mu }A^{\mu }\right] \\
&&-2iqv\left( \partial _{\mu }\xi \right) A^{\mu }+\left\{ q\left[ \eta
\left( \partial _{\mu }\xi \right) -\xi \left( \partial _{\mu }\eta \right) %
\right] A^{\mu }+vq^{2}\left( \eta A_{\mu }A^{\mu }\right) \right. \\
&&\left. +\frac{q^{2}}{2}\left( \xi ^{2}+\eta ^{2}\right) A_{\mu }A^{\mu
}-\lambda \mu \left( \eta ^{3}+\eta \xi ^{2}\right) -\frac{\lambda ^{2}}{4}%
\left( \eta ^{4}+2\eta ^{2}\xi ^{2}+\xi ^{4}\right) \right\} +\frac{\mu
^{2}v^{2}}{4}
\end{eqnarray*}
\end{center}

The particle spectrum consists of

\begin{enumerate}
\item  A field $\eta \;$with mass $\sqrt{2}\mu .$

\item  A vector boson $A_{\mu }$ that has acquired a mass $q\nu >0$ by means
of the VEV.

\item  A massless field $\xi \;$called a Goldstone boson.
\end{enumerate}

However, the Lagrangian above looks dissapointing because of a term of the
form $\left( \partial _{\mu }\xi \right) A^{\mu }\;$which does not have a
clear interpretation in the Feynman formalism. Fortunately, we are able to
remove the unwanted would be Goldstone field out, by exploiting the local
gauge invariance of the Lagrangian. Writing Eq. (\ref{local inv}) in terms
of $\phi _{1}\;$and $\phi _{2}$%
\begin{equation*}
\phi \rightarrow \phi ^{\prime }=e^{i\theta \left( x\right) }\phi =\left[
\phi _{1}\cos \theta \left( x\right) -\phi _{2}\sin \theta \left( x\right) %
\right] +i\left[ \phi _{1}\sin \theta \left( x\right) +\phi _{2}\cos \theta
\left( x\right) \right]
\end{equation*}
where $\theta \left( x\right) \equiv -q\lambda \left( x\right) $,$\;$and
using 
\begin{equation}
\theta \left( x\right) =-\arctan \left( \frac{\phi _{2}\left( x\right) }{%
\phi _{1}\left( x\right) }\right)  \label{gauge1}
\end{equation}
we get $\phi ^{\prime }\;$to be real\footnote{%
Observe that the invariance under a phase rotation of the complex field $%
\phi \rightarrow e^{i\theta }\phi $ is equivalent to the invariance under an
SO(2) rotation of the real and imaginary parts $\phi _{1}\rightarrow \phi
_{1}\cos \theta -\phi _{2}\sin \theta \;;\;\phi _{2}\rightarrow \phi
_{1}\sin \theta +\phi _{2}\cos \theta $}. The gauge field transforms as $%
A_{\mu }^{\prime }(x)=A_{\mu }(x)+\partial _{\mu }\lambda (x).\;$However,
this gauge transformation does not affect the physical content of $A_{\mu
}(x)\;$so we drop the prime notation out from it. Using the local
transformations defined by (\ref{local inv}) and (\ref{gauge1}), (i.e. this
particular \emph{gauge}), the Lagrangian reads 
\begin{eqnarray*}
\pounds &=&\left[ \frac{1}{2}(\partial _{\mu }\eta )(\partial ^{\mu }\eta
)-\mu ^{2}\eta ^{2}\right] +\left[ -\frac{1}{4}F_{\mu \nu }F^{\mu \nu }+%
\frac{q^{2}v^{2}}{2}A_{\mu }A^{\mu }\right] \\
&&+\left\{ q^{2}v\left( \eta A_{\mu }A^{\mu }\right) +\frac{q^{2}}{2}\eta
^{2}A_{\mu }A^{\mu }-\lambda \mu \eta ^{3}-\frac{\lambda ^{2}}{4}\eta
^{4}\right\} +\frac{\mu ^{2}v^{2}}{4}
\end{eqnarray*}
so we have got rid of the massless field $\xi \;$and all its interactions,
especially the ``disgusting'' term $\left( \partial _{\mu }\xi \right)
A^{\mu }.\;$On the other hand, we are left with a massive scalar field $\eta
\;$(a Higgs particle) and a massive four vector field $A_{\mu }\;$(a massive
``photon'').

By making a counting of degrees of freedom we realize that one degree of
freedom has dissapeared (a massless would be Goldstone boson) while another
one has arisen (a longitudinal polarization for the four vector boson i.e.
its mass). Therefore, it is generically said that the photon has ``eaten''
the would be Goldstone boson $\xi \;$in order to acquire mass. This result
is known as the \emph{Higgs mechanism}. Notwithstanding, it worths to
emphasize that as well as the massive vector boson, the Higgs mechanism has
provided us with an additional physical degree of freedom that corresponds
to a scalar field describing the so-called ``Higgs particle''.

We can note that the Higgs mechanism is possible because of the conjugation
of both the local gauge invariance principle and the SSB. For instance, if
we implement a SSB with a global symmetry, what we obtain is a certain
number of (physical) massless Goldstone bosons, it is because with a global
symmetry we do not generate vector bosons that ``eat'' such extra degrees of
freedom. Technically, the number of Goldstone bosons generated after the
symmetry breaking is equal to the number of broken generators (Goldstone
theorem \cite{Goldstone}).

In SM, the Higgs mechanism creates three massive vector bosons\ ($W^{\pm },Z$%
)\ and one massless vector boson (the photon), as well as a Higgs particle
which has not been discovered hitherto.

\section{The Higgs mechanism in the Standard Model\label{SSB SM}}

The SM of particle physics \cite{Salam}, picks up the ideas of local gauge
invariance and SSB to implement a Higgs mechanism. The local gauge symmetry
is $SU\left( 2\right) _{L}\times U\left( 1\right) _{Y}\;$and the SSB obeys
the scheme $SU\left( 2\right) _{L}\times U\left( 1\right) _{Y}\rightarrow
U\left( 1\right) _{Q}\;$where the subscript $L\;$means that $SU\left(
2\right) \;$only acts on left-handed doublets (in the case of fermions), $%
Y\; $is the generator of the original $U\left( 1\right) \;$group, and $Q\;$%
correspond to an unbroken generator (the electromagnetic charge).
Specifically, the symmetry breaking is implemented by introducing a scalar
doublet 
\begin{equation*}
\Phi =\left( 
\begin{array}{c}
\phi ^{+} \\ 
\phi ^{0}
\end{array}
\right) =\left( 
\begin{array}{c}
\phi _{1}+i\phi _{2} \\ 
\phi _{3}+i\phi _{4}
\end{array}
\right)
\end{equation*}
It transforms as an $SU\left( 2\right) _{L}\;$doublet, thus its weak
hypercharge is $Y=1$. In order to induce the SSB the doublet should acquire
a VEV different from zero

\begin{equation}
\langle \Phi \rangle =\left( 
\begin{array}{c}
0 \\ 
v/\sqrt{2}
\end{array}
\right)  \label{VEV SM}
\end{equation}

What is new respect to the toy model above, is that the original local
symmetry $SU\left( 2\right) _{L}\times U\left( 1\right) _{Y}\;$is non abelian%
\footnote{%
A very important consequence of the non-abelianity of the gauge group is the
generation of self interactions among the associated gauge bosons, they
appear when the kinetic term for the gauge bosons (Yang-Mills Lagrangian) is
introduced.}. Its generators are $\tau _{i},Y\;$corresponding to $SU\left(
2\right) _{L}\;$and $U\left( 1\right) _{Y}\;$respectively, the generators $%
\tau _{i}$ are defined as 
\begin{equation*}
\tau _{i}\equiv \frac{\sigma _{i}}{2}\;
\end{equation*}
where $\sigma _{i}\;$are the Pauli matrices. Therefore, such four generators
obey the following lie algebra 
\begin{equation*}
\left[ \tau _{i},\tau _{j}\right] =i\varepsilon _{ij}^{k}\tau _{k}\;\;;\;\;%
\left[ \tau _{i},Y\right] =0
\end{equation*}
When the symmetry is spontaneously broken in the potential (see below) the
doublet acquire a VEV, we can see easily that all generators of the $%
SU\left( 2\right) _{L}\times U\left( 1\right) _{Y}\;$are broken generators 
\begin{eqnarray*}
\tau _{1}\langle \Phi \rangle &=&\frac{1}{2}\left( 
\begin{array}{c}
v/\sqrt{2} \\ 
0
\end{array}
\right) \neq 0\;\;;\;\;\tau _{2}\langle \Phi \rangle =\frac{1}{2}\left( 
\begin{array}{c}
-iv/\sqrt{2} \\ 
0
\end{array}
\right) \neq 0 \\
\tau _{3}\langle \Phi \rangle &=&\frac{1}{2}\left( 
\begin{array}{c}
0 \\ 
-v/\sqrt{2}
\end{array}
\right) \neq 0\;\;;\;\;Y\langle \Phi \rangle =\left( 
\begin{array}{c}
0 \\ 
v/\sqrt{2}
\end{array}
\right) \neq 0
\end{eqnarray*}
However, we can define an unbroken generator by the Gellman-Nijishima
relation 
\begin{equation*}
Q=\left( \tau _{3}+\frac{Y}{2}\right) \;\;;\;\;Q\langle \Phi \rangle =0
\end{equation*}
In such a way that the scheme of SSB is given by $SU\left( 2\right)
_{L}\times U\left( 1\right) _{Y}\rightarrow U\left( 1\right) _{Q}.\;$%
According to the Goldstone theorem, the number of would be Goldstone bosons
generated after the symmetry breaking is equal to the number of broken
generators (which in turn is equal to the number of massive gauge bosons in
the case of local symmetries). Therefore, instead of working with four
broken generators we shall work with three broken generators and one
unbroken generator $Q.\;$This scheme ensures for the photon to remain
massless, while the other three gauge bosons acquire masses.

Let us examine the contributions that the doublet $\Phi \;$introduces in the
SM.

\subsection{The Higgs Potential\label{Higgspot SM}}

The Higgs potential generates the SSB as well as the self interaction terms
of the scalar boson, the most general renormalizable potential invariant
under $SU\left( 2\right) _{L}\times U\left( 1\right) _{Y}\;$is given by 
\textbf{\ } 
\begin{equation}
V(\Phi ^{+}\Phi )=\mu ^{2}(\Phi ^{+}\Phi )+\lambda (\Phi ^{+}\Phi )^{2}
\label{potest}
\end{equation}
where $\mu ^{2}\;$and $\lambda \;$are free parameters of the theory. Since $%
\lambda $ should be positive for the potential to be bounded from below, the
minimization of the potential (\ref{potest}) leads to a SSB when $\mu
^{2}<0\;$with the following scheme $SU\left( 2\right) _{L}\times U\left(
1\right) _{Y}\rightarrow U\left( 1\right) _{Q}$\ where $Q\;$is the
electromagnetic charge. After the SSB the Higgs doublet acquires a VEV as in
Eq. (\ref{VEV SM}) from which the Higgs doublet gives mass to the Higgs
particle and is able to endow the vector bosons and fermions with masses
(see below).

\subsection{The kinetic term\label{kin SM}}

The kinetic term describes the interactions between scalar particles and
vector bosons, and provides the masses for the latter when the Higgs doublet
acquires a VEV. The kinetic Lagrangian reads 
\begin{equation}
\pounds _{kin}=\left( D_{\mu }\Phi \right) \left( D^{\mu }\Phi \right)
^{\dagger }\;\;;\;\;\;D_{\mu }\equiv \partial _{\mu }-\frac{ig^{\prime }}{2}%
YW_{\mu }^{4}-ig\tau _{i}W_{\mu }^{i}  \label{kin SM}
\end{equation}
where$\;W_{\mu }^{i}\;$with$\;i=1,2,3\;$are the four-vector fields (gauge
eigenstates), associated to the three generators $\tau _{i}\;$i.e. the $%
SU\left( 2\right) _{L}\;$symmetry. On the other hand, $W_{\mu }^{4}\;$is the
four-vector field associated to the $Y\;$generator i.e. the $U\left(
1\right) _{Y}\;$symmetry. $g\;$and $g^{\prime }\;$are coupling strengths
associated to $W_{\mu }^{i}\;$and $W_{\mu }^{4}\;$respectively. After
diagonalizing the mass matrix of the gauge bosons we obtain the following
mass eigenstates 
\begin{eqnarray}
W_{\mu }^{\pm } &=&\frac{W_{\mu }^{1}\mp iW_{\mu }^{2}}{\sqrt{2}}%
\;\;\;;\;\;M_{W^{\pm }}^{2}=\frac{1}{4}g^{2}v^{2}  \label{Wmass} \\
M_{Z}^{2} &=&\frac{1}{4}v^{2}(g^{\prime 2}+g^{2})=\;\frac{M_{W}^{2}}{\cos
^{2}\theta _{W}}  \label{Zmass} \\
\left( 
\begin{array}{c}
Z_{\mu } \\ 
A_{\mu }
\end{array}
\right) &=&\left( 
\begin{array}{cc}
\cos \theta _{W} & -\sin \theta _{W} \\ 
\sin \theta _{W} & \cos \theta _{W}
\end{array}
\right) \left( 
\begin{array}{c}
W_{\mu }^{3} \\ 
W_{\mu }^{4}
\end{array}
\right)  \label{ZAmixing}
\end{eqnarray}
and the gauge boson $A_{\mu }\;$(the photon) remains massless, it owes to
the fact that this gauge boson is associated to the unbroken generator $Q\;$%
(the electromagnetic charge) i.e. to the remnant symmetry $U\left( 1\right)
_{Q}$.

\subsection{The Yukawa Lagrangian\label{Yuk SM}}

Finally, we build up a Lagrangian that describes the interaction among the
Higgs bosons and fermions, the resultant $SU\left( 2\right) _{L}\times
U\left( 1\right) _{Y}\;$invariant Lagrangian is called the Yukawa Lagrangian 
\begin{equation}
-\pounds _{Y}=\eta _{ij}^{U}\overline{\Psi }_{L}\widetilde{\Phi }U_{R}+\eta
_{ij}^{D}\overline{\Psi }_{L}\Phi D_{R}+h.c.  \label{Yukawa SM}
\end{equation}
where $\overline{\Psi }_{L}\;$are left-handed fermion doublets, $%
U_{R},D_{R}\;$are the right-handed singlets of the up and down sectors of
quarks\footnote{%
For leptons we have the same structure except that we do not have right
handed singlets of neutrinos (corresponding to the up sector of leptons).}. $%
\eta _{ij}^{U,D}\;$are free parameters that define the vertices and
consequently, the Feynman rules of the Lagrangian where $i,j\;$are family
indices. The Yukawa Lagrangian yields masses to the fermions when the Higgs
doublet acquire VEV.

So in brief, Eqs. (\ref{potest},\ref{kin SM},\ref{Yukawa SM}) describe the
contribution that the Higgs sector gives to the SM.

\chapter{The Two Higgs Doublet Model\label{2HDM}}

\section{Motivation\label{motivation}}

Despite the SM has been very sucessful in describing most of the Elementary
Particles phenomenology, the Higgs sector of the theory remains unknown so
far, and there is not any fundamental reason to assume that the Higgs sector
must be minimal (i.e. only one Higgs doublet). Therefore, we could wonder to
know whether the Higgs sector is not minimal. Of course, evocating arguments
of simplicity we may consider the \emph{next to minimal\ }extension as the
best candidate. The simplest extension compatible with the gauge invariance
is the so called Two Higgs Doublet Model (2HDM), which consists of adding a
second Higgs doublet with the same quantum numbers as the first one.

Another motivation to introduce the second doublet comes from the hierarchy
of Yukawa couplings in the third generation of quarks, the ratio between the
masses of the top and bottom quarks is of the order of $m_{t}/m_{b}\approx
174/5\approx 35.\;$In SM, the masses of both quarks come from the same Higgs
doublet, consequently, it implies a non natural hierarchy between their
corresponding Yukawa couplings. However, if the bottom received its mass
from one doublet (say$\;\Phi _{1}$)$\;$and the top from another\ doublet
(say $\Phi _{2}$), then the hierarchy of their Yukawa couplings could be
more natural if the free parameters of the theory acquired the appropiate
values.

On the other hand, the 2HDM could induce CP violation either explicitly or
spontaneously in the Higgs potential (see section \ref{Higgspot 2HDM}).
However, we shall restrict our discussion on a CP conserving framework. A
recent comprehensive overview of the conditions for the 2HDM to be CP
invariant (or violated) can be found in Ref. \cite{decoupling}.

An extra motivation lies on the study of some rare processes called Flavor
Changing Neutral Currents (FCNC). It is well known that these kind of
processes are severely supressed by experimental data, despite they seem not
to violate any fundamental law of nature. On the other hand, Standard Model
(SM) issues are compatible with experimental constraints on FCNC so far \cite
{Sher91, Cheng Sher}, with the remarkable exception of neutrino oscillations 
\cite{Fukuda}. In the case of the lepton sector this fact is ``explained''
by the implementation of the Lepton Number Conservation (LFC), a new
symmetry that protects phenomenology from these dangerous processes.
However, if we believe that this new symmetry is not exact and we expect to
find out FCNC in near future experiments, SM provides a hopeless framework
since predictions from it, are by far out of the scope of next generation
colliders \cite{SM FC}. This is because in the SM, FCNC are absent in the
lepton sector, and in the quark sector they are prohibited at tree level and
further supressed at one loop by the GIM mechanism \cite{GIM}. So detection
of these kind of events would imply the presence of new Physics effects.
Perhaps the simplest (but not the unique) framework to look for these rare
processes is the Two Higgs Doublet Model (2HDM). Owing to the addition of
the second doublet, the Yukawa interactions lead naturally to tree level
FCNC unless we make additional assumptions. On the other hand, because of
the increasing evidence about neutrino oscillations, especial attention has
been addressed to the Lepton Flavor Violation (LFV) in the neutral leptonic
sector by experiments with atmospheric neutrinos \cite{Fukuda}. Further,
since neutrino oscillations imply LFV in the neutral lepton sector, it is
generally expected to find out LFV processes involving the charged lepton
sector as well, such fact encourage us to study the 2HDM as a possible
source for these FCNC. More details in section \ref{2HDM FCNC} and in
chapter \ref{LFV 2HDM}.

Finally, another motivation to study the 2HDM is the fact\ that some models
have a low energy limit with a non minimal Higgs sector. For instance, at
least two Higgs doublets are necessary in supersymmetric models and the so
called 2HDM type II has the same Yukawa couplings as the Minimal
Supersymmetric Standard Model (MSSM). In particular, if the supersymmetric
particles are heavy enough, the Higgs sector of the MSSM becomes a
constrained 2HDM type II at low energies. SUSY models with two Higgs
doublets could provide solutions to some problems of the SM such as the
Higgs mass behavior at very high scales, the Planck and Electroweak scale
hierarchy, the mass hierarchy among fermion families and the existence of
masses for neutrinos and neutrino oscillations.

With these motivations in mind, let us go to see in what way the terms in
the Higgs sector are modified by the introduction of a new Higgs doublet

\section{The contribution of the Higgs sector in the 2HDM\label{contrib 2HDM}%
}

As explained above, we introduce a new Higgs doublet that is a replication
of the first one, so the Higgs sector includes two Higgs doublets with the
same quantum numbers

\begin{equation}
\Phi _{1}=\left( 
\begin{array}{c}
\phi _{1}^{+} \\ 
\phi _{1}^{0}
\end{array}
\right) \;;\;\Phi _{2}=\left( 
\begin{array}{c}
\phi _{2}^{+} \\ 
\phi _{2}^{0}
\end{array}
\right)  \label{first param}
\end{equation}
with hypercharges $Y_{1}=Y_{2}=1,\;$in general, both doublets could acquire
VEV 
\begin{equation*}
\langle \Phi _{1}\rangle =\frac{v_{1}}{\sqrt{2}}\;\;\;;\;\;\;\langle \Phi
_{2}\rangle =\frac{v_{2}}{\sqrt{2}}e^{i\theta }
\end{equation*}

so it is more convenient to parametrize the doublets in the following way 
\begin{equation}
\Phi _{1}=\left( 
\begin{array}{c}
\phi _{1}^{+} \\ 
\frac{h_{1}+v_{1}+ig_{1}}{\sqrt{2}}
\end{array}
\right) \;;\;\Phi _{2}=\left( 
\begin{array}{c}
\phi _{2}^{+} \\ 
\frac{h_{2}+v_{2}e^{i\theta }+ig_{2}}{\sqrt{2}}
\end{array}
\right)  \label{doub param}
\end{equation}
Now, we examine each part of the Lagrangian that couples to the Higgs
doublets

\subsection{The Higgs potential\label{Higgspot 2HDM}}

Since the Higgs potential is the sector that determines the SSB structure as
well as the Higgs masses, Higgs mass eigenstates and Higgs
self-interactions, we start examining this part of the Higgs sector. Unlike
the SM case, the Higgs potential corresponding to the 2HDM is not unique,
and each potential leads to different Feynman rules, details below.

In order to write the most general renormalizable Higgs potential compatible
with gauge invariance, it is convenient to introduce a basis of hermitian,
gauge invariant operators

\begin{eqnarray*}
\widehat{A} &\equiv &\Phi _{1}^{\dagger }\Phi _{1}\;,\;\widehat{B}\equiv
\Phi _{2}^{\dagger }\Phi _{2},\;\widehat{C}\equiv \frac{1}{2}\left( \Phi
_{1}^{\dagger }\Phi _{2}+\Phi _{2}^{\dagger }\Phi _{1}\right) =\text{Re}%
\left( \Phi _{1}^{\dagger }\Phi _{2}\right) ,\; \\
\widehat{D} &\equiv &-\frac{i}{2}\left( \Phi _{1}^{\dagger }\Phi _{2}-\Phi
_{2}^{\dagger }\Phi _{1}\right) =\text{Im}\left( \Phi _{1}^{\dagger }\Phi
_{2}\right)
\end{eqnarray*}
and write down all possible hermitian bilinear and quartic interactions
compatible with gauge invariance 
\begin{eqnarray}
V_{g}\left( \Phi _{1},\Phi _{2}\right) &=&-\mu _{1}^{2}\widehat{A}-\mu
_{2}^{2}\widehat{B}-\mu _{3}^{2}\widehat{C}-\mu _{4}^{2}\widehat{D}+\lambda
_{1}\widehat{A}^{2}+\lambda _{2}\widehat{B}^{2}+\lambda _{3}\widehat{C}%
^{2}+\lambda _{4}\widehat{D}^{2}  \notag \\
&&+\lambda _{5}\widehat{A}\widehat{B}+\lambda _{6}\widehat{A}\widehat{C}%
+\lambda _{8}\widehat{A}\widehat{D}+\lambda _{7}\widehat{B}\widehat{C}%
+\lambda _{9}\widehat{B}\widehat{D}+\lambda _{10}\widehat{C}\widehat{D}
\label{potH}
\end{eqnarray}
This Lagrangian is much more complex than the SM one given by Eq. (\ref
{potest}), since in the potential (\ref{potH}) we have fourteen free
parameters. As we shall see below, four new Higgses will be generated from
it. The most general 2HDM potential (\ref{potH}), contains interaction
vertices that are independent on the mass matrix and the VEV's.
Nevertheless, these interactions vanish if we assume that the Higgs
potential holds a charge conjugation invariance ($C-$invariance), and the
number of parameters reduces to ten

\begin{eqnarray}
V\left( \Phi _{1},\Phi _{2}\right) &=&-\mu _{1}^{2}\widehat{A}-\mu _{2}^{2}%
\widehat{B}-\mu _{3}^{2}\widehat{C}+\lambda _{1}\widehat{A}^{2}+\lambda _{2}%
\widehat{B}^{2}+\lambda _{3}\widehat{C}^{2}+\lambda _{4}\widehat{D}^{2} 
\notag \\
&&+\lambda _{5}\widehat{A}\widehat{B}+\lambda _{6}\widehat{A}\widehat{C}%
+\lambda _{7}\widehat{B}\widehat{C}  \label{Lag10}
\end{eqnarray}
It is important to say that at this step, charge conjugation invariance is
equivalent to $CP\;$invariance since all fields are scalars\footnote{%
The appearing of pseudoscalar couplings come from the introduction of parity
violation in the theory. It is carried out by introducing left-handed
fermion doublets and right handed fermion singlets which in turn produces
scalar and pseudoscalar couplings, the latter are proportional to $\gamma
_{5}\;$and are responsible for parity violation. Since the Higgs doublets do
not have chirality, their self couplings respect parity and therefore they
behave as scalars. However as we shall see in section (\ref{Yuk 2HDM}) when
the Higgs doublets couple to fermions some pseudoscalar couplings arise and
one Higgs boson behaves as a pseudoscalar.}. Under charge conjugation, a
Higgs doublet $\Phi _{i}\;$of hypercharge $1,\;$transforms as $\Phi
_{i}\rightarrow e^{i\alpha _{i}}\Phi _{i}^{\ast }\;$where the parameters $%
\alpha _{i}\;$are arbitrary. Consequently, under charge conjugation we
obtain $\Phi _{i}^{\dagger }\Phi _{j}\rightarrow e^{i\left( \alpha
_{j}-\alpha _{i}\right) }\Phi _{j}^{\dagger }\Phi _{i}.\;$In particular, if
we choose $\alpha _{i}=\alpha _{j}\;$the operator $\widehat{D}\;$reverse
sign under $C-$conjugation, while the other ones are invariant\footnote{%
Of course, we could have chosen $\alpha _{i}-\alpha _{j}=\pm \pi ,\;$in
whose case the operator $\widehat{C}=\text{Re}\left( \Phi _{1}^{\dagger
}\Phi _{2}\right) \;$is the one that violates charge conservation.
Additionally, any other choice for $\alpha _{i}-\alpha _{j}\;$is possible,
and in general none parameter vanishes. However, taking into account that
these phases must be fixed (though arbitrary), $C-$invariance would impose
relations among the coefficients so that the number of free parameters is
always the same (for instance $\mu _{3}\;$and $\mu _{4}\;$would not be
independent any more).}, leading to the Lagrangian (\ref{Lag10}).

However, Lagrangian (\ref{Lag10}) could induce spontaneous CP violation \cite
{Wolf}. There are two ways of naturally imposing for the minimum of the
potential to be $CP\;$invariant \cite{Velhinho}. The first one consists of
demanding invariance under a$\;Z_{2}$ symmetry where $\Phi _{1}\rightarrow
\Phi _{1},\;\Phi _{2}\rightarrow -\Phi _{2}.\;$The resulting potential, that
will be denoted $V_{A}^{\prime }\;$is 
\begin{equation}
V_{A}^{\prime }=-\mu _{1}^{2}\widehat{A}-\mu _{2}^{2}\widehat{B}+\lambda _{1}%
\widehat{A}^{2}+\lambda _{2}\widehat{B}^{2}+\lambda _{3}\widehat{C}%
^{2}+\lambda _{4}\widehat{D}^{2}+\lambda _{5}\widehat{A}\widehat{B}
\label{VAP}
\end{equation}
and correspond to setting $\mu _{3}^{2}=\lambda _{6}=\lambda _{7}=0\;$in Eq.
(\ref{Lag10}). If we permit a soft breaking term of the form $-\mu _{3}^{2}%
\widehat{C},\;$spontaneous $CP\;$violation occurs \cite{Branco}, in that
case the potential reads 
\begin{equation}
V_{A}=V_{A}^{\prime }-\mu _{3}^{2}\widehat{C}  \label{VA}
\end{equation}

The other potential without spontaneous $CP\;$violation, results from
imposing the global symmetry $\Phi _{2}\rightarrow e^{i\varphi }\Phi _{2}$.\
This potential (called $V_{B}^{\prime }$)\ reads 
\begin{equation}
V_{B}^{\prime }=-\mu _{1}^{2}\widehat{A}-\mu _{2}^{2}\widehat{B}+\lambda _{1}%
\widehat{A}^{2}+\lambda _{2}\widehat{B}^{2}+\lambda _{3}\left( \widehat{C}%
^{2}+\widehat{D}^{2}\right) +\lambda _{5}\widehat{A}\widehat{B}  \label{VBP}
\end{equation}
and is obtained by using $\mu _{3}^{2}=\lambda _{6}=\lambda _{7}=0\;$and $%
\lambda _{3}=\lambda _{4\;}$in Eq. (\ref{Lag10})$.\;$Since we have a global
broken symmetry for the potential$\;V_{B}^{\prime }$, there is an extra
Goldstone boson in the theory.

Additionally, it is customary to allow a soft breaking term $-\mu _{3}^{2}%
\widehat{C}\;$in Lagrangian (\ref{VBP}), obtaining 
\begin{equation}
V_{B}=V_{B}^{\prime }-\mu _{3}^{2}\widehat{C}  \label{VB}
\end{equation}
Neither $V_{A}^{\prime }\;$or $V_{B}\;$have spontaneous $CP\;$violation. In
the case of $V_{B}$,$\;$we end up in the CP conserving case of the scalar
potential considered in the Higgs Hunter's guide (see appendix \ref{minima
of the pot}), which in turn is the one of the MSSM. Both of them contain
seven parameters and lead to different phenomenology, however the kinetic
sector and the Yukawa sector are identical for both potentials. The soft
breaking term in $V_{B}\;$is a quadratic term and consequently it does not
affect the renormalizability of the model, the complete renormalization
scheme for the potential $V_{A}\;$has been accomplished in Ref. \cite{Santos}%
, the results for $V_{B}\;$are similar but changing appropiately the cubic
and quartic scalar vertices.

The Higgs masses and Higgs eigenstates are defined in terms of the
parameters $\mu _{i},\lambda _{i}\;$from the potential, and consequently
depend on the potential chosen (see appendix\ \ref{minima of the pot}). When
the mass matrix described in appendix \ref{minima of the pot} is properly
diagonalized, we get the Higgs masses and Higgs mass eigenstates. From now
on, we will consider the CP conserving case for which both VEV can be taken
real, in that case the Higgs sector consists of the following spectrum, two
Higgs CP-even scalars ($H^{0},h^{0}$), one CP-odd scalar ($A^{0}$), two
charged Higgs bosons ($H^{\pm }$), and the Goldstone bosons ($G^{\pm },G^{0}$%
) corresponding to $W^{\pm },Z\;$respectively. The mass eigenstastes
described above are obtained from the gauge eigenstates defined in (\ref
{doub param}) by the following transformations

\begin{eqnarray}
\left( 
\begin{array}{cc}
\cos \beta & \sin \beta \\ 
-\sin \beta & \cos \beta
\end{array}
\right) \left( 
\begin{array}{c}
\phi _{1}^{+} \\ 
\phi _{2}^{+}
\end{array}
\right) &=&\left( 
\begin{array}{c}
G^{+} \\ 
H^{+}
\end{array}
\right)  \notag \\
\left( 
\begin{array}{cc}
\cos \alpha & \sin \alpha \\ 
-\sin \alpha & \cos \alpha
\end{array}
\right) \left( 
\begin{array}{c}
h_{1} \\ 
h_{2}
\end{array}
\right) &=&\left( 
\begin{array}{c}
H^{0} \\ 
h^{0}
\end{array}
\right)  \notag \\
\left( 
\begin{array}{cc}
\cos \beta & \sin \beta \\ 
-\sin \beta & \cos \beta
\end{array}
\right) \left( 
\begin{array}{c}
g_{1} \\ 
g_{2}
\end{array}
\right) &=&\left( 
\begin{array}{c}
G^{0} \\ 
A^{0}
\end{array}
\right)  \label{mass eigenstates}
\end{eqnarray}

where 
\begin{equation*}
\tan \beta =\frac{v_{2}}{v_{1}}\;,\;\sin \beta =\frac{v_{2}}{\sqrt{%
v_{1}^{2}+v_{2}^{2}}}
\end{equation*}
and $\alpha \;$is the mixing angle for the $CP-$even Higgs bosons, which is
different for each potential (see below). $\tan \beta \;$is a new parameter
that clearly arises from the fact that both Higgs doublets could acquire
VEV. In most of the 2HDM's we shall see that new physics contributions are
very sensitive to it.

In the following we summarize the results obtained in appendix \ref{minima
of the pot}, which are in agreement with \cite{Sher rep, Hunter}.

\textbf{For the potential }$V\;$in Eq. (\ref{Lag10}) the minimum conditions
(Tadpoles at tree level)\ are

\begin{eqnarray*}
0 &=&\allowbreak T_{a}=-\mu _{1}^{2}+\lambda _{1}v_{1}^{2} \\
0 &=&\allowbreak T_{b}=-\mu _{3}^{2}\allowbreak +\frac{\lambda _{6}v_{1}^{2}%
}{2}
\end{eqnarray*}
the Higgs masses and the mixing angle $\alpha $,$\;$are written as

\begin{eqnarray*}
m_{H^{+}}^{2} &=&-\mu _{2}^{2}+\frac{1}{2}\lambda
_{5}v_{1}^{2}\;\;;\;\;m_{A^{0}}=-\allowbreak \mu _{2}^{2}+\frac{1}{2}\left(
\lambda _{4}+\lambda _{5}\right) v_{1}^{2} \\
m_{H^{0},h^{0}} &=&\left( \lambda _{1}+\frac{1}{2}\lambda _{+}\right)
v_{1}^{2}-\frac{1}{2}\mu _{2}^{2}\pm k_{1} \\
k_{1} &=&\sqrt{4\lambda _{1}v_{1}^{2}\left( \lambda _{1}v_{1}^{2}+\mu
_{2}^{2}-v_{1}^{2}\lambda _{+}\right) +\left( \lambda
_{+}^{2}v_{1}^{2}+\lambda _{6}^{2}v_{1}^{2}-2\mu _{2}^{2}\lambda _{+}\right)
v_{1}^{2}+\mu _{2}^{4}} \\
\tan 2\alpha &=&\frac{\lambda _{6}v_{1}^{2}}{\left( \allowbreak 2\lambda
_{1}-\lambda _{+}\right) v_{1}^{2}+\allowbreak \mu _{2}^{2}}
\end{eqnarray*}

\textbf{For the potential }$V_{A}^{\prime }\;$Eq. (\ref{VAP}), the minimum
conditions are 
\begin{eqnarray}
0 &=&T_{1}=v_{1}\left( -\mu _{1}^{2}+\lambda _{1}v_{1}^{2}+\lambda
_{+}v_{2}^{2}\right)  \notag \\
0 &=&T_{2}=v_{2}\left( -\mu _{2}^{2}+\lambda _{2}v_{2}^{2}+\lambda
_{+}v_{1}^{2}\right)  \label{min VA}
\end{eqnarray}
where $\lambda _{+}=\frac{1}{2}\left( \lambda _{3}+\lambda _{5}\right) .\;$%
Relations (\ref{min VA}) lead to the following solutions

i) 
\begin{equation*}
v_{1}^{2}=\frac{\lambda _{2}\mu _{1}^{2}-\lambda _{+}\mu _{2}^{2}}{\lambda
_{1}\lambda _{2}-\lambda _{+}^{2}}\;\;\;;\;\;\;v_{2}^{2}=\frac{\lambda
_{1}\mu _{2}^{2}-\lambda _{+}\mu _{1}^{2}}{\lambda _{1}\lambda _{2}-\lambda
_{+}^{2}}
\end{equation*}
or ii)

\begin{equation*}
v_{2}^{2}=0\;\;\;;\;\;\;v_{1}^{2}=\frac{\mu _{1}^{2}}{\lambda _{1}}
\end{equation*}
for i)\ the masses of the Higgs bosons and the mixing angle $\alpha \;$are
given by 
\begin{eqnarray}
m_{H^{\pm }} &=&-\lambda _{3}\left( v_{1}^{2}+v_{2}^{2}\right)
\;\;\;\;;\;\;\;\;m_{A^{0}}^{2}=\frac{1}{2}\left( \lambda _{4}-\lambda
_{3}\right) \left( v_{1}^{2}+v_{2}^{2}\right)  \notag \\
m_{H^{0},h^{0}}^{2} &=&\lambda _{1}v_{1}^{2}+\lambda _{2}v_{2}^{2}\pm \sqrt{%
\left( \lambda _{1}v_{1}^{2}-\lambda _{2}v_{2}^{2}\right)
^{2}+4v_{1}^{2}v_{2}^{2}\lambda _{+}^{2}}  \notag \\
\tan 2\alpha &=&\frac{2v_{1}v_{2}\lambda _{+}}{\lambda _{1}v_{1}^{2}-\lambda
_{2}v_{2}^{2}}  \label{Higgs mass VAP i}
\end{eqnarray}

and for ii) they are

\begin{eqnarray}
m_{H^{+}}^{2} &=&-\mu _{2}^{2}+\frac{1}{2}\lambda
_{5}v_{1}^{2}\;;\;m_{A^{0}}^{2}=-\allowbreak \mu _{2}^{2}+\frac{1}{2}\left(
\lambda _{4}+\lambda _{5}\right) v_{1}^{2}\;\;;\;m_{H^{0}}^{2}=2\lambda
_{1}v_{1}^{2}\; \\
m_{h^{0}}^{2} &=&-\allowbreak \mu _{2}^{2}+\frac{1}{2}\left( \lambda
_{3}+\lambda _{5}\right) v_{1}^{2}\;\;;\;\;\tan 2\alpha =0
\label{Higgs mass VAP ii}
\end{eqnarray}

\textbf{Finally, for the potential }$V_{B}\;$Eq. (\ref{VB}), the minimum
conditions are 
\begin{equation}
0=T_{1}-\frac{\mu _{3}^{2}}{2}v_{2}\;\;\;\;;\;\;\;0=T_{2}-\frac{\mu _{3}^{2}%
}{2}v_{1}  \label{min VB}
\end{equation}
whose solutions are 
\begin{eqnarray*}
v_{1}^{2} &=&\frac{\lambda _{1}-\lambda _{2}\pm Z_{1}}{2\left( \lambda
_{1}-\lambda _{+}\right) \left( \lambda _{2}-\lambda _{+}\right) } \\
v_{2}^{2} &=&\frac{\lambda _{2}-\lambda _{1}\pm Z_{2}}{2\left( \lambda
_{1}-\lambda _{+}\right) \left( \lambda _{2}-\lambda _{+}\right) } \\
Z_{1} &=&\sqrt{\left( \lambda _{1}-\lambda _{2}\right) ^{2}-4\left( \lambda
_{1}-\lambda _{+}\right) \left( \lambda _{2}-\lambda _{+}\right) \left[
\left( \lambda _{+}v^{2}-\mu _{1}^{2}\right) \left( \lambda _{+}v^{2}-\mu
_{2}^{2}\right) -\frac{1}{4}\mu _{3}^{4}\right] } \\
Z_{2} &=&\sqrt{\left( \lambda _{1}-\lambda _{2}\right) ^{2}-4\left( \lambda
_{2}-\lambda _{+}\right) \left( \lambda _{1}-\lambda _{+}\right) \left[
\left( \lambda _{+}v^{2}-\mu _{2}^{2}\right) \left( \lambda _{1}v^{2}-\mu
_{1}^{2}\right) -\frac{1}{4}\mu _{3}^{4}\right] }
\end{eqnarray*}
The masses and the mixing angle $\alpha $ are given by 
\begin{eqnarray}
m_{H^{\pm }} &=&-\lambda _{3}\left( v_{1}^{2}+v_{2}^{2}\right) +\mu _{3}^{2}%
\frac{v_{1}^{2}+v_{2}^{2}}{v_{1}v_{2}}\;\;\;\;;\;\;\;\;m_{A^{0}}^{2}=\frac{1%
}{2}\mu _{3}^{2}\frac{v_{1}^{2}+v_{2}^{2}}{v_{1}v_{2}}  \notag \\
m_{H^{0},h^{0}}^{2} &=&\lambda _{1}v_{1}^{2}+\lambda _{2}v_{2}^{2}+\frac{1}{4%
}\mu _{3}^{2}\left( \tan \beta +\cot \beta \right)  \notag \\
&&\pm \sqrt{\left[ \lambda _{1}v_{1}^{2}-\lambda _{2}v_{2}^{2}+\frac{1}{4}%
\mu _{3}^{2}\left( \tan \beta -\cot \beta \right) \right] ^{2}+\left(
2v_{1}v_{2}\lambda _{+}-\frac{1}{2}\mu _{3}^{2}\right) ^{2}}  \notag \\
\tan 2\alpha &=&\frac{2v_{1}v_{2}\lambda _{+}-\frac{1}{2}\mu _{3}^{2}}{%
\lambda _{1}v_{1}^{2}-\lambda _{2}v_{2}^{2}+\frac{1}{4}\mu _{3}^{2}\left(
\tan \beta -\cot \beta \right) }\;  \label{Higgs mass VB}
\end{eqnarray}

Observe that a solution with one of the VEV's equal to zero is not possible
for the potential $V_{B}$. As explained before, each potential has different
Feynman rules and consequently, leads to different phenomenology \cite
{fermiophobic}, a complete set of Feynman rules for these potentials could
be found in Refs. \cite{fermiophobic, decoupling}.

The potentials $V_{A}^{\prime }\;$and $V_{B}\;$are different because they
differ in some cubic and quartic interactions. For example, the coupling $%
h^{0}H^{+}H^{-}\;$reveals some subtle aspects of the phenomenology of the
potential; in terms of the $\lambda ^{\prime }s\;$it is given by 
\begin{eqnarray*}
g_{h^{0}H^{+}H^{-}} &=&2v_{2}\lambda _{2}\cos ^{2}\beta \cos \alpha
+v_{2}\lambda _{3}\sin \alpha \cos \beta \sin \beta -v_{1}\lambda _{5}\cos
^{2}\beta \sin \alpha \\
&&-2v_{1}\lambda _{1}\sin ^{2}\beta \sin \alpha +v_{2}\lambda _{5}\sin
^{2}\beta \cos \alpha -v_{1}\lambda _{3}\cos \alpha \cos \beta \sin \beta
\end{eqnarray*}
and coincides for both potentials $V_{A}^{\prime }\;$and $V_{B};\;$it is
because this interaction does not involve $\lambda _{4}$ nor $\mu _{3}\;$%
which are the factors that make the difference between them. However, by
writing the coupling in terms of the Higgs boson masses the result is
different for each potential 
\begin{eqnarray*}
\left( g_{h^{0}H^{+}H^{-}}\right) _{A} &=&\frac{g}{m_{W}}\left[ m_{h^{0}}^{2}%
\frac{\cos \left( \alpha +\beta \right) }{\sin 2\beta }-\left( m_{H^{+}}^{2}-%
\frac{1}{2}m_{h^{0}}^{2}\right) \sin \left( \alpha -\beta \right) \right] \\
\left( g_{h^{0}H^{+}H^{-}}\right) _{B} &=&\frac{g}{m_{W}}\left[ \left(
m_{h^{0}}^{2}-m_{A^{0}}^{2}\right) \frac{\cos \left( \alpha +\beta \right) }{%
\sin 2\beta }\right. \\
&&\left. -\left( m_{H^{+}}^{2}-\frac{1}{2}m_{h^{0}}^{2}\right) \sin \left(
\alpha -\beta \right) \right]
\end{eqnarray*}
we can resolve the puzzle by remembering that what really matters in
perturbative calculations is the position of the minimum of the potential
and the values of the derivatives at that point. Now, the position of the
vacuum (minimum) is different for each potential and the generation of
masses come from the second derivative of the potential evaluated at this
minimum, so the relation among the $\lambda ^{\prime }s\;$and the masses are
different for $V_{A}^{\prime }\;$and $V_{B},\;$thus the coupling $%
h^{0}H^{+}H^{-}$ in terms of physical quantities differs for each potential.
It should be pointed out that even before being able to test the Higgs
bosons self-couplings in order to discriminate among the potentials, we
might see a signature of them in processes with Higgs boson loops, for
example, the decay $h^{0}\rightarrow \gamma \gamma \;$could be very
sensitive to the difference among the self couplings of $V_{A}^{\prime }\;$%
and $V_{B}$.

Finally, it is also important to notice that the symmetry of the potential
ought to be extended to the other Higgs sectors, this fact is particularly
important to write the Yukawa Lagrangian to be discussed in section \ref{Yuk
2HDM}.

\subsection{The kinetic sector\label{kin 2HDM}}

The kinetic Lagrangian (\ref{kin SM})\ of the SM is extended to become 
\begin{equation}
\pounds _{kin}=(D_{\mu }\Phi _{1})^{+}(D^{\mu }\Phi _{1})+(D_{\mu }\Phi
_{2})^{+}(D^{\mu }\Phi _{2})  \label{kin lag}
\end{equation}
where the covariant derivative is defined by Eq. (\ref{kin SM}). This
Lagrangian endows the gauge bosons with mass and provides the interactions
among gauge and Higgs bosons. In contrast to the Higgs potential and the
Yukawa Lagrangian (see sections \ref{Higgspot 2HDM},\ \ref{Yuk 2HDM}), the
kinetic sector is basically unique because of the gauge invariance\footnote{%
However, for some potentials and Yukawa Lagrangians it is possible to rotate
the Doublets such that only one of them acquire VEV, as we will see later.
In that case the kinetic term is basically the same but taking (say) $%
v_{2}=0\;$i.e. $\tan \beta =0.$}. Indeed, we can easily check that the
kinetic Lagrangian described by Eq. (\ref{kin lag}) is already invariant
under charge conjugation as well as under the discrete and global symmetries
described in Sec. (\ref{Higgspot 2HDM}), therefore the imposition of these
symmetries does not produce any difference in the kinetic sector unlike the
case of the potential.

In order to expand the Lagrangian, it is convenient to work on a real
representation for the generators, for which we make the assigment $%
L_{a}=-i\tau _{a}\;$and double the dimension of the representation by means
of the definition

\begin{equation}
\Phi _{k}=\left( 
\begin{array}{c}
Re\phi _{k}^{+}+iIm\phi _{k}^{+} \\ 
Re\phi _{k}^{0}+iIm\phi _{k}^{0}
\end{array}
\right) \rightarrow \left( 
\begin{array}{c}
Re\phi _{k}^{+} \\ 
Im\phi _{k}^{+} \\ 
Re\phi _{k}^{0} \\ 
Im\phi _{k}^{0}
\end{array}
\right) \;\;\;\;\;\;k=1,2\;  \label{doubling}
\end{equation}

From which we find the real representation by looking at the action of the
initial generators (multiplied by $-i$),\ over the two dimensional
representation of the doublet $\Phi _{k}\;$defined in Eq. (\ref{doubling}).
For example, for $\tau _{1}$

\begin{equation}
L_{1}\Phi _{1}=-i\tau _{1}\Phi _{1}=\frac{-i\sigma _{1}}{2}\Phi _{1}=\frac{1%
}{2}\left( 
\begin{array}{c}
Im\phi _{1}^{0}-iRe\phi _{1}^{0} \\ 
Im\phi _{1}^{+}-iRe\phi _{1}^{+}
\end{array}
\right)  \label{def L1}
\end{equation}

And we extend $L_{1}\Phi _{1}\;$with the same correspondence rule

\ \ 
\begin{equation}
L_{1}\Phi _{1}\rightarrow \frac{1}{2}\left( 
\begin{array}{c}
Im\phi _{1}^{0} \\ 
-Re\phi _{1}^{0} \\ 
Im\phi _{1}^{+} \\ 
-Re\phi _{1}^{+}
\end{array}
\right)  \label{find L1}
\end{equation}

Finally, we look for a matrix $L_{1}\;$that acting on (\ref{doubling})
reproduces (\ref{find L1}), we get

\begin{equation}
L_{1}=\frac{1}{2}\left( 
\begin{array}{cccc}
0 & 0 & 0 & 1 \\ 
0 & 0 & -1 & 0 \\ 
0 & 1 & 0 & 0 \\ 
-1 & 0 & 0 & 0
\end{array}
\right)  \label{matrix kin1}
\end{equation}

Similarly we obtain for the other generators

\begin{eqnarray}
L_{2} &=&\frac{1}{2}\left( 
\begin{array}{cccc}
0 & 0 & -1 & 0 \\ 
0 & 0 & 0 & -1 \\ 
1 & 0 & 0 & 0 \\ 
0 & 1 & 0 & 0
\end{array}
\right) \;\;,\;L_{3}=\frac{1}{2}\left( 
\begin{array}{cccc}
0 & 1 & 0 & 0 \\ 
-1 & 0 & 0 & 0 \\ 
0 & 0 & 0 & -1 \\ 
0 & 0 & 1 & 0
\end{array}
\right)  \notag \\
L_{4} &=&\frac{1}{2}\left( 
\begin{array}{cccc}
0 & 1 & 0 & 0 \\ 
-1 & 0 & 0 & 0 \\ 
0 & 0 & 0 & 1 \\ 
0 & 0 & -1 & 0
\end{array}
\right)  \label{matrix kin2}
\end{eqnarray}

From which the covariant derivative in (\ref{kin SM}) reads

\begin{equation}
D_{\mu }=\partial _{\mu }+gL_{i}W_{\mu }^{i}+g^{\prime }L_{4}W_{\mu }^{4}\;
\label{covar double}
\end{equation}

Thus, we can expand the Lagrangian (\ref{kin lag}) by using the four
dimensional representation for each doublet Eq. (\ref{doubling}), and the
four dimensional representation of the generators Eqs. (\ref{matrix kin1}), (%
\ref{matrix kin2}).

\subsubsection{Gauge fields\label{gauge fields}}

The mass terms are obtained from the VEV's, the resulting mass matrix is
given by

\begin{equation}
\frac{1}{2}M_{ab}^{2}W_{\mu }^{a}W_{\mu
}^{b}\;\;;\;\;M_{ab}^{2}=2\sum_{k=1}^{2}(g_{a}L_{a}v_{k})^{\dagger
}(g_{b}L_{b}v_{k})  \label{vector mass}
\end{equation}

Where $a,b=1,2,3,4\;$indicates the gauge bosons (gauge eigenstates)
corresponding to the generators $\tau ^{i},Y\;$respectively. After the
diagonalization of (\ref{vector mass}) the mass terms and eigenstates for
the Gauge bosons read

\begin{eqnarray}
W_{\mu }^{\pm } &=&\frac{W_{\mu }^{1}\mp iW_{\mu }^{2}}{\sqrt{2}}%
\;\;\;;\;\;M_{W^{\pm }}^{2}=\frac{1}{4}g^{2}(v_{1}^{2}+v_{2}^{2})  \label{AW}
\\
M_{Z}^{2} &=&\frac{1}{4}(v_{1}^{2}+v_{2}^{2})(g^{\prime 2}+g^{2})=\;\frac{%
M_{W}^{2}}{\cos ^{2}\theta _{W}}  \label{Z} \\
\left( 
\begin{array}{c}
Z_{\mu } \\ 
A_{\mu }
\end{array}
\right) &=&\left( 
\begin{array}{cc}
\cos \theta _{W} & -\sin \theta _{W} \\ 
\sin \theta _{W} & \cos \theta _{W}
\end{array}
\right) \left( 
\begin{array}{c}
W_{\mu }^{3} \\ 
W_{\mu }^{4}
\end{array}
\right)  \label{WZ}
\end{eqnarray}
where $\theta _{W}\;$is the Weinberg mixing angle. We can realize that the
expressions for the vector bosons masses coincide with the ones in SM if $%
v_{1}^{2}+v_{2}^{2}=v^{2}\;$(where $v\,\;$is the VEV of the Higgs doublet of
the SM), and since $v^{2}=4M_{W}^{2}/g^{2}\;$is a known parameter, we have
at tree level the constraint $v_{1}^{2}+v_{2}^{2}=v^{2}.$

Since the kinetic sector provides the interactions of Higgs bosons with
gauge bosons, let us discuss some interesting features of these interactions.

\subsubsection{Interactions in the kinetic sector\label{kin interac}}

We can obtain the interactions in the kinetic sector by expanding the
Lagrangian (\ref{kin lag}) in terms of the mass eigenstates of Higgs bosons
and Vector bosons Eqs. (\ref{AW}, \ref{WZ}, \ref{mass eigenstates}).\ On the
other hand, it worths to note that some interactions that at first glance
should appear, are absent in this expansion as is the case of $%
A^{0}W^{+}W^{-},\;$and $A^{0}ZZ$.\ Let us discuss shortly the origin of
these missings. First of all we should emphasize that \emph{in the SM, the
discrete symmetries C\ and P are preserved separately when fermions are
absent. }Then, we can assign a unique set of quantum numbers $J^{PC}\;$to
all the bosons of the theory when fermions are ignored. The same argument
holds for the two Higgs doublet model.\ These facts dictate the presence or
missing of some interactions \cite{Hunter}.

First of all, let us examine the assigment of these quantum numbers for the
scalar and vector bosons in the table (\ref{tab:CPnoFer}).

\begin{table}[tbp]
\begin{center}
\begin{tabular}{||l||l||l||l||}
\hline\hline
& $J^{PC}$ &  & $J^{P}$ \\ \hline\hline
$\gamma $ & $1^{--}\;\;\;\;\;\;\;$ & $W^{\pm }$ & $1^{+}$ \\ \hline\hline
$Z$ & $1^{+-}$ & $H^{\pm }$ & $0^{+}$ \\ \hline\hline
$H^{0}$ & $0^{++}$ &  &  \\ \hline\hline
$h^{0}$ & $0^{++}$ &  &  \\ \hline\hline
$A^{0}$ & $0^{+-}$ &  &  \\ \hline\hline
\end{tabular}
\end{center}
\caption{Assigments to the parity ($P$)\ and charge conjugation ($C$),
quantum numbers for the Higgs and vector bosons before the introduction of
fermions. When fermions are absent, $P\;$and $C\;$are conserved separately,
(i.e. in the Higgs potential and kinetic Lagrangian). }
\label{tab:CPnoFer}
\end{table}

The existence of the vertex $ZH^{+}H^{-}\;$indicates that $J^{PC}\left(
Z\right) =1^{+-}$,\ the existence of $H^{0}h^{0}h^{0}\;$in the Higgs
potential says that $J^{PC}\left( H^{0}\right) =0^{++}$; symmetrically the
existence of $h^{0}H^{0}H^{0}\;$tells us that $J^{PC}\left( h^{0}\right)
=0^{++}$. From these quantum number assigments we deduce from the vertex $%
ZH^{0}A^{0}\;$that $J^{PC}\left( A^{0}\right) =0^{+-}$.$\;$

Furthermore, in the 2HDM the imaginary parts combine to produce $A^{0}\;$and 
$G^{0}\;$so both must have the same quantum numbers, consequently $A^{0}\;$%
and $G^{0}$ have the quantum numbers $0^{+-}$. Owing to the assigment $%
J^{PC}\left( A^{0}\right) $ $\;\ =0^{+-}$ the vertices $A^{0}W^{+}W^{-}\;$%
and $A^{0}ZZ\;$are forbidden. We should remark that despite the Higgs $%
A^{0}\;$is usually called a \emph{pseudoscalar\ Higgs,\ }at this step $%
A^{0}\;$is a \emph{scalar }particle. However, we shall see later that $%
A^{0}\;$behaves as a pseudoscalar when it couples to matter (fermions), see
section (\ref{Yuk 2HDM}). Thus, the term pseudoscalar is not so proper for
the Higgs $A^{0}\;$and the denomination as a $CP-$odd Higgs is more
appropiate.

Another way to explain the absence of vertices $A^{0}W^{+}W^{-}\;$and $%
A^{0}ZZ\;$is the following: These interactions come from the kinetic term $%
\left( D_{\mu }\Phi \right) \left( D^{\mu }\Phi \right) ^{\dagger }$ after
replacing one of the $\phi _{i}^{0}$ fields in Eq. (\ref{first param})$\;$by
its VEV, but in a CP\ conserving model the VEV is real while $A^{0}\;$comes
from the imaginary parts of the $\phi _{i}^{0}$ fields as we can see from
Eqs. (\ref{doub param}) and (\ref{mass eigenstates}). Therefore $A^{0}\;$%
cannot be coupled to any massive vector boson. By the same token, the
couplings $\left( H^{0},h^{0}\right) \gamma \gamma \;$are forbidden at tree
level, since $H^{0},h^{0}\;$cannot be coupled at tree level to massless
vector bosons. Finally, another interesting argument is that $%
A^{0}W^{+}W^{-}\;$and $A^{0}ZZ\;$are prohibited by $C-$invariance, however,
when fermions are introduced (by means of loops) $C-$invariance is no longer
valid (though $CP-$invariance still holds in good aproximation).
Consequently, these vertices can appear at one loop level by inserting
fermions into the loops.

So in general, vertices with neutral particles only and one or two photons
clearly vanish at tree level but can be generated at one loop.$\;$And same
for a pair of gluons coupled to any of the neutral Higgs bosons. Gluon gluon
fusion generated by one loop couplings $\left( A^{0},H^{0},h^{0}\right)
gg,\; $is a very important source for Higgs production in Hadron colliders.

Additionally, there are other couplings whose absence can be explained by
the quantum number assigments in table (\ref{tab:CPnoFer}). For instance,
the coupling of $Z\;$to non identical Higgs bosons is allowed only if such
scalars have opposite $CP\;$numbers. Consequently, the couplings$%
\;ZA^{0}H^{0},\;ZA^{0}h^{0}\;$are allowed while $ZH^{0}h^{0}\;$is forbidden.
On the other hand, the couplings of $Z\;$to identical Higgs bosons are\emph{%
\ }forbidden by\emph{\ Bose-Einstein symmetry}. Moreover, $H^{\pm }W^{\mp
}\gamma \;$is prohibited by conservation of electromagnetic current.

As for the vertices $H^{\pm }W^{\mp }Z$, they do not appear at tree level in
the 2HDM. We should note however, that they are not forbidden by these
quantum numbers (they appear in some triplet representations). In the case
of multi-Higgs doublet models (including the 2HDM) they are prohibited at
tree level because of the isospin symmetry. Notwithstanding, such couplings
are allowed at loop levels owing to the breaking of the isospin symmetry by
the loop particles. One loop contributions to the $H^{+}W^{-}Z$ vertex are
studied in \cite{Mendez, KanemuHWZ, Diaz-CruzHWZ} and will be discussed in
section (\ref{phenomenological constraints}).

Finally, we should emphasize that when $CP\;$is violated, many of the
missing vertices described above appear, though they are expected to be
quite suppressed.

Furthermore, it is interesting to notice that the couplings of the kinetic
sector satisfy automatically some tree level unitarity bounds. Since partial
amplitudes cannot grow with energy, cancellations to avoid violation of
unitarity are necessary. For example, in the case of $V_{L}V_{L}\rightarrow
V_{L}V_{L}\;$scattering, cancelation is possible in SM because the vertex is
of the form $g_{\phi ^{0}WW}=gm_{W}.\;$When we have more than one doublet,
this work does not have to be done by only one Higgs boson, instead we have
the sum rule

\begin{equation}
\sum_{i}g_{h_{i}^{0}VV}^{2}=g_{\phi ^{0}VV}^{2}  \label{sum 1}
\end{equation}
where $i\;$labels all the neutral Higgs bosons of the extended Higgs sector,
and $\phi ^{0}$ denotes the SM Higgs. The sum rule (\ref{sum 1}) ensures the
unitarity of $V_{L}V_{L}\rightarrow V_{L}V_{L}.\;$Moreover, to ensure $f%
\overline{f}\rightarrow V_{L}V_{L}\;$unitarity we ought to demand 
\begin{equation}
\sum_{i}g_{h_{i}^{0}VV}g_{h_{i}^{0}f\overline{f}}=g_{\phi ^{0}VV}g_{\phi
^{0}f\overline{f}}  \label{sum 2}
\end{equation}
in such a way that only the contribution of all Higgs bosons cancels the
effect.

In the particular case of the 2HDM, Eq. (\ref{sum 1}) becomes 
\begin{equation}
g_{h^{0}VV}^{2}+g_{H^{0}VV}^{2}=g_{\phi ^{0}VV}^{2}  \label{sum1 2H}
\end{equation}
where $\phi ^{0}VV\;$denotes the coupling of the SM Higgs with two vector
bosons\footnote{%
Remember that the coupling $A^{0}VV\;$does not appear at tree level.}. We
can check that in terms of the mixing angles $\alpha \;$and $\beta \;$we get

\begin{equation}
\frac{g_{h^{0}VV}}{g_{\phi ^{0}VV}}=\sin \left( \beta -\alpha \right) \;\;;\;%
\frac{g_{H^{0}VV}}{g_{\phi ^{0}VV}}=\cos \left( \beta -\alpha \right) \;
\label{relat kin}
\end{equation}
from which the constraint (\ref{sum1 2H}) is obviously accomplished.

A very important phenomenological consequence of (\ref{sum1 2H}) is that the
couplings of$\;h^{0},H^{0}\;$to $VV\;$are supressed respect to SM ones. In
particular, if one of the scalar Higgs bosons decouples at tree level from $%
VV\;$then the Higgs coupling at tree level of the other scalar Higgs is
SM-like. In that case we say that one of the Higgs bosons ``exhausts'' or
``saturates'' the sum rule, this is a natural scenario in the Minimal
Supersymmetric Standard Model (MSSM) in which $\cos \left( \beta -\alpha
\right) \sim 0\;$is expected, so the coupling$\;h^{0}VV\;$almost saturates
the sum rule and is SM like, while the coupling $H^{0}VV\;$is highly
supressed.

It is worthwhile to point out that the existence of V$-$Higgs$-$Higgs
couplings is a new feature of the 2HDM respect to SM. As we shall discuss in
section (\ref{theoretical constraints}), tree level unitarity constraints
involving those kind of couplings are also satisfied.

As for the sum rule (\ref{sum 2}), it involves also the couplings of the
Higgs bosons to fermions, which are model dependent, we shall see in next
section that such rule is also automatically accomplished at tree level by
some sets of Yukawa couplings.

\subsection{The Yukawa Lagrangian\label{Yuk 2HDM}}

The Most general gauge invariant Lagrangian that couples the Higgs fields to
fermions reads 
\begin{eqnarray}
-\pounds _{Y} &=&\eta _{ij}^{U,0}\overline{Q}_{iL}^{0}\widetilde{\Phi }%
_{1}U_{jR}^{0}+\eta _{ij}^{D,0}\overline{Q}_{iL}^{0}\Phi _{1}D_{jR}^{0}+\xi
_{ij}^{U,0}\overline{Q}_{iL}^{0}\widetilde{\Phi }_{2}U_{jR}^{0}+\xi
_{ij}^{D,0}\overline{Q}_{iL}^{0}\Phi _{2}D_{jR}^{0}+  \notag \\
&&\eta _{ij}^{E,0}\overline{l}_{iL}^{0}\Phi _{1}E_{jR}^{0}+\xi _{ij}^{E,0}%
\overline{l}_{iL}^{0}\Phi _{2}E_{jR}^{0}+h.c.,  \label{Yukawa}
\end{eqnarray}
where $\Phi _{1,2}\;$represent the Higgs doublets, $\widetilde{\Phi }%
_{1,2}\equiv i\sigma _{2}\Phi _{1,2}$,$\;\eta _{ij}^{0}\;$and\ $\xi
_{ij}^{0}\;$are non diagonal $3\times 3\;$matrices and $i,j\;$denote family
indices. $D_{R}^{0}\;$refers to the three down-type weak isospin quark
singlets $D_{R}^{0}\equiv \left( d_{R}^{0},s_{R}^{0},b_{R}^{0}\right)
^{T},\;U\;$refers to the three up-type weak isospin quark singlets $%
U_{R}^{0}\equiv \left( u_{R}^{0},c_{R}^{0},t_{R}^{0}\right) ^{T}\;$and $%
E_{R}^{0}\;$to the three charged leptons. Finally, $\overline{Q}_{iL}^{0},\;%
\overline{l}_{iL}^{0}\;$denote the quark and lepton weak isospin left-handed
doublets respectively. The superscript ``$0$''$\;$indicates that the fields
are not mass eigenstates yet.

From now on, I shall restrict the discussion to the quark sector only, since
the extension of the arguments to the lepton sector is straightforward. What
we can see from the model described by Eq. (\ref{Yukawa}) is that in the
most general case, both Higgs bosons couple (and consequently give masses)
to the up and down sectors simultaneously. However, this general case leads
to processes with Flavor Changing Neutral Currents (FCNC) at tree level.
This is basically because by rotating the fermion gauge eigenstates of the
down sector to get the mass eigenstates we are not able to diagonalize both
coupling matrices $\eta ^{D,0},\xi ^{D,0}$ simultaneously\footnote{%
Of course, we still have the possibility of a fine tuning that allows us to
diagonalize both matrices with the same transformation. Although it is not a
very natural assumption, it leads to a model without FCNC different from the
ones obtained from arguments of symmetry.}, the situation is similar for the
up and lepton sectors.

Now, since processes containing FCNC are very strongly supressed
experimentally, especially due to the small $K_{L}-K_{S}\;$mass difference,
these processes were considered dangerous. Consequently, several mechanisms
to avoid them at the tree level were proposed, for instance one possibility
is to consider the exchange of heavy scalar or pseudoscalar Higgs Fields 
\cite{Sher91} or by cancellation of large contributions with opposite sign.
Perhaps the most ``elegant'' or ``natural'' supression mechanism is the one
suggested by Glashow and Weinberg \cite{Glashow}, who implemented a discrete
symmetry that automatically forbids the couplings that generate such rare
decays. Another mechanism was proposed by Cheng and Sher arguing that a
natural value for the FC couplings from different families ought to be of
the order of the geometric average of their Yukawa couplings \cite{Cheng
Sher}, this assumption has been tested by Ref. \cite{Atwood} in the quark
sector. Remarkably, absence of FCNC at tree level is possible even with both
Higgs doublets coupled simultaneously to the up and down sectors; Ref. \cite
{Lin} studies the possibility of suppressing FCNC in a general 2HDM, by
demanding a$\;S_{3}\;$permutation symmetry among the three fermion families,
this $S_{3}\;$permutation symmetry would be broken spontaneously when the
SSB of the gauge symmetry occurs, more details in section \ref{theoretical
constraints}. On the other hand, Ref. \cite{decoupling} shows that the
decoupling limit (to be discussed later on) is a natural scenario to get rid
of tree level FCNC in the general 2HDM.

Let us first discuss the mechanism proposed by Weinberg and Glashow. We have
said that the matrices $\eta ^{U,0},\xi ^{U,0}\;$cannot be diagonalized
simultaneously, and the same applies to the couple of matrices $\eta
^{D,0},\xi ^{D,0}.\;$So it is immediate to realize that we can supress the
FCNC at tree level in Lagrangian (\ref{Yukawa}), if we manage to get rid of
one of the pair of matrices $\left( \eta ^{U,0},\xi ^{U,0}\right) \;$that
couples the up sector to the Higgs doublets, and same for the down sector.
We can achieve that, by implementing the following discrete symmetry 
\begin{eqnarray*}
\Phi _{1} &\rightarrow &\Phi _{1}\;\;\text{and\ \ }\Phi _{2}\rightarrow
-\Phi _{2} \\
D_{jR} &\rightarrow &\mp D_{jR}\;\;\text{and\ \ }U_{jR}\rightarrow -U_{jR}
\end{eqnarray*}
so by demanding invariance under this discrete symmetry we have two cases

\begin{itemize}
\item  When we use $D_{jR}\rightarrow -D_{jR}\;$we should drop out $\eta
_{ij}^{U,0}$and$\;\eta _{ij}^{D,0}.$\ So, $\Phi _{1}$ decouples from the
Yukawa sector and only $\Phi _{2}\;$couples and gives masses to the up and
down sectors. This case is known as the 2HDM type I

\item  When we use $D_{jR}\rightarrow D_{jR}\;$then$\;\eta _{ij}^{U,0}\;$and 
$\xi _{ij}^{D,0}\;$vanish\ and therefore $\Phi _{1}\;$couples and gives
masses to the down sector while $\Phi _{2}\;$couples and gives masses to the
up sector. In this case we call it, the 2HDM type II.
\end{itemize}

In the most general framework of multi-Higgs doublet models, this supression
mechanism acquires the character of a theorem \cite{Glashow}, the theorem of
Glashow and Weinberg says that the FCNC processes mediated by Higgs exchange
are absent at tree level in a general multi-Higgs doublet model if all
fermions of a given electric charge couple to no more than one Higgs
doublet. The Lagrangians type I and II discussed above clearly accomplish
the theorem. It is important to say that we can use the same type of
couplings for both the quark and lepton sectors or on the other hand, we can
treat them asymmetrically giving a total of four different Lagrangians%
\footnote{%
For instance, Ref. \cite{Santos} classified these Yukawa Lagrangians as
models I, II, III, and IV. I prefer to use the most common notation taking
into account that for instance, the leptonic sector could be of type I and
the quark sector of type II.}.

Furthermore, the Yukawa Lagrangians type I and II can also be generated from
a continuous global symmetry. The set of transformations 
\begin{eqnarray*}
\Phi _{1} &\rightarrow &\Phi _{1}\;\;\text{and\ \ }\Phi _{2}\rightarrow
e^{i\varphi }\Phi _{2} \\
D_{jR} &\rightarrow &e^{-i\omega }D_{jR}\;\;\text{and\ \ }U_{jR}\rightarrow
e^{-i\varphi }U_{jR}
\end{eqnarray*}
with $\omega \equiv \varphi ,0$; leads to the models type I and type II,
respectively.

The discrete symmetry arises as a special case of this continuous symmetry
when we fix $\varphi =\pi $. Nevertheless, we should remember that the
discrete symmetry leads to a Higgs potential which is phenomenologically
different from the one generated by the continuous global symmetry as we saw
in Sec. (\ref{Higgspot 2HDM}). Consequently, we should be careful in
choosing the symmetry. For instance, as we saw in Sec. (\ref{Higgspot 2HDM}%
), the potential $V_{B}\;$in Eq. (\ref{VB}) that coincides with the
potential of the Higgs Hunter's Guide and the MSSM, comes from the
imposition of the continuous global symmetry (plus a soft breaking term) and
not from the discrete symmetry\footnote{%
We could argue that the potential $V_{B}\;$i.e. the potential in the Higgs
Hunter's Guide and the MSSM, comes from the discrete symmetry but adding a
fine tuning such that $\lambda _{3}=\lambda _{4}$. Notwithstanding, the
global continuous symmetry generates this potential without the necessity of
any fine tuning.}, while the potential $V_{A}^{\prime }\;$in Eq. (\ref{VAP})
comes from the discrete symmetry.

However, the discrete (or global) symmetry is not compulsory and we should
explore the possibility of having the whole model Eq. (\ref{Yukawa})
including the FCNC, and look for the constraints that the experimental
measurements impose on the region of parameters of the model. When we take
into account all terms in (\ref{Yukawa}), the two Higgs doublets couple and
yield masses to both up-type and down-type fermions, in that case we call it
the 2HDM type III.

From the discussion above, we conclude that the Feynman rules of the Yukawa
Lagrangian are highly model dependent, and could be enhanced or supressed
respect to SM in contrast to the kinetic couplings which are always
supressed. Consequently, the phenomenology is very sensitive to the choice
in the Yukawa sector.

\begin{table}[tbp]
\begin{center}
\begin{tabular}{||l||l||l||l||}
\hline\hline
particle & $J^{PC}$ & particle & $J^{P}$ \\ \hline\hline
$\gamma $ & $1^{--}\;\;\;\;\;\;\;$ & $W^{\pm }$ & $1^{+},1^{-}$ \\ 
\hline\hline
$Z$ & $1^{+-},1^{-+}$ & $H^{\pm }$ & $0^{+},\;0^{-}$ \\ \hline\hline
$H^{0}$ & $0^{++},\;0^{--}$ &  &  \\ \hline\hline
$h^{0}$ & $0^{++},\;0^{--}$ &  &  \\ \hline\hline
$A^{0}$ & $0^{+-},\;0^{-+}$ &  &  \\ \hline\hline
\end{tabular}
\end{center}
\caption{Assigments to the parity ($P$)\ and charge conjugation ($C$),
quantum numbers for the Higgs and vector bosons when fermions are
introduced. After the introduction of fermions, $P\;$and $C\;$are not
conserved separately anymore. However, in good aproximation $CP\;$is still a
good quantum number.}
\label{tab:CPFer}
\end{table}

There are however, some general characteristics related to the quantum
spectrum $J^{PC}\;$for the particles involved in the Yukawa interactions.
When we introduce fermions, $C\;$and $P\;$are not conserved separately any
more, however $CP\;$is still a good symmetry (but not exact). If we assume
CP\ conservation, the Higgs bosons and vectors that were considered above
could be thought as admixtures of two states with well defined $C\;$and $P\;$%
(appropiate linear combinations of them are CP eigenstates, and even we can
introduce a CP deviation $\epsilon $), we show in table (\ref{tab:CPFer})
the $J^{PC}\;$quantum numbers for Higgs and vector bosons when fermions are
introduced i.e. when $C\;$and\ $P\;$are not conserved separately but CP is.

On the other hand, it is well known that for a $f\overline{f}\;$system, the
quantum $C\;$and $P\;$numbers are $P=\left( -1\right) ^{L+1},\;C=\left(
-1\right) ^{L+S}$; where $L,\;S$ denote the total orbital angular momentum
and the total spin of the $f\overline{f}$ system, respectively.$\;$Then, $f%
\overline{f}\;$cannot be coupled to states of the type $0^{--},0^{+-}.\;$%
Therefore, in the couplings $f\overline{f}\left( H^{0},h^{0}\right) \;$the
Higgs bosons\ behave as purely $0^{++}$,$\;$while in $f\overline{f}A^{0}\;$%
the Higgs behaves as purely $0^{-+}$.$\;$Hence, $H^{0},h^{0}\;$couples with
parity $+1$ to matter (fermions) while $A^{0}\;$couples with parity $-1\;$to
fermions, it is because of this fact that the Higgs boson $A^{0}\;$is
usually called a pseudoscalar boson. Obviously, $H^{0},h^{0}\;$continue
being $CP-$even\ and $A^{0}\;$continues being $CP-$odd as long as $CP\;$%
invariance is demanded.

Additionally, there are two interesting gauge frameworks to generate the
Yukawa couplings, the $U-$gauge and the $R-$gauge. Though calculations in
the unitary gauge are simpler because the would be Goldstone bosons are
removed, the usage of the $R-$gauge in which the would be Goldstone bosons
appears explicitly, could be useful if we are interested in calculations
involving longitudinal vector bosons at high energies, since the equivalence
theorem could\ be applied.

Now we shall examine the most interesting choices for the Yukawa Lagrangian
and their features. Once again, for the sake of simplicity, I will restrict
the equations and discussions to the quark sector, nevertheless most of the
results hold for the leptonic sector too, with the proper changes.

\subsubsection{The 2HDM type I\label{Yuk I}}

In the 2HDM type I, only one Higgs doublet (say $\Phi _{2}$), couples to the
fermions, so the Yukawa Lagrangian becomes 
\begin{eqnarray}
-\pounds _{Y}\left( \text{type I}\right) &=&\xi _{ij}^{U,0}\overline{Q}%
_{iL}^{0}\widetilde{\Phi }_{2}U_{jR}^{0}+\xi _{ij}^{D,0}\overline{Q}%
_{iL}^{0}\Phi _{2}D_{jR}^{0}  \notag \\
&&+\text{leptonic\ sector}+h.c.,  \label{type I}
\end{eqnarray}

When we expand in terms of mass eigenstates the Lagrangian reads

\begin{eqnarray}
-\pounds _{Y}\left( \text{type I}\right) &=&\frac{g}{2M_{W}\sin \beta }%
\overline{D}M_{D}^{diag}D\left( \sin \alpha H^{0}+\cos \alpha h^{0}\right) 
\notag \\
&&+\frac{ig\cot \beta }{2M_{W}}\overline{D}M_{D}^{diag}\gamma _{5}DA^{0} 
\notag \\
&&+\frac{g}{2M_{W}\sin \beta }\overline{U}M_{U}^{diag}U\left( \sin \alpha
H^{0}+\cos \alpha h^{0}\right)  \notag \\
&&-\frac{ig\cot \beta }{2M_{W}}\overline{U}M_{U}^{diag}\gamma _{5}UA^{0} 
\notag \\
&&+\frac{g\cot \beta }{\sqrt{2}M_{W}}\overline{U}\left(
KM_{D}^{diag}P_{R}-M_{U}^{diag}KP_{L}\right) DH^{+}  \notag \\
&&-\frac{ig}{2M_{W}}\overline{U}M_{U}^{diag}\gamma _{5}UG_{Z}^{0}+\frac{ig}{%
2M_{W}}\overline{D}M_{D}^{diag}\gamma _{5}DG_{Z}^{0}  \notag \\
&&+\frac{g}{\sqrt{2}M_{W}}\overline{U}\left(
KM_{D}^{diag}P_{R}-M_{U}^{diag}KP_{L}\right) DG_{W}^{+}  \notag \\
&&+\text{leptonic\ sector}+h.c.  \label{type I expand}
\end{eqnarray}

The interactions involving would be Goldstone bosons given above, appears in
the $R-$gauge and are identical in all Yukawa Lagrangians, however they
vanish in the unitary gauge.

Note that since only one Higgs doublet couples to the fermions, the
Lagrangian (\ref{type I}) is SM-like. However, it does not mean that the
Higgs fermion interactions are SM-like, in Eq. (\ref{type I expand}) we see
that all five Higgs bosons and the mixing angles from the 2HDM appears. It
is because when we replace the scalar gauge eigenstates contained in the
doublet by the corresponding mass eigenstates, we are taking into account
the mixing of both doublets coming from the potential to obtain such mass
eigenstates.

On the other hand, the Yukawa couplings in SM are given by 
\begin{equation*}
g_{\phi ^{0}f\overline{f}}=\frac{m_{f}}{v}=\frac{gm_{f}}{2M_{W}}
\end{equation*}
in both the up and down sectors ($f\;$denotes any quark or charged lepton).
It is interesting to analyze the deviations of these couplings respect to
the ones in SM. In order to do it, we calculate the \textbf{relative
couplings\ }(the quotient between the Yukawa couplings of the new physics
and the SM Yukawa couplings). I will follow the notation defined in \cite
{Krawczykg2} for the relative couplings i.e. $\chi _{i}^{h}\equiv
g_{i}^{h}/\left( g_{i}^{\phi ^{0}}\right) _{SM}\;$with ``$i"\;$denoting a
fermion antifermion (or $VV$)\ pair and $``h"\;$represents a generic Higgs
boson. For the Lagrangian\ type I we get in third family notation

\begin{eqnarray}
\chi _{t}^{H^{0}} &:&\frac{\sin \alpha }{\sin \beta }=\cos \left( \beta
-\alpha \right) -\cot \beta \sin \left( \beta -\alpha \right) \;\;  \notag \\
\chi _{b}^{H^{0}} &:&\frac{\sin \alpha }{\sin \beta }=\cos \left( \beta
-\alpha \right) -\cot \beta \sin \left( \beta -\alpha \right)  \notag \\
\chi _{t}^{h^{0}} &:&\frac{\cos \alpha }{\sin \beta }=\sin \left( \beta
-\alpha \right) +\cot \beta \cos \left( \beta -\alpha \right)  \notag \\
\chi _{b}^{h^{0}} &:&\frac{\cos \alpha }{\sin \beta }=\sin \left( \beta
-\alpha \right) +\cot \beta \cos \left( \beta -\alpha \right)  \notag \\
\chi _{t}^{A^{0}} &:&-i\gamma _{5}\cot \beta \;\;\;;\;\;\chi
_{b}^{A^{0}}:i\gamma _{5}\cot \beta  \label{relat SM1}
\end{eqnarray}
There are several interesting features that could be seen above. First, the
relative couplings of a certain Higgs boson to the down-type quarks are
equal to the ones to the up type quarks, only the $A^{0}\;$couplings to the
up and down type quarks differs from a relative sign. Second, from Eqs. (\ref
{relat kin}) and (\ref{relat SM1}) we see that the model type I satisfies
the tree level unitarity bound given in Eq. (\ref{sum 2}). In the case of
minimal mixing in the CP even sector i.e. $\alpha =0,\;H^{0}\;$is decoupled
from the Yukawa sector while the couplings of $h^{0}\;$are maximal (for a
given value of $\tan \beta $).\ In the case of $\alpha =\pi /2,\;$the
behavior is the same but interchanging the roles of both CP even Higgses.
Additionally, the Yukawa couplings for a certain Higgs can be enhanced or
supressed depending on the values of the mixing angles $\alpha ,\beta $.

On the other hand, from (\ref{relat SM1}) we can realize that in model I the
heaviest (lightest) CP even Higgs becomes totally fermiophobic (i.e. all $%
H^{0}\left( h^{0}\right) f\overline{f}$ couplings vanish) if $\alpha
=0\left( \pi /2\right) $. We should bear in mind however, that decays of $%
H^{0}\left( h^{0}\right) \;$to fermions are still possible via $H^{0}\left(
h^{0}\right) \rightarrow W^{\ast }W\left( Z^{\ast }Z\right) \rightarrow 2f%
\overline{f}\;$or $H^{0}\left( h^{0}\right) \rightarrow W^{\ast }W^{\ast
}\left( Z^{\ast }Z^{\ast }\right) \rightarrow 2f\overline{f}$. If the
fermiophobic scenario were accomplished it would be a signature for physics
beyond the SM, and the 2HDM type I would be an interesting candidate. In
addition, we see that if we further assume that $\beta -\alpha =\pi /2\left(
0\right) \;$then $H^{0}\left( h^{0}\right) \;$becomes also bosophobic
(respect to vector bosons, see Eq. (\ref{relat kin}) and ghostphobic, while $%
h^{0}\left( H^{0}\right) \;$acquires SM-like couplings to fermions and
vector bosons. In the latter case, $H^{0}\left( h^{0}\right) \;$always needs
another scalar particle to decay into fermions or vector bosons.

\subsubsection{The 2HDM type II\label{Yuk II}}

In this case, one Higgs\ doublet (e.g. $\Phi _{1}$) couples to the down
sector of fermions while the other Higgs doublet (e.g. $\Phi _{2}$)\ couples
to the up sector.

\begin{equation}
-\pounds _{Y}\left( \text{type II}\right) =\eta _{ij}^{D,0}\overline{Q}%
_{iL}^{0}\Phi _{1}D_{jR}^{0}+\xi _{ij}^{U,0}\overline{Q}_{iL}^{0}\widetilde{%
\Phi }_{2}U_{jR}^{0}+\text{leptonic\ sector}+h.c.  \label{type II}
\end{equation}
In this case the expanded Lagrangian becomes

\begin{eqnarray}
-\pounds _{Y}\left( \text{type II}\right) &=&\frac{g}{2M_{W}\cos \beta }%
\overline{D}M_{D}^{diag}D\left( \cos \alpha H^{0}-\sin \alpha h^{0}\right) 
\notag \\
&&-\frac{ig\tan \beta }{2M_{W}}\overline{D}M_{D}^{diag}\gamma _{5}DA^{0} 
\notag \\
&&+\frac{g}{2M_{W}\sin \beta }\overline{U}M_{U}^{diag}U\left( \sin \alpha
H^{0}+\cos \alpha h^{0}\right)  \notag \\
&&-\frac{ig\cot \beta }{2M_{W}}\overline{U}M_{U}^{diag}\gamma _{5}UA^{0} 
\notag \\
&&-\frac{g}{\sqrt{2}M_{W}}\overline{U}\left[ \left( \cot \beta
M_{U}^{diag}KP_{L}+\tan \beta KM_{D}^{diag}P_{R}\right) \right] DH^{+} 
\notag \\
&&-\frac{ig}{2M_{W}}\overline{U}M_{U}^{diag}\gamma _{5}UG^{0}+\frac{ig}{%
2M_{W}}\overline{D}M_{D}^{diag}\gamma _{5}DG^{0}  \notag \\
&&+\frac{g}{\sqrt{2}M_{W}}\overline{U}\left[ \left(
KM_{D}^{diag}P_{R}-M_{U}^{diag}KP_{L}\right) \right] DG_{W}^{+}  \notag \\
&&+\text{leptonic\ sector}+h.c.  \label{type II expand}
\end{eqnarray}
this is the Lagrangian that is required in the MSSM. Notice that since the
down fermion sector receives its masses on the first Higgs doublet while the
up sector receives its masses from the second one; the hierarchy of the
Yukawa couplings between the top and bottom quarks of the third generation
becomes more natural if $\tan \beta \sim 35$ and $\tan \alpha \sim 1$.$\;$On
the other hand, we can see that the couplings of up type fermions to Higgs
bosons are the same for both models type I and II. It is because in both
lagrangians (\ref{type I}) and (\ref{type II}) the up sector couples to the
same Higgs doublet while the down sector couples to different ones\footnote{%
If we had chosen in the model type II to couple the doublet $\Phi _{1}\;$to
the up sector and $\Phi _{2}\;$to the down sector, the couplings of the down
type fermions would had been the ones that coincide for models type I and
II. It is a fact of convention.}.

Once again, comparison with SM is useful, calculating the relative couplings
in the third generation notation, we get

\begin{eqnarray}
\chi _{t}^{H^{0}} &:&\frac{\sin \alpha }{\sin \beta }=\cos \left( \beta
-\alpha \right) -\cot \beta \sin \left( \beta -\alpha \right)  \notag \\
\chi _{b}^{H^{0}} &:&\frac{\cos \alpha }{\cos \beta }=\cos \left( \beta
-\alpha \right) +\tan \beta \sin \left( \beta -\alpha \right)  \notag \\
\chi _{t}^{h^{0}} &:&\frac{\cos \alpha }{\sin \beta }=\sin \left( \beta
-\alpha \right) +\cot \beta \cos \left( \beta -\alpha \right)  \notag \\
\chi _{b}^{h^{0}} &:&-\frac{\sin \alpha }{\cos \beta }=\sin \left( \beta
-\alpha \right) -\tan \beta \cos \left( \beta -\alpha \right)  \notag \\
\chi _{t}^{A^{0}} &:&-i\gamma _{5}\cot \beta \;\;;\;\;\chi
_{b}^{A^{0}}:-i\gamma _{5}\tan \beta  \label{relat SM2}
\end{eqnarray}
from this results and Eqs. (\ref{relat kin}), we can see the following
interesting relations among the relative couplings in the 2HDM type II 
\begin{eqnarray*}
\left( \chi _{V}^{h^{0}}\right) ^{2}+\left( \chi _{V}^{H^{0}}\right) ^{2}
&=&1 \\
\left( \chi _{u}^{h^{0},H^{0}}-\chi _{V}^{h^{0},H^{0}}\right) \left( \chi
_{V}^{h^{0},H^{0}}-\chi _{d}^{h^{0},H^{0}}\right) &=&1-\left( \chi
_{V}^{h^{0},H^{0}}\right) ^{2}\; \\
\left( \chi _{u}^{h^{0},H^{0}}+\chi _{d}^{h^{0},H^{0}}\right) \chi
_{V}^{h^{0},H^{0}} &=&1+\chi _{u}^{h,H^{0}}\chi _{d}^{h^{0},H^{0}} \\
\left( 1-\frac{m_{h^{0}}^{2}}{2m_{H^{+}}}\right) \chi _{V}^{h^{0}}+\frac{%
m_{h^{0}}^{2}-\mu _{3}^{2}}{2m_{H^{+}}^{2}}\left( \chi _{d}^{h^{0}}+\chi
_{u}^{h^{0}}\right) &=&\chi _{H^{+}}^{h^{0}}
\end{eqnarray*}

From Eqs. (\ref{relat kin}) and (\ref{relat SM2}) we see that the model type
II satisfies the tree level unitarity bound given in Eq. (\ref{sum 2}). The
relative couplings of a certain Higgs to the down-type quarks are different
from the ones to the up type quarks unlike the case of Lagrangian type I .

On the other hand, if $\cos \left( \beta -\alpha \right) \simeq 0,\;$the
couplings of the lightest $CP\;$even Higgs boson $h^{0}\;$are almost
identical to the SM Higgs couplings not only in the Yukawa potential but in
the kinetic sector as well, if we additionally take $H^{0},A^{0},H^{\pm }\;$%
sufficiently heavy while keeping the quartic Higgs self couplings $\lesssim 
\emph{O}\left( 1\right) $, we obtain the decoupling limit in which the low
energy effective theory is the SM \cite{decoupling}. In general, if one
saturates the sume rules for $VV\;$with one scalar, the unitarity sum rules
can be achieved by assuming SM-like couplings of the same Higgs with
fermions. This is a natural scenario in the MSSM. However, it is not the
only way to satisfy the sum rules, for instance the other Higgs (that is
obviously weakly coupled to $VV$) could be strongly coupled to fermions so
that the product $g_{hVV}g_{hf\overline{f}}\;$is still significant.

It is worthwhile to point out that there is a very strong difference between
models I and II, in model II when $\tan \beta >1\;$some couplings (like $At%
\overline{t}$)\ are supressed, and others like $Ab\overline{b}\;$are
enhanced, for instance in $H^{+}\overline{t}b\;$the contributions of the top
(bottom) are suppressed (enhanced). In contrast, model I gives a uniform
supression or enhancement pattern (for fixed $\alpha $).

Moreover, from (\ref{relat SM2}) we can check that model type II does not
exhibit a totally fermiophobic limit for any Higgs boson, unlike the model
type I. For instance, if $\alpha =\pi /2\;$then $h^{0}\left( H^{0}\right) \;$%
become fermiophobic to fermions of up(down)-type, while couplings to the
down(up)-type fermions are maximal, the opposite occurs when $\alpha =0.$
This partially fermiophobic behavior of the model type II is another
important difference with model type I and with SM.

\subsubsection{The 2HDM type III\label{Yuk III}}

The model type III consists of taking into account all terms in Lagrangian (%
\ref{Yukawa}) 
\begin{eqnarray}
-\pounds _{Y} &=&\eta _{ij}^{U,0}\overline{Q}_{iL}^{0}\widetilde{\Phi }%
_{1}U_{jR}^{0}+\eta _{ij}^{D,0}\overline{Q}_{iL}^{0}\Phi _{1}D_{jR}^{0}+\xi
_{ij}^{U,0}\overline{Q}_{iL}^{0}\widetilde{\Phi }_{2}U_{jR}^{0}+\xi
_{ij}^{D,0}\overline{Q}_{iL}^{0}\Phi _{2}D_{jR}^{0}  \notag \\
&&+\text{lepton sector}+h.c.  \label{YukawaIII}
\end{eqnarray}
in the case of the model type III, we are able to make a rotation of the
doublets in such a way that only one of the doublets acquire VEV. Therefore,
we can assume without any loss of generality that $\langle \Phi _{1}\rangle
=v/\sqrt{2},\;\langle \Phi _{2}\rangle =0\;$(see appendices \ref{rotation
Yuk},\ref{rotation pot}).$\;$After expanding the Lagrangian (\ref{YukawaIII}%
) we obtain 
\begin{eqnarray}
-\pounds _{Y}\left( \text{type\ III}\right) &=&\frac{g}{2M_{W}}\overline{D}%
M_{D}^{diag}D\left( \cos \alpha H^{0}-\sin \alpha h^{0}\right)  \notag \\
&&+\frac{1}{\sqrt{2}}\overline{D}\xi ^{D}D\left( \sin \alpha H^{0}+\cos
\alpha h^{0}\right)  \notag \\
&&+\frac{i}{\sqrt{2}}\overline{D}\xi ^{D}\gamma _{5}DA^{0}+\frac{g}{2M_{W}}%
\overline{U}M_{U}^{diag}U\left( \cos \alpha H^{0}-\sin \alpha h^{0}\right) 
\notag \\
&&+\frac{1}{\sqrt{2}}\overline{U}\xi ^{U}U\left[ \sin \alpha H^{0}+\cos
\alpha h^{0}\right] -\frac{i}{\sqrt{2}}\overline{U}\xi ^{U}\gamma _{5}UA^{0}
\notag \\
&&+\overline{U}\left( K\xi ^{D}P_{R}-\xi ^{U}KP_{L}\right) DH^{+}  \notag \\
&&+\text{Goldstone interactions}+\text{leptonic\ sector}+h.c.
\label{YukexpandIII}
\end{eqnarray}
where the Goldstone interactions are the same as in the case of (\ref{type I
expand}) and (\ref{type II expand}).

There are several interesting characteristics:\ the diagonal couplings of
the pseudoscalar Higgs boson and the charged Higgs bosons to fermions are
not proportional to the fermion mass as is the case of models type I and II;
they are proportional to $\xi _{ff}\;$i.e. the diagonal elements of the
mixing matrix $\xi .\;$Additionally, the diagonal couplings involving $CP-$%
even Higgs bosons have one term proportional to the fermion mass and other
term proportional to $\xi _{ff}.\;$Once again we shall calculate the
relative couplings of the model

\begin{eqnarray}
\chi _{t}^{H^{0}} &:&\cos \alpha +\frac{\xi _{tt}\sin \alpha }{\sqrt{2}m_{t}}%
v  \notag \\
\chi _{b}^{H^{0}} &:&\cos \alpha +\frac{\xi _{bb}\sin \alpha }{\sqrt{2}m_{b}}%
v  \notag \\
\chi _{t}^{h^{0}} &:&-\sin \alpha +\frac{\xi _{tt}\cos \alpha }{\sqrt{2}m_{t}%
}v  \notag \\
\chi _{b}^{h^{0}} &:&-\sin \alpha +\frac{\xi _{bb}\cos \alpha }{\sqrt{2}m_{b}%
}v  \notag \\
\chi _{t}^{A^{0}} &:&-\frac{i\xi _{tt}}{\sqrt{2}m_{t}}v\;\;;\;\;\chi
_{b}^{A^{0}}:\frac{i\xi _{bb}}{\sqrt{2}m_{b}}v\;  \label{relative III}
\end{eqnarray}

The presence of diagonal mixing vertices could play a crucial role in
looking for FCNC. At this respect, it is worthwhile to point out that the 
\emph{relative couplings} in model type III are \emph{not universal} because
of the contribution of the factor $\xi _{ff}/m_{f},\;$as a manner of example
the relative couplings $\chi _{t}^{H^{0}},\;\chi _{c}^{H^{0}},\;\chi
_{u}^{H^{0}}\;$are in general different\footnote{%
As we can easily see from Eqs. (\ref{relat SM1}, \ref{relat SM2}) in model
type I, the relative couplings $\chi _{f}^{h}\;$for a certain Higgs $h$, are
the same for all fermions. In model type II, these relative couplings are
equal for all up type fermions and for all down type fermions. In contrast,
in the model type III all these relative couplings are in general, different
each other. This breaking of the universality of the relative couplings in
the model type III owes to the presence of the flavor diagonal contributions 
$\xi _{ii}/m_{i},\;$unless we assume them to be universal as well.}, a
deviation from the universal behavior could be a clear signature of FCNC at
the tree level; even if only diagonal processes are observed. In particular,
we cannot obtain a totally fermiophobic limit of (say) the Higgs $h^{0}\;$%
unless that a very precise pattern of the quotient $\xi _{ff}/m_{f}\;$is
demanded, we can have for instance the possibility of $h^{0}\;$to be
top-phobic\footnote{%
With the top-phobic limit for certain Higgs, we mean that the coupling of
that Higgs to a pair of top quarks vanishes. Nevertheless, the couplings of
such Higgs to a top and any other quark could exist.} but not charm-phobic
or up-phobic, this is due to the lack of universality discussed previously 
\cite{LFVus}. In contrast, model type I, can only be fermiophobic to all
fermions simultaneously \cite{Santos}, while model type II can only be
fermiophobic to all down-type or all up-type fermions simultaneously.
Moreover, as it was explained above, $A^{0}$ couples to a pair of fermions $f%
\overline{f}$ through the matrix element $\xi _{ff}$. Therefore in model
III, if $\xi _{ff}=0\;$for a certain specific fermion, then $A^{0}\;$becomes
fermiophobic to it (and only to it). It is also interesting to see that the
matrix elements $\xi _{ij}\;$modify the charged Higgs couplings as well: $%
g_{H^{+}f_{i}f_{j}}=\left( K_{ik}\xi _{kj}^{D}P_{R}-\xi
_{ik}^{U}K_{kj}P_{L}\right) $, where\ $K\;$is the Kobayashi Maskawa matrix
and $P_{L\left( R\right) }\;$are the left(right) projection operators.$\;$At
this respect there are two features to point out i) the flavor changing
charged currents (FCCC) in the quark sector are modified by the same matrix
that produces FCNC, ii) in the lepton sector FCCC are generated by the same
matrix that generates FCNC \cite{FCCCus}.

Finally, we can see that the tree level unitarity constraints (\ref{sum 2})
impose a condition on the diagonal elements $\xi _{ff}\;$of the mixing
matrix, preventing them to be arbitrarily large.

Since in model III we can find a basis in which we are able to get rid of
one of the VEV's, the mass eigenstates become simpler. By using say $\langle
\Phi _{2}\rangle =0,\;$we obtain the mass eigenstates by setting $\beta =0\;$%
in Eq. (\ref{mass eigenstates}). We shall call it the ``\textbf{fundamental
parametrization}''. In that case, only the doublet $\Phi _{1}\;$generates
the spontaneous symmetry breaking, thus it is the only one that provides
masses for the fermions and vector bosons. The second doublet $\Phi _{2}\;$%
only provides interactions with them.

Additionally, taking $\langle \Phi _{2}\rangle =0\;$and using the basis of
states ($h_{1},h_{2},g_{2},H^{\pm }$) we can see that $\Phi _{1}$
corresponds to the SM doublet and $h_{1}\;$to the SM Higgs field since it
has the same couplings and no interactions with $h_{2},g_{2}$\cite{Froggatt}%
. Consequently, all the new scalar fields belong to the second doublet $\Phi
_{2}.$\ Moreover, $h_{2},g_{2}\;$do not have couplings to gauge bosons of
the form $h_{2}(g_{2})ZZ\;$or $h_{2}(g_{2})W^{+}W^{-}$. Nevertheless, we
ought to remember that $h_{1},h_{2}\;$are not mass eigenstates, they mix
through the angle $\alpha \;$(see Eq. (\ref{mass eigenstates})). However, as
soon as we consider a scenario with $\alpha \approx 0\;$these features
become important since the set ($h_{1},h_{2},g_{2},H^{\pm }$) becomes a mass
eigenstates basis. As we will see later on, when $\alpha =0\;$in the
fundamental parametrization, we arrive to the decoupling limit of the 2HDM.

On the other hand, we can parametrize the Yukawa Lagrangian type III by
preserving both VEV's different from zero, we shall see below that the
relation between the model type III with the models type I or type II is
clearer with this parametrization\cite{LFVIIIus}. In order to distinguish
between the parametrization with only one VEV Eq. (\ref{YukawaIII}) from the
parametrization with two VEV, we write down the latter in the following way 
\begin{eqnarray}
-\pounds _{Y} &=&\widetilde{\eta }_{ij}^{U,0}\overline{Q}_{iL}^{0}\widetilde{%
\Phi }_{1}^{\prime }U_{jR}^{0}+\widetilde{\eta }_{ij}^{D,0}\overline{Q}%
_{iL}^{0}\Phi _{1}^{\prime }D_{jR}^{0}+\widetilde{\xi }_{ij}^{U,0}\overline{Q%
}_{iL}^{0}\widetilde{\Phi }_{2}^{\prime }U_{jR}^{0}+\widetilde{\xi }%
_{ij}^{D,0}\overline{Q}_{iL}^{0}\Phi _{2}^{\prime }D_{jR}^{0}  \notag \\
&&+\text{lepton sector}+h.c.  \label{YukIII}
\end{eqnarray}
where$\;\langle \Phi _{1}^{\prime }\rangle =v_{1},\;\langle \Phi
_{2}^{\prime }\rangle =v_{2}.\;$The relations among $\left( \widetilde{\eta }%
_{ij}^{U,D},\widetilde{\xi }_{ij}^{U,D}\right) \;$and $\left( \eta
_{ij}^{U,D},\xi _{ij}^{U,D}\right) $ are calculated in appendix (\ref
{rotation Yuk}). First we split the Lagrangian (\ref{YukIII}) in two parts

\begin{eqnarray*}
-\pounds _{Y\left( U\right) } &=&\widetilde{\eta }_{ij}^{U,0}\overline{Q}%
_{iL}^{0}\widetilde{\Phi }_{1}U_{jR}^{0}+\widetilde{\xi }_{ij}^{U,0}%
\overline{Q}_{iL}^{0}\widetilde{\Phi }_{2}U_{jR}^{0}+h.c. \\
-\pounds _{Y\left( D\right) } &=&\widetilde{\eta }_{ij}^{D,0}\overline{Q}%
_{iL}^{0}\Phi _{1}D_{jR}^{0}+\widetilde{\xi }_{ij}^{D,0}\overline{Q}%
_{iL}^{0}\Phi _{2}D_{jR}^{0}+\text{lepton sector}+h.c. \\
-\pounds _{Y}\left( \text{type III}\right) &=&-\pounds _{Y\left( U\right)
}-\pounds _{Y\left( D\right) }
\end{eqnarray*}
In order to convert this lagrangian into mass eigenstates we make the
unitary transformations

\begin{eqnarray}
D_{L,R} &=&\left( V_{L,R}\right) D_{L,R}^{0}\;,  \notag  \label{Down transf}
\\
U_{L,R} &=&\left( T_{L,R}\right) U_{L,R}^{0}\;  \label{Up transf}
\end{eqnarray}
from which we obtain the mass matrices. In the parametrization with both VEV
different from zero we get 
\begin{eqnarray}
M_{D}^{diag} &=&V_{L}\left[ \frac{v_{1}}{\sqrt{2}}\widetilde{\eta }^{D,0}+%
\frac{v_{2}}{\sqrt{2}}\widetilde{\xi }^{D,0}\right] V_{R}^{\dagger },  \notag
\label{Masa down} \\
M_{U}^{diag} &=&T_{L}\left[ \frac{v_{1}}{\sqrt{2}}\widetilde{\eta }^{U,0}+%
\frac{v_{2}}{\sqrt{2}}\widetilde{\xi }^{U,0}\right] T_{R}^{\dagger }.
\label{Masa up}
\end{eqnarray}

We can solve for $\widetilde{\xi }^{D,0},\widetilde{\xi }^{U,0}\;$obtaining

\begin{eqnarray}
\widetilde{\xi }^{D,0} &=&\frac{\sqrt{2}}{v_{2}}V_{L}^{\dagger
}M_{D}^{diag}V_{R}-\frac{v_{1}}{v_{2}}\widetilde{\eta }^{D,0}  \notag
\label{Rotation Id} \\
\widetilde{\xi }^{U,0} &=&\frac{\sqrt{2}}{v_{2}}T_{L}^{\dagger
}M_{D}^{diag}T_{R}-\frac{v_{1}}{v_{2}}\widetilde{\eta }^{U,0}
\label{Rotation Iu}
\end{eqnarray}

Let us call the Eqs. (\ref{Rotation Iu}), parametrization of type I.
Replacing them into (\ref{YukIII}) the expanded Lagrangians for up and down
sectors are \cite{LFVIIIus} 
\begin{eqnarray}
-\pounds _{Y\left( U\right) }^{\left( I\right) }\left( \text{type III}%
\right) &=&\frac{g}{2M_{W}\sin \beta }\overline{U}M_{U}^{diag}U\left( \sin
\alpha ^{\prime }H^{0}+\cos \alpha ^{\prime }h^{0}\right)  \notag \\
&&-\frac{1}{\sqrt{2}\sin \beta }\overline{U}\widetilde{\eta }^{U}U\left[
\sin \left( \alpha ^{\prime }-\beta \right) H^{0}+\cos \left( \alpha
^{\prime }-\beta \right) h^{0}\right]  \notag \\
&&-\frac{ig\cot \beta }{2M_{W}}\overline{U}M_{U}^{diag}\gamma _{5}UA^{0}+%
\frac{i}{\sqrt{2}\sin \beta }\overline{U}\widetilde{\eta }^{U}\gamma
_{5}UA^{0}  \notag \\
&&-\frac{g\cot \beta }{\sqrt{2}M_{W}}\overline{U}M_{U}^{diag}KP_{L}DH^{+}+%
\frac{1}{\sin \beta }\overline{U}\widetilde{\eta }^{U}KP_{L}DH^{+}  \notag \\
&&-\frac{ig}{2M_{W}}\overline{U}M_{U}^{diag}\gamma _{5}UG^{0}-\frac{g}{\sqrt{%
2}M_{W}}\overline{U}M_{U}^{diag}KP_{L}DG_{W}^{+}  \notag \\
&&+h.c.  \label{Yukawa 1u}
\end{eqnarray}
\begin{eqnarray}
-\pounds _{Y\left( D\right) }^{\left( I\right) }\left( \text{type III}%
\right) &=&\frac{g}{2M_{W}\sin \beta }\overline{D}M_{D}^{diag}D\left( \sin
\alpha ^{\prime }H^{0}+\cos \alpha ^{\prime }h^{0}\right)  \notag \\
&&-\frac{1}{\sqrt{2}\sin \beta }\overline{D}\widetilde{\eta }^{D}D\left[
\sin \left( \alpha ^{\prime }-\beta \right) H^{0}+\cos \left( \alpha
^{\prime }-\beta \right) h^{0}\right]  \notag \\
&&+\frac{ig\cot \beta }{2M_{W}}\overline{D}M_{D}^{diag}\gamma _{5}DA^{0}-%
\frac{i}{\sqrt{2}\sin \beta }\overline{D}\widetilde{\eta }^{D}\gamma
_{5}DA^{0}  \notag \\
&&+\frac{g\cot \beta }{\sqrt{2}M_{W}}\overline{U}KM_{D}^{diag}P_{R}DH^{+}-%
\frac{1}{\sin \beta }\overline{U}K\widetilde{\eta }^{D}P_{R}DH^{+}  \notag \\
&&+\frac{ig}{2M_{W}}\overline{D}M_{D}^{diag}\gamma _{5}DG^{0}+\frac{g}{\sqrt{%
2}M_{W}}\overline{U}KM_{D}^{diag}P_{R}DG_{W}^{+}  \notag \\
&&+\text{leptonic\ sector}+h.c.  \label{Yukawa 1d}
\end{eqnarray}
where $K$ is the CKM matrix, $\widetilde{\eta }^{U(D)}=T_{L}(V_{L})%
\widetilde{\eta }^{U(D),0}T_{R}^{\dagger }(V_{R})^{\dagger }$ and similarly
for $\widetilde{\xi }^{U(D)}$. The superindex $\left( I\right) $ refers to
the parametrization type I. In order to make the expansion above, we have
used the Eqs. (\ref{mass eigenstates}) but replacing $\alpha \rightarrow
\alpha ^{\prime }\;$since the notation $\alpha \;$will be reserved for the
mixing angle between CP even scalars in the fundamental parametrization (see
appendix \ref{rotation Yuk}). The leptonic sector is obtained from the
equation (\ref{Yukawa 1d}) replacing the down-type (up-type) quarks by the
charged leptons (neutrinos).

It is easy to check that if we add (\ref{Yukawa 1u}) and (\ref{Yukawa 1d})
we obtain a Lagrangian consisting of the one in the 2HDM type I \cite{Hunter}%
, plus some FC interactions, i.e., $-\pounds _{Y(U)}^{(I)}-\pounds
_{Y(D)}^{(I)}$. Therefore, we obtain the Lagrangian of type I from Eqs. (\ref
{Yukawa 1u}) and (\ref{Yukawa 1d}) by setting $\widetilde{\eta }^{D}=%
\widetilde{\eta }^{U}=0.$

On the other hand, from (\ref{Masa down}), we can also solve for $\widetilde{%
\eta }^{D,0},\widetilde{\eta }^{U,0}$ instead of $\widetilde{\xi }^{D,0},%
\widetilde{\xi }^{U,0}$, to get 
\begin{eqnarray}
\widetilde{\eta }^{D,0} &=&\frac{\sqrt{2}}{v_{1}}V_{L}^{\dagger
}M_{D}^{diag}V_{R}-\frac{v_{2}}{v_{1}}\widetilde{\xi }^{D,0}  \notag
\label{Rotation IId} \\
\widetilde{\eta }^{U,0} &=&\frac{\sqrt{2}}{v_{1}}T_{L}^{\dagger
}M_{U}^{diag}T_{R}-\frac{v_{2}}{v_{1}}\widetilde{\xi }^{U,0}
\label{Rotation IIu}
\end{eqnarray}
which we call parametrization of type II. Replacing them into (\ref{YukIII})
the expanded Lagrangians for up and down sectors become \cite{LFVIIIus}

\begin{eqnarray}
-\pounds _{Y\left( U\right) }^{\left( II\right) }\left( \text{type III}%
\right) &=&\frac{g}{2M_{W}\cos \beta }\overline{U}M_{U}^{diag}U\left( \cos
\alpha ^{\prime }H^{0}-\sin \alpha ^{\prime }h^{0}\right)  \notag \\
&&+\frac{1}{\sqrt{2}\cos \beta }\overline{U}\widetilde{\xi }^{U}U\left[ \sin
\left( \alpha ^{\prime }-\beta \right) H^{0}+\cos \left( \alpha ^{\prime
}-\beta \right) h^{0}\right]  \notag \\
&&+\frac{ig\tan \beta }{2M_{W}}\overline{U}M_{U}^{diag}\gamma _{5}UA^{0}-%
\frac{i}{\sqrt{2}\cos \beta }\overline{U}\widetilde{\xi }^{U}\gamma
_{5}UA^{0}  \notag \\
&&+\frac{g\tan \beta }{\sqrt{2}M_{W}}\overline{U}M_{U}^{diag}KP_{L}DH^{+}-%
\frac{1}{\cos \beta }\overline{U}\widetilde{\xi }^{U}KP_{L}DH^{+}  \notag \\
&&-\frac{ig}{2M_{W}}\overline{U}M_{U}^{diag}\gamma _{5}UG^{0}-\frac{g}{\sqrt{%
2}M_{W}}\overline{U}M_{U}^{diag}KP_{L}DG_{W}^{+}  \notag \\
&&+h.c.  \label{Yukawa 2u}
\end{eqnarray}
\begin{eqnarray}
-\pounds _{Y\left( D\right) }^{\left( II\right) }\left( \text{type\ III}%
\right) &=&\frac{g}{2M_{W}\cos \beta }\overline{D}M_{D}^{diag}D\left( \cos
\alpha ^{\prime }H^{0}-\sin \alpha ^{\prime }h^{0}\right)  \notag \\
&&+\frac{1}{\sqrt{2}\cos \beta }\overline{D}\widetilde{\xi }^{D}D\left[ \sin
\left( \alpha ^{\prime }-\beta \right) H^{0}+\cos \left( \alpha ^{\prime
}-\beta \right) h^{0}\right]  \notag \\
&&-\frac{ig\tan \beta }{2M_{W}}\overline{D}M_{D}^{diag}\gamma _{5}DA^{0}+%
\frac{i}{\sqrt{2}\cos \beta }\overline{D}\widetilde{\xi }^{D}\gamma
_{5}DA^{0}  \notag \\
&&-\frac{g\tan \beta }{\sqrt{2}M_{W}}\overline{U}KM_{D}^{diag}P_{R}DH^{+}+%
\frac{1}{\cos \beta }\overline{U}K\widetilde{\xi }^{D}P_{R}DH^{+}  \notag \\
&&+\frac{ig}{2M_{W}}\overline{D}M_{D}^{diag}\gamma _{5}DG^{0}+\frac{g}{\sqrt{%
2}M_{W}}\overline{U}KM_{D}^{diag}P_{R}DG_{W}^{+}  \notag \\
&&+\text{leptonic\ sector}+h.c.  \label{Yukawa 2d}
\end{eqnarray}

The superindex $(II)\;$refers to the parametrization type II. Moreover, if
we add the Lagrangians (\ref{Yukawa 1u}) and (\ref{Yukawa 2d}) we find the
Lagrangian of the 2HDM type II \cite{Hunter} plus some FC interactions, $%
-\pounds _{Y(U)}^{(I)}-\pounds _{Y(D)}^{(II)}$. Similarly like before,
Lagrangian type II is obtained setting $\widetilde{\xi }^{D}=\widetilde{\eta 
}^{U}=0$,$\;$in the total Lagrangian $-\pounds _{Y(U)}^{(I)}-\pounds
_{Y(D)}^{(II)}$.\ Therefore, Lagrangian type III can be written as the model
type II plus FC interactions if we use the parametrization type I in the up
sector and the parametrization type II in the down sector, it is valid since 
$\widetilde{\xi }^{U}\;$and $\widetilde{\xi }^{D}\;$are independent each
other and same to $\widetilde{\eta }^{U,D}.$

Moreover, we can build two additional parametrizations by adding $-\pounds
_{Y(U)}^{(II)}-\pounds _{Y(D)}^{(II)}\;$and $-\pounds _{Y(U)}^{(II)}-\pounds
_{Y(D)}^{(I)}$. The former corresponds to a model type I plus FC
interactions but interchanging the role of the doublets $\Phi
_{1}\leftrightarrow \Phi _{2};\;$the latter gives a model type II plus FC
interactions with the same interchange. On the other hand, terms involving
would-be Goldstone bosons are the same in all parametrizations in the
R-gauge, while in the unitary gauge they vanish \cite{Hunter, LFVIIIus}.

Finally, as it is demostrated in appendix (\ref{rotation Yuk}) the FC
matrices $\widetilde{\eta }^{U,D},\widetilde{\xi }^{U,D}\;$are related to
the FC matrices $\eta ^{U,D},\xi ^{U,D}\;$in the fundamental parametrization
Eq. (\ref{YukexpandIII}) through the following formulae (see Eqs. (\ref
{parametlink})), here we shall write them in terms of $\tan \beta $%
\begin{eqnarray}
\widetilde{\eta }^{U,D} &=&\frac{M_{U,D}}{v}\sqrt{\frac{2}{1+\tan ^{2}\beta }%
}-\frac{\tan \beta }{\sqrt{1+\tan ^{2}\beta }}\xi ^{U,D}  \notag \\
\widetilde{\xi }^{U,D} &=&\left( \sqrt{\frac{2}{1+\tan ^{2}\beta }}\right) 
\frac{M_{U,D}}{v}\tan \beta +\frac{1}{\sqrt{1+\tan ^{2}\beta }}\xi ^{U,D} 
\notag \\
\eta ^{U,D} &=&\frac{\sqrt{2}}{v}M_{U,D}  \notag \\
\;\xi ^{U,D} &=&\left( \sqrt{1+\tan ^{2}\beta }\right) \widetilde{\xi }%
^{U,D}-\frac{\sqrt{2}\tan \beta }{v}M_{U,D}  \label{parametlink2}
\end{eqnarray}

The development of these new parametrizations of the 2HDM type III,
facilitates the comparison of it with the models type I and type II. In
addition, one of these parametrizations has been used by Haber and Gunion 
\cite{decoupling} to study the decoupling limit of the general two Higgs
doublet model. Further, in Sec. (\ref{bounds on Higgs}) we shall obtain some
limits on the free parameters of the model in the framework of one of these
parametrizations.

\chapter{Present Status of the general Two Higgs Doublet Models\label%
{status 2HDM}}

\section{Theoretical constraints\label{theoretical constraints}}

The most important theoretical constraints come from the $\rho \;$parameter
and the unitarity limits. In SM, $\rho =1$\ at tree level. On the other
hand, in models beyond the standard model values around $\rho \simeq 1\;$are
expected too. In arbitrary Higgs multiplet representations the value of $%
\rho \;$at tree level is given by\cite{Hunter} 
\begin{equation*}
\rho \equiv \frac{M_{W}^{2}}{M_{Z}^{2}\cos ^{2}\theta _{W}}=\frac{%
\sum_{T_{i},Y_{i}}\left[ 4T_{i}(T_{i}+1)-Y_{i}^{2}\right] \left|
V_{T_{i}Y_{i}}\right| ^{2}C_{T_{i}Y_{i}}}{\sum_{T_{i},Y_{i}}2Y_{i}^{2}\left|
V_{T_{i}Y_{i}}\right| ^{2}}\;,
\end{equation*}
where$\;V_{T_{i}Y_{i}}=\;<\Phi (T_{i}Y_{i})>_{0}\;$defines the VEV of the
neutral Higgses, $Y_{i}$ is the hypercharge value of the multiplet number $%
i.\;T_{i}$ is the $SU(2)_{L\;\;}$isospin i.e. $T_{i}(T_{i}+1)$ are the
eigenvalues of the generators:\ $\sum_{j=1}^{3}\left( T_{i}\right) _{jL}^{2}$%
. Finally, the coefficients $C_{T_{i}Y_{i}}\;$are defined as

\begin{equation*}
C_{T_{i}Y_{i}}=\left\{ 
\begin{array}{c}
1,\;\text{if\thinspace }Y_{i}\neq 0 \\ 
\frac{1}{2},\ \text{if\ }T_{i}\;\text{is integer and}\ Y_{i}=0.
\end{array}
\right.
\end{equation*}
in the case of the 2HDM, we have $Y_{1}=Y_{2}=1,%
\;C_{T_{1}Y_{1}}=C_{T_{2}Y_{2}}=1,\;T_{1}=T_{2}=1/2.\;$From which we can see
that the relation $\rho =1\;$is maintained at tree level.

In SM the one loop correction to the $\rho \;$parameter \cite{Einborn} is
dominated by the top quark loop 
\begin{equation*}
\Delta \rho =\frac{3G_{F}m_{t}^{2}}{8\sqrt{2}\pi ^{2}}
\end{equation*}
by contrast, the contributions coming from the first family are given by
mild logarithmic terms. This fact is related to the breaking of a global $%
SU\left( 2\right) _{V}\;$symmetry \cite{custodial}. This isospin symmetry
``protects'' the relation $\rho =1\;$at next to leading order. The first
generation of quarks preserves in good approximation such ``custodial\
symmetry''. In contrast, the third generation of quarks breaks the symmetry
strongly producing a very sizeable one loop contribution\ \cite{custodial}.
Constraints coming from radiative corrections to $\rho \;$in the 2HDM will
be discussed later on.

Other important theoretical constraints come from unitarity. As it was
mentioned in section (\ref{kin interac}), the unitarity bound coming from $%
V_{L}V_{L}\rightarrow V_{L}V_{L}\;$scatterings is accomplished if

\begin{equation}
\sum_{i}g_{h_{i}^{0}VV}^{2}=g_{\phi ^{0}VV}^{2}  \label{unit1}
\end{equation}
where $i\;$labels all the neutral Higgs bosons of the 2HDM and $\phi ^{0}\;$%
refers to the SM Higgs. Additionally, we saw that the requirement of
unitarity for the $f\overline{f}\rightarrow V_{L}V_{L}\;$scattering leads to 
\begin{equation}
\sum_{i}g_{h_{i}^{0}VV}g_{h_{i}^{0}f\overline{f}}=g_{\phi ^{0}VV}g_{\phi
^{0}f\overline{f}}  \label{unit2}
\end{equation}

As we saw in Sec. (\ref{kin interac}), the condition (\ref{unit1}) is
accomplished for the general 2HDM since the couplings of any Higgs to $VV\;$%
are universal. In contrast, the second condition (\ref{unit2}) is
accomplished automatically by the 2HDM type I and II, but not by the type
III one, in the latter case this condition provides a constraint for the
diagonal terms of the mixing matrix as it was already mentioned.

On the other hand, unitarity bounds of processes involving $V-Higgs-Higgs\;$%
couplings determines the form of such couplings. For instance, unitarity
constraints \ in $H^{+}W^{-}\rightarrow H^{+}W^{-}\;$and $A^{0}Z\rightarrow
A^{0}Z\;$leads to the following sum rule \cite{Hunter}.

\begin{equation}
g_{H^{+}W^{-}h^{0}}^{2}+g_{H^{+}W^{-}H^{0}}^{2}=\left(
g_{h^{0}A^{0}Z}^{2}+g_{H^{0}A^{0}Z}^{2}\right) \cos ^{2}\theta _{W}=\frac{%
g^{2}}{4}  \label{sum rule 3}
\end{equation}
Moreover, unitarity of $A^{0}Z\rightarrow W^{+}W^{-}\;$requires a precise
ratio between $g_{H^{0}A^{0}Z}\;$and $g_{h^{0}A^{0}Z}^{2}$\cite{Hunter}.\
Similarly, unitarity of $H^{+}W^{-}\rightarrow W^{+}W^{-}\;$determines a
relation among the couplings $g_{H^{+}W^{-}h^{0}}\;$and $g_{H^{+}W^{-}H^{0}}$%
. Combining the sum rule (\ref{sum rule 3}) with the unitarity bounds from$%
\;A^{0}Z\rightarrow W^{+}W^{-}\;$and $H^{+}W^{-}\rightarrow W^{+}W^{-}$,\
the following values for the $V-Higgs-Higgs\;$couplings are obtained 
\begin{eqnarray*}
g_{h^{0}A^{0}Z} &=&\frac{g}{2\cos \theta _{W}}\cos \left( \beta -\alpha
\right) \;\;\;\;\;\;\;;\;\;\;\;\;g_{H^{+}W^{-}h^{0}}=\frac{g}{2}\cos \left(
\beta -\alpha \right) \\
g_{H^{0}A^{0}Z} &=&\frac{g}{2\cos \theta _{W}}\cos \left( \beta -\alpha
\right)
\end{eqnarray*}

They coincide with the values gotten by direct expansion of the kinetic
Lagrangian, and tree level unitarity is also automatically satisfied by $%
V-Higgs-Higgs\;$couplings.

On the other hand, tree level unitarity bounds in the Higgs potential can
also be derived from the two body scattering of scalars $S_{1}S_{2}%
\rightarrow S_{3}S_{4}.\;$Tree level unitarity constraints for the potential 
$V_{A}^{\prime }\;$in Eq. (\ref{VAP}) have been calculated \cite{Kanemura}
obtaining strong upper limits on some free parameters of the Higgs potential
with exact $Z_{2}\;$invariance. For instance, Ref. \cite{Kanemura} obtains
the following useful bound

\begin{equation*}
m_{h^{0}}\leq \sqrt{\frac{16\pi \sqrt{2}}{3G_{F}}\cos ^{2}\beta
-m_{H^{0}}^{2}\cot ^{2}\beta }
\end{equation*}

Notwithstanding, Akeroyd \emph{et. al.} \cite{Akeroyd} showed that by
introducing the soft breaking term $\mu _{3}^{2}$ (from which we arrive to
the potential $V_{A}\;$Eq.(\ref{VA})) such bounds are weaken considerably.
Additionally, Ref. \cite{Akeroyd} also shows that the inclusion of charged
states not included in \cite{Kanemura} changes significantly the bounds on $%
m_{H^{\pm }}\;$and $m_{A^{0}}$.$\;$In the absence of the soft breaking term,
general unitarity bounds are $m_{H^{\pm }}\lesssim 691\;$GeV$%
,\;m_{A^{0}}\lesssim 695\;$GeV$,\;m_{h^{0}}\lesssim 435\;$GeV, \ $%
m_{H^{0}}\lesssim 638\;$GeV.$\;$These bounds are obtained for relatively
small values of $\tan \beta \;$($\tan \beta \approx 0.5$). For large values
of it the upper bound is smaller, but only $h^{0}\;$is very sensitive to the
variation in $\tan \beta ,\;$especially for small values of the soft
breaking term. For $\mu _{3}^{2}=0,\;$the sensitivity is very strong, but
for large values of $\mu _{3}^{2}\;$e.g.\ $\mu _{3}^{2}=15,\;$the plot $%
m_{h^{0}}\;$vs$\;\tan \beta \;$becomes a horizontal line predicting $%
m_{h^{0}}\lesssim 670\;$GeV. The relaxation of the strong correlation
between $m_{h^{0}}\;$and $\tan \beta \;$could be a mechanism to distinguish
between the potential with the exact symmetry and with the softly broken
symmetry.

Unfortunately, no general unitarity bounds have been calculated for the
potentials $V\;$Eq. (\ref{Lag10}) and $V_{B}\;$Eq.$\;$(\ref{VB})$.\;$However
in the fermiophobic limit of model I, a tree level unitarity bound can be
easily obtained for $V_{B}$ \cite{fermiophobic} 
\begin{equation}
m_{h^{0}}^{2}=m_{A^{0}}-2\left( \lambda _{+}-\lambda _{1}\right) v^{2}\cos
^{2}\beta  \label{unitarity VB}
\end{equation}
we see that if we assume $\beta \approx \pi /2\;$then $m_{h^{0}}^{2}\approx
m_{A^{0}}^{2}.\;$As for the potential $V$,$\;$Ref. \cite{decoupling}
estimates that in the decoupling limit the introduction of unitarity implies
for the heavy Higgs bosons to be nearly degenerate.

However, it should be pointed out that tree level unitarity is only a guide,
its violation could imply a strongly interacting Higgs sector. Nevertheless,
in that case perturbative arguments become suspect.

It has also been mentioned that flavor changing neutral currents impose
strong constraints on the 2HDM, since the introduction of the second doublet
produces them automatically. These rare processes can be suppressed on
either theoretical or phenomenological grounds. The most common theoretical
mechanism consists of the imposition of a $Z_{2}\;$symmetry (or a global $%
U\left( 1\right) $ symmetry) as dicussed in section (\ref{Yuk 2HDM})
generating the models type I and II. Nevertheless, there are also some
theoretical considerations that permits us to avoid FCNC in the type III
model. Two interesting scenarios for a flavor conserving type III model
deserves special attention. 1) A model type III with a permutation symmetry
among the fermion families \cite{Lin}, and 2) the decoupling limit of the
2HDM type III \cite{decoupling}. Let us discuss them briefly.

The motivation for introducing the $S_{3}\;$permutation symmetry, comes from
the similarity among the fermion families. Since there are not fermion
masses before the SSB of the gauge symmetry, it is logical to assume that
before the SSB the three generations of fermions are indistinguishable i.e.
they hold an $S_{3}\;$symmetry. After the SSB of the gauge symmetry,
fermions acquire masses i.e. distinguibility, thus the $S_{3}\;$symmetry
should be spontaneously broken as well. The implementation of such
permutation symmetry conduces automatically to the supression of FCNC at
tree level \cite{Lin}, but also leads to a 3$\times $3\ unit CKM matrix.
Notwithstanding, Ref. \cite{Lin} showed that by introducing an appropiate
soft breaking term, we could still get a flavor conserving type III model
while obtaining a reasonable CKM matrix.

As for the decoupling limit, Ref. \cite{decoupling} shows that this
particular scenario automatically suppress both FCNC and CP violating
couplings in the 2HDM type III. More details about the decoupling limit will
be discussed in Sec. (\ref{the decoupling limit}).

Furthermore, renormalization group equations could be utilized to obtain
some useful constraints. For instance, we can demand that the self
interacting parameters in the Higgs potential remain perturbative up to a
large scale (typically the grand unification scale i.e. $\Lambda _{GUT}\sim
10^{16}$GeV), and also that the vacuum of the potential remains stable up to
such large scale. In SM such bounds have been studied at first by Cabibbo 
\emph{et. al.} \cite{Cabibbo} and Lindner \cite{Lindner}. The criterium of
perturbativity provides an upper limit of the SM Higgs mass of about 180 GeV
when the cut-off scale is taken to be $\Lambda _{GUT}\sim 10^{16}$GeV.
Meanwhile, vacuum stability provides a lower limit of about 130 GeV \cite
{vacuum stability}, for the same cut-off. On the other hand, if the cut-off
scale is extended up to the Planck scale $\Lambda _{Planck}\sim 10^{19}\;$%
GeV, the allowed interval becomes $145\;$GeV$\lesssim m_{h}\lesssim 175$ GeV 
\cite{Kanemu}, therefore perturbativity and vacuum stability requirements
gives a rather strong allowed interval for the Higgs mass in SM. As for the
2HDM, Ref. \cite{vacuum stability} has examined the constraints coming from
perturbativity and vacuum stability up to the grand unification scale in the
type II version without the soft breaking term in the potential. Such
reference found that these criteria impose an upper limit for the charged
Higgs of $m_{H^{\pm }}\lesssim 150\;$GeV which is in conflict with the lower
bound for it $m_{H^{\pm }}\gtrsim 500\;$GeV\footnote{%
At the time of ref \cite{vacuum stability}, the lower bound was
significantly smaller $m_{H^{+}}\gtrsim 165\;$GeV, but enough for the
conflict to arise.}. Consequently, the 2HDM type II (without the soft
breaking term) cannot be valid up to the unification scale. In contrast,
model I is not excluded by these criteria, since the experimental lower
limit for $m_{H^{\pm }}\;$is significantly milder.

We should remark however that bounds from perturbativity and vacuum
stability from the electroweak (EW) scale up to any cut-off scale is only a
guidance, if they are not accomplished it merely means that the model is
only valid up to a lower scale (or that it belongs to a non-perturbative
regime). Indeed, we can reverse the problem and check until what scale might
our model be valid. For instance, Ref. \cite{Kanemu} have calculated lower
and upper bounds of $m_{h^{0}}$ in the context of the 2HDM types I and II
but including the soft breaking term in the potential. These bounds are
estimated as a function of the cutoff scale and also of the soft breaking
term of the discrete symmetry, requiring that the running coupling constants
neither blow up nor fall down below $\Lambda $. Comparing with the
corresponding SM constraints, the upper bound does not change so much but
the lower bound is substantially reduced. In particular, in the decoupling
limit with a large soft breaking term, the upper bound for the 2HDM for $%
\Lambda _{Planck}=10^{19}$ GeV, and $m_{t}=175$ GeV, is about $175$ GeV,
which is basically the same as the corresponding SM upper bound. By
contrast, the corresponding lower bound for the 2HDM is about $100$ GeV
while the SM bound is about $145$ GeV. In the region of small soft breaking
term the lower and upper bounds depend on the soft-breaking mass and when
the soft breaking term vanishes, the lower bound no longer appears \cite
{Kanemu}.

Finally, perturbativity could be also examined at a fixed scale (typically
the electroweak one) to bound the free parameters of the model at that
scale. For example, perturbative constraints at the EW scale \cite
{perturbative} on the $\lambda _{i}\;$restrict the allowed values of $\tan
\beta $,\ from these grounds it is found that $\tan \beta \geq 30\;$is
strongly disfavoured \cite{Akeroyd}.

\section{Phenomenological constraints\label{phenomenological constraints}}

Let us discuss radiative corrections to $\rho \;$first.\ There are different
renormalization schemes, Marciano and Sirlin \cite{Sirlin ro} maintain the
relation $m_{W}=m_{Z}\cos \theta _{W}\;$and radiative corrections appear in
the relation between $G_{F}\;$and\ $m_{W}.\;$In the Veltman scheme \cite
{Veltman ro}, the free parameters of the electroweak theory are $g,\;\sin
\theta _{W}\;$and $m_{W}.\;$The effective one loop value for $m_{Z}^{2}$\ is
extracted from $\overline{\vartheta }_{\mu }e\;$and $\vartheta _{\mu }%
\overline{e}\;$at one loop, and writing them in the tree level form with a
corrected $m_{Z}\;$value. From the Veltman scheme the one loop correction $%
\Delta \rho \;$is written as 
\begin{equation}
m_{W}^{2}\Delta \rho =A_{WW}\left( k^{2}=0\right) -\cos ^{2}\theta
_{W}A_{ZZ}\left( k^{2}=0\right)  \label{delta ro}
\end{equation}
$A_{VV}\;$are the propagator corrections at zero momentum transfer.

In the $R-$gauge, the terms that could be quadratically dependent on the
Higgs boson masses come from diagrams with either two Higgs bosons or a
Higgs boson and a Goldstone boson, as well as associated tadpole type
diagrams.

The contribution of the Higgs sector in the $R-gauge\;$in the 2HDM type II
is given by \cite{Hunter} 
\begin{eqnarray}
A_{WW}^{HH}\left( 0\right) -\cos ^{2}\theta _{W}A_{ZZ}^{HH}\left( 0\right)
&=&\frac{g^{2}}{64\pi ^{2}}\left\{ F_{\Delta \rho }\left(
m_{H^{+}}^{2},m_{A^{0}}^{2}\right) +\left[ F_{\Delta \rho }\left(
m_{H^{+}}^{2},m_{H^{0}}^{2}\right) \right. \right.  \notag \\
&&\left. -F_{\Delta \rho }\left( m_{A^{0}}^{2},m_{H^{0}}^{2}\right) \right]
\sin ^{2}\left( \beta -\alpha \right)  \notag \\
&&+\left[ F_{\Delta \rho }\left( m_{H^{+}}^{2},m_{h^{0}}^{2}\right) \right. 
\notag \\
&&\left. \left. -F_{\Delta \rho }\left( m_{A^{0}}^{2},m_{h^{0}}^{2}\right) 
\right] \cos ^{2}\left( \beta -\alpha \right) \right\}  \notag \\
A_{WW}^{HG}\left( 0\right) -\cos ^{2}\theta _{W}A_{ZZ}^{HG}\left( 0\right)
&=&\frac{g^{2}}{64\pi ^{2}}\cos ^{2}\left( \beta -\alpha \right) \left[
F_{\Delta \rho }\left( m_{W}^{2},m_{H^{0}}^{2}\right) \right.  \notag \\
&&-F_{\Delta \rho }\left( m_{W}^{2},m_{h^{0}}^{2}\right) -F_{\Delta \rho
}\left( m_{Z}^{2},m_{H^{0}}^{2}\right)  \notag \\
&&\left. -F_{\Delta \rho }\left( m_{Z}^{2},m_{H^{0}}^{2}\right) +F_{\Delta
\rho }\left( m_{Z}^{2},m_{h^{0}}^{2}\right) \right]  \label{A ro}
\end{eqnarray}
where $F_{\Delta \rho }\left( m_{1}^{2},m_{2}^{2}\right) \;$is defined as

\begin{equation}
F_{\Delta \rho }\left( m_{1}^{2},m_{2}^{2}\right) \equiv \frac{1}{2}\left(
m_{1}^{2}+m_{2}^{2}\right) -\frac{m_{1}^{2}m_{2}^{2}}{m_{1}^{2}-m_{2}^{2}}%
\ln \frac{m_{1}^{2}}{m_{2}^{2}}  \label{function ro}
\end{equation}
Notwithstanding, we should be careful in removing the contribution from the
SM Higgs, a natural approach could be to take $h^{0}\;$as the SM Higgs with
modified couplings, and extract the contribution of $h^{0}\;$with SM
couplings.

Since $m_{W}\sim m_{Z}\;$the HG\ (Higgs-Goldstone) contribution is very
small.\ While the HH\ contribution could be large depending upon the mixing
angles and masses. From expressions (\ref{delta ro}-\ref{function ro}) it is
easy to check that $\Delta \rho \;$depends basically on the splitting
between $m_{H^{+}}\;$and$\;m_{A^{0}}\;$as well as from the splitting of $%
m_{H^{+}}\;$with the CP even Higgses. Combined experimental measurements on
neutral currents of $\sin ^{2}\theta _{W},\;m_{Z}\;$and $m_{W},\;$imply that
the charged Higgs mass cannot be much larger than the neutral Higgs mass
(except if the Higgs is SM like). There are some bounds about the splitting
between $m_{H^{0}}\;$and $m_{H^{+}}.\;$In SUSY a very natural scenario
consists of $m_{H^{0}}\sim m_{A^{0}}\sim m_{H^{+}}\;$with a SM like $h^{0}\;$%
Higgs.\ The term$\;F_{\Delta \rho }\left( m_{H^{+}}^{2},m_{A^{0}}^{2}\right)
\;$vanishes in this case and the other terms tend to be cancelled by the
small splitting of $m_{W}\sim m_{Z},\;$then these contributions are expected
to be small in SUSY scenarios. Notice that $A_{ZZ}\left( 0\right) \;$tends
to cancel the otherwise large $A_{WW}\left( 0\right) \;$contribution.

For the model type I, the contribution to $\Delta \rho \;$has been estimated
in the unitary gauge \cite{Denner} 
\begin{eqnarray*}
\Delta \rho &=&\frac{1}{16\pi ^{2}v^{2}}\left[ \sin ^{2}\left( \beta -\alpha
\right) F\left( m_{H^{\pm }},m_{A^{0}}^{2},m_{H^{0}}^{2}\right) \right. \\
&&\left. +\cos ^{2}\left( \beta -\alpha \right) F\left( m_{H^{\pm
}},m_{A^{0}}^{2},m_{h^{0}}^{2}\right) \right] \\
F\left( a,b,c\right) &\equiv &a+\frac{bc}{b-c}\ln \frac{b}{c}-\frac{ab}{a-b}%
\ln \frac{a}{b}-\frac{ac}{a-c}\ln \frac{a}{c}
\end{eqnarray*}

In the case of model type I, if a light $h^{0}\;$Higgs still exists, the
variation of $\Delta \rho \;$with $h^{0}\;$is soft, but it depends on $%
m_{A^{0}}\;$and $m_{A^{0}}-m_{H^{\pm }}\;$(or $m_{H^{\pm }}$ and $%
m_{A^{0}}-m_{H^{\pm }}$). Once again, $\Delta \rho \;$can be used to
restrict the splitting between the charged Higgs and neutral Higgs bosons.
Ref \cite{Logan} shows that in the general 2HDM when $m_{A^{0}}\;$is small,
the contribution to $\Delta \rho \;$can grow quadratically with the other
Higgs masses, so in order to keep $\Delta \rho \;$small, a correlation among
the Higgs boson masses to cancel this large loop contributions should be
assumed.\ Ref \cite{Larios} presents constraints\ in the framework of the
general 2HDM with a very light CP-odd scalar, these bounds are shown in a $%
m_{H}-m_{H^{+}}\;$plane for three different values of $\sin ^{2}\left( \beta
-\alpha \right) ,\;$constraining strongly the splitting between such scalars.

Other observables useful to constrain the 2HDM are $Br\left( b\rightarrow
s\gamma \right) ,\;$the$\;Z\rightarrow b\overline{b}\;$hadronic decay
branchig ratio $\left( R_{b}\right) ,\;$the forward backward asymmetry of
the bottom quark in $Z\;$decays$\;\left( A_{b}\right) $;$\;$and the muon
anomalous magnetic moment of the muon, $a_{\mu }.\;$The effectiveness of
each observable to constrain the model depend on the specific scenario, for
instance, when $\tan \beta \;$is of the order of 1\ the asymmetry $A_{b}\;$%
is less effective to constrain in the $m_{H^{+}}-\tan \beta \;$plane than $%
R_{b}\;$\cite{Larios}. The anomalous magnetic moment of the muon $a_{\mu }\;$%
is a very important parameter to restrict any physics beyond the SM, in
particular the 2HDM. Since much of our analysis is based on this observable,
we shall revise its present status in section \ref{survey g-2}. For now, I
just say that its present world average experimental value is given by \cite
{g2exp} 
\begin{equation*}
a_{\mu }^{\text{exp}}\equiv \frac{\left( g-2\right) _{\mu }^{\text{exp}}}{2}%
=11\;659\;203\;\left( 8\right) \times 10^{-10}
\end{equation*}
and the SM estimation has been calculated recently by a variety of authors 
\cite{g2th}, though their estimations are rather different, all of them
coincide in the fact that the SM value based on $e^{+}e^{-}\rightarrow $%
hadrons data, is roughly $3\sigma \;$below the present world average
measurement (see details in section \ref{survey g-2}). Thus, the theoretical
and experimental current estimations seem to show the necessity of new
physics. The fact that $\Delta a_{\mu }\;$should be positive at 95\% CL,
impose strong restrictions to any physics beyond the SM.

On the other hand, bounds on the 2HDM might also be extracted from the
non-observation of direct Higgs production and/or decays. For instance, in
scenarios with a very light CP odd Higgs boson constraints can be derived
from $Br\left( \Upsilon \rightarrow A^{0}\gamma \right) ,\;Z\rightarrow
AAA,\;f\left( f^{\prime }\right) \overline{f}\rightarrow Z\left( W^{\pm
}\right) AA\rightarrow Z\left( W^{\pm }\right) +``\gamma \gamma "\;$and $f%
\overline{f}\rightarrow H^{+}H^{-}\rightarrow Z\left( W^{+}W^{-}\right)
+``\gamma \gamma "\;$\cite{Larios}.\ Similar decays with $A^{0}\;$replaced
by $h^{0}\;$can be used to bound scenarios with a very light CP even Higgs
boson.\ For intermediate Higgs bosons, limits could be derived from the non
observations of processes like$\;e^{+}e^{-}\rightarrow Z^{\ast }\rightarrow
Zh^{0}$, and in the heavy regime from processes like $h^{0}\rightarrow
ZZ\rightarrow $leptons.

Further, as well as constraining the model it is necessary to explore the
possibility to produce the Higgs bosons from colliders. The expected
observed production rate is usually very model dependent and, of course
depend also on the experimental luminosity and detector efficiency.

At $e^{+}e^{-}\;$machines production mechanisms depend upon substantial $VV$
couplings. The most advantageous scenario for these machines would be a
roughly equally $VV\;$couplings for both scalars, if one of the Higgses tend
to saturate these couplings (as in MSSM) then the discovery of the weak
coupled Higgs would be problematical in $e^{+}e^{-}\;$machines. $A^{0}\;$is
particularly problematical in these colliders since it is not coupled to $%
VV,\;$the main production mode that is available is $e^{+}e^{-}\rightarrow
Z^{\ast }\rightarrow A^{0}h^{0}\;$or $A^{0}H^{0}.\;$The cross sections of
these processes are proportional to $\cos ^{2}\left( \beta -\alpha \right)
,\;\sin ^{2}\left( \beta -\alpha \right) \;$for $h^{0},H^{0}\;$respectively.
Detection might be possible if $m_{A^{0}}+m_{h^{0}}\;$(or $%
m_{A^{0}}+m_{H^{0}}$)\ is not too large compared to the machine energy, and
the vertex $ZA^{0}\left( h^{0}\right) H^{0}\;$is a saturating coupling.
Further, if $h^{0},A^{0}\;$are sufficiently light, they may be produced with
a $Z\;$on shell and the production rate would be higher. With sufficiently
high energy, the heavy states can be produced via $e^{+}e^{-}\rightarrow
H^{+}H^{-}\;$or $e^{+}e^{-}\rightarrow H^{0}A^{0}.\;$We should notice that
modes in which one or both Higgses go to $ee\;$are difficult to detect
because of the QED background. Prospects for detection of a single Higgs in $%
e^{+}e^{-}\;$colliders are $e^{+}e^{-}\rightarrow Z\rightarrow Zh^{0}\left(
A^{0}\right) \;$where one of the $Z^{\prime }s\;$is off shell. If $%
m_{h^{0}}>2m_{Z}\;$``the gold-plated'' detection mode is $h^{0}\rightarrow
ZZ\rightarrow $leptons\ (for $m_{h^{0}}\gtrsim 130\;$GeV, $h^{0}\rightarrow
ZZ^{\ast }\;$could be a signature). In the intermediate mass Higgs regimes
other decays are required \cite{Sopczak}. With increasing center of mass
energy,$\;W^{+}W^{-}\;$fusion begins to be the dominant mechanism.

In hadron colliders, the dominant mechanism for Higgs production is the $%
gg\; $fusion through a top quark loop. Important prospects for single Higgs
production are $gg\rightarrow A^{0},h^{0},H^{0}\;$and $gb\rightarrow
H^{-}t.\;$An interesting method could be from $gg\rightarrow Q\overline{Q}%
^{\prime }\left( H^{0},A^{0},H^{\pm }\right) \;$where $Q\;$is a heavy quark (%
$b\;$or $t$)\ and the Higgs decays into a Heavy quark pair. Another
interesting source is the quarkonium decay to $h\gamma \;$which could be
enhanced or supressed according to the model. Indirect signatures could be
given by $^{3}S_{1}\left( t\overline{t}\right) \rightarrow b\overline{b},\;$%
the SM contributions come from $W\;$exchange, but in two doublets a\ $%
H^{+}\; $could be placed wherever the $W^{+}\;$is. So alterations of the
rate could be detected especially if $\tan \beta <1\;$so that the $t%
\overline{b}H^{+}\;$coupling is enhanced. Many other quarkonium decays can
be a direct source for Higgs boson pairs. Some of them take into account
exotic quarks with new CKM elements.

Another possibility is the $\gamma \gamma \;$collider mode via $\gamma
\gamma \;$fusion (dominated by $\rho W^{-}\;$or $t\overline{t}\;$loop),
depending on the mass of the Higgs in the final state decay.

On the other hand, muon muon colliders offer interesting perspectives for
Higgs discovery, because Higgs bosons $s-$channels are enhanced respect to $%
e^{+}e^{-}\;$colliders, though still weak (proportional to the muon mass)
but production can be significant if the collider is run on the Higgs
resonance $\sqrt{s}=m_{H}$.\ Further, they could be made feasible to reach
energy in the multi-TeV regime. Many references have address the perpective
of looking for the Higgs bosons in this kind of machines \cite{muon collider}%
. In particular, the fourth of Refs. \cite{muon collider} shows that in the
large $\tan \beta $ regime, production of a single charged or pseudoscalar
Higgs in association with a gauge boson is possible in muon colliders (i.e. $%
\mu ^{+}\mu ^{-}\rightarrow H^{\pm }W^{\mp },A^{0}Z$) with sizeable cross
sections whose anologies at $e^{+}e^{-}$ would be very small.

As for $e^{+}e^{-}\;$machines, LEP have obtained limits on the charged Higgs
mass from $e^{+}e^{-}\rightarrow \gamma ^{\ast },Z^{\ast }\rightarrow
H^{+}H^{-}.\;$It worths to point out that far from the $Z\;$mass shell
electromagnetic interaction become dominant, thus the bound on $m_{H^{+}}\;$%
obtained from this process is nearly model independent, LEP excludes a
region of $2m_{\tau }\lesssim m_{H^{+}}\lesssim 80.5GeV.\;$The region $%
m_{H^{+}}\lesssim 2m_{\tau }\;$is excluded by the non observation of $%
B\rightarrow H^{+}X_{c}.\;$In this region they look for $H^{+}\rightarrow
\tau \vartheta _{\tau }\;$when $m_{H^{+}}>m_{\tau }.\;$

Some other limits are gotten from some rare decays like $b\rightarrow
s\gamma ,\;b\rightarrow sg,\;K\rightarrow \pi \vartheta \overline{\vartheta }%
.\;$In particular, for the model type II a lower bound of $%
m_{H^{+}}\geqslant 500\;$GeV\ at 95\% CL is reported based on NLO analysis
of the $b\rightarrow s\gamma \;$data \cite{Gambino}. Further, as it was
mentioned above constraints on the $m_{H}-m_{H^{+}}\;$plane are obtained
based on $\Delta \rho \;$and on the $m_{H^{+}}-\tan \beta $ plane from $%
R_{b}\;$and $A_{b}\;$\cite{Logan, Larios}.

The top and charged Higgs phenomenologies are intimately related. Since the
present upper bound of $m_{H^{+}}\geq 500$GeV$\;$for model II seems to show
that $m_{H^{+}}>m_{t}+m_{b}\;$we shall examine this assumption first. In\
that case two important scenarios emerge. In the first, $%
m_{H^{+}}<m_{W}+m_{h^{0}}\;$so\ that $H^{+}\rightarrow W^{+}h^{0}\;$is
forbidden, thus $H^{+}\rightarrow t\overline{b}\;$dominates fermion modes.
But if $H^{+}\rightarrow W^{+}h^{0}\;$is allowed there is a competition. We
get

\begin{equation*}
\frac{BR\left( H^{+}\rightarrow W^{+}h^{0}\right) }{BR\left(
H^{+}\rightarrow t\overline{b}\right) }=\frac{2\cos ^{2}\left( \beta -\alpha
\right) p_{W}^{2}m_{H^{+}}^{2}}{3p_{\overline{b}}M}
\end{equation*}
\begin{equation*}
M\equiv \left[ \left( m_{t}^{2}\cot ^{2}\beta +m_{b}^{2}\tan ^{2}\beta
\right) \left( m_{H^{+}}^{2}-m_{t}^{2}-m_{b}^{2}\right) -4m_{t}^{2}m_{b}^{2}%
\right]
\end{equation*}
for $H^{0}\;$we replace $\cos \left( \beta -\alpha \right) \rightarrow \sin
\left( \beta -\alpha \right) .\;$So for large $m_{H^{+}},\;H^{+}\rightarrow
W^{+}h^{0}\;$could be important or even dominant for $\cos \left( \beta
-\alpha \right) \;$not too small, because of the availability of
longitudinal $W^{\prime }$s,\ if\ we assume saturation of $h^{0}\;$i.e. $%
\cos ^{2}\left( \beta -\alpha \right) =1,\;$we see that for large $%
m_{H^{+}}\;$the mechanism $H^{+}\rightarrow W^{+}h^{0}\;$would be dominant.
However for masses accesible to the LEP II\ $H^{+}\rightarrow W^{+}h^{0}\;$%
is smaller than $H^{+}\rightarrow t\overline{b}.\;$Additionally,$\;$since
the condition $m_{t}<m_{H^{+}}+m_{b}$, is obviously accomplished and taking
into account that $t\rightarrow W^{+}b\;$is the dominant top decay, the off
shell contribution of the charged Higgs is very small even for light
Higgses; in that case it could be interesting to examine branchings of the
type $BR\left( t\rightarrow H^{\ast +}b\rightarrow \tau ^{+}\vartheta _{\tau
}b\right) $.\ On the other hand, if $m_{H^{+}}<m_{t}+m_{b}\;$and $%
m_{H^{+}}<m_{W}+m_{h^{0}}\;$(as it is still possible in models I and III)
only $H^{+}\rightarrow \tau ^{+}\vartheta _{\tau }\;$and $H^{+}\rightarrow c%
\overline{s}\;$are relevant. So we examine the ratio of their branchings, in
model I this ratio is independent of $\tan \beta $, while in model type II
it yields 
\begin{equation*}
\frac{BR\left( H^{+}\rightarrow \tau ^{+}\vartheta _{\tau }\right) }{%
BR\left( H^{+}\rightarrow c\overline{s}\right) }=\frac{p_{\tau }}{3p_{c}}%
\frac{m_{\tau }^{2}\tan ^{2}\beta \cos ^{2}\theta _{c}\left(
m_{H^{+}}^{2}-m_{\tau }^{2}\right) }{K}
\end{equation*}
\begin{equation*}
K\equiv \left( m_{s}^{2}\tan ^{2}\beta +m_{c}^{2}\cot ^{2}\beta \right)
+\left( m_{H^{+}}^{2}-m_{c}^{2}-m_{s}^{2}\right) -4m_{c}^{2}m_{s}^{2}\tan
\beta \cot \beta
\end{equation*}
in model I it is $\sim 30\%\;$and independent of $\tan \beta $. We should
remember that small values of $\tan \beta \;$with small values of $%
m_{H^{+}}\;$are probably excluded by $B_{d}^{0}-\overline{B}_{d}^{0}\;$%
mixing. Now, if $t\rightarrow H^{+}b\;$is allowed, it is very likely that $%
H^{+}$ mass is sufficiently large to produce $H^{+}\rightarrow W^{+}h^{0}.\;$%
Since $H^{+}\rightarrow t\overline{b}\;$is not allowed then $%
H^{+}\rightarrow W^{+}h^{0}\;$will be dominant because of the $%
H^{+}W^{-}h^{0}\;$coupling ($\left( g/2\right) \cos \left( \beta -\alpha
\right) $), as long as it is not very supressed. So a possible scenario for $%
H^{-}\;$discovery could be $t\overline{t}\;$production followed by the
conventional $t\rightarrow W^{+}b\;$and $\overline{t}\rightarrow H^{-}%
\overline{b}\rightarrow W^{-}h^{0}\overline{b}.\;$Since the $h^{0}\;$is most
likely to decay to $b\overline{b},\;$we would have to distinguish between a $%
W^{+}b\overline{b}b\;$final state from a $W^{+}b\;$final state.

Furthermore, the loop induced $H^{+}\rightarrow W^{+}Z$ decay, though highly
suppress in most of the parameter space, can be significantly enhanced owing
to non-decoupling effects of Heavy Higgs bosons on the $H^{\pm }W^{\mp }Z$
vertices. As we explained in Sec. (\ref{kin interac}), these vertices are
forbidden at tree level in multi-Higgs doublet models and in particular in
the 2HDM because of the isospin symmetry, but might arise at loop levels due
to the breaking of the isospin symmetry through the loop particles. The
quark loop contributions for the $H^{\pm }W^{\mp }Z$ couplings have been
studied in \cite{Mendez}. The full one loop calculation in the 2HDM was
carried out in \cite{KanemuHWZ}, finding that the inclusion of
non-decoupling heavy Higgs modes with large mass splitting between $H^{+}$
and $A^{0}$ could yield a substantial enhancement for the decay width $%
H^{+}\rightarrow W^{+}Z$ due to the strong breakdown of the custodial $%
SU\left( 2\right) _{V}$ invariance \cite{custodial} in the Higgs sector.
Ref. \cite{KanemuHWZ}, found that the branching ratio for this decay could
reach values up to $10^{-2}\sim 10^{-1}$ for $m_{H^{+}}=300$ GeV, giving a
detectable mode at LHC or future $e^{+}e^{-}$ colliders. On the other hand,
Diaz-Cruz \emph{et. al.} \cite{Diaz-CruzHWZ} studied the same decay with an
effective Lagrangian approach, finding a branching ratio enhancement up to
about $10^{-1}$. Additionally, Ref. \cite{Diaz-CruzHWZ} also studies the
rare decays $H^{+}\rightarrow W^{+}\gamma $, and$\;H^{+}\rightarrow
W^{+}h^{0}$ with an effective Lagrangian approach, finding that they can
have ratios of order $10^{-5}$ and $O\left( 1\right) $ respectively.
Moreover, the relative behavior among the three decays $H^{+}\rightarrow
W^{+}\left( \gamma ,Z,h^{0}\right) $ could give clues about the underlying
Higgs structure \cite{Diaz-CruzHWZ}.

Since detection of a charged Higgs would be an unambiguous signal of physics
beyond the SM, many estimations of the potential of either hadron or
electron positron colliders to discover such Higgs boson and its properties
have been carried out. The production of charged Higgs at the Tevatron and
the LHC proceeds via the partonic processes: $gb\rightarrow tH^{+}\;$and $%
gg\rightarrow tbH^{+}\;$\cite{Plehn}. Other partonic processes could
generate pair production of charged Higgs bosons: $gg(q\overline{q}%
)\rightarrow H^{+}H^{-}$\cite{Eichten} and $gg(b\overline{b})\rightarrow
H^{\pm }W^{\mp }$ \cite{Dicus}, notwithstanding $H^{\pm }W^{\mp }\;$%
production lead to a much smaller rate at Hadron colliders. At\ $%
e^{+}e^{-}\; $colliders, the main source is $e^{+}e^{-}\rightarrow \gamma
^{\ast },Z^{\ast }\rightarrow H^{+}H^{-}.\;$If the collider energy is not
enough for pair production the associated $H^{\pm }W^{\mp }\;$production
becomes dominant if kinematically accesible. This latter possibility has
been studied in Ref. \cite{Brein}, where $H^{\pm }W^{\mp }\;$production is
analized in both hadron and $e^{+}e^{-}\;$colliders in the framework of the
MSSM and the general 2HDM. At hadron colliders $H^{\pm }W^{\mp }$ final
states arise via $gg\;$fusion and $b\overline{b}\;$annihilation, Ref. \cite
{Brein} found that hadronic cross section in the general 2HDM type II, can
be much larger than in the MSSM (about a factor 500-1000!), thus, its
discovery at hadron colliders could in addition provide a distinctive
signature of the Higgs sector it belongs to. In contrast, for $e^{+}e^{-}\;$%
colliders the MSSM signals are considerably higher than for the 2HDM type II
(up to two orders of magnitude).

On the other hand, Kanemura \emph{et. al.} \cite{Kanemura Moretti}, showed
that there are only two channels that offers an opportunity for single $%
H^{\pm }\;$detection at $e^{+}e^{-}\;$colliders for $m_{H^{\pm }}\gtrsim 
\sqrt{s}/2,\;$when $\sqrt{s}=500\;$GeV, (1) the one loop $H^{\pm }W^{\mp }\;$%
production discussed above, and (2) the tree level scattering $%
e^{+}e^{-}\rightarrow \tau ^{-}\overline{\nu }_{\tau }H^{+},\tau ^{+}\nu
_{\tau }H^{-}$. The latter is relevant in the large $\tan \beta \;$region
while the former is relevant in the small $\tan \beta \;$regime. Since the
rate of process (1) is rather poor \cite{Kanemura Moretti}, Ref. \cite
{Moretti} resort to the dominant decay channel $H^{+}\rightarrow t\overline{b%
}$. Such process has as the main background source, the SM process $%
e^{+}e^{-}\rightarrow t\overline{t}\rightarrow \tau ^{-}\overline{\nu }%
_{\tau }t\overline{b}$. It was found that for $1\;$and $5\;$ab$^{-1}\;$of
accumulated luminosity, neither evidence ($\gtrsim 3\sigma $) nor discovery (%
$\gtrsim 5\sigma $)\ of charged Higgs bosons is possible if $m_{H^{\pm
}}\gtrsim \sqrt{s}/2$,\ whereas if $m_{H^{\pm }}\lesssim \sqrt{s}/2\;$the
signal should be easily observed.

Further, many of the technics to look for the SM Higgs boson, can be
extrapolated to the exploration for neutral Higgs bosons of extended Higgs
sectors, if a neutral CP even scalar were discovered a precision test of its
couplings to fermions and vector bosons, as well as its self couplings,
could give us information on the underlying Higgs structure. Linear
colliders (LC) will explore the Higgs structure in case of discovery by
precise measurements of the Higgs couplings to fermions and vector bosons.
However, the structure of the Higgs potential can only be revealed by
measuring the Higgs self couplings. The trilinear Higgs coupling $\lambda
_{h^{0}h^{0}h^{0}}\;$can be directly measured from $e^{+}e^{-}\rightarrow
Z^{\ast }\rightarrow Zh^{0}h^{0}\;$and $e^{+}e^{-}\rightarrow W^{+\ast
}W^{-\ast }\nu \overline{\nu }\rightarrow h^{0}h^{0}\nu \overline{\nu }$,
when the Higgs boson is light \cite{Djouadi1}. Ref. \cite{Battaglia} shows
that at $e^{+}e^{-}\;$colliders with the energy of $500\;$GeV (3TeV) and the
integrated luminosity of $1\;ab^{-1}\left( 5\;ab^{-1}\right) $,\ $\lambda
_{h^{0}h^{0}h^{0}}\;$can be measured by about 20\% (7\%)\ accuracy for $%
m_{h}\sim 120\;$GeV. It worths to remark that even in the decoupling limit
in which the couplings of $h^{0}\;$to fermions and vector bosons are
SM-like, the one loop corrections to the trilinear self coupling $\lambda
_{h^{0}h^{0}h^{0}}\;$can undergo a significant deviation from the SM
behavior in both the MSSM and the general 2HDM \cite{Shingo}. In particular,
they find that quantum corrections can be of the order of 100\% in the
general 2HDM due to the quartic mass terms of heavier Higgses. In the case
of the MSSM they find that the deviation can exceed 5\% in light stop
scenarios.

The $\tau \overline{\tau }\;$decay mode \cite{Richter} is a promising
discovery channel for $A^{0}\;$and $H^{0}\;$at the CERN LHC for large $\tan
\beta $. On the other hand, the potential of the muon pair channel for MSSM
Higgs bosons was studied in \cite{Richter, Stepanov}. Ref. \cite{Kao} focus
on the potentiality to look for neutral Higgs bosons of the MSSM via $%
pp\rightarrow b\overline{b}H\rightarrow b\overline{b}\mu \overline{\mu }+X\;$%
at the LHC. He estimated that such process could serve to discover $A^{0}\;$%
and $H^{0}\;$at the LHC with an integrated luminosity of 30 fb$^{-1}\;$if $%
m_{A^{0}}\lesssim 300\;$GeV. At a higher luminosity of 300 fb$^{-1}$, the
discovery reach is not expanded much. In addition, the inclusive final state 
$H\rightarrow \mu \overline{\mu }$ could allow the discovery of $A^{0}\;$and 
$H^{0}$ at the LHC with an integrated luminosity of 30 fb$^{-1}\;$if $%
m_{A}\lesssim 450\;$GeV. At a higher luminosity of 300 fb$^{-1}$, the
discovery region in $m_{A^{0}}$ increases significantly to $m_{A}\lesssim
650\;$GeV for $\tan \beta =50$.

In this section we have examined the most hopeful channels for possible
direct and indirect detection of the 2HDM Higgs sector. However, the
importance of this channels depend strongly on the specific scenario, in
what follows we describe some interesting scenarios of the general 2HDM.

\subsection{The fermiophobic limit\label{the fermiophobic limit}}

If we take into account the relative Yukawa couplings given in Eqs. (\ref
{relat SM1}) we can see that in model I the heaviest (lightest) CP even
Higgs becomes totally fermiophobic (i.e. all $hf\overline{f}\;$couplings
vanish) if $\alpha =0\left( \pi /2\right) $. In contrast, model type II does
not exhibit a totally fermiophobic limit for any Higgs boson, see Eqs. (\ref
{relat SM2}); for instance, if $\alpha =\pi /2\;$then $h^{0}\left(
H^{0}\right) \;$become fermiophobic to fermions of up(down)-type, while
couplings to the down(up)-type fermions are maximal, the opposite occurs
when $\alpha =0.\;$On the other hand, in model III, there is still a
possible fermiophobic scenario, nevertheless in order for a certain Higgs
boson to be totally fermiophobic, a non natural fine tuning must be
accomplished\footnote{%
There are other scenarios with fermiophobic Higgs bosons such as the Higgs
triplet model of Refs. \cite{Higgstriplet}, and some models with
extradimensions \cite{lightHiggsSher}. Ref. \cite{lightHiggsSher} shows that
the latter can exhibit a more natural fermiophobic limit with less fine
tuning.}; see details in section (\ref{Yuk III}).

Consequently, only the model type I provides a natural totally fermiophobic
framework. The fermiophobic limit for $h^{0}$ $\left( \alpha =\pi /2\right)
\;$in the model I can be obtained in the potential $V_{A}^{\prime }\;$in two
ways: $\lambda _{+}=0,\;$or $v_{1}=0\;;\;$the latter assumption leads to a
massless $h^{0}.\;$In potential $V_{B}\;$there is only one way: $%
2v_{1}v_{2}\lambda _{+}=\frac{1}{2}\mu _{3}^{2}.\;$We have assumed $%
v_{1}<v_{2}\;$but of course we can interchange the role of each VEV. Despite 
$h^{0}\;$becomes totally fermiophobic by setting $\alpha =\pi /2;\;h^{0}\;$%
can still decay into fermions by the channels $h^{0}\rightarrow W^{\ast
}W\left( Z^{\ast }Z\right) \rightarrow 2\overline{f}f\;$or by means of
decays of the type $h^{0}\rightarrow \overline{f}f\;$through scalar and
vector boson loops; the dominant fermionic decay in the fermiophobic limit
is $h^{0}\rightarrow b\overline{b}.$

If we additionally assume $\alpha -\beta \equiv \delta =0\;$then $h^{0}\;$%
becomes bosophobic and ghostphobic as well. Besides, in this scenario $%
A^{0},H^{\pm }\;$are also fermiophobic. In that case, $h^{0}\;$needs another
scalar particle to be able to decay. Moreover, in the 2HDM type I with these
assumptions, $H^{0}\;$acquires the couplings of the SM Higgs. In contrast,
the signature of $h^{0}\;$is model dependent.

Br\"{u}cher and Santos \cite{fermiophobic} shows that in the fermiophobic
limit the most important three level decays for $h^{0}\;$are 
\begin{equation*}
h^{0}\rightarrow WW,ZZ,ZA,WH,AA,H^{+}H^{-}
\end{equation*}
and the most important one loop decays are 
\begin{equation*}
h^{0}\rightarrow \gamma \gamma ,Z\gamma ,b\overline{b}
\end{equation*}
and decays into fermions via virtual vector bosons. The dominance of any of
them depend on the parameters of the model, especially on $\delta \equiv
\alpha -\beta =\pi /2-\beta \;$see details in Ref. \cite{fermiophobic}. The
fact to emphasize is that the branching fractions depend strongly on the
potential chosen and that $h^{0}\rightarrow b\overline{b}\;$is the fermionic
dominant decay. In addition, the decay $h^{0}\rightarrow \gamma \gamma \;$is
always important for light $h^{0}\;$(and sometimes for intermediate $h^{0}$)$%
,\;$but for $h^{0}\;$sufficiently heavy it is suppressed.

As for the pseudoscalar, below the $Zh^{0}\;$and the $W^{\pm }H^{\mp }\;$%
thresholds\ $A^{0}$ decays mainly into fermions, but if $\delta \rightarrow
0,\;A^{0}\;$becomes stable or decay outside the detector; thus the only
signature would be missing energy or momentum. Over the threshold of $%
Zh^{0}\;$or $W^{\pm }H^{\mp }\;$the pseudoscalar Higgs boson$\;$decays
inside the detector. Finally, the signature of $A^{0}\;$is the same for both 
$V_{A}^{\prime }\;$and $V_{B}.$

Now for the charged Higgs boson, if $m_{H^{\pm }}$ is below $m_{W^{\pm }}\;$%
the decays are fermionic and independent of $\delta ,\;$but if $\delta \;$%
decreases the branching is unchanged while the total decay width decreases
(so $H^{\pm }\;$becomes stable), if $m_{H^{\pm }}\geq m_{W^{\pm }}\;$then\ $%
H^{\pm }\rightarrow W^{\pm }\gamma \;$becomes dominant for tiny $\delta \;$%
but for large values the signature is again fermionic, but as soon as $%
H^{\pm }\rightarrow W^{\pm }A^{0}(h^{0})\;$is possible the sum of both
becomes dominant, except in the case in which $m_{H^{\pm }}\geq
m_{t}+m_{b}\; $in whose case $H^{+}\rightarrow t\overline{b}\;$will be
dominant. In the case of potential $V_{B}\;$below the $W^{\pm }\gamma \;$and
above the $W^{\pm }h^{0}\;$or\ $W^{\pm }A^{0}\;$thresholds, the situation is
the same as in the potential $V_{A}.\;$But in the regime of $H^{\pm
}\rightarrow W^{\pm }\gamma \;$dominance, the situation depends strongly on
the choice of parameters; for instance, when $\delta \rightarrow 0\;$the
unitarity constraint for $V_{B}\;$given by Eq. (\ref{unitarity VB}) predicts
a degeneracy between $m_{A^{0}}\;$and $m_{h^{0}}$; such degeneracy$\;$can
suppress $H^{\pm }\rightarrow W^{\pm }\gamma \;$to a few percent respect to
the fermionic decays even above the $m_{H^{\pm }}\geq m_{W^{\pm }}\;$%
threshold unlike the case of $V_{A}^{\prime }.$

Finally, $H^{0}\;$is not fermiophobic in this scenario so the one loop
decays are not important, the dominant decay is obviously $H^{0}\rightarrow b%
\overline{b}\;$and $H^{0}\rightarrow WW\;$above the $WW\;$threshold like in
SM, this situation is the same in both potentials. A difference in signature
can only be detected by purely scalar decay modes.

From the experimental point of view, LEP and Tevatron have looked for
fermiophobic Higgs bosons. The main channel at LEP has been $%
e^{+}e^{-}\rightarrow h^{0}Z$, $h^{0}\rightarrow \gamma \gamma $ obtaining a
lower bound of about $100\;$GeV \cite{LEPfermiophobic}, the channels $%
e^{+}e^{-}\rightarrow h^{0}A^{0}$, $h^{0}\rightarrow \gamma \gamma $ and $%
h^{0}\rightarrow WW^{\ast }$ have also been considered \cite{LEPfermiophobic}%
. As for Tevatron run I, the mechanism used is $qq^{\prime }\rightarrow
V^{\ast }\rightarrow h^{0}V$, $h^{0}\rightarrow \gamma \gamma $ and a lower
limit of about 80 GeV was obtained \cite{TeVfermiophobic}. Nevertheless,
such limits were obtained assuming that the couplings of the type $%
g_{h^{0}VV}$ are of the same order as the ones in SM. Notwithstanding, in
the limit of fermiophobia we can see that the relative couplings $\chi
_{V}^{h^{0}}$ behave like 
\begin{equation*}
\chi _{V}^{h^{0}}\sim \cos ^{2}\beta =\frac{1}{1+\tan ^{2}\beta }
\end{equation*}
therefore, in scenarios with large values of $\tan \beta $ the couplings $%
g_{h^{0}VV}$ might be highly suppress and the bounds described above would
not be valid. If it is the case, a light Higgs boson ($h^{0}<<100$ GeV)
could have eluded the LEP and Tevatron constraints and may also escape
detection at Tevatron run II \cite{fermiophobicMDiaz}. Owing to it, Akeroyd
and M. A. Diaz \cite{fermiophobicMDiaz}, have proposed new production
mechanisms at Tevatron run II, they are: (1) $q\overline{q}\rightarrow
\gamma ^{\ast },Z^{\ast }\rightarrow H^{+}H^{-}$; (2) $qq^{\prime
}\rightarrow W^{\ast }\rightarrow H^{\pm }h^{0}$; (3) $qq^{\prime
}\rightarrow W^{\ast }\rightarrow H^{\pm }A^{0}$; (4) $q\overline{q}%
\rightarrow Z^{\ast }\rightarrow A^{0}h^{0}.$

The subsequent decays $H^{\pm }\rightarrow h^{0}W^{\ast }$, $%
A^{0}\rightarrow Z^{\ast }h^{0},\;h^{0}\rightarrow \gamma \gamma $ would
give rise to $\gamma \gamma \gamma \gamma \;$final states very easy to
measure accurately. In the fermiophobic scenarios, these new channels are
effective even when $h^{0}VV$ are very suppressed \cite{fermiophobicMDiaz}.
These mechanisms are complementary to the traditional one $qq^{\prime
}\rightarrow Vh^{0}$ which could suffer a strong suppression in realistic
models. Note that gluon gluon fussion is not relevant in this framework and
that the four additional channels proposed are based on the couplings $%
g_{HHV}$.

\subsection{The decoupling limit\label{the decoupling limit}}

The discussion of the decoupling limit will be given based on the following
parametrization of the most general CP invariant potential\cite{decoupling} 
\begin{eqnarray*}
V &=&m_{11}^{2}\Phi _{1}^{\dagger }\Phi _{1}+m_{22}^{2}\Phi _{2}^{\dagger
}\Phi _{2}-\left[ m_{12}^{2}\Phi _{1}^{\dagger }\Phi _{2}+h.c.\right] +\frac{%
1}{2}\lambda _{1}\left( \Phi _{1}^{\dagger }\Phi _{1}\right) ^{2} \\
&&+\frac{1}{2}\lambda _{2}\left( \Phi _{2}^{\dagger }\Phi _{2}\right)
^{2}+\lambda _{3}\left( \Phi _{1}^{\dagger }\Phi _{1}\right) \left( \Phi
_{2}^{\dagger }\Phi _{2}\right) +\lambda _{4}\left( \Phi _{1}^{\dagger }\Phi
_{2}\right) \left( \Phi _{2}^{\dagger }\Phi _{1}\right) \\
&&+\left\{ \frac{1}{2}\lambda _{5}\left( \Phi _{1}^{\dagger }\Phi
_{2}\right) ^{2}+\left[ \lambda _{6}\left( \Phi _{1}^{\dagger }\Phi
_{1}\right) +\lambda _{7}\left( \Phi _{2}^{\dagger }\Phi _{2}\right) \right]
\Phi _{1}^{\dagger }\Phi _{2}+h.c\right\} \;.
\end{eqnarray*}

In principle $m_{12}^{2},\;\lambda _{5},\;\lambda _{6}\;$and $\lambda _{7}\;$%
can be complex, but for simplicity all of them are chosen real. The
exploration of the decoupling limit is inspired in the idea that\ the
standard model could be an effective theory embedded in a more fundamental
structure characterized by a scale $\Lambda \;$which is supposed to be much
larger than the ElectroWeak Symmetry Breaking (EWSB) scale, $v=246GeV.\;$In
particular, many models end up in a low energy model with a non-minimal
Higgs sector, as is the case of the MSSM whose low energy Higgs sector
correspond to that of a 2HDM.

Based on this idea, there are two important scenarios to differenciate in
the 2HDM, in the first one there is not a low energy theory containing only
one light Higgs boson i.e. no decoupling limit exists. In the second case,
the lightest CP even Higgs boson is much lighter than the other Higgs bosons
whose masses are of the order of a new scale $\Lambda _{2HDM}.\;$In
particular, if $\Lambda _{2HDM}>>v\;$and all dimensionless Higgs
self-coupling parameters accomplish the condition $\lambda _{i}\lesssim
O\left( 1\right) ,\;$then all couplings of the light Higgs are SM-like,\ the
deviations lie on the order of $O\left( v^{2}/\Lambda _{2HDM}^{2}\right) ,\;$%
this is called the decoupling limit. What we do in an effective theory is to
integrate out the heavy modes; if we believe that the SM is an effective
theory of a 2HDM then one of the doublets should be integrated out. Since
only one Higgs boson must exist in the low energy theory then the mass of
the light Higgs should be of the order of EWSB scale\ while the heavy
scalars masses should be of the order of $\Lambda _{2HDM}$. And as the low
energy theory only contain one doublet, then $h^{0}\;$should be
indistinguishable from the SM Higgs. For some choices of the scalar
potential, no decoupling limit exists. Haber and Gunion \cite{decoupling}
has established that \emph{no decoupling limit exists if and only if }$%
\lambda _{6}=\lambda _{7}=0\;$\emph{in the basis where }$m_{12}^{2}=0.\;$%
Thus, the absence of a decoupling limit implies a discrete symmetry for the
scalar potential (this symmetry could be hidden in other bases). The low
energy effective theory reduces the potential to the potential of the SM
with all the SM constraints.

When we assume that $m_{h^{0}}<<m_{H^{0}},m_{A^{0}},m_{H^{\pm }}\;$one can
simply impose tree level unitarity constraints to the low energy effective
scalar theory i.e. to the SM in this case. However, at one loop, the heavier
scalars can contribute by virtual exchanges so the restrictions on the self
couplings involves all Higgs bosons, to maintain unitarity and
perturbativity the splitting of the squared masses among the heavier states
should be of order $O\left( v^{2}\right) \;$so they should be rather
degenerate \cite{decoupling}.

In MSSM the decoupling limit is achieved by setting $m_{A^{0}}>>m_{Z}.\;$%
However, one loop effects mediated by loops of SUSY particles can generate a
deviation from the SM expectations (e.g. light squarks contributions to $%
h^{0}\rightarrow \gamma \gamma ,gg$),$\;$so to avoid significant SUSY
contributions we should assume heavy SUSY particles (say of order 1TeV). If
the latter condition is fulfilled, $h^{0}\;$is SM-like even at one loop
level. The leading one loop radiative correction to $g_{h^{0}b\overline{b}%
}\; $is of order $O\left( m_{Z}^{2}\tan \beta /\left( m_{A^{0}}^{2}\right)
\right) \;$and formally decouples in the regime in which $%
m_{A^{0}}^{2}>>m_{Z}^{2}\tan \beta \;$which is more demanding in the case of
large $\tan \beta \;$regime. This behavior is called delay decoupling in
Ref. \cite{delaydecoupling}, this phenomenon can also occur in the general
2HDM, with tree level couplings.

In general, the amplitudes of all loop induced processes which involve $%
h^{0}\;$and SM particles as external states approaches the SM values
whenever $\lambda _{i}\lesssim O\left( 1\right) \;$and $m_{A^{0}}\rightarrow
\infty ,\;$this is guaranteed by the applequist-Carazone decoupling theorem.

After describing the framework of the decoupling limit, phenomenological
consequences must be discussed. For instance, in the most general Yukawa
Lagrangian (model type III) FCNC are generated through the mixing matrices $%
\xi _{ij}^{U,D}$and $\eta _{ij}^{U,D}.\;$These matrices are in general
complex and non-diagonal; thus, they could be also a source for CP violating
couplings between the neutral Higgs bosons and fermions. Notwithstanding, in
the decoupling limit (which correspond to the condition $\cos \left( \beta
-\alpha \right) =0\;$in the Yukawa Lagrangian),$\;$both non-diagonal and CP
violating couplings of $h^{0}\;$vanish (but not for $H^{0},A^{0}$)\ and the $%
h^{0}\;$couplings to fermions reduces to the SM ones. However, FCNC and CP
violating processes mediated by $H^{0}\;$or $A^{0}\;$are supressed by their
square masses (relative to$\;v$), due to the propagator supression, and
since $\cos \left( \beta -\alpha \right) \simeq O\left(
v^{2}/m_{A}^{2}\right) \;$the suppression factor is roughly the same for $%
h^{0},H^{0},A^{0}.\;$Therefore, the decoupling limit is a natural mechanism
to suppress Higgs mediated FCNC and to suppress Higgs mediated CP violating
couplings in the most general 2HDM. Additionally, all vertices containing at
least one vector boson and exactly one of the heavy Higgs states are
proportional to $\cos \left( \beta -\alpha \right) \;$and hence vanish in
the decoupling limit.

Nevertheless, Ref. \cite{decoupling} shows that there are some alternative
scenarios in which $\cos \left( \beta -a\right) =0$,$\;$but all Higgs bosons
masses are of $O\left( v\right) ,\;$although $h^{0}\;$is SM-like, there is
not an effective theory with only one light Higgs. Radiative corrections
introduce in this case significant deviation respect to the tree-level $%
h^{0}\;$behavior.

The scenario in which $\sin \left( \beta -\alpha \right) =0$,$\;$is usually
called the non-decoupling regime, since $h^{0}\;$couplings$\;$deviate
maximally from SM, while $H^{0}\;$ones are SM-like.

Now the following question arise, if only one CP even Higgs boson were
discovered, how could we distinguish whether this scalar belongs to the
minimal Higgs sector of the SM or to a non minimal Higgs sector?. Far from
the decoupling limit, the deviation from the coupling of the Higgs boson to
fermions and gauge bosons (respect to the ones in SM) could provide the
answer. However, in the decoupling regime couplings of this Higgs boson are
similar to the SM ones and the discrimination becomes problematic; in that
case there are two alternatives: (1) Direct production of heavy states and
(2) High precision measurements of the couplings of the light Higgs to look
even for tiny deviations.

As for the first alternative, at the LHC $gg\rightarrow A^{0},H^{0}\;$and $%
gb\rightarrow H^{-}t$ signatures could be suppressed in the decoupling
scenario. In addition, the branching $H^{0}\rightarrow ZZ\;$which is the
gold plated in SM is almost absent in this regime. An interesting method
could be from $gg\rightarrow Q\overline{Q}^{\prime }\left(
H^{0},A^{0},H^{\pm }\right) \;$where $Q\;$is a heavy quark ($b\;$or $t$)\
and the Higgs decays into a Heavy quark pair because these couplings are not
suppressed in the decoupling limit. In $e^{+}e^{-}\;$colliders with
sufficiently high energy the heavy states can be produced via$%
\;e^{+}e^{-}\rightarrow H^{+}H^{-}\;$or $e^{+}e^{-}\rightarrow H^{0}A^{0}\;$%
without a rate suppression,$\;$the$\;$mechanism $e^{+}e^{-}\rightarrow
Z\rightarrow h^{0}A^{0}\;$discussed in section (\ref{phenomenological
constraints}) is supressed by $\cos \left( \beta -\alpha \right) $. These
production mechanisms requires very high $\sqrt{s}\;$since they involve a
pair of heavy states. Another possibility is to consider the production of
one heavy state in association with light states (the light Higgs and the
top quark are considered light states), however this option seems to be
hopeless\cite{decoupling}.

On the other hand, for the second alternative, one of the most promising
channels seems to be $Br(h^{0}\rightarrow \overline{b}b).\;$Additionally, $%
\gamma \gamma \;$colliders could play an important role by either measuring
the $h^{0}\rightarrow \gamma \gamma \;$coupling with sufficient precision or
directly producing $A^{0}\;$and/or\ $H^{0}\;$in $\gamma \gamma \;$fusion,\
only the latter is viable in the decoupling regime. Finally, as it was
pointed out by Kanemura \emph{et. al.} \cite{Shingo}, another promising
coupling to distinghish the underlying Higgs structure could be the
trilinear self coupling $\lambda _{h^{0}h^{0}h^{0}}\;$which has been
estimated to be very sensitive to the introduction of new physics in both
the 2HDM type II and the MSSM, see discussion in Sec. (\ref{phenomenological
constraints}).

\subsection{Scenarios with a light Higgs boson\label{light Higgs}}

As we saw above, SM estimation of the muon anomalous magnetic moment differs
from the experimental measurement by about $3\sigma \;$deviations. In the SM
the Higgs contribution to $a_{\mu }\;$is negligible since it is proportional
to the factor $m_{\mu }^{2}/m_{h}^{2},\;$and the present SM Higgs mass limit
provided by LEP is$\;m_{h}\gtrsim 114.4\;$GeV. If we believe that an
extended Higgs sector provides the necessary enhancement to get a value of
the $\left( g-2\right) _{\mu }\;$compatible with experimental measurement,
we should examine possible factors to generate such enhancement. In a
general 2HDM with no FCNC, it might occur due to two factors. First, an
increment on the $h^{0}\mu ^{+}\mu ^{-}\;$coupling proportional to $\tan
\beta $, and second owing to the possibility of a light Higgs boson arising
from the suppression of the vertex $h^{0}ZZ\;$from which the LEP bound could
be avoided. The total enhancement at one loop level would be of the order of 
\cite{Haberg2} 
\begin{equation}
\frac{m_{\mu }^{2}}{m_{h^{0}}^{2}}\tan ^{2}\beta \ln \left( \frac{m_{\mu
}^{2}}{m_{h^{0}}^{2}}\right)  \label{enhancement}
\end{equation}
The couplings of $h^{0}ZZ\;$and $h^{0}A^{0}Z\;$for the 2HDM read

\begin{equation*}
g_{h^{0}ZZ}=\frac{gm_{Z}\sin \left( \beta -\alpha \right) }{\cos \theta _{W}}%
\;\;;\;g_{h^{0}AZ}=\frac{g\cos \left( \beta -\alpha \right) }{2\cos \theta
_{W}}
\end{equation*}
so we can evade the LEP constraint by assuming $\sin \left( \beta -\alpha
\right) \simeq 0,\;$i.e. a non decoupling scenario, in that case, a light CP
even scalar can be considered. Exploration of light scalar and pseudoscalar
modes has been studied by a variety of authors \cite{lightHiggsSher, Larios,
Haberg2, Krawczykg2}. In general, one light Higgs boson is necessary in the
2HDM type I and II, in order to get a sufficient value of $\Delta a_{\mu }\;$%
to be compatible with present experimental measurement. M. Krawczyk \cite
{Krawczykg2} explores the possibility of having a light scalar or
pseudoscalar Higgs boson in the framework of the 2HDM (II), considering only
the case in which the soft breaking term $\mu _{3}^{2}\;$vanishes.$\;$The
LEP data allows the existence of a light neutral Higgs in the 2HDM, but not
two of them could be light simultaneously \cite{Chankowski, Abbiendi,
Bambade, OPAL}, the constraints are given in a $\left(
m_{h^{0}},m_{A^{0}}\right) \;$plane\ and they indicate roughly that $%
m_{h^{0}}+m_{A^{0}}\gtrsim 90\;$GeV. On the other hand, the Bjorken process
gives an upper limit on the relative coupling $\chi _{V}^{h^{0}}$\ \cite
{Krawczykg2},$\;$for instance, it is found that at 95\% CL,$\;\chi
_{V}^{h^{0}}<<1$ for $m_{h^{0}}\lesssim 50$ GeV$.\;$Moreover, the Yukawa
couplings $\chi _{d}^{h^{0},A^{0}}\;$are constrained by low energy data
coming from $\Upsilon \rightarrow h^{0}\left( A^{0}\right) \gamma \;$\cite
{Keh, Prades} while LEP do it for masses $\gtrsim 4\;$GeV$.\;$Additionally,
the analysis of the decay $Z\rightarrow h^{0}\left( A^{0}\right) \gamma \;$%
at LEP \cite{Chankowski} gives both the upper and lower limits for $\left|
\chi _{d}\right| .$

Assuming that the light Higgs ($h^{0}\;$or $A^{0}$), is the only one
contributing at one loop to the muon anomalous magnetic moment, the one loop
diagram gives a positive (negative) contribution for $h^{0}\left(
A^{0}\right) .\;$However, calculation of the two loops diagram with a
charged Higgs in one of the loops produce a contribution that is positive
(negative) for $A^{0}\left( h^{0}\right) .\;$Taking into account both
contributions Ref. \cite{Krawczykg2} found that the scalar Higgs gives a
positive (negative) contribution for $m_{h^{0}}\;$below (above) 5 GeV, while
the pseudoscalar gives a positive (negative) contribution for $m_{h^{0}}\;$%
above (below) 3 GeV.

Constraints from $\left( g-2\right) _{\mu }\;$\cite{g2th},$\;$current upper
limits of LEP from the Yukawa processes \cite{OPAL}, lower limits from $%
Z\rightarrow h^{0}\left( A^{0}\right) \gamma \;$\cite{Chankowski}, upper
90\% CL limits from $\Upsilon \rightarrow h^{0}\left( A^{0}\right) \gamma \;$%
and upper limits from the Tevatron \cite{Roco} are combined to obtain
exclusion regions in $m_{h^{0}}-\tan \beta \;$and $m_{A^{0}}-\tan \beta \;$%
planes. The two loop analysis of \cite{Krawczykg2} shows that light scalars
are excluded at 95\% CL, but pseudoscalar masses between $\sim 25\;$GeV$\;$%
and\ $70\;$GeV\ with $25\lesssim \tan \beta \lesssim 115\;$are still allowed.

An alternative study of very light pseudoscalar scenarios (of the order of $%
0.2\;$GeV) has been carried out in Ref. \cite{Larios}, they show that in the
2HDM type I and II this very light CP-odd scalar can be compatible with the $%
\rho \;$parameter, Br$\left( b\rightarrow s\gamma \right)
,\;R_{b},\;A_{b},\;\left( g-2\right) _{\mu }\;Br\left( \Upsilon \rightarrow
A^{0}\gamma \right) ,\;$and direct searches via the Yukawa process at LEP.
For a mass of around $m_{A^{0}}\sim 0.2\;GeV,\;$they got that $\tan \beta
\sim 1\;$and that $A^{0}\;$behaves as a CP odd fermiophobic scalar, decaying
predominantly into a $\gamma \gamma \;$pair. The second of Refs. \cite
{Larios} found that a very light CP odd Scalar can be either allowed or
excluded depending on the statistic used for the Hadronic contributions for $%
a_{\mu }$.\ Therefore, a better stabilization of the Hadronic contributions
for $a_{\mu }^{SM}\;$is necessary to elucidate this point.

Recently, Ref. \cite{lightHiggsSher} showed that additional constraints on
light pseudoscalars belonging to a 2HDM are obtained from $K$ and $B$
decays. They showed that masses between 100 MeV and 200 MeV might evade such
stringent constraints owing to a possible cancellation in the decay
amplitude. In that case, the coupling of $A^{0}$ to quarks is strong and may
produce sufficient pseudoscalar states via photoproduction. This kind of
test could be performed by the Jefferson laboratory and Ref. \cite
{lightHiggsSher} shows the perspective for detection of this modes in this
laboratory.

\subsection{The 2HDM with FCNC\label{2HDM FCNC}}

\subsubsection{FCNC, general framework}

Flavor Changing Neutral Currents (FCNC) are processes highly suppressed by
some underlying principle still unknown, despite they seem not to violate
any fundamental law of nature. On the other hand, Standard Model (SM)
observables are compatible with experimental constraints on FCNC so far,
with the remarkable exception of neutrino oscillations \cite{Fukuda}. In the
case of the lepton sector, the high supression of FCNC suggested by direct
search is ``explained'' by the implementation of the Lepton Number
Conservation (LFC), a new symmetry that protects the phenomenology from
these dangerous processes. Notwithstanding, the observation of oscillation
of neutrinos \cite{Beshtoev} coming from the sun and the atmosphere, seem to
indicate the existence of such rare couplings. Neutrino oscillations can be
explained by introducing mass terms for these particles, in whose case the
mass eigenstates are different from the interaction eigenstates. It should
be emphasized that the oscillation phenomenon implies that the lepton family
number is violated, and such fact leads us in turn to consider the existence
of physics beyond the standard model (SM), because in SM neutrinos are
predicted to be massless and lepton flavor violating mechanisms are
basically absent. These considerations motivate the study of scenarios with
Lepton Flavor Violation (LFV).

The original motivation for the introduction of neutrino oscillations comes
from the first experiment designed to measure the flux of solar neutrinos 
\cite{Davis}, such measurement was several times smaller than the value
expected from the standard solar model, so Ref. \cite{Pontecorvo} suggested
the neutrino oscillation mechanism as a possible explanation of the neutrino
deficit problem. In addition, models of neutrino oscillations in matter \cite
{Mikheyev} arose to solve the neutrino deficit confirmed by SuperKamiokande 
\cite{Hirata}. Since then, further evidence of solar neutrino oscillations
has been found by SuperKamiokande and SNO \cite{Kameda}. Besides, this
phenomenon can be inferred from experiments with atmospheric neutrinos as
well \cite{Fukuda}.

On the other hand, since neutrino oscillations imply LFV in the neutral
lepton sector, it is generally expected to find out LFV processes involving
the charged lepton sector as well. Experiments to look directly for LFV have
been performed for many years, all with null results so far, those searches
for these processes have only provided some upper limits, some examples are $%
\mu -e\;$conversion in nuclei \cite{doh}, $\mu \rightarrow eee\;$\cite{bell}$%
,\;\mu \rightarrow e\gamma \;$\cite{bolton}$,\;$and $\mu ^{-}\rightarrow \nu
_{e}e^{-}\overline{\nu }_{\mu }$\cite{Okada}. In addition, the search for
LFV can also be made by analyzing semileptonic decays, upper bounds from LFV
meson decays have been estimated, some examples are$\;K_{L}^{0}\rightarrow
\mu ^{+}e^{-}$ \cite{arisaka}, and $K_{L}^{0}\left( K^{+}\right) \rightarrow
\pi ^{0}\left( \pi ^{+}\right) \mu ^{+}e^{-}$ \cite{plb}.

As for experimental perspectives, a muon muon collider could provide very
interesting new constraints on FCNC, for example $\mu \mu \rightarrow \mu
\tau (e\tau )\;$ mediated by Higgs exchange \cite{muon collider} which test
the mixing between the second and third generations. Additionally, the muon
collider could be a Higgs factory and it is well known that the Higgs sector
is crucial for FCNC.\ Several authors have studied the potentiality of $\mu
\mu \;$colliders as a Higgs factory and as a source of Higgs mediated FCNC 
\cite{muon collider}.

Other interesting perspectives consist of improving the sensitivity of
experiments that look for LFV processes. For example, in the case of $\mu
\rightarrow e\gamma $, the present upper bound of its branching ratio comes
from MEGA collaboration ($1.2\times 10^{-11}$) \cite{MEGAKit}. A new
experiment is under construction at PSI whose aim is to improve present
bound by about three orders of magnitude \cite{PSIKit}. On the other hand,
the SINDRUM II experiment at PSI provides the current upper bound of $\mu
-e\;$conversion branching ratio ($6.1\times 10^{-13}$) \cite{SINDRUMKit}.
The MECO experiment at BNL \cite{MECOKit} is planned to analyse such
conversion in aluminum nuclei with a sensitivity below $10^{-16}$. In
addition, the PRISM project \cite{PRISMKit} aims to improve upper limits for 
$\mu \rightarrow e\gamma $, and $\mu -e\;$conversion in nuclei, by one or
two orders of magnitude.

Another posible source to study LFV is by means of lepton flavor violating
decays of a Higgs, in Higgs factories. In particular, the Fermilab Tevatron
and the CERN Large Hadron Collider have the potentiality to study LFV Higgs
boson decays such as $h\rightarrow \tau \mu $ \cite{Kao, flavoredHiggs}.

From the theoretical point of view, the phenomenon of LFV has been widely
studied in different scenarios such as Two Higgs Doublet Models,
Supersymmetry, Grand Unification, effective theories, leptoquark models,
technicolor, Superstrings and SM extensions with heavy neutrinos \cite{nosno}%
. For instance, LFV processes in SU(5) SUSY models with right-handed
neutrinos have been examined based on recent results of neutrino oscillation
experiments \cite{Baek}. On the other hand, Ref. \cite{LFVMasina} discuss
how the supersymmetric SM with a see-saw mechanism could face the flavor
problem; in particular, current constraints on the leptonic soft
supersymmetry breaking terms and possible improvement on those constraints
are studied in this reference. Moreover, Grimus \emph{et. al.} \cite{Lavoura}
explores the generation of LFV by using a multi-Higgs doublet model with
additional right-handed neutrinos for each lepton generation, finding a
non-decoupling behavior of some LFV amplitudes respect to the right-handed
neutrino masses.

Further, Babu and Pakvasa \cite{BabuPakvasa}, studied the $\Delta L=2$
process $\mu ^{+}\rightarrow e^{+}\overline{\vartheta }_{e}\overline{%
\vartheta }_{i}$ ($i=e,\mu ,\tau $), in an effective Lagrangian approach
showing that it can explain the neutrino anomaly reported by the LSND
experiment. Ref. \cite{BabuPakvasa}, found two effective operators in SM
that lead to such decays and no other processes. These operators arise from
integrating out scalar fields with LFV interactions.

In addition, Cvetic \emph{et. al.}\cite{Cvetic}, studied lepton flavor
violation in tau decays induced by heavy Majorana neutrinos within two
models (1) the SM with additional right-handed heavy Majorana neutrinos \
which is a typical see-saw model; and (2) the SM with left-handed and
right-handed neutral singlets. The first of these models predicts very small
branching ratios of the LFV tau decays considered, in most of the parameter
space. Unlike the second of these models, which might show large enough
branching ratios for such decays to be tested in near future experiments.

Now, since there are interesting perspectives to improve the experimental
sensitivity for the LFV processes $\mu \rightarrow e\gamma $, and $\mu
\rightarrow e\;$conversion in nuclei. It might also be interesting to check
for theoretical predictions for them coming from different scenarios. Both
processes might occur in many models with LFV. Nevertheless, the simple
see-saw neutrino model does not induce experimentally observable rates for
the $\mu \rightarrow e\gamma \;$decay \cite{mue Kitano}. But in the case of
SUSY models the rate for this decay can be significant owing to the
production of LFV processes by means of one loop diagrams of sleptons.
Particularly interesting are SUSY models with $R-$parity broken because LFV
interactions could arise even at tree level \cite{SUSYRKit}. The branching
ratios of LFV processes has been calculated for a variety of models such as
SUSY-GUTs and SUSY models with right-handed neutrinos \cite{manyKit}. These
references show that the branching ratios for $\mu \rightarrow e\gamma $,
and $\mu -e$ conversion could get values near to the observable threshold in
forthcoming experiments.

The $\mu -e$ conversion rate in nuclei, has been calculated by a variety of
authors with a variety of approaches \cite{mue authors}. In particular,
Kitano \emph{et. al.} \cite{mue Kitano} have evaluated the $\mu -e\;$%
conversion by using the method of Czarnecki \emph{et. al. }\cite{mue authors}%
. These calculations were carried out for nuclei in a wide range of atomic
numbers with an effective theory approach. In general they found that the
conversion branching ratios grow along with the atomic number ($Z$) up to $%
Z\sim 30$, those branching ratios become the largest for $Z\sim 30-60$, and
decrease for heavy nuclei $Z\gtrsim 60$. Though this tendency is general,
the $Z\;$dependence on conversion rates is significantly different for
several LFV couplings. Therefore, the atomic number dependence of the
conversion branching ratio could be useful to discriminate among the
theoretical models with LFV. Additionally, the conversion rate also depends
on proton and neutron densities for each nucleus, Ref. \cite{mue Kitano}
uses several models for proton and neutron density distributions (see
appendix A on that reference), finding a reasonable agreement among them.

In general, many extensions of the SM permit FCNC at tree level. The
introduction of new representations of fermions different from doublets
produce them by means of the Z-coupling \cite{2}. In addition, they are
generated at tree level in the Yukawa sector by adding a second doublet to
the SM \cite{Wolf}. Such couplings also appear in SUSY theories with
R-parity broken \cite{R1}, because FCNC coming from R-parity violation
generates massive neutrinos and neutrino oscillations \cite{Kaustubh}. In
this work we shall concentrate on the FCNC arising from the general 2HDM.
For other scenarios with FCNC, I refer the reader to the literature \cite
{ZepedaCoti, 2, R1}.

On the other hand, scenarios with FCNC in the lepton sector automatically
generates FCNC in the quark sector as well, experimental bounds on FCNC in
the quark sector come from $\Delta F=2$ processes, rare B-decays, $%
Z\rightarrow \overline{b}b$ and the $\rho $-parameter \cite{Atwood}.
Reference \cite{Atwood} also explored the implications of FCNC at tree level
for $e^{+}e^{-}(\mu ^{+}\mu ^{-})\rightarrow t\overline{c}+\overline{t}c$, $%
t\rightarrow c\gamma (Z,g)$, $D^{0}-\overline{D}^{0}$ and $B_{s}^{0}-%
\overline{B}_{s}^{0}$. Moreover, there are other important processes
involving FCNC. For instance, the decay $B^{-}(D^{-})\rightarrow K^{-}\mu
^{+}\tau ^{-}\;$ which depends on $\mu -\tau \;$mixing and vanishes in the
SM.\ Another one is $B^{-}(D^{-})\rightarrow K^{-}\mu ^{+}e^{-}\;$ whose
form factors have been calculated in \cite{Sher91}, \cite{nosotros}.

As explained above, FCNC in the lepton sector are basically absent in the
SM. On the other hand, in the quark sector FCNC are very tiny because they
are prohibited at the tree level and are further suppressed at one loop by
the GIM mechanism \cite{GIM}. Consequently, SM provides a hopeless framework
for this rare processes since predictions from it, are by far out of the
scope of next generation colliders \cite{Atwood}. So detection of these kind
of events in near future experiments in either the lepton or quark sector,
would imply the existence of new Physics.

Since FCNC are strongly constrained by experimental data, several mechanism
to avoid potentially larger contribution to these exotic processes have been
developed. For instance, Glashow and Weinberg \cite{Glashow} proposed a
discrete symmetry in the Two Higgs Doublet Model (2HDM) which forbids the
couplings that generate such rare decays, hence they do not appear at tree
level. This discrete symmetry led to the so called 2HDM type I and II
discussed in section (\ref{Yuk 2HDM}). Another possibility is to consider
heavy exchange of scalar or pseudoscalar Higgs fields \cite{Sher91} or by
cancellation of large contributions with opposite sign. Another mechanism
was proposed by Cheng and Sher arguing that a natural value for the FC
couplings from different families should be of the order of the geometric
average of their Yukawa couplings \cite{Cheng Sher}. On the other hand, the
possibility of having a flavor conserving type III model at the tree level
has been considered in Ref. \cite{Lin} by demanding an $S_{3}\;$permutation
symmetry on the fermion families, see details in Sec. (\ref{theoretical
constraints}). Finally, as dicussed in Sec. (\ref{theoretical constraints}),
the decoupling limit provides another natural scenario for suppression of
both FCNC and CP violating couplings in the 2HDM type III.

\subsubsection{FCNC in the 2HDM\label{FCNC in the 2HDM}}

Although FCNC in SM are very suppressed because of the GIM mechanism, FC
transitions involving down-type quarks get enhanced because of the presence
of a top quark in the loop. In contrast, transitions involving up-type
quarks are usually very small. Therefore in SM processes like $b\rightarrow
\left( s,d\right) \gamma ,\;K-\overline{K}^{0},\;B-\overline{B}^{0}$; are
more interesting than for example $t\rightarrow c\left( \gamma ,Z\right)
,\;D-\overline{D}^{0}.\;$However, by introducing FCNC at tree level (as is
the case of the 2HDM type III) transitions like $t\rightarrow c\left( \gamma
,Z\right) \;$could be enhanced too, giving a clear signal of new physics 
\cite{Atwood, tcgammaus}.

In $e^{+}e^{-}$ colliders a very interesting FCNC reaction is $%
e^{+}e^{-}\rightarrow t\overline{c}.\;$Its signature would be very clear and
could compensate the lower statistics respect to hadron colliders. This
process can proceeds by a Higgs in the $s-$channel at the tree level, and at
one loop via corrections to the $Ztc,\;\gamma tc\;$vertices. The former is
negligible and only the latter have to be considered. Defining 
\begin{equation*}
R^{tc}\equiv \frac{\sigma \left( e^{+}e^{-}\rightarrow t\overline{c}+%
\overline{t}c\right) }{\sigma \left( e^{+}e^{-}\rightarrow \gamma ^{\ast
}\rightarrow \mu ^{+}\mu ^{-}\right) }
\end{equation*}
and using the following parametrization for the mixing matrix elements 
\begin{equation*}
\xi _{ij}=\lambda _{ij}\frac{\sqrt{m_{i}m_{j}}}{v}
\end{equation*}
Atwood \emph{et. al.} \cite{Atwood} found that assuming $\lambda
_{tt}=\lambda _{ct}\equiv \lambda $, values of $R^{tc}/\lambda ^{4}\;$up to $%
10^{-5}\;$are reached at $\sqrt{s}=400-500\;GeV\;$and that not much is
gained for larger values of $\sqrt{s}.$

The decay $t\rightarrow c\gamma \;$is highly supress in SM since the GIM
mechanism supresses it strongly owing to the smallness of the quarks running
into the loop (down type quarks) and also because of the large tree level
rate $t\rightarrow bW$.\ This process as well as the processes $t\rightarrow
c\left( Z,g\right) \;$can be substantially enhanced in model type III
respect to SM and also respect to type I and type II \cite{Atwood, tcgammaus}%
.

Due to the great agreement between $SM\;$and\ experimental determination of
the mass difference in the $K-\overline{K}^{0},\;$and $B_{d}-\overline{B}%
_{d}^{0}\;$systems, any tree level contribution from elementary flavor
changing couplings needs to be strongly suppressed. This was the original
motivation to impose the discrete symmetry given in Ref. \cite{Glashow} on
the 2HDM. If we were to work on a model with FCNC a very usual assumption is
that $\xi _{ij}=\lambda _{ij}\frac{\sqrt{m_{i}m_{j}}}{v}\;$\cite{Cheng Sher}.%
$\;$If we assume that all $\lambda _{ij}\;$are equal in the quark sector
then $\Delta F=2\;$mixing imposes severe constraints to the common $\lambda
.\;$If on the other hand, we assume that the mixings involving the first
family are negligible then these processes do not place further constraints
on the remaining $FC\;$couplings, allowing quite strong mixing between the
second and third family \cite{Atwood}. Finally, further constraints coming
from $R_{b}\;$seems to indicate that $\lambda _{bb}>>1,\lambda _{tt}<<1\;$%
and $\lambda _{sb}>>1,\lambda _{ct}<<1.\;$i.e. only the down sector of the
second and third family is strongly mixed.\ Disregarding constraints coming
from $R_{b}\;$the latter assumptions are not neccesary and the assumption of
negligible mixing with the first family is compatible with constraints given
by processes like $Br\left( B\rightarrow X_{s}\gamma \right) \;$and the $%
\rho \;$parameter. In general, the mixing between the second and third
generation of down-type quarks stays almost unconstrained, for example the
contribution of new Physics for $\Gamma \left( b\rightarrow sc\overline{c}%
\right) \;$is appreciably smaller than SM one for $\lambda _{sb}<40,\;$i.e.
in a large range of values.

As for the lepton sector, FCNC (i.e. LFV processes) in extended Higgs
sectors have been analized by several authors \cite{Sher91, vacuum
stability, g2us, LFVus, FCCCus, KangLee}. Nie and Sher examined the
contribution on $a_{\mu }$ from Higgs mediated FCNC coming from a very
general model with extended Higgs sector. They found that those
contributions can be significantly enhanced if the mass of the scalar is
light enough \cite{vacuum stability}, and obtained a bound on $\xi _{\mu
\tau }$ from $a_{\mu }$ by assuming that the heavier generations have larger
flavor changing couplings\ \cite{Cheng Sher}. Kang and Lee \cite{KangLee},
got constraints on some LFV couplings ($\xi _{\tau \tau }$, $\xi _{e\tau }$,$%
\;\xi _{\mu \tau }$,$\;\xi _{\mu \mu }$,$\;\xi _{e\mu }$,$\;\xi _{ee}$), in
the framework of a general extended Higgs sector by means of the $g-2\;$muon
factor and several LFV decays. These limits are gotten by setting the
lightest scalar mass as $m_{h^{0}}=100\;$GeV, $1000\;$GeV, and assuming that
the other Higgs bosons are decoupled. On the other hand, Ref. \cite{LFVus}
provides bounds on the mixing vertices ($\xi _{\tau \tau }$, $\xi _{e\tau }$,%
$\;\xi _{\mu \tau }$,$\;\xi _{\mu \mu }$) in the framework of the 2HDM type
III, by a similar approach but sweeping a wide region of the free parameters
of the model. For instance, since the contribution of the scalars and the
pseudoscalar have opposite signs, those limits could change significantly if
the pseudoscalar Higgs is not decoupled. Specifically, Ref. \cite{LFVus}
shows that lower limits on $\xi _{\mu \tau }^{2}\;$are obtained by using
light values of the pseudoscalar Higgs boson, while upper limits are found
by using scenarios with very heavy pseudoscalar, details in next chapter.
Finally, Diaz \emph{et. al.} \cite{FCCCus} have found the constraint $\left|
\xi _{\mu e}/m_{H^{+}}\right| \leq 3.8\times 10^{-3}GeV^{-1}$, based on the
flavor changing charged current decay $\mu \rightarrow \nu _{e}e\overline{%
\nu }_{\mu }$. Despite the bound is not so strong, it should be pointed out
that it is independent on the other parameters of the model, so improvements
on the upper limit of $\mu \rightarrow \nu _{e}e\overline{\nu }_{\mu }\;$%
could provide very useful information on $\xi _{\mu e}$.

\chapter{Flavor Changing Neutral Currents in the Lepton Sector of the 2HDM
(III)\label{LFV 2HDM}}

Having motivated the study of Higgs mediated FCNC in the previous chapter,
we shall concentrate on the lepton sector of the 2HDM type III henceforth.
Now, since neutrino oscillations seem to indicate the existence of Lepton
Flavor Violation (LFV) in the neutral lepton sector; we shall consider the
possibility of having LFV couplings in the form of either FCNC in the
charged sector, or Flavor Changing Charged Currents (FCCC) between the
charged and neutral sector.

For easy reference, I write the Yukawa Lagrangian type III Eq. (\ref
{YukexpandIII}) in the fundamental parametrization, for the lepton sector
only 
\begin{eqnarray}
-\pounds _{Y} &=&\overline{E}\left[ \frac{g}{2M_{W}}M_{E}^{diag}\right]
E\left( \cos \alpha H^{0}-\sin \alpha h^{0}\right)  \notag \\
&&+\frac{1}{\sqrt{2}}\overline{E}\xi ^{E}E\left( \sin \alpha H^{0}+\cos
\alpha h^{0}\right)  \notag \\
&&+\overline{\vartheta }\xi ^{E}P_{R}EH^{+}+\frac{i}{\sqrt{2}}\overline{E}%
\xi ^{E}\gamma _{5}EA^{0}+h.c.  \label{Yuk lepton}
\end{eqnarray}

In order to get information on LFV couplings in the 2HDM type III, we shall
use the following observables: the anomalous magnetic moment of the muon$%
\;\left( g-2\right) _{\mu }\;$and the LFV decays: $\mu \rightarrow e\gamma
,\;\mu \rightarrow eee,\;\tau \rightarrow eee,\,\tau \rightarrow \mu \mu \mu
,\;\tau \rightarrow e\gamma ,\;\tau \rightarrow \mu \gamma ,\;\mu
^{-}\rightarrow \nu _{e}e^{-}\overline{\nu }_{\mu }.$

In general, the muon anomalous magnetic moment $a_{\mu }$,$\;$is one of the
most important observables to constrain contributions coming from new
Physics. Since, it is perhaps the most important source to obtain our
bounds, we shall make a brief survey on the present status of its
experimental and theoretical estimations.

\section{Brief survey on $\left( g-2\right) _{\protect\mu }$\label%
{survey g-2}}

The first study of $g-$factors\ of subatomic particles was realized by Stern
in 1921 \cite{Stern}, the famous Stern Gerlach experiment was done in 1924 
\cite{Gerlach} the conclusion was that the magnetic moment of the electron
was one Bohr magneton i.e. that the gyromagnetic ratio $g\;$was 2. This
gyromagnetic factor is the proportionality constant between the magnetic
moment and the spin, 
\begin{equation*}
\overrightarrow{\mu }=g\left( \frac{e}{2m}\right) \overrightarrow{s}
\end{equation*}
the discovery that $g_{e}\neq 2\;$\cite{Nafe} and the calculation by Swinger 
\cite{Schwinger} was one of the great success of quantum electrodynamics.
Additionally the long lifetime of the muon allows a precision measurement of
its anomalous magnetic moment at ppm level. The anomalous muon magnetic
moment is given by 
\begin{equation*}
\mu _{\mu }=\left( 1+a_{\mu }\right) \frac{e\hbar }{2m_{\mu }}\;\;\;\;a_{\mu
}\equiv \frac{g-2}{2}
\end{equation*}
A series of experiments to evaluate $a_{\mu }\;$were carried out at CERN 
\cite{CERN79}. Nowadays, an ongoing experiment at BNL has been running with
much higher statistics and a very stable magnetic field in its storage ring 
\cite{Marciano}. By combining CERN and BNL statistics the present world
average result is \cite{g2exp} 
\begin{equation*}
a_{\mu }^{\text{exp}}\equiv \frac{\left( g-2\right) _{\mu }^{\text{exp}}}{2}%
=11\;659\;203\;\left( 8\right) \times 10^{-10}
\end{equation*}
the ultimate goal of the E821 experiment at BNL is to reduce present
uncertainty by a factor of two.

In the framework of the SM, theoretical predictions of $a_{\mu }\;$comes
from QED, electroweak, and hadronic contributions 
\begin{equation*}
a_{\mu }=a_{\mu }^{QED}+a_{\mu }^{EW}+a_{\mu }^{Had}
\end{equation*}

$a_{\mu }^{QED}$ constitute the bulk of the SM contributions ($%
11\;658\;470.57\times 10^{-10}$) and its uncertainty is small $\left(
2.9\times 10^{-11}\right) \;$\cite{Kinoshita},$\;$the EW contribution is
small ($152.0\times 10^{-11}$) and it is known with a similar accuracy $%
\left( 4.0\times 10^{-11}\right) \;$\cite{Czarnecki}, noteworthy, the EW
contribution has the same order of magnitude as the current experimental
uncertainty. Hadronic contributions are the second largest ones $\left( \sim
7000\times 10^{-11}\right) \;$\cite{lightby, g2th} and their uncertainty
constitutes the bulk of the uncertainty contributions\ $\left( \sim
70.7\times 10^{-11}\right) $.

As for the hadronic contributions, the most important ones in terms of both
contributions to $a_{\mu }\;$and its uncertainties are the leading vacuum
polarization terms, they have been estimated recently by a set of authors 
\cite{g2th}. The first of Refs. \cite{g2th} has used the most recent data on
hadron production from $e^{+}e^{-}\;$collisions from BES, CMD-2, and SND 
\cite{CMD}. The HMNT group (see fourth of Refs. \cite{g2th}) has made
estimations based on both inclusive and exclusive analysis of $%
e^{+}e^{-}\rightarrow $hadrons. Combining such exclusive and inclusive sets
of data, they obtain a more optimum estimate, and evaluate QCD sum rules in
order to solve a discrepancy between the inclusive and exclusive analysis.
In addition, the second and third Refs. \cite{g2th} presents an analysis
based on $e^{+}e^{-}$ data as well as $\tau \;$hadronic decays data. All of
Refs. \cite{g2th} found that analysis based on $e^{+}e^{-}\rightarrow $%
hadrons,\ predicts a SM value which is roughly $3\sigma \;$below the present
world average measurement, however, the second and third of Refs. \cite{g2th}
have showed that data based on hadronic $\tau \;$decays only differs with
the experimental world average by about $1.6\sigma $. Stabilization of the
predictions coming from $e^{+}e^{-}\;$based data and $\tau \;$decays data is
necessary in order to say something definitive. For our purposes we shall
use the estimation given by Jegerlehner \cite{g2th} whose analysis is based
on $e^{+}e^{-}\;$data by taking the most recent results from CMD-2 \cite{CMD}%
. Furthermore, other important contribution is the light by light scattering
whose sign has been corrected recently \cite{lightby}.

On the other hand, it is important to emphasize that despite the measurement
of the muon anomalous magnetic moment of the muon is about 180 times less
accurate than the electron anomalous magnetic moment; the former is more
sensitive to new physics by a factor of $\sim \left( m_{\mu
}^{2}/m_{e}^{2}\right) ,\;$this factor more than compensates the difference
in precision. Therefore, $a_{\mu }\;$is more important than $a_{e}\;$in
looking for new physics.

In constraining the 2HDM type III we shall use the estimation for $\Delta
a_{\mu }\;$at 95\% CL reported by \cite{Krawczykg2} 
\begin{equation}
9.38\times 10^{-10}\leq \Delta a_{\mu }\leq 51.28\times 10^{-10}
\label{Maria K}
\end{equation}
which in turn are based on preliminary results by Jegerlehner \cite{g2th}.

\section{The $\left( g-2\right) _{\protect\mu }\;$factor in the 2HDM type III%
\label{g-2 type III}}

The first one loop electroweak corrections for $a_{\mu }\;$coming from new
Physics ($\Delta a_{\mu }$) were calculated in Refs. \cite{Jackiw}. For $%
\Delta a_{\mu }$\ the integral expression \cite{Hunter} is given by

\begin{eqnarray}
\Delta a_{\mu } &=&\sum_{S}I_{S}+\sum_{P}I_{P}\;,  \notag \\
I_{S\left( P\right) } &=&C_{S\left( P\right) }^{2}\frac{m_{\mu }^{2}}{8\pi
^{2}}\int_{0}^{1}\frac{x^{2}\left( 1-x\pm m_{\tau }/m_{\mu }\right) }{m_{\mu
}^{2}x^{2}+\left( m_{\tau }^{2}-m_{\mu }^{2}\right) x+M_{S\left( P\right)
}^{2}\left( 1-x\right) }dx  \label{aNP}
\end{eqnarray}
where $I_{S\left( P\right) }\;$is an integral involving an Scalar
(Pseudoscalar) Higgs boson with mass $M_{S\left( P\right) }$,\ and $%
C_{S\left( P\right) }\;$is the corresponding coefficient in the Yukawa
Lagrangian Eq. (\ref{Yuk lepton}). If we assume that $m_{\mu }^{2}<<m_{\tau
}^{2}\;$and $m_{\mu }^{2}<<m_{h^{0},H^{0},A^{0}}^{2}\;$in the calculation of
(\ref{aNP}), the one loop contribution from all neutral Higgs bosons reads 
\cite{g2us} 
\begin{eqnarray}
\Delta a_{\mu } &=&\frac{m_{\mu }m_{\tau }}{16\pi ^{2}}\left\{ \left[
\sum_{S}C_{S}^{2}\left( F\left( M_{S}\right) -\frac{m_{\mu }}{3m_{\tau }}%
G\left( M_{S}\right) \right) \right] \right. +  \notag \\
&&\left. \left[ \sum_{P}C_{P}^{2}\left( F\left( M_{P}\right) +\frac{m_{\mu }%
}{3m_{\tau }}G\left( M_{P}\right) \right) \right] \right\}  \label{aNPaprox}
\end{eqnarray}
where 
\begin{eqnarray}
G\left( M\right) &\equiv &\left[ \frac{2+3\widehat{M}^{2}+6\widehat{M}%
^{2}\ln \left( \widehat{M}^{2}\right) -6\widehat{M}^{4}+\widehat{M}^{6}}{%
M^{2}\left( 1-\widehat{M}^{2}\right) ^{4}}\right]  \notag \\
F\left( M\right) &\equiv &\frac{\left[ 3+\widehat{M}^{2}\left( \widehat{M}%
^{2}-4\right) +2\ln \left( \widehat{M}^{2}\right) \right] \widehat{M}}{%
M^{2}\left( 1-\widehat{M}^{2}\right) ^{3}}  \notag \\
\widehat{M} &\equiv &\frac{m_{\tau }}{m_{H}}  \label{GFM}
\end{eqnarray}
We have neglected the contribution of the charged Higgs boson because of two
reasons \cite{g2us}: on one hand this contribution involves the neutrino
mass and on the other hand, the CERN $e^{+}e^{-}\;$collider LEP bound on its
mass is $m_{H^{+}}\geq 80.5\;$GeV.

\section{Lepton Flavor Violating decays\label{LFV decays}}

As we explained below, some lepton flavor violating decays are useful to
constrain the 2HDM type\ III. The relevant expressions of the lepton decays
involved at leading order are given by \cite{LFVus}:

\begin{eqnarray}
\Gamma \left( \tau \rightarrow l\gamma \right) &=&\xi _{l\tau }^{2}\frac{%
G_{F}\alpha _{em}m_{\tau }^{5}}{4\pi ^{4}\sqrt{2}}R\left(
m_{H^{0}},m_{h^{0}},m_{A^{0}},\alpha ,\xi _{\tau \tau }\right) \;,  \notag \\
\Gamma \left( \mu \rightarrow e\gamma \right) &=&\xi _{\mu \tau }^{2}\xi
_{e\tau }^{2}\frac{\alpha _{em}m_{\tau }^{4}m_{\mu }}{16\pi ^{4}}S\left(
m_{H^{0}},m_{h^{0}},m_{A^{0}},\alpha ,\xi _{\tau \tau }\right) \;.
\label{tamufot}
\end{eqnarray}
where $l\equiv e,\mu \;$denotes a light charged lepton. In addition, we have
defined 
\begin{eqnarray}
R\left( m_{H^{0}},m_{h^{0}},m_{A^{0}},\alpha ,\xi _{\tau \tau }\right)
&=&\left| \left( m_{\tau }\sin 2\alpha +\frac{\sqrt[4]{2}\xi _{\tau \tau
}\sin ^{2}\alpha }{\sqrt{G_{F}}}\right) \frac{\ln \left[ m_{H^{0}}/m_{\tau }%
\right] }{m_{H^{0}}^{2}}\right.  \notag \\
&-&\left. \left( m_{\tau }\sin 2\alpha -\frac{\sqrt[4]{2}\xi _{\tau \tau
}\cos ^{2}\alpha }{\sqrt{G_{F}}}\right) \frac{\ln \left[ m_{h^{0}}/m_{\tau }%
\right] }{m_{h^{0}}^{2}}\right.  \notag \\
&-&\left. \frac{2\sqrt[4]{2}\xi _{\tau \tau }}{\sqrt{G_{F}}}\frac{\ln \left[
m_{A^{0}}/m_{\tau }\right] }{m_{A^{0}}^{2}}\right| ^{2}\;,  \notag \\
S\left( m_{H^{0}},m_{h^{0}},m_{A^{0}},\alpha ,\xi _{\tau \tau }\right)
&=&\left| \sin ^{2}\alpha \frac{\ln \left[ m_{H^{0}}/m_{\tau }\right] }{%
m_{H^{0}}^{2}}+\cos ^{2}\alpha \frac{\ln \left[ m_{h^{0}}/m_{\tau }\right] }{%
m_{h^{0}}^{2}}\right.  \notag \\
&+&\left. \frac{\ln \left[ m_{A^{0}}/m_{\tau }\right] }{m_{A^{0}}^{2}}%
\right| ^{2}\;.  \label{R}
\end{eqnarray}
On the other hand, the expression for a lepton $L\;$going to three leptons
of the same flavor $l\;$is given by 
\begin{eqnarray}
\Gamma \left( L\rightarrow \overline{l}ll\right) &=&\frac{m_{L}^{5}}{2048\pi
^{3}}\left[ \sin 2\alpha \sqrt{\frac{G_{F}}{\sqrt{2}}}\left( \frac{1}{%
m_{H^{0}}^{2}}-\frac{1}{m_{h^{0}}^{2}}\right) m_{l}\right.  \notag \\
&+&\left. \xi _{ll}\left( \frac{\sin ^{2}\alpha }{m_{H^{0}}^{2}}+\frac{\cos
^{2}\alpha }{m_{h^{0}}^{2}}-\frac{1}{m_{A^{0}}^{2}}\right) \right] ^{2}\;.
\label{Llll}
\end{eqnarray}

\begin{figure}[tbp]
\resizebox{\textwidth}{!}{
\rotatebox{0}{\includegraphics[height=5cm]{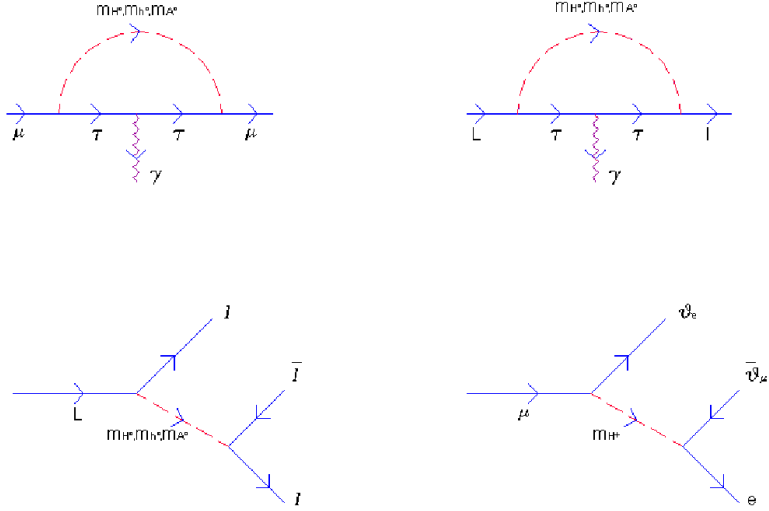}}}
\caption{Leading order diagrams for the processes:\ $\left( g-2\right) _{%
\protect\mu }\;$(top left), $L\rightarrow l\protect\gamma \;$(top right), $%
L\rightarrow \overline{l}ll\;$(bottom left), and $\protect\mu %
^{-}\rightarrow \protect\vartheta _{e}\overline{\protect\vartheta }_{\protect%
\mu }e^{-}$ (bottom right). In the two former, we have neglected the
contribution of the charged Higgs and a neutrino into the loop. }
\label{fig:alldiagrams}
\end{figure}

Finally, the decay width for$\;\mu ^{-}\rightarrow \nu _{e}e^{-}\overline{%
\nu }_{\mu }$, reads 
\begin{equation}
\Gamma \left( \mu ^{-}\rightarrow \nu _{e}e^{-}\overline{\nu }_{\mu }\right)
=\frac{m_{\mu }^{5}}{24\,576\pi ^{3}}\left( \frac{\xi _{\mu e}}{m_{H^{+}}}%
\right) ^{4}  \label{muneutrino}
\end{equation}
the latter is a process with Flavor Changing Charged Currents (FCCC)
mediated by a charged Higgs boson as we see in figure (\ref{fig:alldiagrams}%
).

In calculating $\Delta a_{\mu }\;$and the leptonic decays of the type $%
L\rightarrow l\gamma $, we neglect the contribution of light leptons into
the loop, so only the contribution with a tao into the loop is considered,
see figure (\ref{fig:alldiagrams}).

\section{Obtaining the bounds for the flavor changing vertices\label%
{bounds for LFV}}

From the previous section, we see that the free parameters that we are
involved with, are: the three neutral Higgs boson masses $\left(
m_{h^{0}},m_{H^{0}},m_{A^{0}}\right) $,\ the mixing angle $\alpha $,$\;$and
some flavor changing vertices $\xi _{ij}$. Based on the present bounds from
LEP2 we shall assume that $m_{h^{0}}\approx 115\;$GeV. In addition, we shall
assume that $m_{A^{0}}\gtrsim m_{h^{0}}$, both assumptions will be held
throughout this section. Now, since we are going to consider plots in the $%
\xi _{ij}-m_{A^{0}}\;$plane, we should manage to use appropiate values of $%
\left( m_{H^{0}},\alpha \right) \;$in order to sweep a wide region of
parameters. To sweep a reasonable set of this pair of parameters we utilize
for $m_{H^{0}}\;$values of the order of $115\;$GeV (light), $300\;$GeV
(intermediate), and very large masses (heavy). As for the angle $\alpha $,\
we consider values of $\alpha =0\;$(minimal mixing), $\alpha =\pi /4\;$%
(intermediate mixing), and $\alpha =\pi /2$\ (maximal mixing). It can be
seen that all possible combinations of $\left( m_{H^{0}},\alpha \right) $
could be done by considering five cases 1) when $m_{H^{0}}\simeq 115\;$GeV;$%
\;$ 2) when $m_{H^{0}}\simeq 300\;$GeV$\;$and $\alpha =\pi /2;\;$ 3) when $%
m_{H^{0}}\;$is very large and $\alpha =\pi /2$; 4) when $m_{H^{0}}\simeq
300\;$GeV$\;$and $\alpha =\pi /4$;$\;$5) when $m_{H^{0}}\;$is\ very\ large,
and $\alpha =\pi /4$.$\;$In all these cases the mass of the pseudoscalar
will be considered in the range $115\;$GeV$\lesssim m_{A^{0}}$.

It is worthwhile to remark that these five cases cover all possible
combinations of$\;\left( m_{H^{0}},\alpha \right) $, with the three
different values for $m_{H^{0}}$ ($115$\ GeV, $300\;$GeV, and very heavy)
and the three different values for $\alpha \;\left( 0,\;\pi /4,\;\pi
/2\right) $, though at the first glance some cases seem to miss. For
example, in the first case we have $m_{H^{0}}\simeq m_{h^{0}}$. In this
scenario,$\;$all the processes that we calculate become independent on $%
\alpha \;$and therefore no distinction is neccesary among the three
different values of it. The cases a) $m_{H^{0}}=300\;$GeV, with $\alpha =0$;$%
\;$and b) $m_{H^{0}}\;$very large, with $\alpha =0$;$\;$are also included in
case 1, because when $\alpha =0\;$the processes in the previous section does
not depend on $m_{H^{0}}$.

\begin{table}[tbp]
\begin{center}
\centering
\vskip0.1 in 
\begin{tabular}{||l||l|l|l||}
\hline\hline
& Bounds$\;$on$\;\xi _{\mu \tau }^{2}$ & Bounds$\;$on$\;\xi _{e\tau }^{2}\xi
_{\mu \tau }^{2}$ & Bounds$\;$on$\;\xi _{e\tau }^{2}$ \\ \hline\hline
case 1 & $7.62\times 10^{-4}\lesssim \xi _{\mu \tau }^{2}\lesssim
\allowbreak 8.31\times 10^{-3}$ & $\xi _{e\tau }^{2}\xi _{\mu \tau
}^{2}\lesssim 7.33\times 10^{-18}$ & $\xi _{e\tau }^{2}\lesssim 4.82\times
10^{-15}$ \\ \hline
case 2 & $1.29\times 10^{-3}\lesssim \xi _{\mu \tau }^{2}\lesssim
\allowbreak 4.42\times 10^{-2}$ & $\xi _{e\tau }^{2}\xi _{\mu \tau
}^{2}\lesssim 2.24\times 10^{-16}$ & $\xi _{e\tau }^{2}\lesssim 2.77\times
10^{-14}$ \\ \hline
case 3 & $1.53\times 10^{-3}\lesssim \xi _{\mu \tau }^{2}$ & $\xi _{e\tau
}^{2}\xi _{\mu \tau }^{2}\lesssim 2.24\times 10^{-16}$ & $\xi _{e\tau
}^{2}\lesssim 2.76\times 10^{-14}$ \\ \hline
case 4 & $9.57\times 10^{-4}\lesssim \xi _{\mu \tau }^{2}\lesssim
\allowbreak 1.40\times 10^{-2}$ & $\xi _{e\tau }^{2}\xi _{\mu \tau
}^{2}\lesssim 2.10\times 10^{-17}$ & $\xi _{e\tau }^{2}\lesssim 8.22\times
10^{-15}$ \\ \hline
case\ 5 & $1.02\times 10^{-3}\lesssim \xi _{\mu \tau }^{2}\lesssim
\allowbreak 1.66\times 10^{-2}$ & $\xi _{e\tau }^{2}\xi _{\mu \tau
}^{2}\lesssim 2.93\times 10^{-17}$ & $\xi _{e\tau }^{2}\lesssim 9.65\times
10^{-15}$ \\ \hline\hline
\end{tabular}
\end{center}
\caption{Constraints on the mixing parameters $\protect\xi _{\protect\mu 
\protect\tau }^{2}$,\ $\protect\xi _{e\protect\tau }^{2}\protect\xi _{%
\protect\mu \protect\tau }^{2}\;$and $\protect\xi _{e\protect\tau }^{2}\;$%
for the five cases mentioned in the text. The two former are generated from $%
\Delta a_{\protect\mu }\;$and $\Gamma \left( \protect\mu \rightarrow e%
\protect\gamma \right) \;$respectively, while the latter comes from the
combination of the lower limit on $\protect\xi _{\protect\mu \protect\tau
}^{2}\;$and the upper bound on $\protect\xi _{e\protect\tau }^{2}\protect\xi %
_{\protect\mu \protect\tau }^{2}$.}
\label{tab:mutao}
\end{table}

The first bounds come from $\Delta a_{\mu }$. We use the estimated value of
it,$\;$given by \cite{Krawczykg2} at 95\% C.L. Eq. (\ref{Maria K}). Since $%
\Delta a_{\mu }\;$in Eq. (\ref{Maria K}) is positive, lower and upper bounds
for the FC vertex $\xi _{\mu \tau }\;$can be gotten at 95\%\ C.L. The
results are indicated in table (\ref{tab:mutao}) column 1. The lower bounds
in each case are obtained when $m_{A^{0}}\approx 115\;$GeV, and using the
minimum value of $\Delta a_{\mu }$ in Eq. (\ref{Maria K}), while the upper
bounds are obtained when $A^{0}\;$is very heavy, and using the maximum value
of $\Delta a_{\mu }$ in Eq. (\ref{Maria K}). From these results a quite
general and conservative allowed interval can be extracted \cite{LFVus} 
\begin{equation}
7.62\times 10^{-4}\lesssim \xi _{\mu \tau }^{2}\lesssim 4.44\times 10^{-2}.
\label{genmutao}
\end{equation}

Furthermore, upper bounds for the product $\xi _{\mu \tau }^{2}\xi _{e\tau
}^{2}$ are obtained from the expression of the decay width $\Gamma \left(
\mu \rightarrow e\gamma \right) $ in Eq. (\ref{tamufot}) and from the
experimental upper limit $\Gamma \left( \mu \rightarrow e\gamma \right) \leq
3.6\times 10^{-30}$ GeV\ \cite{data particle}. The most general upper bounds
are obtained for $A^{0}\;$very heavy.$\;$The results are shown in table (\ref
{tab:mutao}) column 2. From this table we infer that quite generally the
bound is \cite{LFVus} 
\begin{equation}
\xi _{e\tau }^{2}\xi _{\mu \tau }^{2}\lesssim 2.24\times 10^{-16}
\end{equation}

Moreover, combining these upper limits with the lower bounds on $\xi _{\mu
\tau }^{2}$ given in the first column of table (\ref{tab:mutao}), we find
upper limits on $\xi _{e\tau }^{2}.$\ The results appear on table (\ref
{tab:mutao}) third column, and the general bound can be written as 
\begin{equation}
\xi _{e\tau }^{2}\lesssim 2.77\times 10^{-14}\;.  \label{genemutao}
\end{equation}
Noteworthy, these constraints predict a strong hierarchy between the mixing
elements $\xi _{\mu \tau }\;$and $\xi _{e\tau }\;$i.e. $\left| \xi _{e\tau
}\right| <<\left| \xi _{\mu \tau }\right| $\ and they differ by at least
five orders of magnitude.

\begin{figure}[tbph]
\centerline{\hbox{ \hspace{0.2cm}
    \includegraphics[width=6.5cm]{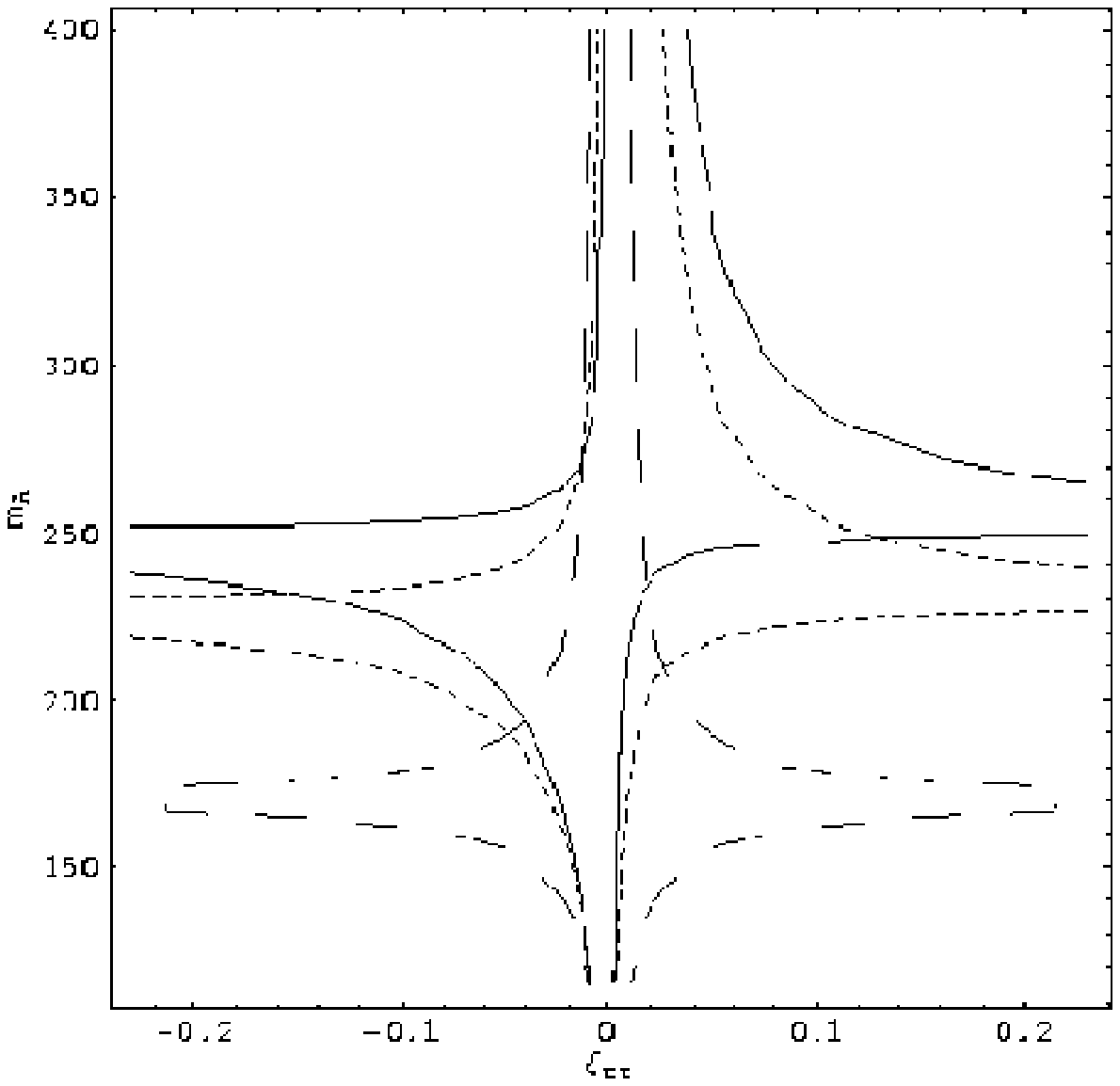}
    \hspace{0.3cm}
    \includegraphics[width=6.5cm]{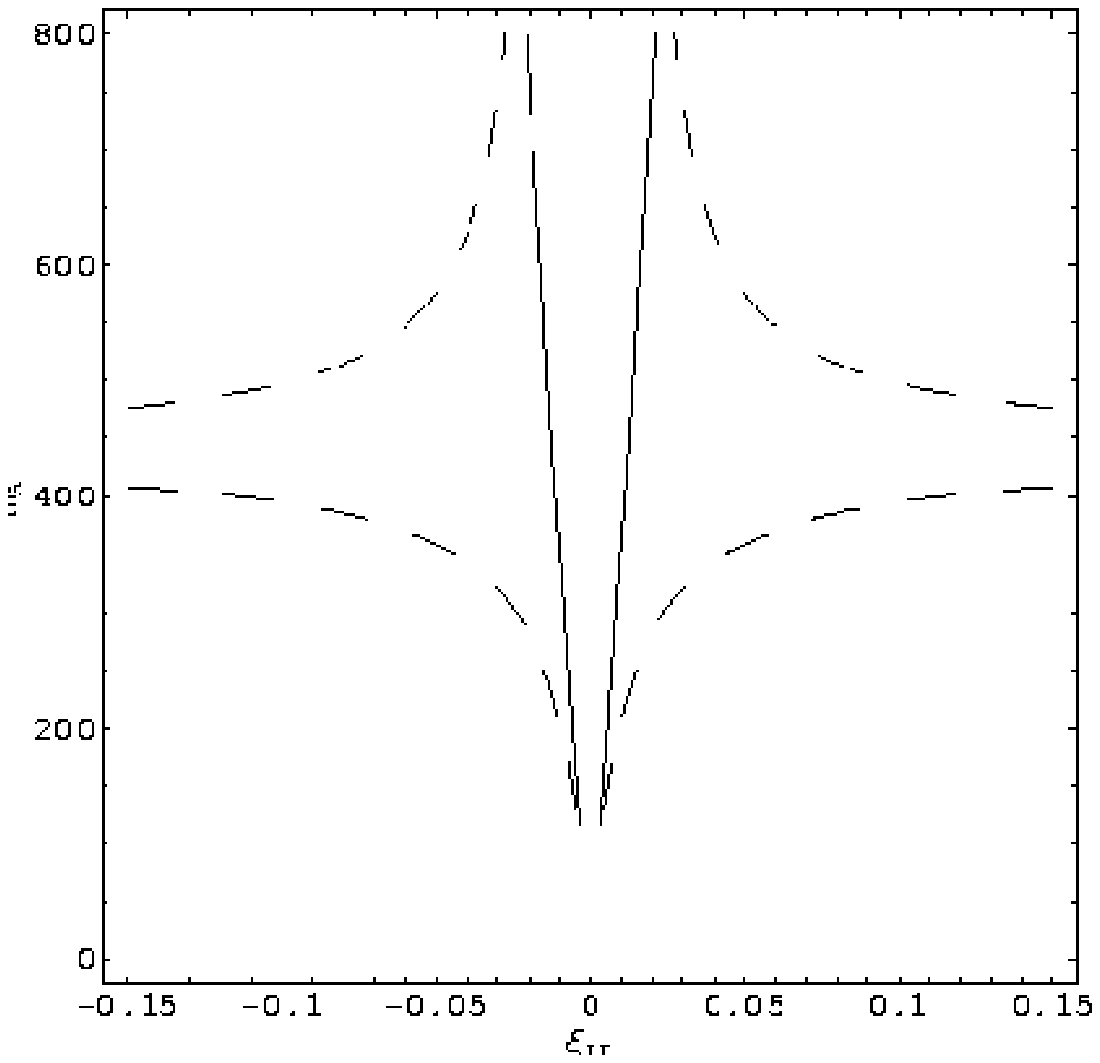}
    }
  }
\caption{Contourplots in the $\protect\xi _{\protect\tau \protect\tau
}-m_{A^{0}}\;$plane for each of the five cases cited in the text. On left:
case 1 corresponds to the long-dashed line, case 4 to the short-dashed line,
and case 5 to the solid line. On right: case 2 corresponds to dashed line,
and case 3 is solid line.}
\label{fig:taotao}
\end{figure}
\begin{figure}[tbph]
\centerline{\hbox{ \hspace{0.2cm}
    \includegraphics[width=6.5cm]{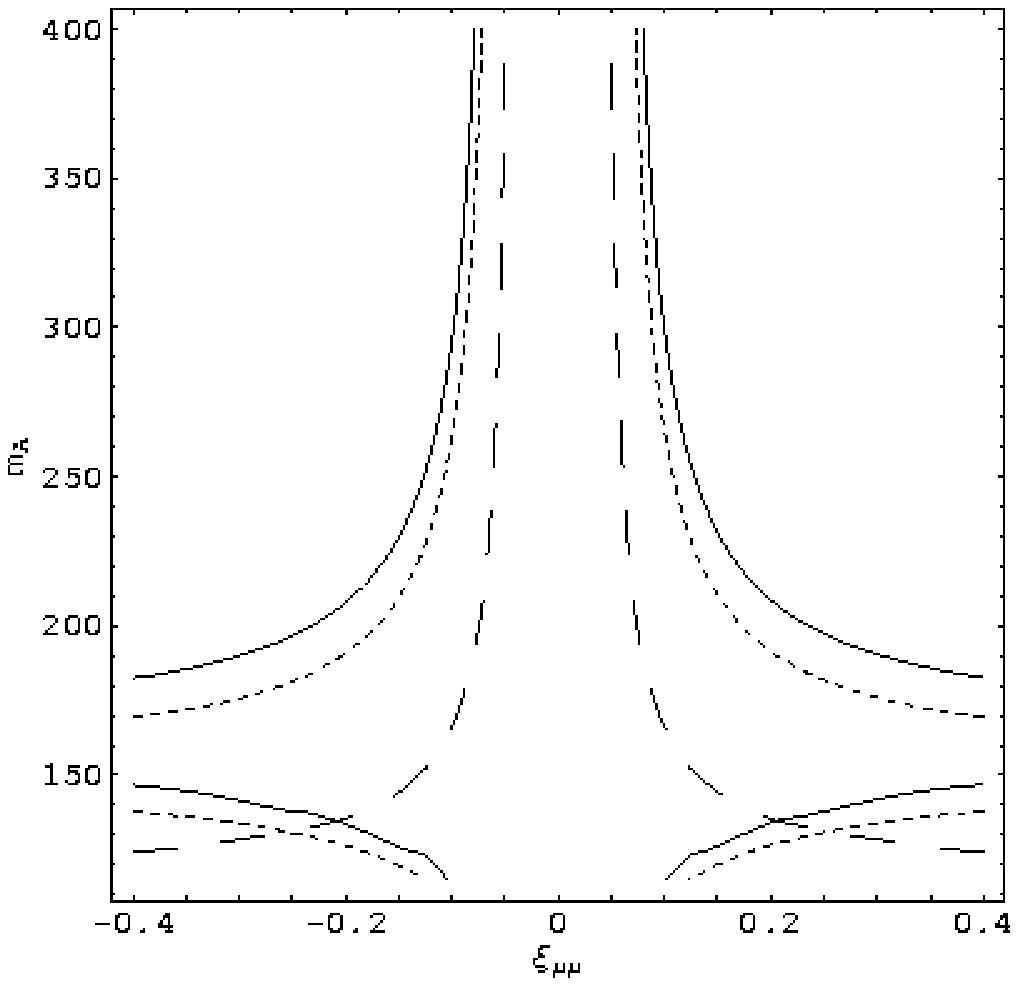}
    \hspace{0.3cm}
    \includegraphics[width=6.5cm]{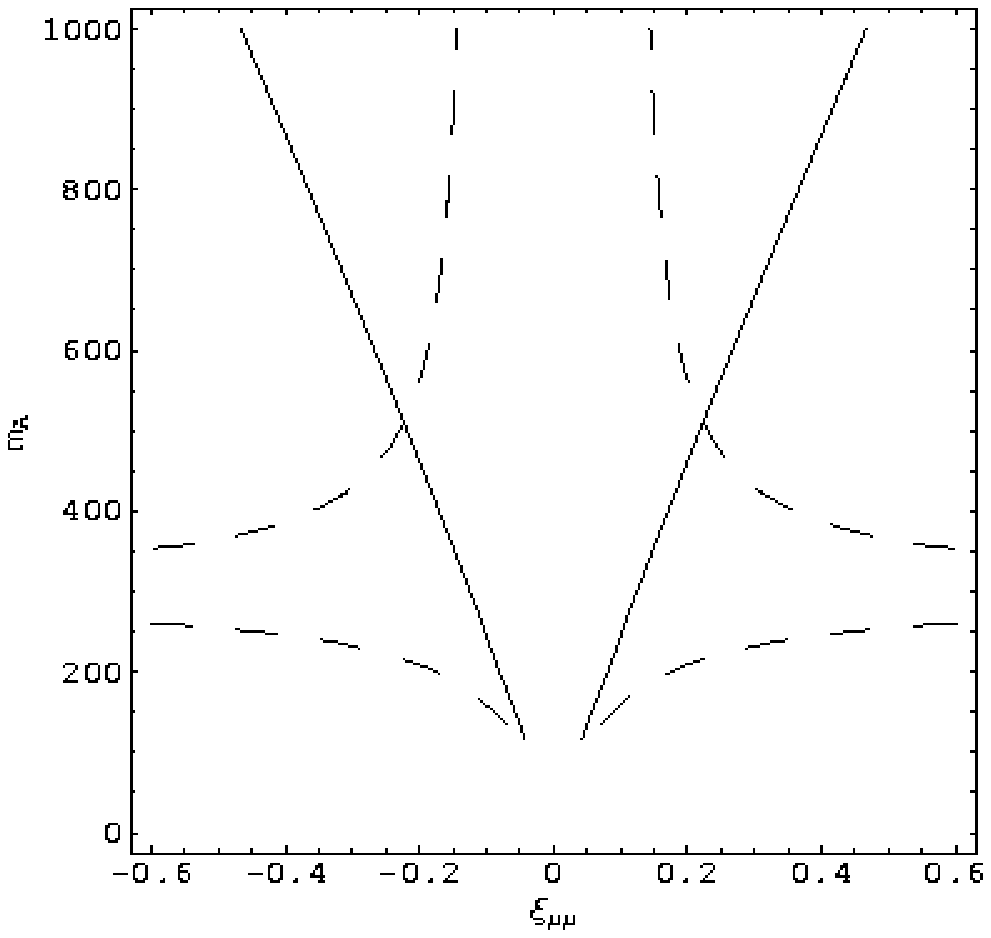}
    }
  }
\caption{Contourplots in the $\protect\xi _{\protect\mu \protect\mu
}-m_{A^{0}}\;$plane for each of the five cases cited in the text. On left:
case 1 corresponds to the long-dashed line, case 4 to the short-dashed line,
and case 5 to the solid line. On right: case 2 corresponds to dashed line,
and case 3 is solid line.}
\label{fig:mumu}
\end{figure}

On the other hand, from Eq. (\ref{tamufot}) we see that the decay widths $%
\Gamma \left( \tau \rightarrow \mu \gamma \right) $ and $\Gamma \left( \tau
\rightarrow \mu \mu \mu \right) $ depend on two mixing vertices $\xi _{\mu
\tau }^{2},\xi _{\tau \tau }\;$and $\xi _{\mu \tau }^{2},\xi _{\mu \mu }$
respectively. Then, we can get conservative constraints on the diagonal
mixing vertices $\xi _{\tau \tau }$, $\xi _{\mu \mu }\;$by using once again
the lower bounds on $\xi _{\mu \tau }^{2}\;$obtained from $\Delta a_{\mu }$.$%
\;$Since $\Gamma \left( \tau \rightarrow \mu \gamma \right) $, and $\Gamma
\left( \tau \rightarrow \mu \mu \mu \right) \;$are rather complicate
functions of $\xi _{\tau \tau }$, and $\xi _{\mu \mu }\;$respectively, we
present these constraints in the form of contourplots in the $m_{A^{0}}-\xi
_{\tau \tau }\;$plane and the $m_{A^{0}}-\xi _{\mu \mu }\;$plane, figures 
\ref{fig:taotao} and \ref{fig:mumu}, each one for the five cases.

We can check that for each contourplot there is a value of $m_{A^{0}}\;$for
which $\xi _{\tau \tau }\;$or$\;\xi _{\mu \mu }\;$stays unconstrained, and
they are shown in tables \ref{tab:taotao}, and \ref{tab:mumu} respectively.
Additionally, the bounds for $\xi _{\tau \tau },\;\xi _{\mu \mu }\;$when $%
m_{A^{0}}$ is very large$\;$and when $m_{A^{0}}\approx 115\;$GeV are also
included in tables \ref{tab:taotao}, \ref{tab:mumu} respectively.$\;$We see
that the general constraints for $\xi _{\tau \tau }$ read 
\begin{eqnarray}
-1.8\times 10^{-2} &\lesssim &\xi _{\tau \tau }\lesssim 2.2\times 10^{-2}%
\text{ , for }m_{A^{0}}\text{ very large}  \notag \\
-1.0\times 10^{-2} &\lesssim &\xi _{\tau \tau }\lesssim 1.0\times 10^{-2}%
\text{ , for }m_{A^{0}}\approx 115\text{ GeV}
\end{eqnarray}
These constraints are valid for all cases, except for the third case with $%
m_{A^{0}}\;$very large, since in that scenario $\xi _{\tau \tau }\;$remains
unconstrained. Now, for $\xi _{\mu \mu }\;$the general bounds read 
\begin{eqnarray}
\left| \xi _{\mu \mu }\right| &\lesssim &0.12\text{,\ for }m_{A^{0}}\text{
very large}  \notag \\
\left| \xi _{\mu \mu }\right| &\lesssim &0.13\text{, for }m_{A^{0}}\approx
115\text{ GeV}
\end{eqnarray}
Once again these constraints are not valid for the third case when $%
m_{A^{0}}\;$is very large, but are valid in all the other cases.

We should remember that the diagonal mixing vertices generate differences
among the relative couplings in model type III, and that this lack of
universality could be a distinctive signature for it. \ This fact might
leads us to scenarios in which a certain Higgs boson (say $h^{0}$) may be
tau-phobic\footnote{%
With the tau-phobic limit for certain Higgs, we mean that the coupling of
that Higgs to a pair of tau leptons vanishes. Nevertheless, the couplings of
such Higgs to a tau and any other lepton could exist.} but not muon-phobic
or electron-phobic (see section \ref{Yuk III}).

\begin{table}[tbp]
\begin{center}
\centering
\vskip0.1 in 
\begin{tabular}{||l||l|l|l||}
\hline\hline
& 
\begin{tabular}{l}
Values of $m_{A^{0}}\;$for \\ 
$\xi _{\tau \tau }\;$unconstrained
\end{tabular}
& 
\begin{tabular}{l}
$\xi _{\tau \tau }\;$intervals for \\ 
$m_{A^{0}}\;$very large
\end{tabular}
& 
\begin{tabular}{l}
$\xi _{\tau \tau }\;$intervals \\ 
for $m_{A^{0}}\approx 115\;$GeV
\end{tabular}
\\ \hline
case 1 & $m_{A^{0}}\approx 170\;$GeV & $-0.0072\lesssim \xi _{\tau \tau
}\lesssim 0.0072$ & $-0.010\lesssim \xi _{\tau \tau }\lesssim 0.010$ \\ 
\hline
case 2 & $m_{A^{0}}\approx 440\;$GeV & $-0.018\lesssim \xi _{\tau \tau
}\lesssim 0.018$ & $-0.0043\lesssim \xi _{\tau \tau }\lesssim 0.0043$ \\ 
\hline
case 3 & ----------------- & unconstrained & $-0.0036\lesssim \xi _{\tau
\tau }\lesssim 0.0036$ \\ \hline
case 4 & $m_{A^{0}}\approx 228\;$GeV & $-0.0075\lesssim \xi _{\tau \tau
}\lesssim 0.022$ & $-0.0094\lesssim \xi _{\tau \tau }\lesssim 0.0036$ \\ 
\hline
case 5 & $m_{A^{0}}\approx 250\;$GeV & $0.00024\lesssim \xi _{\tau \tau
}\lesssim 0.021$ & $-0.0093\lesssim \xi _{\tau \tau }\lesssim 0.0026$ \\ 
\hline\hline
\end{tabular}
\end{center}
\caption{Bounds extracted from the contourplots shown in Fig. \ref
{fig:taotao}. The first column indicates the values of $m_{A^{0}}\;$for
which $\protect\xi _{\protect\tau \protect\tau }\;$stays unconstrained for
each of the five cases. The second and third columns show the allowed
intervals on $\protect\xi _{\protect\tau \protect\tau }\;$when $m_{A^{0}}\;$%
is very large and when $m_{A^{0}}\approx 115$ GeV respectively.}
\label{tab:taotao}
\end{table}

\begin{table}[tbp]
\begin{center}
\centering
\vskip0.1 in 
\begin{tabular}{||l||l|l|l||}
\hline\hline
& 
\begin{tabular}{l}
Values of $m_{A^{0}}\;$for \\ 
$\xi _{\mu \mu }\;$unconstrained
\end{tabular}
& 
\begin{tabular}{l}
$\xi _{\mu \mu }\;$intervals \\ 
for $m_{A}>>m_{h^{0}}$%
\end{tabular}
& 
\begin{tabular}{l}
$\xi _{\mu \mu }\;$intervals \\ 
for $m_{A^{0}}\approx 115\;$GeV
\end{tabular}
\\ \hline\hline
case 1 & $m_{A^{0}}\approx 115\;$GeV & $\left| \xi _{\mu \mu }\right|
\lesssim 0.043$ & unconstrained \\ \hline
case 2 & $m_{A^{0}}\approx 300\;$GeV & $\left| \xi _{\mu \mu }\right|
\lesssim 0.12$ & $\left| \xi _{\mu \mu }\right| \lesssim 0.055$ \\ \hline
case 3 & -------------------- & unconstrained & $\left| \xi _{\mu \mu
}\right| \lesssim 0.043$ \\ \hline
case 4 & $m_{A^{0}}\approx 152\;$GeV & $\left| \xi _{\mu \mu }\right|
\lesssim 0.058$ & $\left| \xi _{\mu \mu }\right| \lesssim 0.13$ \\ \hline
case 5 & $m_{A^{0}}\approx 163\;$GeV & $\left| \xi _{\mu \mu }\right|
\lesssim 0.061$ & $\left| \xi _{\mu \mu }\right| \lesssim 0.11$ \\ 
\hline\hline
\end{tabular}
\end{center}
\caption{Bounds extracted from the contourplots shown in Fig. \ref{fig:mumu}%
. The first column indicates the values of $m_{A^{0}}\;$for which $\protect%
\xi _{\protect\mu \protect\mu }\;$stays unconstrained for each one of the
five cases. The second and third columns show the allowed intervals on $%
\protect\xi _{\protect\mu \protect\mu }\;$when $m_{A^{0}}\;$is very large$\;$%
and when $m_{A^{0}}\approx 115\;$GeV$\;$respectively.}
\label{tab:mumu}
\end{table}
We also ought to remember that in model type III, the pseudoscalar Higgs
boson couples to fermions only by means of the matrix element $\xi _{ff},\;$%
while in models I and II it couples through the mass of the corresponding
fermion. Therefore in model III, if $\xi _{ff}=0\;$for a certain specific
fermion, then $A^{0}\;$becomes fermiophobic to it (and only to it).

At this step, it is important to remark that our bounds are compatible with
all tau-phobic limits for each Higgs boson. It is easy to check that $%
h^{0}\; $becomes tau-phobic if $\xi _{\tau \tau }=0\;$for $\alpha =0$, and
if $\xi _{\tau \tau }\simeq 0.01\;$for $\alpha =\pi /4;\;$for $\alpha =\pi
/2\;$there is not tau-phobic limit.$\;$On the other hand, $H^{0}\;$becomes
tau-phobic if $\xi _{\tau \tau }=0\;$for $\alpha =\pi /2$, and if $\xi
_{\tau \tau }\simeq -0.01\;$for $\alpha =\pi /4;\;$and there is no
tau-phobic limit for $\alpha =0.\;$The tau-phobic limit for the pseudoscalar
is the simplest one, $\xi _{\tau \tau }=0.\;$Finally, if $\xi _{\tau \tau
}=0\;$and $\alpha =0\left( \pi /2\right) \;$then $A^{0}\;$and $h^{0}\left(
H^{0}\right) \;$become tau-phobic simultaneously.

On the other hand, bounds given above could in contrast produce a
significant enhancement to the Yukawa couplings. As a matter of example, in
the case 5 with $\alpha =\pi /4\;$using $\xi _{\tau \tau }\approx 0.1\;$%
\footnote{%
In the case 5 both values $\xi _{\tau \tau }=0.01\;$and\ $\xi _{\tau \tau
}=0.1\;$are allowed at least for $m_{A^{0}}$\ around $250\;$GeV, as it is
shown in Fig. (\ref{fig:taotao}). However, for a very light ($%
m_{A^{0}}\approx 115\;$GeV), or a very heavy pseudoscalar Higgs boson ($%
m_{A^{0}}>>m_{h^{0}}$), constraints on $\xi _{\tau \tau }\;$are considerably
stronger in all cases, see table \ref{tab:taotao}.} we see that the
contribution coming from the term proportional to $\xi _{\tau \tau },\;$is
about 10 times larger in magnitude than the contribution coming from the
term proportional to the tau mass, for both the $H^{0}\;$and $h^{0}\;$%
couplings. So it is even possible for the couplings of the $CP-$even Higgs
bosons to be overwhelmed by the $\xi _{\tau \tau }\;$contribution. Of
course, since the bounds for $\xi _{\mu \mu }\;$are weaker than the
constraints on $\xi _{\tau \tau }\;$as we can see from Fig. \ref{fig:mumu}
and table \ref{tab:mumu}, we can also have highly supressed or highly
enhanced Yukawa couplings to a pair of muon fermions.

Based on the bounds obtained above, we are able to estimate upper limits for
some leptonic decays by using the expressions (\ref{tamufot},\ref{R},\ref
{Llll}). We shall assume that $\left| \xi _{\tau \tau }\right| \lesssim 0.1$%
, and that$\;\left| \xi _{ee}\right| \lesssim 0.1$;$\;$from these
assumptions we can infer the following upper bounds \cite{LFVus} 
\begin{eqnarray}
\Gamma \left( \tau \rightarrow e\gamma \right) &\lesssim &1.5\times
10^{-27}\;,  \notag \\
\Gamma \left( \tau \rightarrow eee\right) &\lesssim &5\times 10^{-29}.
\end{eqnarray}
If we compare with the current experimental upper bounds, \cite{data
particle} 
\begin{eqnarray*}
\Gamma \left( \tau \rightarrow e\gamma \right) &\leq &6.12\times 10^{-18} \\
\Gamma \left( \tau \rightarrow eee\right) &\leq &6.57\times 10^{-18}
\end{eqnarray*}
$\;$ We see that these decays are predicted to be very far from the reach of
next generation experiments in the context of the 2HDM type III, unless that 
$\xi _{\tau \tau }$, and/or $\xi _{ee}\;$acquire unexpectedly large values,
this remarkably strong supression owes to the dependence of these decays on
the mixing vertex $\xi _{e\tau }$. For the sake of comparison, Ref. \cite
{Cvetic} has obtained upper limits for these decays in the context of two
models with heavy Majorana neutrinos: (I) the SM with additional
right-handed heavy Majorana neutrinos i.e. a typical see-saw type model and
(II) the standard model with right-handed and left-handed neutral singlets.
The first of these models predicts very small decay widths for LFV processes
in most of the parameter space, but the second one might show large enough
decay widths. Namely, the upper limits in the context of the second of these
models read \cite{Cvetic} 
\begin{eqnarray}
\Gamma \left( \tau \rightarrow e\gamma \right) &\lesssim &\allowbreak
2.27\times 10^{-20}\text{\ GeV}\;\text{,}  \notag \\
\Gamma \left( \tau \rightarrow eee\right) &\lesssim &2.\,\allowbreak
27\times 10^{-21}\;\text{GeV.}
\end{eqnarray}
which are not so far from the present experimental threshold and could be
tested by near future experiments, unlike the case of the 2HDM (III). Such
fact could be useful to discriminate between these models.

On the other hand, the decay $\mu ^{-}\rightarrow \nu _{e}e^{-}\overline{\nu 
}_{\mu }$ provides information about the mixing between the first and second
lepton family. The leptonic processes analyzed so far, involves FCNC
mediated by neutral Higgs bosons. Instead, the process $\mu ^{-}\rightarrow
\nu _{e}e^{-}\overline{\nu }_{\mu }$, involves Flavor Changing Charged
Currents mediated by a charged Higgs boson. Nevertheless, from Lagrangian (%
\ref{Yuk lepton}) we see that the matrix elements $\xi _{ij}^{E}\;$that
generates FCNC automatically generates FCCC which are strongly suppress in
the leptonic sector. Consequently, by constraining flavor changing charged
currents we are indirectly constraining flavor changing neutral currents as
well. In the case of the 2HDM (III), interactions involving FCCC at tree
level only contains the contribution from the charged Higgs boson, reducing
the free parameters to manage. Thus, from the decay width $\Gamma \left( \mu
\rightarrow \nu _{e}e^{-}\overline{\nu }_{\mu }\right) \;$we are able to
extract a bound for the quotient $\xi _{e\mu }/m_{H^{+}}\;$independent on
the other free parameters of the model. Taking the analytical expression for
the decay width Eq. (\ref{muneutrino}) and the current experimental upper
bound for it \cite{data particle} 
\begin{equation}
\Gamma \left( \mu ^{-}\rightarrow \nu _{e}e^{-}\overline{\nu }_{\mu }\right)
\leq 3.\,\allowbreak 6\times 10^{-21}\text{\ GeV},
\end{equation}
the following constraint is gotten 
\begin{equation}
\left| \frac{\xi _{e\mu }}{m_{H^{+}}}\right| \leq 3.8\times 10^{-3}\text{\
GeV}^{-1}.
\end{equation}
Despite this constraint is not so strong, it is interesting since it does
not depend on the other free parameters of the model, because the
calculation does not involve neutral Higgs bosons nor mixing angles. This is
a good motivation to improve the experimental upper limit for processes
involving FCCC in the leptonic sector.

\section{Bounds on the Higgs boson masses\label{bounds on Higgs}}

In the previous section, we obtained constraints for the mixing vertices
based on some sets of values for the Higgs masses. Conversely, we can assume
a set of values for the flavor changing vertices and try to get constraints
on the Higgs boson masses. We shall do it in the framework of the new
parametrization developed in Ref. \cite{LFVIIIus} and in section (\ref{Yuk
III}) of this work. We should be very careful in interpreting such results
because a different basis has been used for the Yukawa Lagrangian. For
instance, we shall show allowed regions in the $m_{A^{0}}-\tan \beta \;$%
plane by using fixed values for $\widetilde{\xi }_{\mu \tau
},\;m_{H^{0}},\;m_{h^{0}}$.$\;$However, we ought to bear in mind that $%
\widetilde{\xi }_{\mu \tau }\;$depends explicitly on $\tan \beta \;$as we
see in Eqs. (\ref{parametlink2}), checking Eqs. (\ref{parametlink2}) we see
that the only way to keep $\widetilde{\xi }_{\mu \tau }\;$fixed with
changing $\tan \beta $,$\;$is by varying the $\xi _{\mu \tau }\;$parameter
of the fundamental parametrization; applying the last of Eqs. (\ref
{parametlink2})\ to the vertex$\;\xi _{\mu \tau }\;$we get 
\begin{equation}
\xi _{\mu \tau }=\left( \sqrt{1+\tan ^{2}\beta }\right) \widetilde{\xi }%
_{\mu \tau }  \label{inkmutao}
\end{equation}
so by using a fixed value of $\widetilde{\xi }_{\mu \tau }\;$and varying $%
\tan \beta \;$(say\ from $0\;$to $N$),$\;$what we are really doing from the
point of view of the fundamental parametrization, is sweeping $\xi _{\mu
\tau }\;$from $\widetilde{\xi }_{\mu \tau }\;$to $\sqrt{1+N^{2}}\widetilde{%
\xi }_{\mu \tau }\;$.$\;$In addition, we should take into account that $%
\alpha ^{\prime }=\alpha +\beta $;\ Therefore, if we want to keep $\alpha
^{\prime }$ constant as well,\ $\alpha $ should be varied accordingly.
Consequently, to vary $\tan \beta $ while keeping fixed $\widetilde{\xi }%
_{\mu \tau }$ and $\alpha ^{\prime }$, it is necessary to vary the
parameters $\xi _{\mu \tau }$ and $\alpha $ in the fundamental
parametrization\footnote{%
We can assume another point of view in which $\tan \beta $ and $\alpha
^{\prime }$ are changed in such a way that only $\xi _{\mu \tau }$ varies in
the fundamental parametrization. But it implies to work with an additional
variable in the non-trivial parametrization.}. It should be pointed out
then, that changing $\tan \beta $ would imply a correlation among the
variation of $\alpha $ and $\xi _{\mu \tau }$ which is not a necessary
condition. A more appropiate picture would be fixing $\tan \beta $ (the
basis) and use the other parameters as variables. Notwithstanding, we shall
show some plots in the $m_{A^{0}}-\tan \beta $ plane, assuming that
correlation.

The possibility of plotting in a $m_{A^{0}}-\tan \beta \;$plane for the
model type III, facilitates the comparison of the model type III with the
models type I or II.

With this clarification we proceed to constrain the FC vertex involving the
second and third charged leptonic sector by using the estimated value for $%
\Delta a_{\mu }^{NP}$. Additionally, we get lower bounds on the Pseudoscalar
Higgs mass by taking into account the lower experimental value of $\Delta
a_{\mu }^{NP}\;$ at $95\%$ CL reported in \cite{Krawczykg2} and making
reasonable assumptions on the FC vertex.

For easy reference, I include the Yukawa Lagrangian type III for the
leptonic sector by using parametrizations of type I and II explained in
section (\ref{Yuk III}). 
\begin{eqnarray}
-\pounds _{Y\left( E\right) }^{\left( I\right) } &=&\frac{g}{2M_{W}\sin
\beta }\overline{E}M_{E}^{diag}E\left( \sin \alpha ^{\prime }H^{0}+\cos
\alpha ^{\prime }h^{0}\right)  \notag \\
&+&\frac{ig}{2M_{W}}\overline{E}M_{E}^{diag}\gamma _{5}EG^{0}+\frac{ig\cot
\beta }{2M_{W}}\overline{E}M_{E}^{diag}\gamma _{5}EA^{0}  \notag \\
&-&\frac{1}{\sqrt{2}\sin \beta }\overline{E}\widetilde{\eta }^{E}E\left[
\sin \left( \alpha ^{\prime }-\beta \right) H^{0}+\cos \left( \alpha
^{\prime }-\beta \right) h^{0}\right]  \notag \\
&-&\frac{i}{\sqrt{2}\sin \beta }\overline{E}\widetilde{\eta }^{E}\gamma
_{5}EA^{0}+h.c.  \label{leptonic 1ad}
\end{eqnarray}
where the superindex $(I)\;$refers to the parametrization type I. It is easy
to check that Lagrangian (\ref{leptonic 1ad}) is just the one in the 2HDM
type I, plus some FC interactions. Therefore, we obtain the Lagrangian of
the 2HDM type I from Eq. (\ref{leptonic 1ad}) by setting $\widetilde{\eta }%
^{E}=0.\;$In this case, it is clear that when $\tan \beta \rightarrow 0$
then $\widetilde{\eta }^{E}$ should go to zero, in order to have a finite
contribution for FCNC at the tree level.

In the parametrization of type II, the Yukawa Lagrangian reads 
\begin{eqnarray}
-\pounds _{Y(E)}^{(II)} &=&\frac{g}{2M_{W}\cos \beta }\overline{E}%
M_{E}^{diag}E\left( \cos \alpha ^{\prime }H^{0}-\sin \alpha ^{\prime
}h^{0}\right)  \notag \\
&+&\frac{ig}{2M_{W}}\overline{E}M_{E}^{diag}\gamma _{5}EG^{0}-\frac{ig\tan
\beta }{2M_{W}}\overline{E}M_{E}^{diag}\gamma _{5}EA^{0}  \notag \\
&+&\frac{1}{\sqrt{2}\cos \beta }\overline{E}\widetilde{\xi }^{E}E\left[ \sin
\left( \alpha ^{\prime }-\beta \right) H^{0}+\cos \left( \alpha ^{\prime
}-\beta \right) h^{0}\right]  \notag \\
&+&\frac{i}{\sqrt{2}\cos \beta }\overline{E}\widetilde{\xi }^{E}\gamma
_{5}EA^{0}+h.c.  \label{leptonic 2ad}
\end{eqnarray}
The Lagrangian (\ref{leptonic 2ad}) coincides with the one of the 2HDM type
II \cite{Hunter}, plus some FC interactions. So, the Lagrangian of the 2HDM
type II is obtained setting $\widetilde{\xi }^{E}=0.\;$In this case it is
clear that when $\tan \beta \rightarrow \infty $, then $\widetilde{\xi }^{E}$
should go to zero, in order to have a finite contribution for FCNC at the
tree level.

We now use the expression for $\Delta a_{\mu }\;$from either Eq. (\ref{aNP})
or Eq. (\ref{aNPaprox}) to get bounds on the Higgs boson masses.
Additionally, we can notice that in Eq. (\ref{GFM}) for Higgs bosons heavier
than 100 GeV we find 
\begin{equation*}
\frac{m_{\mu }}{3m_{\tau }}G\left( m_{H_{i}}\right) <<F\left(
m_{H_{i}}\right)
\end{equation*}
and the contribution of the factor involving $G\left( m_{H_{i}}\right) $ is
negligible \cite{g2us}. However, we put it for completeness.

\begin{figure}[tbp]
\resizebox{\textwidth}{!}{
\rotatebox{0}{\includegraphics{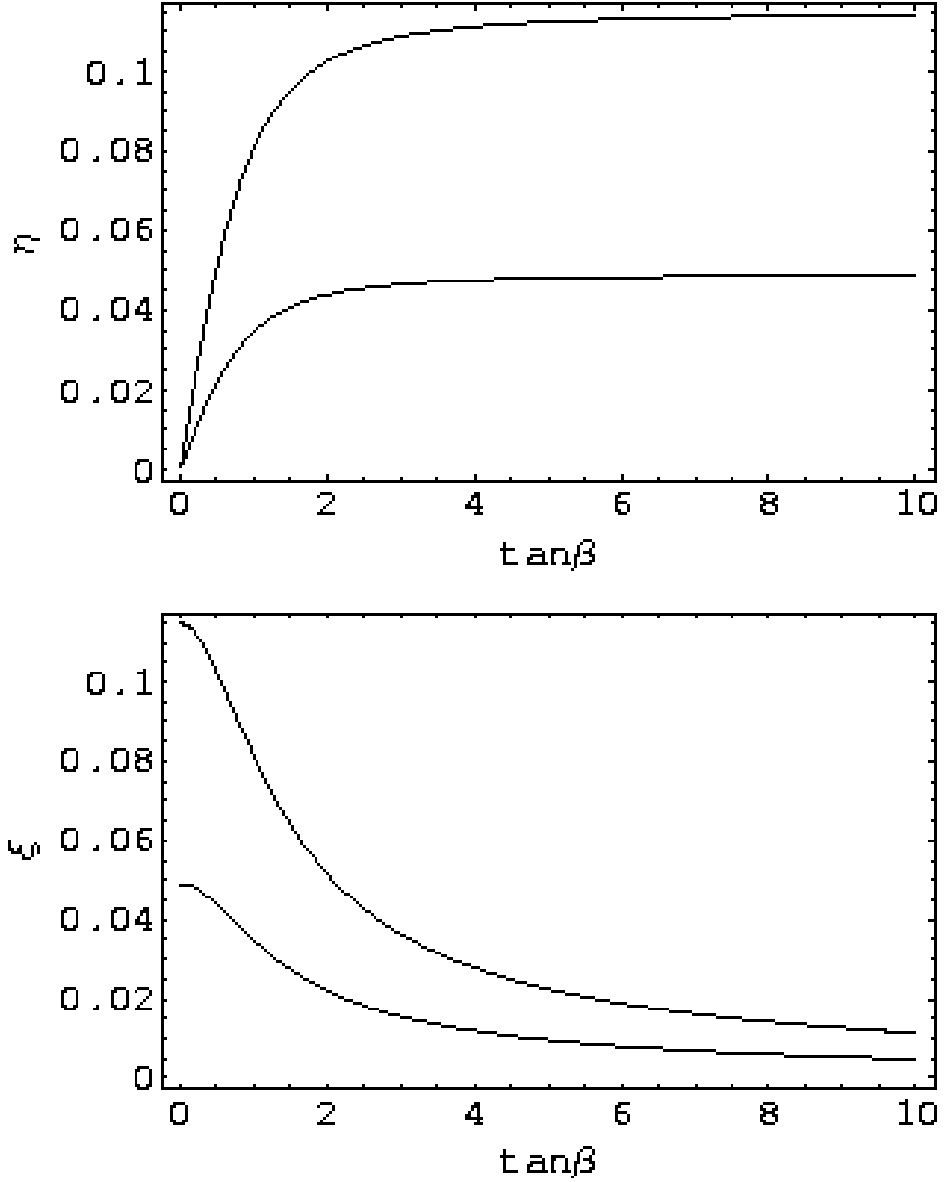}}}
\caption{ Lower and upper bounds for $\widetilde{\protect\eta }_{\protect\mu 
\protect\tau }\left( \widetilde{\protect\xi }_{\protect\mu \protect\tau
}\right) \;$vs\ tan$\protect\beta ,\;$for parametrizations I and II using $%
m_{h^{0}}=m_{H^{0}}=150$ GeV and $m_{A^{0}}\rightarrow \infty .\;$}
\label{fig:figg-21}
\end{figure}

In Ref. \cite{g2us} the estimated value for $\Delta a_{\mu }\;$has been
taken from \cite{Marciano}. In this section we are going to use the basic
procedure developed in Ref. \cite{g2us}, but using a more updated value for $%
\Delta a_{\mu }\;$i.e.\ the one reported in Ref. \cite{Krawczykg2}, instead
of the one reported in \cite{Marciano}. We get some lower and upper bounds
on the mixing vertex $\widetilde{\eta }\left( \widetilde{\xi }\right) _{\mu
\tau }\;$for the parametrizations of type I (II). In Fig. (\ref{fig:figg-21}%
), we display lower and upper bounds for the FC vertices as a function of $%
\tan \beta $ for both types of parametrizations with $%
m_{h^{0}}=m_{H^{0}}=150 $ GeV and $m_{A^{0}}\rightarrow \infty $. In the
first case, with parametrization type I, the allowed region for $\widetilde{%
\eta }_{\mu \tau } $ is $0.045\lesssim \widetilde{\eta }_{\mu \tau }\lesssim
0.12$ for large values of $\tan \beta $. Meanwhile, for parametrization type
II, the allowed region for small $\tan \beta $ is the same. From the first
of Figs. (\ref{fig:figg-21}) we can see that when $\tan \beta \rightarrow 0$%
, $\widetilde{\eta }_{\mu \tau }\;$goes to zero as expected. Similarly, the
second of these figures, shows that when $\tan \beta \rightarrow \infty $, $%
\widetilde{\xi }_{\mu \tau }\;$goes to zero as expected too.

In addition, the Figs. (\ref{fig:figg-21}) show us that the FC vertices are
basis dependent. We can realize that by varying\ $\tan \beta $, what we are
doing is changing the basis. Figs. (\ref{fig:figg-21}) show that for
different values of $\tan \beta $ (i.e.\ different bases) we get different
values for the lower and upper limits on $\widetilde{\eta }_{\mu \tau
}\left( \widetilde{\xi }_{\mu \tau }\right) $.

\begin{figure}[tbp]
\resizebox{\textwidth}{!}{
\includegraphics{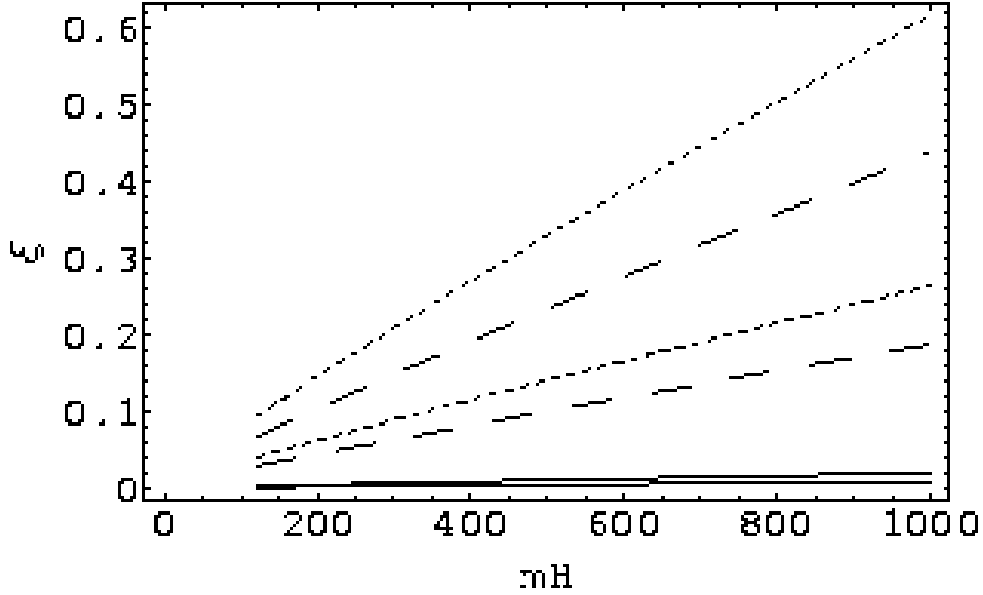}}
\caption{Lower and upper bounds for $\widetilde{\protect\xi }_{\protect\mu 
\protect\tau }$ vs \ $m_{H^{0}},\;$for parametrization of type II, taking$%
\;m_{h^{0}}=m_{H^{0}}\;$and $m_{A^{0}}\rightarrow \infty $, the pair of
dotted lines correspond to $\tan \protect\beta =0.1,\;$the dashed lines are
for $\tan \protect\beta =1$,\ and the solid lines are for $\tan \protect\beta
=30.$}
\label{fig:figg-22}
\end{figure}

In figure (\ref{fig:figg-22}), we show lower and upper bounds for the FC
vertex as a function of $m_{H^{0}}$ for parametrization of type II when $%
m_{h^{0}}=m_{H^{0}}$ and $m_{A^{0}}\rightarrow \infty $. We see that the
larger value for $\tan \beta $ the smaller value of $\widetilde{\xi }_{\mu
\tau }$ . We only consider the case of parametrization type II because there
is a complementary behaviour between both parametrizations as could be seen
in figure (\ref{fig:figg-21}). In particular, for $\tan \beta =1,$ the
behaviour of the bounds for both parametrizations is the same. Since in this
case we are fixing $\tan \beta $, we are fixing the basis and the
interpretation is clear. It should be emphasized that according to Eq. (\ref
{inkmutao}) for different values of $\tan \beta ,\;$the same value of $%
\widetilde{\xi }_{\mu \tau }\;$correspond to different values of $\xi _{\mu
\tau }\;$in the fundamental parametrization. Thus, we see again that the
values of the FC vertices are basis dependent.

Observe that according to its current estimated value, $\Delta a_{\mu }^{NP}$
is positive definite, and it is a new feature from most updated results \cite
{g2th}. On the other hand, the Feynman rules from (\ref{leptonic 1ad}) and (%
\ref{leptonic 2ad}), show that the Scalar (Pseudoscalar) contribution to $%
\Delta a_{\mu }^{NP}\;$Eqs. (\ref{aNP}, \ref{aNPaprox}) is positive
(negative). Such fact permits us to impose lower bounds on the pseudoscalar
Higgs mass, by using the lower limit\ for $\Delta a_{\mu }^{NP}$. However,
it is very important to clarify that for very light Pseudoscalars (i.e. $%
m_{A^{0}}<<100$ GeV), the contributions up to two loops become important 
\cite{Krawczykg2}. Further, the diagrams at two loops for the scalars and
the pseudoscalar could have opposite signs respect to the contribution at
one loop. Consequently, the lower bounds shown here are not valid in the
very light pseudoscalar regime, and they are only valid if we assume that $%
m_{A^{0}}\gtrsim 100$ GeV.

\begin{figure}[tbp]
\resizebox{\textwidth}{!}
{\includegraphics{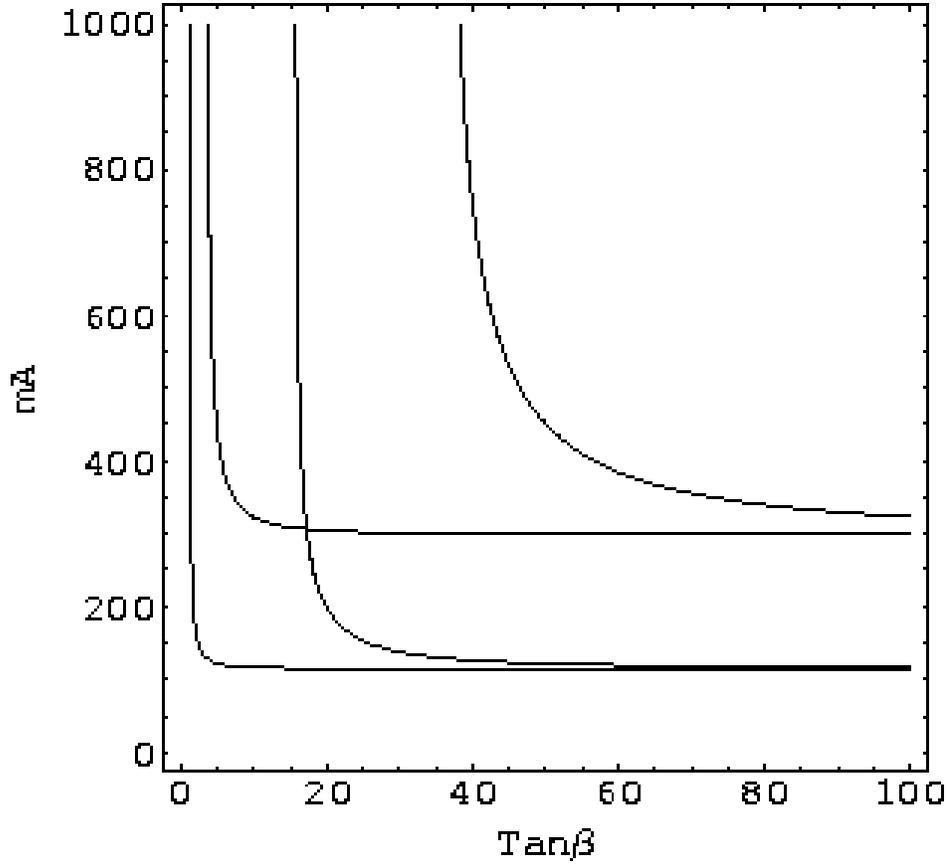}}
\caption{Contourplot of $m_{A^{0}}\;$vs$\;\tan \protect\beta \;$using
parametrization type II and assuming $m_{h^{0}}=m_{H^{0}}$. From left to
right: (1) $m_{h^{0}}=115$ GeV$,\;\protect\xi _{\protect\mu \protect\tau
}=2.5\times 10^{-2}$ (2)\ $m_{h^{0}}=300$ GeV$,\;\protect\xi _{\protect\mu 
\protect\tau }=2.5\times 10^{-2}$ (3)\ $m_{h^{0}}=115$ GeV$,\;\protect\xi _{%
\protect\mu \protect\tau }=2.5\times 10^{-3}$ (4)\ $300$ GeV$,\;\protect\xi
_{\protect\mu \protect\tau }=2.5\times 10^{-3}$.}
\label{fig:figg-23}
\end{figure}

\begin{figure}[tbp]
\resizebox{\textwidth}{!}
{\includegraphics{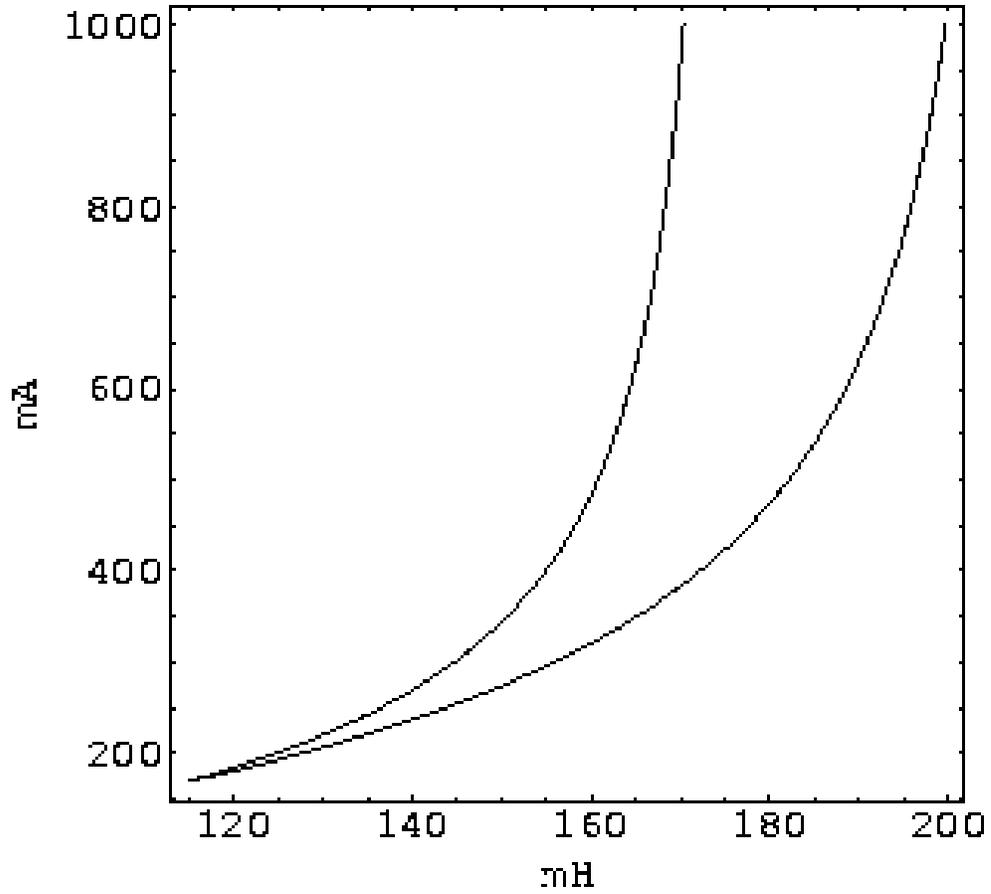}}
\caption{Contourplot of$\;m_{A^{0}}\;$vs$\;m_{H^{0}}\;$setting $\tan \protect%
\beta =2$, the line above correspond to $\widetilde{\protect\xi }_{\protect%
\mu \protect\tau }=2.5\times 10^{-2}\;$and $m_{h^{0}}=m_{H^{0}}$, while the
line below correspond to$\;\widetilde{\protect\xi }_{\protect\mu \protect\tau
}=2.5\times 10^{-2}\;$for $m_{h^{0}}=115\;$GeV.}
\label{fig:figg-24}
\end{figure}

\begin{figure}[tbp]
\resizebox{0.62\textwidth}{!}{
\rotatebox{0}{\includegraphics{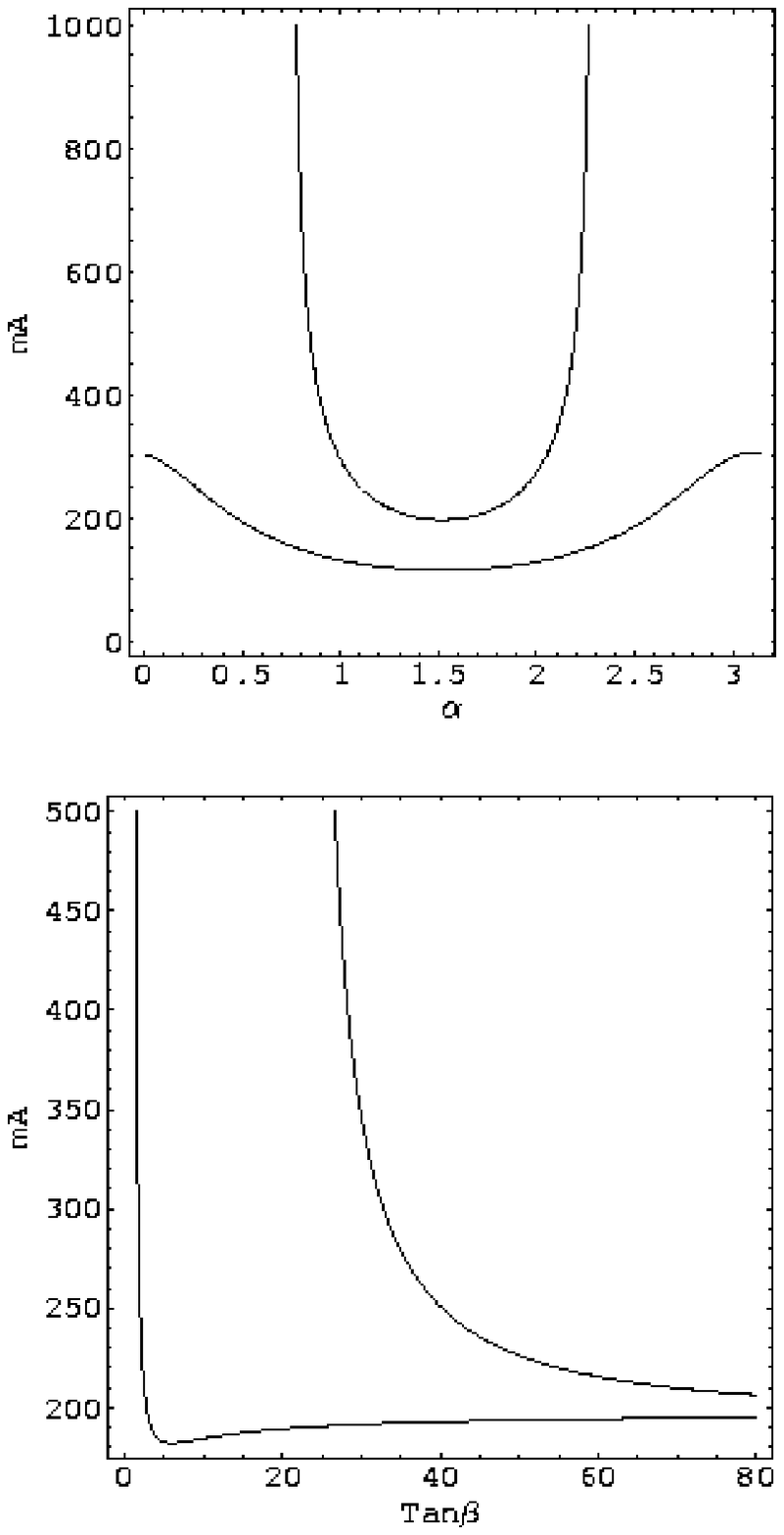}}
}
\caption{{\protect\small (top) Contourplot of }$m_{A^{0}}\;${\protect\small %
vs\ }$\protect\alpha ^{\prime },\;${\protect\small for parametrization type
II, with }$m_{h^{0}}=115${\protect\small \ GeV,}$\;m_{H^{0}}=300$%
{\protect\small \ GeV and }$\tan \protect\beta =20${\protect\small . Line
above corresponds to }$\widetilde{\protect\xi }_{\protect\mu \protect\tau
}=2.5\times 10^{-3}${\protect\small \ ,\ and line below corresponds to }$%
\widetilde{\protect\xi }_{\protect\mu \protect\tau }=2.5\times 10^{-2}.\;$%
{\protect\small (bottom) Contourplot of }$m_{A^{0}}\;${\protect\small vs\ }$%
\tan \protect\beta \;${\protect\small for }$m_{h^{0}}=115${\protect\small \
GeV, }$m_{H^{0}}=300${\protect\small \ GeV, }$\protect\alpha =\protect\pi
/6,\;${\protect\small and for parametrization type II. Line above
corresponds to }$\widetilde{\protect\xi }_{\protect\mu \protect\tau
}=2.5\times 10^{-3}${\protect\small , and line below corresponds to }$%
\widetilde{\protect\xi }_{\protect\mu \protect\tau }=2.5\times 10^{-2}.\;$}
\label{fig:figg-25}
\end{figure}

Now, to take into account the experimental value, we should make a
supposition about the value of the FC vertex. A reasonable assumption
consists of taking the geometric average of the Yukawa couplings \cite{Cheng
Sher} i.e. $\widetilde{\eta }\left( \widetilde{\xi }\right) _{\mu \tau
}\approx 2.5\times 10^{-3}$. Additionally, we shall use also the values $%
\widetilde{\eta }\left( \widetilde{\xi }\right) _{\mu \tau }\approx
2.5\times 10^{-2}$ and$\;\widetilde{\eta }\left( \widetilde{\xi }\right)
_{\mu \tau }\approx 2.5\times 10^{-4}\;$which are one order of magnitude
larger and smaller than the former. Notwithstanding, we should remember that
these values are basis dependent. Therefore, in this work we shall take the
values yielded above for the FC vertex among the second and third family by
using different bases, i.e. different values of $\tan \beta $. Using these
suppositions and the experimental value for $\Delta a_{\mu }^{NP}$ reported
in Ref. \cite{Krawczykg2} we get lower bounds for $m_{A^{0}}\;$ and they are
plotted in figures (\ref{fig:figg-23}-\ref{fig:figg-25}).

Figure \ref{fig:figg-23} displays $m_{A^{0}}\;$vs$\;\tan \beta \;$utilizing
the parametrization type II and using $\widetilde{\xi }_{\mu \tau
}=2.5\times 10^{-2}$, and $\widetilde{\xi }_{\mu \tau }=2.5\times 10^{-3}$.
Additionally it is assumed that $m_{h^{0}}=m_{H^{0}}\;$with $%
m_{h^{0}}=115,\;300$ GeV\footnote{%
By using $\widetilde{\xi }_{\mu \tau }=2.5\times 10^{-4}$ we still have an
allowed region but it appears only for $\tan \beta \gtrsim 155$ and for $%
\tan \beta \gtrsim 360$ when $m_{h^{0}}=115$\ GeV and when $m_{h^{0}}=300$\
GeV respectively. The behavior of those curves is similar to the others and
is not shown in Fig. \ref{fig:figg-23}.}. According to the interpretation
explained above, since we are using $0\leq \tan \beta \leq 100$, it is
equivalent to use a range of $\xi _{\mu \tau }\;$from$\;\widetilde{\xi }%
_{\mu \tau }\;$to$\;100\widetilde{\xi }_{\mu \tau }\;$in the fundamental
parametrization. It could be seen that in the limit of large $\tan \beta ,\;$%
the lower limit reduces to$\;m_{A^{0}}\approx m_{h^{0}}$. The same behavior
can be seen in parametrization type I but the bound $m_{A^{0}}\approx
m_{h^{0}}$ is gotten in the limit of small $\tan \beta .$ We also see that
the smaller value of $\widetilde{\xi }_{\mu \tau }\;$the stronger lower
limit for $m_{A^{0}}$.

As a proof of consistency, we can see that for instance in the case $%
m_{h^{0}}=m_{H^{0}}=115$ GeV with $\widetilde{\xi }_{\mu \tau }=2.5\times
10^{-3}$ the allowed region starts at about $\tan \beta \approx 13$. If we
calculate the value of $\xi _{\mu \tau }$ in the fundamental parametrization
corresponding to $\tan \beta \approx 13$ we obtain from Eq. (\ref{inkmutao})
that $\xi _{\mu \tau }^{2}\approx 1\times 10^{-3}$, in rough agreement with
the minimum allowed value for $\xi _{\mu \tau }^{2}$ reported in table (\ref
{tab:mutao}) for the case 1\footnote{%
The difference owes to the fact that here we have used the approximate
expression for $\Delta a_{\mu }$ given in Eq. (\ref{aNPaprox}), instead of
the Eq. (\ref{aNP}) which was the one used to obtain the table (\ref
{tab:mutao}). Additional differences could be traced to the errors involved
in the calculation of the contourplots. A more careful numerical analysis
could improve the agreement, but it is beyond the purpose of the present
work.}.

Figure \ref{fig:figg-24} shows $m_{A^{0}}\;$vs$\;m_{H^{0}}\;$with $%
\widetilde{\xi }_{\mu \tau }=2.5\times 10^{-2}\;$and $\tan \beta =2$, using $%
m_{h^{0}}=m_{H^{0}}\;$and$\;m_{h^{0}}=115$ GeV. With these settings, the
values $\widetilde{\xi }_{\mu \tau }=2.5\times 10^{-3}$ and $\widetilde{\xi }%
_{\mu \tau }=2.5\times 10^{-4}$ are excluded.

Once again, we can see the consistency with the analysis in Sec. (\ref
{bounds for LFV}) noting that for $\widetilde{\xi }_{\mu \tau }=2.5\times
10^{-3}\;$and $\widetilde{\xi }_{\mu \tau }=2.5\times 10^{-4}$ with $\tan
\beta =2$, we get $\xi _{\mu \tau }^{2}=3.\,\allowbreak 125\times 10^{-5}\;$%
and $\xi _{\mu \tau }^{2}=3.\,\allowbreak 125\times 10^{-7}$ respectively.
Looking at the case 1 in table (\ref{tab:mutao}) we see that these values
are not allowed, and hence they are excluded in Fig. \ref{fig:figg-24}. By
contrast, the value $\widetilde{\xi }_{\mu \tau }=2.5\times 10^{-2}\;$with $%
\tan \beta =2$, leads to$\;\xi _{\mu \tau }^{2}=$ $\allowbreak
3.\,\allowbreak 1\times 10^{-3}$ which is clearly allowed according to the
table (\ref{tab:mutao}) in the case 1.

In figure \ref{fig:figg-25} we suppose that $m_{h^{0}}=115$ GeV,$%
\;m_{H^{0}}=300$ GeV. The figure above shows the sensitivity of lower bounds
on $m_{A^{0}}\;$with the mixing angle $\alpha ^{\prime },\;$for
parametrization type II, taking $\tan \beta =20$. The value $\widetilde{\xi }%
_{\mu \tau }=2.5\times 10^{-4}\;$is excluded again. The constraints are very
sensitive to the $\alpha ^{\prime }\;$mixing angle for $\widetilde{\xi }%
_{\mu \tau }=2.5\times 10^{-3}\;$but less sensitive for $\widetilde{\xi }%
_{\mu \tau }=2.5\times 10^{-2}$. The figure below shows $m_{A^{0}}\;$vs\ $%
\tan \beta \;$for $m_{h^{0}}=115$ GeV, $m_{H^{0}}=300$ GeV, $\alpha ^{\prime
}=\pi /6,\;$ for parametrization type II and considering the $\widetilde{\xi 
}_{\mu \tau }=2.5\times 10^{-2}$ and $\widetilde{\xi }_{\mu \tau }=2.5\times
10^{-3}$.$\;$The value $\widetilde{\xi }_{\mu \tau }=2.5\times 10^{-4}$ is
excluded. The $m_{A^{0}}$ lower asymptotic limit for large $\tan \beta \;$is
approximately $m_{h^{0}}$.

\chapter{Addendum: Top-squark searches at the Fermilab Tevatron in models of
low-energy supersymmetry breaking}

\section{General framework}

This chapter describes the work developed at the Fermi National Accelerator
Laboratory, along with M. Carena, D. Choudhoury, H. E. Logan, and C.E.M.
Wagner \cite{stop}. It is about the perspectives for detection of the
lightest top squark at the Tevatron collider in the case in which such stop
comes from a supersymmetric model with low energy supersymmetry breaking
pattern. Since it is an addendum, I will provide only a very brief survey on
the framework of supersymmetric models and supersymmetry breaking schemes.
For more details I refer the reader to the literature \cite{primer, SUSYTeV}

One of the greatest problems of the SM is the so called ``naturalness
problem''. The radiative corrections of the Higgs mass squared depend
quadratically on the cut off scale energy used. If one hopes to have a Higgs
mass of the order of the EWSB scale, it is required a very precise fine
tuning (of about one part in 10$^{16}$), in order to get the cancellation
necessary at the Planck scale, the apparent necessity of such extremely
accurate cancellation is known as the naturalness problem. On the other
hand, it is well known that such radiative corrections with fermions and
bosons into the loops have opposite signs. Consequently, if we had a fermion
of the same mass for each boson and vice versa, the necessary cancellation
would occur. Supersymmetry provides a framework in which each SM particle
posseses a superpartner i.e. a particle with the same mass and quantum
numbers under the gauge symmetry but with different spin. Therefore,
imposition of SUSY provides an elegant solution to the naturalness problem.

Supersymmetry consists of an extension to the Poincar\'{e} group by the
introduction of new generators. The algebra of the group is determined by a
set of conmuting and anticonmuting relations among the poincar\'{e}
generators and the new generators. The new supersymmetric generators permits
to relate fermion fields with boson fields providing a unified description
of them. In order to keep the supersymmetry, the spectrum of the SM should
be doubled arising the so called superpartners.

Nevertheless, since the superpartners have not been detected yet, they
cannot have the same mass as their corresponding SM partners. Consequently,
supersymmetry must be broken in such a way that the superpartners should be
heavier than their partners. However, such breaking have to preserve the
cancellation of quadratic divergencies that was one of the original
motivations, this fact imposes constraints on the splittings among the
partner and superpartner masses, in general such splitting cannot be larger
than a few TeV. As a consequence, some of the lightest superpartners could
have masses within the reach of accelerators like the Tevatron or the LHC.
As we shall see later, the lightest stop (i.e. one of the the superpartners
of the top quark), is a good candidate to be one of the lightest
supersymmetric particles. Before discussing that point, let us describe
briefly the particle content of the Minimal Supersymmetric Standard Model
(MSSM), and some breaking schemes of supersymmetric models.

\subsection{The MSSM}

In the MSSM, the particle spectrum of the SM is doubled. On the other hand,
in order to get cancellation of anomalies and preserve SUSY, two Higgs
doublets must be introduced. The Higgs spectrum is similar to the one of the
general 2HDM i.e. two CP-even scalars ($H^{0},h^{0}$), one CP-odd scalar ($%
A^{0}$),\ and two charged Higgs bosons ($H^{\pm }$). The Higgs bosons and
the SM particles have superpartners with the same quantum numbers under the
gauge symmetry but with different spin. The superpartners of the gauge
bosons (gauginos) are spin 1/2\ particles denoted as photinos ($\widetilde{%
\gamma }$),$\;$Winos ($\widetilde{W}$), Zinos ($\widetilde{Z}$)$\;$and
gluinos ($\widetilde{g}$). The fermionic superpartners of the Higgs bosons
(Higgsinos) are $\widetilde{H}_{1},\widetilde{H}_{2},\widetilde{H}^{\pm }$.
Owing to the EWSB, the Higgsinos and the gauginos of the electroweak
symmetry can be mixed to give the mass eigenstates, arising two charged
Dirac fermions called charginos $\widetilde{\chi }^{\pm }$, and four neutral
majorana fermions (the neutralinos $\widetilde{\chi }_{1-4}^{0}$). By
contrast, the gaugino of the color symmetry (gluino $\widetilde{g}$) does
not mix with Higgsinos nor electroweak gauginos, because it belongs to an
unbroken symmetry. Finally, the spin-$0\;$partners of the fermion fields
(the sfermions) are the squarks $\widetilde{q}$, charged$\;$sleptons $%
\widetilde{l}\;$and sneutrinos\ $\widetilde{\nu }$. There are two
superpartners associated to each quark or charged slepton, one for each
chirality, the left-handed sfermions transform as $SU\left( 2\right) _{L}\;$%
doublets while right-handed ones are $SU\left( 2\right) _{L}\;$singlets,
there are only left-handed sneutrinos because of the lack of right-handed
partners in the SM. The gluino is a $SU\left( 3\right) _{C}\;$octet while
the squarks are $SU\left( 3\right) _{C}\;$triplets.

As well as the interactions of SUSY particles with SM particles, the
Lagrangian of Supersymmetry should contain some soft SUSY breaking terms.
The term ``soft'' means that these terms breaks SUSY and generates
splittings among the SUSY particles and their partners but maintaining the
gauge symmetry and the cancellation of the quadratic divergencies mentioned
above. The soft SUSY breaking parameters are mass terms for gauginos and
sfermions as well as trilinear scalar couplings. The number of those kind of
terms depend on the specific SUSY breaking scheme.

The chargino and neutralino masses and mixing angles are determined by the
gauge boson masses $M_{W}$, $M_{Z}$,$\;$the parameter $\tan \beta ,\;$the
SUSY Higgsino mass parameter\ $\mu \;$and the two soft breaking parameters
associated to the $SU\left( 2\right) _{L}\;$gaugino mass ($M_{1}$)$\;$and
the $U\left( 1\right) \;$gaugino mass\ ($M_{2}$), evaluated at the EW scale.
The neutralino mass matrix in the $\widetilde{B}-\widetilde{W}^{3}-%
\widetilde{H}_{1}-\widetilde{H}_{2}\;$basis is written as: 
\begin{eqnarray}
\mathbf{M}_{N} &=&\left( 
\begin{array}{cc}
\mathbf{M}_{i} & \mathbf{Z} \\ 
\mathbf{Z}^{T} & \mathbf{M}_{\mu }
\end{array}
\right) \;;\;\mathbf{M}_{i}=\left( 
\begin{array}{cc}
M_{1} & 0 \\ 
0 & M_{2}
\end{array}
\right) \;;\;\mathbf{M}_{\mu }=\left( 
\begin{array}{cc}
0 & -\mu \\ 
-\mu & 0
\end{array}
\right)  \notag \\
\mathbf{Z} &=&\left( 
\begin{array}{cc}
-M_{Z}\cos \beta \sin \theta _{W} & M_{Z}\sin \beta \sin \theta _{W} \\ 
M_{Z}\cos \beta \cos \theta _{W} & -M_{Z}\sin \beta \cos \theta _{W}
\end{array}
\right)  \label{neutralino matrix}
\end{eqnarray}
The values of the parameters $M_{1},\;M_{2},\;\mu ,\;\tan \beta \;$%
determines the Higgsino and gaugino content of the neutralinos. There are
some interesting limits in which the Higgsino and gaugino components become
simpler. As a manner of example, if $\left| \mu \right| >>M_{Z}\;$and $%
M_{1},M_{2}\approx M_{Z}$, with $M_{1}<M_{2}$, the lightest neutralinos are
gaugino-like and the heaviest are Higgsino like, the eigenvalues of the
matrix $\mathbf{M}_{N}\;$in Eq. (\ref{neutralino matrix}) yield the
following spectrum 
\begin{equation*}
M_{\widetilde{\chi }_{1}^{0}}\approx M_{1}\;;\;M_{\widetilde{\chi }%
_{2}^{0}}\approx M_{2}\;;\;M_{\widetilde{\chi }_{3}^{0}}\approx M_{%
\widetilde{\chi }_{4}^{0}}\approx \left| \mu \right|
\end{equation*}
In particular, we shall work under the assumption that the lightest
neutralino is a pure bino $\widetilde{\chi }_{1}^{0}\approx \widetilde{B}$,$%
\;$details later on. The Higgsino and gaugino composition affects strongly
the couplings of the neutralinos to gauge bosons and sfermions leading to
subtantial differences in production and decay rates.

Similarly, the mass matrix for the two charginos can be written in the $%
\widetilde{W}^{+}-\widetilde{H}^{+}$ basis, and the production and decay
rates depend on their gaugino and Higgsino composition as well.

As for the squarks and sleptons, their mass eigenstates are in general
combinations of left-handed and right-handed components. The soft SUSY
breaking sfermion mass parameters are strongly constrained by the
suppression of FCNC, requiring that (1) the soft SUSY breaking sfermion mass
matrix be diagonal and degenerate or (2) the masses of the first and second
generation sfermions be very large.

The mixing among left and right-handed sfermions depends on the mass of the
corresponding SM fermion. So those mixings are negligible for squarks and
sleptons of the first and second generations but could be significant for
the fermions of the third generation. In particular, the mass matrix for the
stops in the $\widetilde{t}_{L},\;\widetilde{t}_{R}\;$basis reads 
\begin{equation*}
M_{\widetilde{t}}^{2}=\left( 
\begin{array}{cc}
m_{Q_{3}}^{2}+m_{t}^{2}+D_{\widetilde{t}_{L}} & m_{t}\left( A_{t}-\frac{\mu 
}{\tan \beta }\right) \\ 
m_{t}\left( A_{t}-\frac{\mu }{\tan \beta }\right) & 
m_{U_{3}}^{2}+m_{t}^{2}+D_{\widetilde{t}_{R}}
\end{array}
\right)
\end{equation*}
where $m_{Q_{3}}^{2},\;A_{t}\;$are SUSY breaking parameters and 
\begin{eqnarray*}
D_{\widetilde{f}_{L}} &=&M_{Z}^{2}\cos \left( 2\beta \right) \left(
T_{3f}-Q_{f}\sin ^{2}\theta _{W}\right) \\
D_{\widetilde{f}_{R}} &=&M_{Z}^{2}\cos \left( 2\beta \right) Q_{f}\sin
^{2}\theta _{W}
\end{eqnarray*}
Unless there is a cancellation of the factor $\left( A_{t}-\frac{\mu }{\tan
\beta }\right) $, the mixing will be significant because of the large top
mass. This large mixing in turn produces a large splitting between the
lightest ($\widetilde{t}_{1}$)\ and the heaviest ($\widetilde{t}_{2}$)\ top
squarks. This fact makes the lightest stop a good candidate to be one of the
lightest superparticles. In this work we shall assume that the stop is the
next-to-next-to lightest sparticle, details below.

Another important issue on supersymmetric models is the $R-$parity, such
discrete symmetry is defined as\ $R=\left( -1\right) ^{2S+3B+L}$, where $S\;$%
is the particle spin, $B\;$the baryon number, and $L\;$the lepton number. It
can be checked that this quantum number is $R=1$ for the SM particles and $%
R=-1$ for the sparticles, if this symmetry were conserved, a SUSY particle
could not decay into just SM particles. Specifically, under $R-$parity
conservation each vertex should contain an even number of sparticles. An
inmediate consequence is that the lightest supersymmetric particle (LSP)
must be absolutely stable in the case of $R-$parity conservation. Thus, from
the point of view of collider physics the track of such LSP would be missing
energy and momentum. The violation of $R-$parity conduces to the violation
of lepton and/or baryon number. We shall restrict to the case of $R-$parity
conservation.

\subsection{SUSY breaking schemes}

As we have explained above, supersymmetry must be broken and there is a
variety of breaking mechanisms. We shall discuss briefly two schemes:
supergravity and gauge mediated susy breaking models.

In supergravity (SUGRA) it is assumed the existence of additional
superfields (the hidden sector) that couples to the MSSM particles by means
of gravitational interactions. After the SUSY breaking, some components of
the hidden sector acquire VEV, the soft SUSY breaking terms arise from the
interaction among the MSSM superfields with the components of the hidden
sector that acquire a VEV, from which it follows that the strength of these
SUSY breaking terms is proportional to those VEV divided by the Planck
scale. Despite the number of SUSY breaking terms is enormous in those kind
of models, in minimal SUGRA they are reduced considerably because the MSSM
sparticles couples universally to the hidden sector. This universality is
extended to the MSSM mass pattern at a scale of the order of $M_{\text{Planck%
}}\left( \sim 10^{19}\text{GeV}\right) $ or at a scale of the order of GUT $%
\left( \sim 10^{16}\text{GeV}\right) $; at such a high scale the scalars
(Higgs bosons and sfermions) are assumed to have a common mass $m_{0}$, all
gauginos (Bino, Wino, and gluino) have a common mass $m_{1/2}$, and all
trilinear couplings have a common strength as well ($A_{0}$). All these
values at GUT or Planck scale can be run by Renormalization Group Equations
(RGE) to obtain their values at EW scale. Evolution of the mass pattern by
RGE shows that the lightest two neutralinos and the lightest chargino tend
to be gaugino like. In addition, squarks are in general heavier than
sleptons. Finally, the lightest neutralino seems to be a good candidate to
be the LSP, except in the cases in which $m_{0}\;$or $m_{1/2}\;$are very
small, in those cases the best candidates become the sneutrino and the
gluino respectively.

As for the scale of supersymmetry breaking, in SUGRA models it is expected
to be very high ($\sim 10^{11}$GeV if we expect the SUSY masses to lie at
the TeV scale) as we shall discuss in next section. Thus, SUGRA are in
general models with high energy supersymmetry breaking scale.

Another interesting scheme is the gauge mediated SUSY breaking, in which the
SUSY breaking terms are generated from gauge interactions. In these models
the masses of sfermions with the same quantum numbers under the gauge group
i.e. with the same gauge couplings are predicted to be degenerate, in this
way FCNC are naturally suppressed. As we shall see in next section, in Gauge
Mediated Spontaneously Breaking (GMSB) models the scale of SUSY breaking is
much smaller than the Planck or GUT scale avoiding corrections at those
scales to the degeneracy. The existence of heavy messenger superfields is
assumed. The breaking of SUSY occurs in a hidden sector which also couples
to the messenger superfields, thus the fermion components of the messenger
superfields acquire a commom mass $M\;$and the scalar components acquire a
commom mass $M\sqrt{1\pm x}\;$where $x\;$is a dimensionless parameter
related to the breaking scale. The gauginos and sfermions masses receive
radiative contributions from the messenger fields that generates a splitting
among them and their corresponding SM partners. Gaugino and scalar masses
lie roughly on the same order of magnitude. Moreover, after evolving by RGE,
sfermions with the same quantum numbers continue degenerate if we ignore the
effect of Yukawa couplings\footnote{%
In this work we assume that there are ten degenerate squarks corresponding
to the five light quarks. Only the top squarks are considered to have an
splitting between them, owing to the large Yukawa coupling of the top quark.}%
, therefore the mass hierarchy is directly related to the gauge coupling
strength; the gauge coupling pattern also determines the gaugino mass
hierarchy. In such models the gravitino might be very light, and it is a
good candidate to be the LSP, in that case it plays a crucial role in low
energy phenomenology. In this work the gravitino will be the LSP.

\section{Introduction}

The Standard Model with a light Higgs boson provides a very good description
of all experimental data. The consistency of the precision electroweak data
with the predictions of the Standard Model suggests that, if new physics is
present at the weak scale, it is most probably weakly interacting and
consistent with the presence of a light Higgs boson in the spectrum.
Extensions of the Standard Model based on softly broken low energy
supersymmetry (SUSY) \cite{HaberKaneNilles} provide the most attractive
scenarios of physics beyond the Standard Model fulfilling these properties.
If the supersymmetry breaking masses are $\mathrel{\raise.3ex\hbox{$<$%
\kern-.75em\lower1ex\hbox{$\sim$}}}\mathcal{O}(1\;\mathrm{TeV})$,
supersymmetry stabilizes the hierarchy between the Planck scale $M_{P}$ and
the electroweak scale. Furthermore, the minimal supersymmetric extension of
the Standard Model (MSSM) significantly improves the precision with which
the three gauge couplings unify and leads to the presence of a light Higgs
boson with a mass below about 128 GeV \cite{lightHiggsmass}\footnote{%
This upper limit for the lightest CP even Higgs boson could be moved up in
some non-minimal supersymmetric versions. For instance, Ref. \cite{SU3U1us}
shows that in the supersymmetric version of the SU(3)$\times $U(1)\ gauge
model in which the Higgs triplets are included in the lepton superfields,
the upper limit for the lightest CP even Higgs boson can be shifted up to
about 140 GeV.}.

Perhaps the most intriguing property of supersymmetry is that local
supersymmetry naturally leads to the presence of gravity (supergravity). In
the case of local supersymmetry, the Goldstino provides the additional
degrees of freedom necessary to make the gravitino a massive particle~\cite
{Fayet}. In the simplest scenarios, the gravitino mass $m_{\tilde G}$ is
directly proportional to the square of the supersymmetry breaking scale $%
\sqrt{F_{\mathrm{SUSY}}}$: 
\begin{equation}
m_{\tilde{G}} \simeq F_{\mathrm{SUSY}}/\sqrt{3} M_P \ ,  \label{eq:mgrav}
\end{equation}
where $M_P$ denotes the Planck mass.

The relation between the supersymmetry breaking scale $\sqrt{F_{\mathrm{SUSY}%
}}$ and the masses of the supersymmetric partners depends on the specific
supersymmetry breaking mechanism. In general, the superpartner masses $M_{%
\mathrm{SUSY}}$ are directly proportional to $F_{\mathrm{SUSY}}$ and
inversely proportional to the messenger scale $M_m$ at which the
supersymmetry breaking is communicated to the visible sector: 
\begin{equation}
M_{\mathrm{SUSY}} \simeq C_{M} \frac{F_{\mathrm{SUSY}}}{M_m},
\label{eq:Msusy}
\end{equation}
where $C_{M}$ is the characteristic strength of the coupling between the
messenger sector and the visible one. If the breakdown of supersymmetry is
related to gravity effects, $M_m$ is naturally of the order of the Planck
scale and $C_M$ is of order one; hence, for $\sqrt{F_{\mathrm{SUSY}}} \sim
10^{11}$ GeV, $M_{\mathrm{SUSY}}$ is naturally at the TeV scale. In gauge
mediated scenarios (GMSB)~\cite{gms,DineLESB}, instead, the couplings $C_M$
are associated with the Standard Model gauge couplings (times a loop
suppression factor), so that a $F_{\mathrm{SUSY}}/M_m \overset{<}{{}_\sim}
100$ TeV yields masses of the order of 100 GeV for the lighter Standard
Model superpartners.

When relevant at low energies, the gravitino interactions with matter are
well described through the interactions of its spin 1/2 Goldstino component~ 
\cite{Fayet}. The Goldstino has derivative couplings with the visible sector
with a strength proportional to $1/F_{\mathrm{SUSY}}$. In scenarios with a
high messenger scale, of order $M_P$, Eqs.~\ref{eq:mgrav} and \ref{eq:Msusy}
imply that the gravitino has a mass of the same order as the other SUSY
particles, and its interactions are extremely weak. In such scenarios, the
gravitino plays no role in the low-energy phenomenology. However, in low
energy supersymmetry breaking scenarios such as GMSB in which the messenger
scale is significantly lower than the Planck scale, the supersymmetry
breaking scale is much smaller. Typical values in the GMSB case are $M_m
\sim 10^5 - 10^8$ GeV, leading to a supersymmetry breaking scale $\sqrt{F_{%
\mathrm{SUSY}}}$ roughly between $10^5$ and a few times $10^6$ GeV. The
gravitino then becomes significantly lighter than the superpartners of the
quarks, leptons and gauge bosons, and its interaction strength is larger. As
the lightest supersymmetric particle (LSP), the gravitino must ultimately be
produced at the end of all superparticle decay chains if $R$-parity is
conserved (for an analysis of the case of $R$-parity violation see, Ref.~ 
\cite{rpgm}).

Depending on the strength of the gravitino coupling, the decay length of the
next-to-lightest supersymmetric particle (NLSP) can be large (so that the
NLSP is effectively stable from the point of view of collider phenomenology;
this occurs when $\sqrt{F_{\mathrm{SUSY}}} \mathrel{\raise.3ex\hbox{$>$%
\kern-.75em\lower1ex\hbox{$\sim$}}} 1000$ TeV), intermediate (so that the
NLSP decays within the detector giving rise to spectacular displaced vertex
signals; this occurs when 1000 TeV $\mathrel{\raise.3ex\hbox{$>$\kern-.75em%
\lower1ex\hbox{$\sim$}}} \sqrt{F_{\mathrm{SUSY}}} \mathrel{\raise.3ex%
\hbox{$>$\kern-.75em\lower1ex\hbox{$\sim$}}} 100$ TeV), or microscopic (so
that the NLSP decays promptly; this occurs when $\sqrt{F_{\mathrm{SUSY}}} %
\mathrel{\raise.3ex\hbox{$<$\kern-.75em\lower1ex\hbox{$\sim$}}} 100$ TeV)~ 
\cite{SUSYRun2,at2,KaneTev}. The decay branching fractions of the Standard
Model superpartners other than the NLSP into the gravitino are typically
negligible. However, if the supersymmetry breaking scale is very low, $\sqrt{%
F_{\mathrm{SUSY}}} \ll 100$ TeV (corresponding to a gravitino mass $\ll 1$
eV), then the gravitino coupling strength can become large enough for
superpartners other than the NLSP to decay directly into final states
containing a gravitino \cite{KaneTev,Zwirner}. In any case, the
SUSY-breaking scale must be larger than the mass of the heaviest
superparticle; an approximate lower bound of $\sqrt{F_{\mathrm{SUSY}}}
\simeq 1$ TeV corresponds to a gravitino mass of about $10^{-3}$ eV.

In many models, the lightest amongst the supersymmetric partners of the
Standard Model particles is a neutralino, $\tilde \chi^0_1$. The partial
width for $\tilde \chi^0_1$ decaying into the gravitino and an arbitrary SM
particle $X$ is given by 
\begin{equation}
\Gamma(\tilde{\chi}_1^0 \to X \tilde{G}) \simeq K_X N_X \frac{m_{\tilde
\chi_1^0}}{96 \pi} \left(\frac{m_{\tilde \chi_1^0}} {\sqrt{M_P m_{\tilde{G}}}%
}\right)^4 \left( 1 - \frac{m_X^2}{m_{\tilde \chi_1^0}^2} \right)^4,
\label{rategm}
\end{equation}
where $K_X$ is a projection factor equal to the square of the component in
the NLSP of the superpartner $\tilde X$, and $N_X$ is the number of degrees
of freedom of $X$. If $X$ is a photon, for instance, $N_X = 2$ and 
\begin{equation}
K_X = \left| N_{11} \cos\theta_W + N_{12} \sin\theta_W \right|^2,
\end{equation}
where $N_{ij}$ is the mixing matrix connecting the neutralino mass
eigenstates to the weak eigenstates in the basis $\tilde{B}, \tilde{W}, 
\tilde{H}_1, \tilde{H}_2$.

If the neutralino has a significant photino component, it will lead to
observable decays into photon and gravitino. Since the heavier
supersymmetric particles decay into the NLSP, which subsequently decays into
photon and gravitino, supersymmetric particle production will be
characterized by events containing photons and missing energy. This is in
contrast to supergravity scenarios, where, unless very specific conditions
are fulfilled~\cite{radneu,m2eqm1}, photons do not represent a
characteristic signature. The presence of two energetic photons plus missing
transverse energy provides a distinctive SUSY signature with very little
Standard Model background.

It might be argued that a $\tilde \chi^0_1$, decaying with a large branching
ratio into photons, would be severely constrained by LEP data. However, such
bounds are extremely model-dependent. For example, there exists no
tree-level coupling between a photon (or $Z$) and either two binos or two
neutral winos. Since the bino is associated with the smallest of all gauge
interactions and, in addition, its mass is more strongly renormalized
downward at smaller scales compared to the wino mass, in many models the
lightest neutralino has a significant bino component. Therefore, if the NLSP
is approximately a pure bino, the bounds on its mass depend strongly on the
selectron mass (since, at LEP, pair-production of binos or neutral winos
could occur through $t$-channel exchange of selectrons). For selectron
masses below 200 GeV, the present bound on such a neutralino is
approximately 90 GeV~\cite{LEPSUSYWG} (the bound weakens with increasing
selectron mass). As emphasized before, once produced, such a neutralino
would decay via $\tilde \chi^0_1 \to \gamma \tilde G$. If $\tilde \chi^0_1$
is heavy enough, then the decays $\tilde \chi^0_1 \to Z \tilde G$ and $%
\tilde \chi^0_1 \to h^0 \tilde G$ are also allowed; however, the decay
widths into these final states are kinematically suppressed compared to the $%
\gamma \tilde G$ final state and will be important only if either the
photino component of $\tilde \chi^0_1$ is small or if $\tilde \chi^0_1$
itself is significantly heavier than $Z$ and $h^0$ \cite{KaneTev}.

In most SUSY models, it is natural for the lighter top squark, $\tilde t_1$,
to be light compared to the other squarks. In general, due to the large top
Yukawa coupling, there is a large mixing between the weak eigenstates $%
\tilde t_L$ and $\tilde t_R$, which leads to a large splitting between the
two stop mass eigenstates. In addition, even if all squarks have a common
mass at the messenger scale, the large top Yukawa coupling typically results
in the stop masses being driven (under renormalization group evolution) to
smaller values at the weak scale. An extra motivation to consider light
third generation squarks comes from the fact that light stops, with masses
of about or smaller than the top quark mass, are demanded for the
realization of the mechanism of electroweak baryogenesis within the context
of the MSSM \cite{EWBG}.

In this chapter, we examine, in detail, the production and decay of top
squarks at Run II of the Tevatron collider in low-energy SUSY breaking
scenarios wherein the lightest neutralino is the NLSP and decays promptly
into $\gamma \tilde G$. We also investigate the production and decay of the
other squarks, and provide an estimate of the reach of Run II of the
Tevatron in the heavy gluino limit. We work in the context of a general SUSY
model in which the SUSY particle masses are \textit{not} constrained by the
relations predicted in the minimal GMSB models. We assume throughout that
the gravitino coupling is strong enough (or, equivalently, that the scale of
SUSY breaking is low enough) that the NLSP decays promptly. This implies an
upper bound on the supersymmetry breaking scale of a few tens to a few
hundred TeV~\cite{at2,KaneTev,SUSYRun2}, depending on the mass of the NLSP.
Our analysis can be extended to higher supersymmetry breaking scales for
which the NLSP has a finite decay length, although in this case some signal
will be lost on account of the NLSP decaying outside the detector. At least
50\% of the diphoton signal cross section remains though for NLSP decay
lengths $c\tau \mathrel{\raise.3ex\hbox{$<$\kern-.75em\lower1ex\hbox{$%
\sim$}}} 40$ cm~\cite{SUSYRun2}; this corresponds to a supersymmetry
breaking scale below a few hundred to about a thousand TeV, depending on the
mass of the NLSP. Moreover, the displaced vertex associated with a finite
decay length could be a very good additional discriminator for the signal.
Thus, in totality, our choice is certainly not an overly optimistic one.

This chapter is organized as follows. In Sec.~\ref{sec:light_stop}, we
outline models of low-energy supersymmetry breaking wherein the lighter stop
is the lightest sfermion and, moreover, is lighter than the charginos as
well as the gluino. In the following section, we review the stop pair
production cross section at the Tevatron. In Sec.~\ref{sec:literature}, we
summarize previous studies of stop production and decay at Run II of the
Tevatron. This is followed, in Sec.~\ref{sec:stopBRs}, by a discussion of
the SUSY parameter space and the relative partial widths of the various stop
decay modes. In Sec.~\ref{sec:signals}, we describe the signal for each of
the stop decay modes considered. We describe the backgrounds and the cuts
used to separate signal from background in each case, and give signal cross
sections after cuts. This gives the reach at Run~II. We also note the
possibility that stops can be produced in the decays of top quarks. In Sec.~%
\ref{sec:10squarks} we consider the production and decay of 10 degenerate
squarks that are the supersymmetric partners of the five light quarks.
Finally, we summarize our results (along with the results concerning the
previous chapters of this Ph. D. thesis) in chapter~\ref{sec:conclusions}.


\section{Light Stop in Low Energy Supersymmetry Breaking Models}

\label{sec:light_stop} As discussed above, the mass of the gravitino as well
as its interaction strength are governed by the supersymmetry breaking scale 
$\sqrt{F_{\mathrm{SUSY}}}$. Low energy supersymmetry breaking models are
defined as those obtained for low values of $\sqrt{F_{\mathrm{SUSY}}}$ ($%
\overset{<}{{}_{\sim }}10^{6}$~GeV) and hence resulting in a gravitino
lighter than a few keV. Apart from evading cosmological problems~\cite
{Moroi:1993mb}, a further striking consequence of such models is that
sparticle decays into gravitinos may occur at scales that may be of interest
for collider phenomenology. As is apparent from Eq. (\ref{eq:Msusy}), in
such models the messenger mass scale $M_{m}$ should be smaller than $10^{9}$
GeV.

In this chapter, we are interested in the presence of light stops in low
energy supersymmetry breaking scenarios. The motivation is simple: assuming
that the supersymmetry breaking masses are flavor independent, that is the
left- and right-handed squark masses of the three generations are the same
at the messenger scale, the lighter stop turns out to be the lightest of all
the squarks. The reasons are twofold. On the one hand, there are
renormalization group effects induced by top-quark Yukawa interactions that
tend to reduce the stop mass scale compared to the other squark masses. On
the other, there are non-trivial mixing effects that tend to push the
lightest stop mass down compared to the overall left- and right-squark
masses. For similar reasons, it is also natural to assume that the lightest
stop will be lighter than the gluino.

There is a further motivation behind our analysis, namely that of
baryogenesis. A crucial requirement for electroweak baryogenesis scenarios
within the Minimal Supersymmetric Extension of the Standard Model is the
presence of a light stop with mass of the order of, or smaller than, the top
quark mass. Since this scenario does not depend on the nature of
supersymmetry breaking, it is very important to develop strategies to look
for light top squarks in all their possible decay modes, in particular in
those related to the possibility of low energy supersymmetry breaking.

In general, light stops can induce large corrections to the precision
electroweak parameter $\Delta \rho $ \cite
{Barbieri:1983wy,Alvarez-Gaume:1983gj,Chankowski:1993eu}, unless they are
mainly right-handed or there is some correlation between the masses and
mixing angles in the stop and sbottom sector \cite{Carena:1997xb}. The most
natural way of suppressing potentially large contributions to the
rho-parameter is to assume that there is a large hierarchy between the
supersymmetry breaking mass parameters for the left- and right-handed stops.
The simplest and most efficient way of doing this is to assume that
sparticles which are charged under the weak gauge interactions acquire large
supersymmetry breaking masses. Observe that, under this assumption, the
charged wino will also be naturally heavier than the lightest stop and, for
simplicity, we will assume that both charginos are heavier than the particle
under study. We will further assume that the superpartner of the $U(1)_{Y}$
gauge boson, the so-called bino, is the lightest standard model superpartner.

The above mentioned properties are naturally obtained in simple extensions
of the minimal gauge mediated model. Indeed, let us assume that there are $N$
copies of messengers, which transform as complete representations (${} + 
\bar{}$) of $SU(5)$ and hence do not spoil the unification relations. Under
the Standard Model gauge group, some of these fields will then transform as
left-handed lepton multiplets ($W_i$, $i = 1 \dots N$) and their mirror
partners ($\bar{W}_i$), while the others would transform as a right-handed
down quark multiplet ($C_i$) and its mirror partner ($\bar{C}_i$). We shall
assume that $N \leq 4$ in order to keep the gauge couplings weak up to the
grand unification scale~\cite{Martin:1996zb}. Let us further assume that $%
W_i $ and $C_i$ couple to two different singlet fields $S_{2,3}$, with 
\begin{equation}
\langle S_j \rangle = S_j + \theta^2 F_j \ ,
\end{equation}
and, for simplicity, assume that all couplings are of order one and that $%
F_j \ll S_j^2$. We will further assume that $S_1 \simeq S_2 \simeq S$, where 
$S$ characterizes the messenger scale.

In this simple case, the masses of the gauginos at the messenger scale will
be given by 
\begin{eqnarray}
M_3 & = & N \frac{\alpha_3}{4\pi} \Lambda_3 \ ,  \notag \\
M_2 & = & N \frac{\alpha_2}{4\pi} \Lambda_2 \ ,  \notag \\
M_1 & = & N \frac{\alpha_1}{4\pi} \left( \frac{2 \; \Lambda_3}{5} + \frac{3
\; \Lambda_2}{5} \right) \ ,
\end{eqnarray}
where $\Lambda_3 \approx F_3/S_3$ and $\Lambda_2 \approx F_2/S_2$~\cite
{Martin:1996zb,carlos}. Now, it is easy to see that for $\Lambda_2 > 3
\Lambda_3$, the weak gaugino can have a mass similar to that of the gluino
(or be even heavier), while the bino is still much lighter than the wino
(about three times lighter than the wino at the messenger scale).

The squared scalar masses at the messenger scale are also proportional to
the number of messengers $N$ 
\begin{eqnarray}
m_Q^2 & = & 2 N \left[\sum_{a = 2,3} C_Q^a \left(\frac{\alpha_a}{4\pi}%
\right)^2 \Lambda_a^2 + C_Q^1 \left(\frac{\alpha_1}{4\pi}\right)^2 \left( 
\frac{2}{5} \Lambda_3^2 + \frac{3}{5} \Lambda_2^2 \right) \right] \ ,
\end{eqnarray}
where, for a particle transforming in the fundamental representation of $%
SU(n)$, $C_Q^n = (n^2 - 1)/(2n)$, while $C_Q^1 = 3/5 (Q - T_3)^2$. The above
quoted masses should be renormalized to the weak scale. The details of this
procedure in a general gauge mediated model have been given by one of us in
Ref.~\cite{carlos}. We shall not repeat these expressions here. For our
purpose, it suffices to stress some important details.

As we mentioned before, after renormalization and mixing effects are added,
the right-handed stop is the lightest squark in the spectrum. The large
value of the left-handed stop mass increases the negative Yukawa dependent
radiative corrections to the right-handed stop mass. Therefore, for $\Lambda
_{2}$ larger than a few times $\Lambda _{3}$ and $N>1$, the lightest stop is
lighter than the gluino and the wino. Even for $N=1$ this tends to be true,
for not too small values of the messenger scale ($S_{3}\overset{>}{{}_{\sim }%
}10^{7}$ GeV). Notice that, for $\Lambda _{2}$ larger by an order of
magnitude than $\Lambda _{3}$ and/or large values of $N$, the right-stop
(or, in extreme cases the right-sbottom) becomes the lightest standard model
superpartner. We shall thus concentrate on cases where $\Lambda _{2}$,
though larger than $\Lambda _{3}$, is still of the same order.

The sleptons do not play a relevant role in stop decays as long as the
charginos are heavier than the lightest stop. One has only to ensure that
the lightest slepton does not become lighter than the bino. The hierarchy of
the right-handed slepton mass and the bino mass is approximately given by 
\begin{equation}
m_{\tilde{e}_{R}}^{2}\approx \frac{2M_{1}^{2}}{N}r_{1}\ ,
\end{equation}
where $r_{1}$ is the appropriate renormalization group factor relating the
masses at the messenger scale to the masses at the weak scale. This factor
is about 1.5--1.7~\cite{Martin:1996zb} in the region of interest for this
work and therefore we are led to conclude that, as long as $N\leq 3$, the
bino is the next-to-lightest supersymmetric particle. Observe that this
constraint on $N$ is somewhat weaker than in the minimal model, due to the
appearence of the factor $3/5$ in front of the dominant $\Lambda _{2}$
contribution.

In all of the above, we have not discussed the process of radiative
electroweak symmetry breaking and the determination of the $\mu$ parameter.
This is due to the fact that in the minimal gauge mediated models of the
kind we described above, the generation of a proper value of the parameter $%
B \mu$ requires the presence of new physics that necessarily modifies the
value of the Higgs mass parameters~\cite{Dvali:1996cu}. It is, therefore,
justified to treat $\mu$ as an independent parameter in such models.

To summarize, we have seen that a mild modification of the simplest gauge
mediated models leads to a model where the stop is lighter than all the
other squarks and also lighter than the weak and strong gauginos, while the
bino remains the next-to-lightest supersymmetric particle. The unification
relations are preserved as long as the number of messenger families is less
than or equal to 3. The only condition for this to happen is that the
effective scale $\Lambda_2$ is a few times larger than $\Lambda_3$, but
still of the same order of magnitude. The relation between the stop mass and
the bino mass will be governed by the hierarchy between $\Lambda_2$ and $%
\Lambda_3$. This modification of the minimal gauge mediated models is well
justified in order to get consistency of a light stop with electroweak
precision observables. Some of the sleptons might be lighter than the stop,
but, as long as the charginos remain heavy, they have no impact on the stop
phenomenology.

Let us stress that the above should be considered only as a simple example
in which a light stop, mainly right-handed, can appear in the spectrum in
low energy supersymmetry breaking models. As the nature of supersymmetry
breaking is unknown, so is the exact spectrum. We only make the simplifying
assumptions that the left- and right-handed sfermions receive an
approximately common mass at the messenger scale and that the gaugino masses
are of the same order of magnitude as the squark masses. The truly defining
feature of our assumption is that the wino and the gluino are heavier than
the squarks and that the bino is the next-to-lightest supersymmetric
particle. Although the left-handed squarks tend to be heavier than the
right-handed ones, the exact spectrum obtained in the simple model detailed
above does not differ markedly from the simplified spectrum that we choose
to work with. In this kind of models, apart from the somewhat lighter stops,
there will be ten squarks approximately degenerate in mass, which will
predominantly decay into a quark and a bino, which will subsequently decay
into photons (or $Z$ bosons) and missing energy.

\section{Top squark production at Tevatron Run~II}

\label{sec:production}

Top squarks are produced at hadron colliders overwhelmingly via the strong
interaction, so that the tree-level cross sections are model independent and
depend only on the stop mass. The production modes of the lightest stop, $%
\tilde{t}_{1}$, at the Tevatron are $q\bar{q}\rightarrow \tilde{t}_{1}\tilde{%
t}_{1}^{\ast }$ and $gg\rightarrow \tilde{t}_{1}\tilde{t}_{1}^{\ast }$. The
cross sections for these processes are well known at leading order (LO) \cite
{Harrison,DEQ}, and the next-to-leading order (NLO) QCD and SUSY-QCD
corrections have been computed \cite{PROSPINOstops} and significantly reduce
the renormalization scale dependence. The NLO cross section is implemented
numerically in PROSPINO \cite{PROSPINO,PROSPINOstops}.

\begin{figure}[h]
\resizebox{0.77\textwidth}{!}{
\rotatebox{270}{\includegraphics{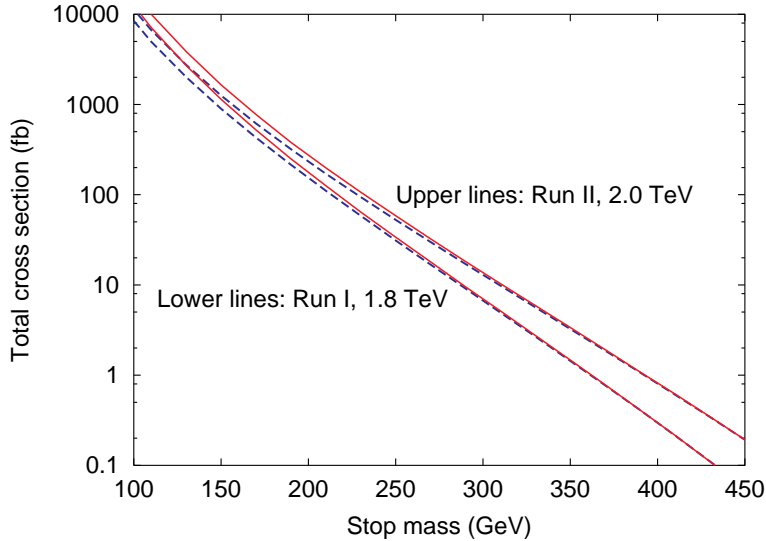}}}
\caption{LO (dashed) and NLO (solid) cross sections for stop pair production
in $p\bar{p}$ collisions at Tevatron Run I (1.8 TeV) and Run II (2.0 TeV),
from PROSPINO~\protect\cite{PROSPINO,PROSPINOstops}. Cross sections are
evaluated at the scale $\protect\mu =m_{\tilde{t}}$. }
\label{fig:xsecmass}
\end{figure}

We generate stop events using the LO cross section evaluated at the scale $%
\mu = m_{\tilde t}$, improved by the NLO K-factor~\footnote{%
Although gluon radiation at NLO leads to a small shift in the stop $p_T$
distribution to lower $p_T$ values~\cite{PROSPINO,PROSPINOstops}, we do not
expect this shift to affect our analysis in any significant way.} obtained
from PROSPINO \cite{PROSPINO,PROSPINOstops} (see Fig.~\ref{fig:xsecmass}).
The K-factor varies between 1 and 1.5 for $m_{\tilde t}$ decreasing from 450
to 100 GeV. We use the CTEQ5 parton distribution functions \cite{CTEQ5}. We
assume that the gluino and the other squarks are heavy enough that they do
not affect the NLO cross section. This is already the case for gluino and
squark masses above about 200 GeV \cite{PROSPINOstops}.

Top squarks can also be produced via cascade decays of heavier
supersymmetric particles, with a highly model dependent rate. To be
conservative, we assume that the masses of the heavier supersymmetric
particles are large enough that their production rate at Tevatron energies
can be neglected.


\section{Previous studies of stops at Run II}

\label{sec:literature}

A number of previous studies have considered the prospects for stop
discovery at Run~II of the Tevatron, which we summarize here. In general,
the most detailed SUSY studies have been done in the context of
supergravity; in this case SUSY is broken at the Planck scale so that the
gravitino plays no role in the collider phenomenology. Then the lightest
neutralino is the LSP and ends all superparticle decay chains. The reach of
the Tevatron for a number of stop decay modes has been analyzed in Refs.~ 
\cite{Baer,Demina}. The signal depends on the decay chain, which in turn
depends on the relative masses of various SUSY particles. For sufficiently
heavy stops, the decay $\tilde t \to t \tilde \chi_1^0$ will dominate. This
channel is of limited use at Run II because the stop pair production cross
section falls rapidly with increasing stop mass, and this channel requires $%
m_{\tilde t} > m_t + m_{\tilde \chi_1^0}$, which is quite heavy for the
Tevatron in the case of minimal supergravity.\footnote{%
In low-energy SUSY breaking scenarios, however, the mass range $m_{\tilde t}
> m_t + m_{\tilde \chi_1^0}$ is interesting at Tevatron energies because the
distinctive signal allows backgrounds to be reduced to a very low level, as
we will show.} For lighter stops, if a chargino is lighter than the stop
then $\tilde t \to b \tilde \chi_1^+$ tends to dominate (followed by the
decay of the chargino). The details of the signal depend on the Higgsino
content of $\tilde \chi_1^+$. If $\tilde \chi_1^+$ has a mass larger than $%
m_{\tilde{t}} - m_b$, the previous decay does not occur and the three-body
decay $\tilde t \to b W^+ \tilde \chi_1^0$ dominates; this decay proceeds
through the exchange of a virtual top quark, chargino, or bottom squark. If
the stop is too light to decay into an on-shell $W$ boson and $\tilde
\chi_1^0$, then the flavor-changing decay $\tilde t \to c \tilde \chi_1^0$
tends to dominate. Finally, if a sneutrino or slepton is light, then $\tilde
t \to b \ell^+ \tilde \nu_{\ell}$ or $\tilde t \to b \tilde \ell^+
\nu_{\ell} $, respectively, will occur (followed by the decays of the
slepton or the sneutrino, if it is not the LSP). At Run~II with 2 (20) fb$%
^{-1}$ of total integrated luminosity, in the context of minimal
supergravity one can probe stop masses up to 160 (200) GeV in the case of
the flavor changing decay, while stop masses as high as 185 (260) GeV can be
probed if the stop decays into a bottom quark and a chargino~\cite{Demina}.
A similar reach holds in the case of a light sneutrino~\cite{Demina} and in
the case of large $\tan\beta$ when the stop can decay into $b \tau \nu
\tilde \chi_1^0$ \cite{Djouadi}.

One can also search for stops in the decay products of other SUSY particles 
\cite{Demina}. In top decays, the process $t \to \tilde t \tilde g$ is
already excluded because of the existing lower bound on the gluino mass~%
\footnote{%
The bound on the gluino mass is, however, model dependent. Under certain
conditions, an allowed window exists for gluino masses below the gauge boson
masses~\cite{stuart,lightsb}.}. Other possibilities are $\tilde \chi^- \to b
\tilde t^*$ and $\tilde g \to t \tilde t^*$. Finally, the decays $\tilde b
\to \tilde t W^-$ and $\tilde t H^-$ have to compete with the preferred
decay, $\tilde b \to b \tilde \chi_1^0$, and so may have small branching
ratios (depending on the masses of $\tilde t$, $\tilde b$, $\tilde \chi_1^0$
and $H^-$). The signals for these processes at the Tevatron Run~II have been
considered in minimal supergravity in Ref.~\cite{Demina}.

If R-parity violation is allowed, then single stop production can occur via
the fusion of two quarks. Single stop production is kinematically favored
compared to stop pair production and offers the opportunity to measure
R-parity violating couplings. This has been considered for the Tevatron in
Ref.~\cite{Bergerstop}, which showed that the stop could be discovered at
Run II for masses below about 400 GeV provided that the R-parity violating
coupling $\lambda^{\prime\prime} > 0.02 - 0.1$ and that the stop decays via $%
\tilde t_1 \to b \tilde \chi_1^+$ (followed by $\tilde \chi_1^+ \to l^+
\nu_l \tilde \chi_1^0$).

Relatively few studies have been done in the context of low-energy SUSY
breaking with a gravitino LSP. A study of GMSB signals performed as part of
the Tevatron Run~II workshop \cite{SUSYRun2} considered the decays of
various SUSY particles as the NLSP. As discussed before, the NLSP in such
models will decay directly to the gravitino and Standard Model particles. If
the stop is the NLSP in such a model, then it will decay via $\tilde t \to
t^{(*)} \tilde G \to b W^+ \tilde G$ (for $m_{\tilde t} > m_b + m_W$). Note
that because $\tilde G$ is typically very light in such models, $m_{\tilde
t} > m_W + m_b$ is sufficient for this decay to proceed with an on-shell $W$
boson. Ref.~\cite{Demina} found sensitivity at Run~II to this decay mode for
stop masses up to 180 GeV with 4 fb$^{-1}$. This stop decay looks very much
like a top quark decay; nevertheless, even for stop masses near $m_t$, such
stop decays can be separated from top quark decays at the Tevatron using
kinematic correlations among the decay products \cite{Peskin}.

Finally, Ref.~\cite{KaneTev} considered general GMSB signals at the Tevatron
of the form $\gamma \gamma \not\!\!E_T + X$. The authors of Ref.~\cite
{KaneTev} provide an analysis of the possible bounds on the stop mass coming
from the Run I Tevatron data. They analyze the stop decay mode into a charm
quark and a neutralino, and also possible three body decays, by scanning
over a sample of models. They conclude that stop masses smaller than 140 GeV
can be excluded already by the Run I Tevatron data within low energy
supersymmetry breaking models independent of the stop decay mode, assuming
that $m_{\tilde \chi_1^0} > 70$ GeV.


\section{Top squark decay branching ratios}

\label{sec:stopBRs}

The decay properties of the lighter stop depend on the supersymmetric
particle spectrum. Of particular relevance are the mass splittings between
the stop and the lightest chargino, neutralino and bottom squark. In our
analysis, we assume that the charginos and bottom squarks are heavier than
the lighter stop, so that the on-shell decays $\tilde t \to \tilde b W$ and $%
\tilde t \to \tilde \chi^+ b$ are kinematically forbidden. Then the details
of the stop decay depend on the mass splitting between the stop and the
lightest neutralino. If $m_{\tilde t} < m_W + m_b + m_{\tilde \chi_1^0}$,
two decay modes are kinematically accessible:

\begin{enumerate}
\item  the flavor-changing (FC) two-body decay $\tilde{t}\rightarrow c\tilde{%
\chi}_{1}^{0}$. This two-body decay proceeds through a flavor-changing loop
involving $W^{+}$, $H^{+}$ or $\tilde{\chi}^{+}$ exchange or through a
tree-level diagram with a $\tilde{t}-\tilde{c}$ mixing mass insertion;

\item  the four-body decay via a virtual $W$ boson, $\tilde{t}\rightarrow
W^{+\ast }b\tilde{\chi}_{1}^{0}\rightarrow jjb\tilde{\chi}_{1}^{0}$ or $\ell
\nu b\tilde{\chi}_{1}^{0}$~\cite{4body}.
\end{enumerate}

The branching ratio of the stop decay into charm and neutralino strongly
depends on the details of the supersymmetry breaking mechanism. In models
with no flavor violation at the messenger scale, the whole effect is induced
by loop effects and receives a logarithmic enhancement which becomes more
relevant for larger values of the messenger mass. In the minimal
supergravity case, the two-body FC decay branching ratio tends to be the
dominant one. In the case of low energy supersymmetry breaking, this is not
necessarily the case. Since the analysis of the four-body decay process is
very similar to the three-body decay described below for larger mass
splittings between the stop and the lighter neutralino, here we shall
analyze only the case in which the two-body FC decay $\tilde t \to c \tilde
\chi^0_1$ is the dominant one whenever $m_{\tilde t} < m_W + m_b + m_{\tilde
\chi_1^0}$.

For larger mass splittings, so that $m_W + m_b + m_{\tilde \chi^0_1} <
m_{\tilde t} < m_t + m_{\tilde \chi^0_1}$, the three-body decay $\tilde t
\to W^+ b \tilde \chi_1^0$ becomes accessible, with $\tilde \chi^0_1 \to
\gamma \tilde G$. This stop decay proceeds through a virtual top quark,
virtual charginos, or virtual sbottoms. Quite generally, this 3-body decay
will dominate over the 2-body FC decay in this region of phase space.

For still heavier stops, $m_{\tilde t} > m_t + m_{\tilde \chi_1^0}$, the
two-body tree-level decay mode $\tilde t \to t \tilde \chi_1^0$ becomes
kinematically accessible and will dominate. Although the 3-body and 2-body
FC decays are still present, their branching ratios are strongly suppressed.

\begin{figure}[h]
\resizebox{0.77\textwidth}{!}{
\rotatebox{270}{\includegraphics{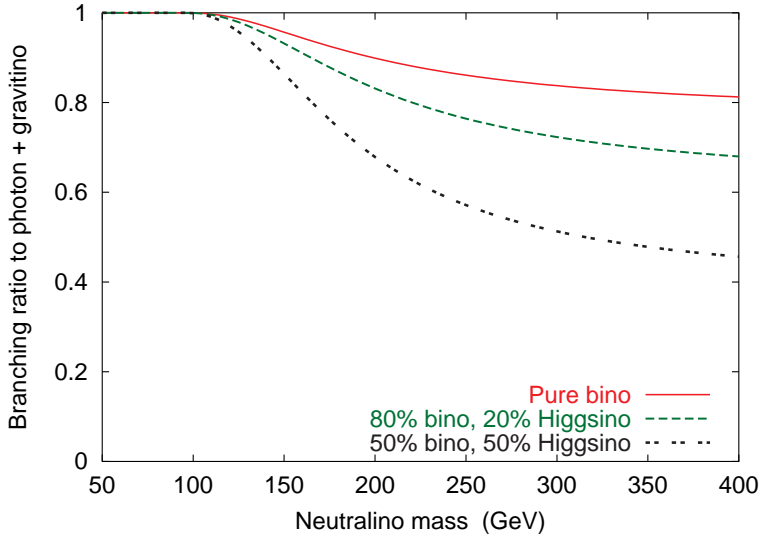}}
}
\caption{Branching ratio for the decay of the lightest neutralino into a
photon and a gravitino as a function of the neutralino mass. Shown are the
branching ratio if the neutralino is a pure bino (solid line) and for 20\%
and 50\% Higgsino admixtures (long and short dashed lines, respectively)
assuming that $m_{h^{0}}=120$ GeV and that the other MSSM Higgs bosons are
very heavy. For the Higgsino admixture in the lightest neutralino, we choose
the $\tilde{H}_{1}$--$\tilde{H}_{2}$ mixing so that the field content is
aligned with that of $h^{0}$ and the longitudinal component of the $Z$ boson
in order to minimize the branching ratio to photons. }
\label{fig:binoBR}
\end{figure}

Let us emphasize that, since the bino is an admixture of the zino and the
photino, a pure bino neutralino can decay into either $\gamma \tilde G$ or $%
Z \tilde G$ (see Eq.~\ref{rategm}). If the lightest neutralino is a mixture
of bino and wino components, then the relative zino and photino components
can be varied arbitrarily, leading to a change in the relative branching
ratios to $\gamma \tilde G$ and $Z \tilde G$. If the lightest neutralino
contains a Higgsino component, then the decay to $h^0 \tilde G$ is also
allowed. We show in Fig.~\ref{fig:binoBR} the branching ratio of the
lightest neutralino into $\gamma \tilde G$ as a function of its mass and
Higgsino content.

\section{Top squark signals in low-energy SUSY breaking}

\label{sec:signals}

As discussed in the previous section, the decay properties of the lighter
stop depend primarily on the mass splitting between the stop and the
lightest neutralino. In this section, we proceed with the phenomenological
analysis of the signatures of top squark production associated with the
different decay modes. In Sec.~\ref{sec:2body} we shall analyze the
signatures associated with the two-body FC decay, which after the neutralino
decay leads to $\tilde t \to c \gamma \tilde G$. In Sec.~\ref{sec:3body} we
shall analyze the signatures associated with the three-body decay, which
after neutralino decay leads to $\tilde t \to b W^+ \gamma \tilde G$. The
two-body decay $\tilde t \to t \tilde \chi_1^0$, which typically dominates
for $m_{\tilde t} > m_t + m_{\tilde \chi_1^0}$, leads to the same final
state as the three-body decay and therefore the discussion of this case will
be included in Sec.~\ref{sec:3body}. Finally, in Sec.~\ref{sec:topdecay} we
consider the possibility that stops are produced in the decays of top quarks.


\subsection{Two-body FC decay: $\tilde t \to c \protect\gamma \tilde G$}

\label{sec:2body}

With the stop undergoing the aforementioned decay, the final state consists
of a pair each of charm jets, photons and (invisible) gravitinos. As we will
show below, the backgrounds to this process are small enough that we need
not require charm tagging. Thus the signal consists of 
\begin{equation*}
2 (\mathrm{jets}) + 2 \gamma + \not\!\!E_T.
\end{equation*}
The selection criteria we adopt are:

\begin{enumerate}
\item  each event must contain two jets and two photons, each of which
should have a minimum transverse momentum ($p_{T}>20$ GeV) and be contained
in the pseudorapidity range $-2.5<\eta <2.5$;

\item  the jets and the photons should be well separated from each other;
namely, 
\begin{equation*}
\delta R_{jj}>0.7,\quad \delta R_{\gamma \gamma }>0.3,\quad \delta
R_{j\gamma }>0.5
\end{equation*}
where $\delta R^{2}=\delta \eta ^{2}+\delta \phi ^{2}$, with $\delta \eta $ (%
$\delta \phi $) denoting the difference in pseudorapidity (azimuthal angle)
of the two entities under consideration;

\item  the invariant mass of the two jets should be sufficiently far away
from the $W$- and $Z$-masses:\newline
$m_{jj}\notin $ (75 GeV, 95 GeV);

\item  each event should be associated with a minimum missing transverse
momentum ($\not{\!}\!p_{T}>30$ GeV).
\end{enumerate}

The photons and jets in signal events tend to be very central; in
particular, reducing the pseudorapidity cut for the two photons to $%
-2.0<\eta <2.0$ would reduce the signal by less than about 3\%. (This change
would reduce the background by a somewhat larger fraction.) Apart from
ensuring observability, these selection criteria also serve to eliminate
most of the backgrounds, which are listed in Table \ref{tab:2bodybg}.

\begin{table}[tbp]
\begin{center}
\begin{tabular}{|l|r|r|}
\hline
Background & Cross section after cuts & after $\gamma$ ID \\ \hline\hline
$jj\gamma\gamma Z$, $Z \to \nu \bar \nu$ & $\sim 0.2$ fb & $\sim 0.13$ fb \\ 
$j j \gamma \nu \bar \nu$ + $\gamma$ radiation & $\sim 0.002$ fb & $\sim
0.001$ fb \\ \hline
$b \bar b \gamma\gamma$, $c \bar c \gamma\gamma$ & $\mathrel{\raise.3ex%
\hbox{$<$\kern-.75em\lower1ex\hbox{$\sim$}}} 0.1$ fb & $\sim 0.06$ fb \\ 
\hline
$jj \gamma \gamma$ & $\sim 0.2$ fb & $\sim 0.13$ fb \\ \hline\hline
Backgrounds with fake photons: &  &  \\ \hline
$jj(ee \to \gamma\gamma)$ & $\sim 5 \times 10^{-4}$ fb & $\sim 5 \times
10^{-4}$ fb \\ \hline
$jj\gamma(j \to \gamma)$ & $\sim 0.8$ fb & $\sim 0.8$ fb \\ \hline
$jj(jj \to \gamma\gamma)$ & $\sim 0.8$ fb & $\sim 0.8$ fb \\ \hline\hline
Total & $\sim 2$ fb & $\sim 2$ fb \\ \hline
\end{tabular}
\end{center}
\caption{Backgrounds to $\tilde t \tilde t^* \to jj\protect\gamma\protect%
\gamma \not\!\!E_T$. The photon identification efficiency is taken to be $%
\protect\epsilon_{\protect\gamma} = 0.8$ for each real photon. See text for
details.}
\label{tab:2bodybg}
\end{table}

A primary source of background is the SM production of $j j \gamma \gamma
\nu_i \bar \nu_i$ where the jets could have arisen from either quarks or
gluons in the final state of partonic subprocesses. A full diagrammatic
calculation would be very computer-intensive and is beyond the scope of this
work. Instead, we consider the subprocesses that are expected to contribute
the bulk of this particular background, namely $p \bar p \rightarrow 2 j + 2
\gamma + Z + X$ with the $Z$ subsequently decaying into neutrinos. These
processes are quite tractable and were calculated with the aid of the
helicity amplitude program \textsc{Madgraph}~\cite{Madgraph}. On imposition
of the abovementioned set of cuts, this background at the Run II falls to
below $\sim 0.2$ fb.

An independent estimate of the $jj\gamma \gamma \nu _{i}\bar{\nu}_{i}$
background may be obtained through the consideration of the single-photon
variant, namely $jj\gamma \nu _{i}\bar{\nu}_{i}$ production, a process that 
\textsc{Madgraph} can handle. After imposing the same kinematic cuts (other
than requiring only one photon) as above, this process leads to a cross
section of roughly 0.2 fb. Since the emission of a second hard photon should
cost us a further power of $\alpha _{\mathrm{em}}$, the electromagnetic
coupling constant, this background falls to innocuous levels. We include
both this estimate and the one based on $Z$ production from the previous
paragraph in Table~\ref{tab:2bodybg}. Though a naive addition of both runs
the danger of overcounting, this is hardly of any importance given the
overwhelming dominance of one.

A second source of background is $b\bar b\gamma\gamma$ ($c \bar
c\gamma\gamma $), with missing transverse energy coming from the
semileptonic decay of one or both of the $b$ ($c$) mesons. In this case,
though, the neutrinos tend to be soft due to the smallness of the $b$ and $c$
masses. Consequently, the cut on $\not\!\!p_T$ serves to eliminate most of
this background, leaving behind less than 0.1 fb. This could be further
reduced (to $\mathrel{\raise.3ex\hbox{$<$\kern-.75em\lower1ex\hbox{$\sim$}}}
0.001$ fb) by vetoing events with leptons in association with jets. However,
such a lepton veto would significantly impact the selection efficiency of
our signal, which contains $c \bar c$, thereby reducing our overall
sensitivity in this channel.

A third source of background is $j j \gamma \gamma$ production, in which the
jet (light quark or gluon) and/or photon energies are mismeasured leading to
a fake $\not\!\!p_T$. To simulate the effect of experimental resolution, we
use a (very pessimistic) Gaussian smearing: $\delta E_j / E_j = 0.1 + 0.6 / 
\sqrt{E_j (\mathrm{GeV})}$ for the jets and $\delta E_\gamma / E_\gamma =
0.05 + 0.3 / \sqrt{E_\gamma (\mathrm{GeV})}$ for the photons.\footnote{%
We have also performed similar smearing for the other background channels as
well as for the signal. However, there it hardly is of any importance as far
as the estimation of the total cross section is concerned.} While the
production cross section is much higher than that of any of the other
backgrounds considered, the ensuing missing momentum tends to be small; in
particular, our cut on $\not\!\!E_T$ reduces this background by almost 99\%.%
\footnote{%
This reduction factor is in rough agreement with that found for $\gamma + j$
events in CDF Run I data in Ref.~\cite{gammabCDF}.} On imposition of our
cuts, this fake background is reduced to $\sim 0.2$ fb.\footnote{%
We note that in the squark searches in supergravity scenarios the signal is $%
j j \not\!\!E_T$; in this case a similar background due to dijet production
with fake $\not\!\!p_T$ is present. This background is larger by two powers
of $\alpha$ than the $jj\gamma\gamma$ background, yet it can still be
reduced to an acceptable level by a relatively hard cut on $\not\!\!p_T$
(see, \textit{e.g.}, Ref.~\cite{Demina}).}

Finally, we consider the instrumental backgrounds from electrons or jets
misidentified as photons. Based on Run I analyses \cite{RunIee} of electron
pair production and taking the probability for an electron to fake a photon
to be about 0.4\%, we estimate the background due to electrons faking
photons to be of order $5 \times 10^{-4}$ fb. More important is the
background in which a jet fakes a photon. Based on a Run I analysis \cite
{gammabCDF} and taking the probability for a jet to fake a photon to be
about 0.1\%, we estimate the background due to $jjj\gamma$ in which one of
the jets fakes a second photon to be about 0.8 fb. Similarly, we estimate
that the background due to $jjjj$ in which two of the jets fake photons is
somewhat smaller; to be conservative we take it to be of the same order, 
\textit{i.e.}, 0.8 fb.

\begin{figure}[!h]
\vspace*{-4ex} 
\resizebox{0.8\textwidth}{!}{
\includegraphics*[0,0][550,550]{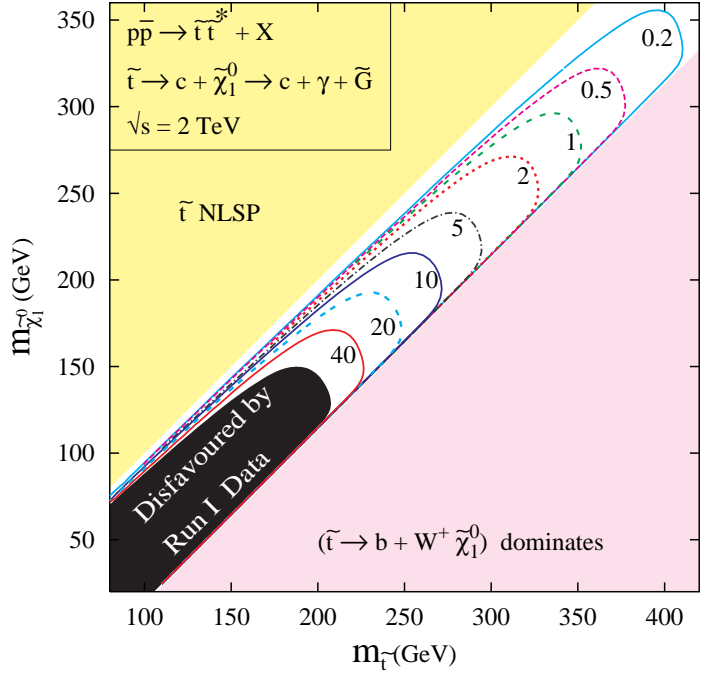}
}
\caption{Cross sections in fb for stop pair production in Run II with $%
\tilde t \to c \protect\gamma \tilde G$, after cuts. No efficiencies have
yet been applied. The black area is excluded by non-observation of $jj%
\protect\gamma\protect\gamma\not\!\!E_T$ events in Run I. }
\label{fig:2bodyxsec}
\end{figure}
Having established that the backgrounds are small, let us now turn to the
signal cross section that survives the cuts. In Fig.~\ref{fig:2bodyxsec}, we
present these as contours in the $m_{\tilde t}$--$m_{\tilde \chi_1^0}$
plane. We assume that the branching ratio of $\tilde t \to c \tilde \chi_1^0$
dominates in the region of parameter space that we consider here. The
branching ratio of $\tilde \chi_1^0 \to \gamma \tilde G$ is taken from Fig.~%
\ref{fig:binoBR} assuming that $\tilde \chi_1^0$ is a pure bino.\footnote{%
If $\tilde \chi_1^0$ is not a pure bino, its branching ratio to $%
\gamma\tilde G$ will typically be reduced somewhat when its mass is large
(see Fig.~\ref{fig:binoBR}). This will lead to a reduction of the signal
cross section at large $m_{\tilde \chi_1^0}$ by typically a few tens of
percent.} All our plots are made before detector efficiencies are applied.
With a real detector, each photon is identified with about $%
\epsilon_{\gamma} = 80$\% efficiency; thus the cross sections shown in Fig.~%
\ref{fig:2bodyxsec} must be multiplied by $\epsilon_{\gamma}^2 = 0.64$ in
order to obtain numbers of events.\footnote{%
The diphoton trigger efficiency is close to 100\%, so we neglect it here.}
While the production cross sections are independent of the neutralino mass,
the decay kinematics have a strong dependence on $m_{\tilde \chi_1^0}$. For
a given $m_{\tilde t}$, a small mass splitting between $m_{\tilde t}$ and $%
m_{\tilde \chi_1^0}$ would imply a soft charm jet, which would often fail to
satisfy our selection criteria. This causes the gap between the cross
section contours and the upper edge of the parameter space band that we are
exploring in Fig.~\ref{fig:2bodyxsec}. This is further compounded at large $%
m_{\tilde \chi_1^0}$ by the fact that a large neutralino mass typically
implies a smaller $\tilde \chi_1^0 \to \gamma \tilde G$ branching fraction
(see Fig.\ref{fig:binoBR}). On the other hand, a small $m_{\tilde \chi_1^0}$
results in reduced momenta for the gravitino and the photon, once again
resulting in a loss of signal; however, this is important only for $%
m_{\tilde \chi_1^0} \mathrel{\raise.3ex\hbox{$<$\kern-.75em\lower1ex\hbox{$%
\sim$}}} 70$ GeV, which is not relevant in this search channel.

The signal cross sections are fairly substantial. In particular, the dark
area in Fig.~\ref{fig:2bodyxsec} corresponds to a signal cross section of 50
fb or larger at Run I of the Tevatron. Taking into account the
identification efficiency of 64\% for the two photons, such a cross section
would have yielded 3 signal events in the 100 pb$^{-1}$ collected in Run I
over a background of much less than 1 event. Run I can thus exclude this
region at 95\% confidence level. In particular, we conclude that Run I data
excludes stop masses up to about 200 GeV, for large enough mass splitting
between the stop and the neutralino. When the stop-neutralino mass splitting
is small (\textit{i.e.}, less than about $10-20$ GeV), the charm quark jets
become too soft and the signal efficiency decreases dramatically. For
comparison, a D{\O} search \cite{DZerosquark} for inclusive $p \bar p \to
\tilde \chi^0_2 + X$ with $\tilde \chi_2^0 \to \gamma \tilde \chi_1^0$ in
the context of minimal supergravity yields a limit on the production cross
section of about 1 pb for parent squark masses of order 150-200 GeV.
Interpreting this in terms of stop pair production with $\tilde \chi_2^0 \to
\gamma \tilde \chi_1^0$ reidentified as $\tilde \chi_1^0 \to \gamma \tilde G$
increases the signal efficiency by a factor of $\sim 2.7$ because every SUSY
event now contains two photons \cite{DZerosquark}; the D{\O} analysis then
yields a stop mass bound of about 180 GeV, in rough agreement with our
result.\footnote{%
Note, however, that the non-negligible mass of the LSP of about 35 GeV
assumed in Ref.~\cite{DZerosquark} leads to kinematics that differ
significantly from those in our analysis, in which the gravitino is
essentially massless.} In addition, Ref.~\cite{KaneTev} projected a Run I
exclusion of stops in this decay channel for masses below about 160-170 GeV,
again in rough agreement with our result.

To claim a discovery at the $5\sigma$ level, one must observe a large enough
number of events that the probability for the background to fluctuate up to
that level is less than $5.7 \times 10^{-7}$. Because the number of expected
background events in this analysis is small, we use Poisson statistics to
find the number of signal events required for a $5\sigma$ discovery. Taking
the total background cross section to be 2 fb from Table \ref{tab:2bodybg},
we show in Table~\ref{tab:2bodyevents} the expected maximum stop discovery
mass reach at Tevatron Run~II for various amounts of integrated luminosity.%
\footnote{%
The cross sections and numbers of events required for discovery are quoted
in terms of integrated luminosities at a single detector. If data from the
CDF and D{\O} detectors are combined, the integrated luminosity of the
machine is effectively doubled.} In particular, with 4 fb$^{-1}$ a stop
discovery can be expected in this channel if $m_{\tilde t} < 285$ GeV, with $%
S/B$ of more than 2/1.\footnote{%
For comparison, in the case of minimal supergravity a reach of $m_{\tilde t}
< 180$ GeV can be expected in the $\tilde t \to c \tilde \chi_1^0$ channel
with 4 fb$^{-1}$ at Tevatron Run II; the same reach is obtained in
low-energy SUSY breaking scenarios in which the stop is the NLSP rather than
the neutralino~\cite{Demina}. In both of these cases, the signal consists of 
$jj\not\!\!E_T$, with no photons in the final state.} Including the effects
of mixing in the composition of the lightest neutralino, a 50\% reduction in
the $\tilde \chi_1^0 \to \gamma \tilde G$ branching ratio compared to the
pure bino case would reduce the stop mass reach by only about 20 GeV. For
such a reduction to occur in the relevant neutralino mass range of about $%
200 - 250$ GeV, the lightest neutralino would have to be less than half bino.

\begin{table}[!h]
\vspace*{2ex}
\par
\begin{center}
\begin{tabular}{|c|c|c|c|c|}
\hline
$\int\mathcal{L}$ & B & S for a 5$\sigma$ discovery & $\sigma_S \times
\epsilon_{\gamma}^2$ & Maximum stop mass reach \\ \hline
2 fb$^{-1}$ & 4 & 14 & 7.0 fb & 265 GeV \\ 
4 fb$^{-1}$ & 8 & 18 & 4.5 fb & 285 GeV \\ 
15 fb$^{-1}$ & 30 & 31 & 2.1 fb & 310 GeV \\ 
30 fb$^{-1}$ & 60 & 42 & 1.4 fb & 325 GeV \\ \hline
\end{tabular}
\end{center}
\caption{ Number of signal events (S) required for a 5$\protect\sigma$ stop
discovery at Tevatron Run~II in the $jj\protect\gamma\protect\gamma %
\not\!\!E_T$ channel and the corresponding signal cross section after cuts
and efficiencies and maximum stop mass reach. We assume a photon
identification efficiency of $\protect\epsilon_{\protect\gamma} = 0.80$. The
number of background events (B) is based on a background cross section of 2
fb from Table~\ref{tab:2bodybg}. }
\label{tab:2bodyevents}
\end{table}


\subsection{Three-body decay: $\tilde t \to b W^+ \protect\gamma \tilde G$}

\label{sec:3body}

For a large enough splitting between the stop and neutralino masses, the
signature of stop pair production would be\footnote{%
The backgrounds are again small enough after cuts in this channel that we do
not need to tag the $b$ quarks. In the case of a discovery, one could
imagine tagging the $b$ quarks and reconstructing the $W$ bosons in order to
help identify the discovered particle.} $jjWW\gamma\gamma\not\!\!E_T$. In
this analysis, we consider only the dominant hadronic decay mode of both $W$
bosons.

This decay mode of the stop proceeds via three Feynman diagrams, involving
an intermediate off-shell top quark, chargino or sbottom. We use the full
decay matrix elements as given in Ref.~\cite{Porod}. Although this
introduces several additional parameters into the analysis, a few
simplifying assumptions may be made without becoming too model dependent.
For example, assuming that the lightest stop is predominantly the
superpartner of the right-handed top-quark ($\tilde t_R$), eliminates the
sbottom exchange diagram altogether. Even if the stop contains a mixture of $%
\tilde t_R$ and $\tilde t_L$, under our assumption that the lighter stop is
the next-to-lightest Standard Model superpartner, the sbottom exchange
diagram will be suppressed by the necessarily larger sbottom mass. As for
the chargino exchange, the wino component does not contribute for a $\tilde
t_R$ decay. Thus the chargino contribution is dominated by its Higgsino
component. Furthermore, if we concentrate on scenarios with large values of
the supersymmetric mass parameter $\mu$ and of the wino mass parameter (in
which case the charginos are heavy and the neutralino is almost a pure
bino), the chargino exchange contribution is also suppressed and the
dominant decay mode is via the diagram involving an off-shell top quark. To
simplify our numerical calculations, we have then assumed that only this
diagram contributes to the stop decay matrix element. We have checked for a
few representative points though, that the inclusion of the sbottom and
chargino diagrams does not significantly change the signal efficiency after
cuts as long as we require that $m_{\tilde b}$, $m_{\tilde \chi^+} >
m_{\tilde t}$.

As the $W$ bosons themselves decay, it might be argued that their
polarization information needs to be retained. However, since we do not
consider angular correlations between the decay products, this is not a
crucial issue; the loss of such information at intermediate steps in the
decay does not lead to a significant change in the signal efficiency after
cuts. This is particularly true for the hadronic decay modes of the $W$, for
which the profusion of jets frequently leads to jet overlap, thereby
obscuring detailed angular correlations. It is thus safe to make the
approximation of neglecting the polarization of the $W$ bosons in their
decay distributions, and we do so in our analysis.

Before we discuss the signal profile and the backgrounds, let us elaborate
on the aforementioned jet overlapping. With six quarks in the final state,
some of the resultant jets will very often be too close to each other to be
recognizable as coming from different partons. We simulate this as follows.
We count a final-state parton (quark or gluon) as a jet only if it has a
minimum energy of 5 GeV and lies within the pseudorapidity range $-3 < \eta
< 3$. We then merge any two jets that fall within a $\delta R$ separation of
0.5; the momentum of the resultant jet is the sum of the two momenta. We
repeat this process iteratively, starting with the hardest jet. Our
selection cuts are then applied to the (merged) jets that survive this
algorithm.

The signal thus consists of: 
\begin{equation*}
n \; \mathrm{jets} + 2 \gamma + \not\!\!E_T \qquad (n \leq 6)
\end{equation*}
Hence, all of the SM processes discussed in the previous section yield
backgrounds to this signal when up to four additional jets are radiated.
Now, the radiation of each hard and well separated jet suppresses the cross
section by a factor of order $\alpha_s \simeq 0.118$. Then, since the $%
jj\gamma\gamma \not\!\!E_T$ backgrounds are already quite small after the
cuts applied in the previous section (see Table~\ref{tab:2bodybg}), the
backgrounds with additional jets are expected to be still smaller.~\footnote{%
For the backgrounds in which one or both of the photons are faked by
misidentified jets, we have taken into account the larger combinatoric
factor that arises when more jets are present to be misidentified.}

There exists a potential exception to the last assertion, namely the
background due to $t \bar t \gamma \gamma$ production. To get an order of
magnitude estimate, the cross section for $t \bar t$ production at Tevatron
Run~II is 8 pb \cite{topxsec}. If both of the photons are required to be
energetic and isolated, we would expect a suppression by a factor of order $%
\alpha^2_{\mathrm{em}}$, leading to a cross section of the order of 0.5 fb.
This cross section is large enough that we need to consider it carefully.

If both $W$ bosons were to decay hadronically, then a sufficiently large
missing transverse energy can only come from a mismeasurement of the jet or
photon energies. As we have seen in the previous section, this missing
energy is normally too small to pass our cuts, thereby suppressing the
background. If one of the $W$ bosons decays leptonically, however, it will
yield a sizable amount of $\not\!\!E_T$. This background can be largely
eliminated by requiring that no lepton ($e$ or $\mu$) is seen in the
detector. This effectively eliminates the $W$ decays into $e$ or $\mu$ or
decays to $\tau$ followed by leptonic $\tau$ decays; the remaining
background with hadronic $\tau$ decays is naturally quite small without
requiring additional cuts. Considerations such as these lead us to an
appropriate choice of criteria for an event to be selected:

\begin{enumerate}
\item  At least four jets, each with a minimum transverse momentum $%
p_{Tj}>20 $ GeV and contained in the pseudorapidity interval of $-3<\eta
_{j}<3$. Any two jets must be separated by $\delta R_{jj}>0.7$. As most of
the signal events do end up with 4 or more energetic jets (the hardest jets
coming typically from the $W$ boson decays), this does not cost us in terms
of the signal, while reducing the QCD background significantly. In addition,
the $t\bar{t}\gamma \gamma $ events with both $W$'s decaying leptonically
are reduced to a level of order $10^{-4}$ fb by this requirement alone.

\item  Two photons, each one with $p_{T\gamma }>20$ GeV and pseudorapidity $%
-2.5<\eta _{\gamma }<2.5$. The two photons must be separated by at least $%
\delta R_{\gamma \gamma }>0.3$.

\item  Any photon-jet pair must have a minimum separation of $\delta
R_{j\gamma }>0.5$.

\item  A minimum missing transverse energy $\not{\!}\!E_{T}>30$ GeV.

\item  The event should not contain any isolated lepton with $p_{T}>10$ GeV
and lying within the pseudorapidity range $-3<\eta <3$.
\end{enumerate}

As in the previous section, the cut on $\not\!\!E_T$ serves to eliminate
most of the background events with only a fake missing transverse momentum
(arising out of mismeasurement of jet energies). In association with the
lepton veto, it also eliminates the bulk of events in which one of the $W$
bosons decays leptonically (including the $\tau$ channel). A perusal of
Table~\ref{tab:3bodybg}, which summarizes the major backgrounds after cuts,
convinces us that the backgrounds to this channel are very small, in fact
much smaller than those for the previous channel.

\begin{table}[h]
\vspace*{2ex}
\par
\begin{center}
\begin{tabular}{|l|r|r|}
\hline
Background & Cross section after cuts & after $\gamma$ ID \\ \hline\hline
$(jj\gamma\gamma Z$, $Z \to \nu \bar \nu) + 2j$ & $\sim 0.003$ fb & $\sim
0.002$ fb \\ 
$j j \nu \bar \nu \gamma \gamma +2j$ & $\sim 3 \times 10^{-5}$ fb & $\sim 2
\times 10^{-5}$ fb \\ \hline
$(b \bar b \gamma\gamma$, $c \bar c \gamma\gamma) + 2j$ & $\sim 0.001$ fb & $%
\sim 0.0006$ fb \\ \hline
$jj \gamma \gamma + 2j$ & $\mathrel{\raise.3ex\hbox{$<$\kern-.75em\lower1ex%
\hbox{$\sim$}}} 0.003$ fb & $\mathrel{\raise.3ex\hbox{$<$\kern-.75em%
\lower1ex\hbox{$\sim$}}} 0.002$ fb \\ \hline
$t \bar t \gamma\gamma$, $WW \to jjjj$ & $\mathrel{\raise.3ex\hbox{$<$%
\kern-.75em\lower1ex\hbox{$\sim$}}} 10^{-4}$ fb & $\mathrel{\raise.3ex%
\hbox{$<$\kern-.75em\lower1ex\hbox{$\sim$}}} 10^{-4}$ fb \\ 
$t \bar t \gamma\gamma$, $WW \to jj \ell \nu$, $\ell = e$, $\mu$, or $\tau
\to \ell$ & $\sim 0.001$ fb & $\sim 0.0006$ fb \\ 
$t \bar t \gamma\gamma$, $WW \to jj \tau \nu$, $\tau \to j$ & $%
\mathrel{\raise.3ex\hbox{$<$\kern-.75em\lower1ex\hbox{$\sim$}}} 0.01$ fb & $%
\mathrel{\raise.3ex\hbox{$<$\kern-.75em\lower1ex\hbox{$\sim$}}} 0.006$ fb \\ 
\hline\hline
Backgrounds with fake photons: &  &  \\ \hline
$jj(ee \to \gamma\gamma) + 2j$ & $\sim 7\times 10^{-6}$ fb & $\sim 7\times
10^{-6}$ fb \\ \hline
$jj\gamma(j \to \gamma) + 2j$ & $\sim 0.02$ fb & $\sim 0.02$ fb \\ \hline
$jj(jj \to \gamma\gamma) + 2j$ & $\sim 0.03$ fb & $\sim 0.03$ fb \\ 
\hline\hline
Total & $\mathrel{\raise.3ex\hbox{$<$\kern-.75em\lower1ex\hbox{$\sim$}}}
0.07 $ fb & $\sim 0.06$ fb \\ \hline
\end{tabular}
\end{center}
\caption{ Backgrounds to $\tilde{t}\tilde{t}^{\ast }\rightarrow jjWW\protect%
\gamma \protect\gamma \not\!\!E_T$\ with both $W$ bosons decaying
hadronically. The photon identification efficiency is taken to be $\protect%
\epsilon _{\protect\gamma }=0.8$ for each real photon. See text for details.}
\label{tab:3bodybg}
\end{table}

\begin{figure}[!h]
\vspace*{-8ex} 
\resizebox{0.8\textwidth}{!}{
\includegraphics*[0,0][550,550]{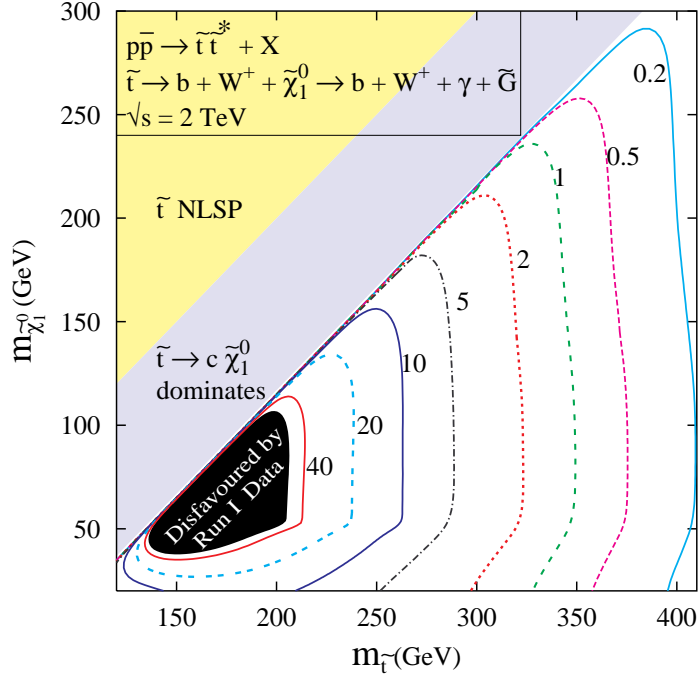}
}
\caption{Cross section in fb for stop pair production with $\tilde t \to b W 
\protect\gamma \tilde G$, after cuts. Both $W$ bosons are assumed to decay
hadronically. The black area is excluded by non-observation of $jjWW\protect%
\gamma\protect\gamma\not\!\!E_T$ events in Run I. }
\label{fig:3bodyxsec}
\end{figure}

The signal cross section after cuts (but before photon identification
efficiencies) is shown in Fig.~\ref{fig:3bodyxsec} as a function of the stop
and $\tilde \chi_1^0$ masses. We assume that the branching ratio of $\tilde
t \to b W \tilde \chi_1^0$ dominates in the region of parameter space under
consideration. The branching ratio of $\tilde \chi_1^0 \to \gamma \tilde G$
is again taken from Fig.~\ref{fig:binoBR} assuming that $\tilde \chi_1^0$ is
a pure bino. The signal efficiency after cuts is about 45\%. For small
neutralino masses ($m_{\tilde \chi_1^0} \mathrel{\raise.3ex\hbox{$<$%
\kern-.75em\lower1ex\hbox{$\sim$}}} 50$ GeV), though, both the photons and
the gravitinos ($\not\!\!E_T$) tend to be soft, leading to a decrease in the
signal efficiency. For small stop masses (as well as for large stop masses
when the stop-neutralino mass difference is small), on the other hand, the
jets are soft leading to a suppression of the signal cross section after
cuts. As one would expect, both of these effects are particularly pronounced
in the contours corresponding to large values of the cross section. The
additional distortion of the contours for large neutralino masses can, once
again, be traced to the suppression of the $\tilde \chi_1^0 \to \gamma
\tilde G$ branching fraction (see Fig.~\ref{fig:binoBR}). 

The non-observation of $jjWW\gamma\gamma\not\!\!E_T$ events at Run I of the
Tevatron already excludes the region of parameter space shown in black in
Fig.~\ref{fig:3bodyxsec}. As in the previous section, in this excluded
region at least 3 signal events would have been produced after cuts and
detector efficiencies in the 100 pb$^{-1}$ of Run I data, with negligible
background. In particular, Run I data excludes stop masses below about 200
GeV, for neutralino masses larger than about 50 GeV.

We show in Table~\ref{tab:3bodyevents} the expected maximum stop discovery
mass reach at Tevatron Run II for various amounts of integrated luminosity.%
\footnote{%
Again, if data from the CDF and D{\O} detectors are combined, the integrated
luminosity of the machine is effectively doubled.} In particular, with 4 fb$%
^{-1}$, a stop discovery can be expected in this channel if $m_{\tilde t} <
320$ GeV.\footnote{%
For comparison, in the case of minimal supergravity a reach of $m_{\tilde t}
< 190$ GeV can be expected in the $\tilde t \to b W \tilde \chi_1^0$ channel
with 4 fb$^{-1}$ at Tevatron Run II \cite{Demina}.} If the lightest
neutralino is not a pure bino, the reach at large neutralino masses will be
reduced. However, the maximum stop mass reach quoted here will not be
affected, because it occurs for $m_{\tilde \chi_1^0} \sim 50 - 100$ GeV; in
this mass range the lightest neutralino decays virtually 100\% of the time
to $\gamma \tilde G$ due to the kinematic suppression of all other possible
decay modes, unless its photino component is fine-tuned to be tiny.

Including a separate analysis of stop production and decay with one or more
of the $W$ bosons decaying leptonically would yield an increase in the
overall signal statistics; however, we do not expect this increase to
dramatically alter the stop discovery reach.

\begin{table}[!h]
\vspace*{2ex}
\par
\begin{center}
\begin{tabular}{|c|c|c|c|c|}
\hline
$\int\mathcal{L}$ & B & S for a $5\sigma$ discovery & $\sigma_S \times
\epsilon_{\gamma}^2$ & Maximum stop mass reach \\ \hline
2 fb$^{-1}$ & 0.1 & 5 & 2.5 fb & 300 GeV \\ 
4 fb$^{-1}$ & 0.2 & 6 & 1.5 fb & 320 GeV \\ 
15 fb$^{-1}$ & 0.9 & 8 & 0.53 fb & 355 GeV \\ 
30 fb$^{-1}$ & 1.8 & 10 & 0.33 fb & 375 GeV \\ \hline
\end{tabular}
\end{center}
\caption{ Number of signal events (S) required for a $5\protect\sigma$ stop
discovery at Tevatron Run II in the $jjWW\protect\gamma\protect\gamma %
\not\!\!E_T$ channel and the corresponding signal cross section after cuts
and efficiencies and maximum stop mass reach. We take $\protect\epsilon_{%
\protect\gamma} = 0.80$. The number of background events (B) is based on a
background cross section of 0.06 fb from Table~\ref{tab:3bodybg}. }
\label{tab:3bodyevents}
\end{table}


\subsection{Stop production in top quark decays}

\label{sec:topdecay}

If $m_{\tilde t} + m_{\tilde \chi_1^0} < m_t$, then stops can be produced in
the decays of top quarks. As we will explain here, most of the parameter
space in this region is excluded by the non-observation of stop events via
direct production or in top quark decays at Run I of the Tevatron. However,
some interesting parameter space for this decay remains allowed after Run I,
especially if the lighter stop is predominantly $\tilde t_L$.

For $m_{\tilde t} < m_b + m_W + m_{\tilde \chi_1^0}$, so that the stop
decays via $\tilde t \to c \tilde \chi_1^0$, the region in which $t \to
\tilde t \tilde \chi_1^0$ is possible is almost entirely excluded by the
limit on stop pair production at Run I, as shown in Fig.~\ref{fig:2bodyxsec}%
. A sliver of parameter space in which the stop-neutralino mass splitting is
smaller than about 10 GeV remains unexcluded. For $m_{\tilde t} > m_b + m_W
+ m_{\tilde \chi_1^0}$, so that the stop decays via $\tilde t \to b W \tilde
\chi_1^0$, the signal efficiency in the search for direct stop production is
degraded for light neutralinos with masses below about 50 GeV and for stops
lighter than about 150 GeV. This prevents Run I from being sensitive to stop
pair production in the region of parameter space in which top quark decays
to stops are possible with the stops decaying to $b W \tilde \chi_1^0$, as
shown in Fig.~\ref{fig:3bodyxsec}. In what follows, we focus on this latter
region of parameter space.

As discussed before, if the lightest neutralino is mostly bino, the
constraints on its mass are model-dependent. The constraints from Tevatron
Run I are based on inclusive chargino and neutralino production~\cite
{KaneTev,DZeroino} under the assumption of gaugino mass unification; the
cross section is dominated by production of $\tilde{\chi}_{1}^{\pm }\tilde{%
\chi}_{1}^{\mp }$ and $\tilde{\chi}_{1}^{\pm }\tilde{\chi}_{2}^{0}$. If the
assumption of gaugino mass unification is relaxed, then Run I puts no
constraint on the mass of $\tilde{\chi}_{1}^{0}$. At LEP, while the pair
production of a pure bino leads to an easily detectable diphoton signal, it
proceeds only via $t$-channel selectron exchange. The mass bound on a bino $%
\tilde{\chi}_{1}^{0}$ from LEP thus depends on the selectron mass \cite
{LEPSUSYWG}. In particular, for selectrons heavier than about 600 GeV, bino
masses down to 20 GeV are still allowed by the LEP data. If $\tilde{\chi}%
_{1}^{0}$ contains a Higgsino admixture, it couples to the $Z$ and can be
pair-produced at LEP via $Z$ exchange. For an NLSP with mass between 20 and
45 GeV, as will be relevant in our top quark decay analysis, the LEP search
results limit the $\tilde{H}_{2}$ component to be less than 1\%. Such a
small $\tilde{H}_{2}$ admixture has no appreciable effect on the top quark
partial width to $\tilde{t}\tilde{\chi}_{1}^{0}$. We thus compute the
partial width for $t\rightarrow \tilde{t}\tilde{\chi}_{1}^{0}$ assuming that
the neutralino is a pure bino. Taking the lighter stop to be $\tilde{t}_{1}=%
\tilde{t}_{L}\cos \theta _{\tilde{t}}+\tilde{t}_{R}\sin \theta _{\tilde{t}}$%
, we find, 
\begin{equation}
\Gamma (t\rightarrow \tilde{t}_{1}\tilde{B})=\left[ \frac{4}{9}\sin
^{2}\theta _{\tilde{t}}+\frac{1}{36}\cos ^{2}\theta _{\tilde{t}}\right] 
\frac{\alpha }{\cos ^{2}\theta _{W}}\frac{E_{\tilde{B}}}{m_{t}}\sqrt{E_{%
\tilde{B}}^{2}-m_{\tilde{B}}^{2}},  \label{eq:topwidth}
\end{equation}
where $E_{\tilde{B}}=(m_{t}^{2}+m_{\tilde{B}}^{2}-m_{\tilde{t}}^{2})/2m_{t}$%
. The numerical factors in the square brackets in Eq.~\ref{eq:topwidth} come
from the hypercharge quantum numbers of the two stop electroweak
eigenstates. Clearly, the partial width is maximized if the lighter stop is
a pure $\tilde{t}_{R}$ state; it drops by a factor of 16 if the lighter stop
is a pure $\tilde{t}_{L}$ state. In any case, the branching ratio for $%
t\rightarrow \tilde{t}\tilde{B}$ does not exceed 6\% for $m_{\tilde{t}}>100$
GeV and $m_{\tilde{\chi}_{1}^{0}}>20$ GeV.\footnote{%
For neutralino masses below $m_{Z}$, as are relevant here, the branching
ratio into $\gamma \tilde{G}$ \ is virtually 100\% (see Fig.~\ref{fig:binoBR}%
), almost independent on the neutralino composition.}

\begin{figure}[h]
\resizebox{0.8\textwidth}{!}{\rotatebox{270}{\includegraphics{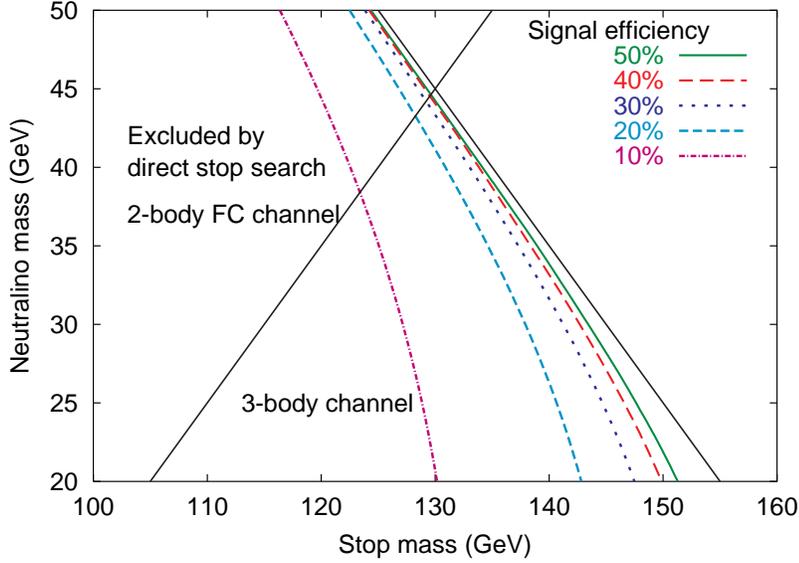}}}
\caption{95\% confidence level exclusion limits for top quark decays to
stops from Run I for various signal efficiencies, in the case that $\tilde{t}%
_{1}=\tilde{t}_{R}$, which gives the largest event rates. The area to the
left of the curves is excluded. The case $\tilde{t}_{1}=\tilde{t}_{L}$ gives
no exclusion for the signal efficiencies considered and is not shown here.
The solid line from upper left to lower right is the kinematic limit for $%
m_{t}=175$ GeV. The solid line from upper right to lower left separates the
regions in which the 2-body FC decay and the 3-body decay of the stop
dominate.}
\label{fig:topRun1}
\end{figure}

The signal from top quark pair production followed by one top quark decaying
as in the Standard Model and the other decaying to $\tilde{t}\tilde{\chi}%
_{1}^{0}$, followed by the stop 3-body decay and the neutralino decays to $%
\gamma \tilde{G}$, is $bbWW\gamma \gamma \not\!\!E_T$. This signal is the
same (up to kinematics) as that from stop pair production in the 3-body
decay region. As in the case of stop pair production followed by the 3-body
decay, we expect that the background to this process can be reduced to a
negligible level. Run I can then place 95\% confidence level exclusion
limits on the regions of parameter space in which 3 or more signal events
are expected after cuts and efficiencies are taken into account. Using the
Run I top quark pair production cross section of 6 pb \cite{topxsec} and a
total luminosity of 100 pb$^{-1}$, we compute the number of signal events as
a function of the stop and neutralino masses and the stop composition,
assuming various values of the signal efficiency after cuts and detector
efficiencies. If $\tilde{t}_{1}=\tilde{t}_{R}$, then Run I excludes most of
the parameter space below the kinematic limit for this decay even for fairly
low signal efficiency $\sim 20$\%, as shown in Fig.~\ref{fig:topRun1}. If $%
\tilde{t}_{1}=\tilde{t}_{L}$, on the other hand, the signal cross section is
much smaller and Run I gives no exclusion unless the signal efficiency is
larger than 75\%, which would already be unfeasible including only the
identification efficiencies for the two photons; even 100\% signal
efficiency would only yield an exclusion up to $m_{\tilde{t}}\simeq 118$ GeV.

\begin{figure}[h]
\resizebox{0.8\textwidth}{!}{\rotatebox{270}{\includegraphics{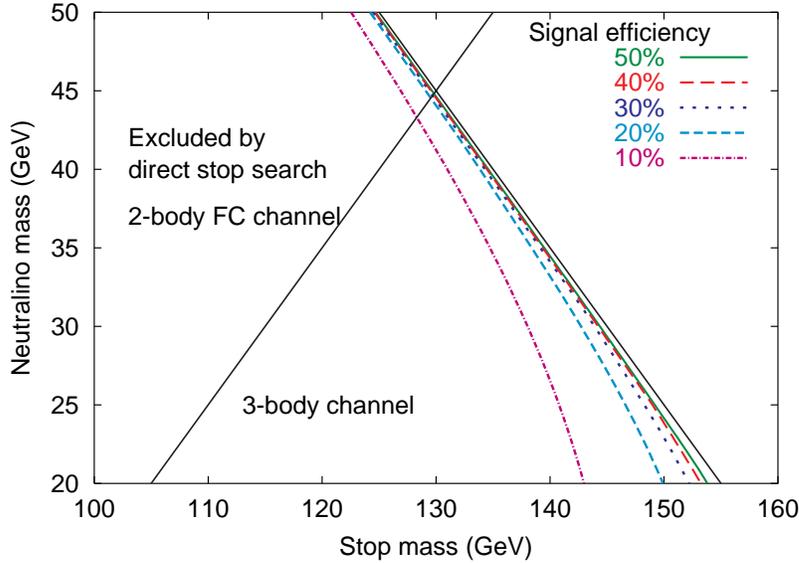}}}
\caption{$5\protect\sigma $ discovery contours for top quark decays to stops
at Run II with 4 fb$^{-1}$ for various signal efficiencies, in the case that 
$\tilde{t}_{1}=\tilde{t}_{L}$, which gives the smallest event rates. The
case $\tilde{t}_{1}=\tilde{t}_{R}$ would be discovered virtually up to the
kinematic limit even with 10\% signal efficiency, and is not shown here. The
solid lines are as in Fig.~\ref{fig:topRun1}.}
\label{fig:topRun2}
\end{figure}

At Run II, the top quark pair production cross section is 8 pb \cite{topxsec}
and the expected total luminosity is considerably higher. This allows top
quark decays to $\tilde{t}\tilde{\chi}_{1}^{0}$ to be detected for stop and
neutralino masses above the Run I bound. For $\tilde{t}_{1}=\tilde{t}_{R}$,
top quark decays to stops will be probed virtually up to the kinematic
limit, even with low signal efficiency $\sim $10\% and only 2 fb$^{-1}$ of
integrated luminosity. If $\tilde{t}_{1}=\tilde{t}_{L}$, so that the signal
event rate is minimized, top quark decays to stops would be discovered up to
within 10 GeV of the kinematic limit for signal efficiencies $%
\mathrel{\raise.3ex\hbox{$>$\kern-.75em\lower1ex\hbox{$\sim$}}}20$\% and 4 fb%
$^{-1}$ of integrated luminosity (see Fig.~\ref{fig:topRun2}). In this
region of parameter space, stops would also be discovered in Run II with
less than 2 fb$^{-1}$ via direct stop pair production (see Fig.~\ref
{fig:3bodyxsec}). 


\section{Ten degenerate squarks}

\label{sec:10squarks}

Having concentrated until now on the production and decay of the top squark,
let us now consider the other squarks. In the most general MSSM, the
spectrum is of course quite arbitrary. However, low energy constraints~\cite
{fcnc} from flavor changing neutral current processes demand that such
squarks be nearly mass degenerate, at least those of the same chirality.
Interestingly, in many theoretical scenarios, such as the minimal gauge
mediated models, this mass degeneracy between squarks of the same chirality
happens naturally; in addition the mass splitting between the left-handed
and right-handed squarks associated with the five light quarks turns out to
be small. For simplicity, then, we will work under the approximation that
all of these 10 squarks (namely, $\tilde u_{L,R}$, $\tilde d_{L,R}$, $\tilde
c_{L,R}$, $\tilde s_{L,R}$ and $\tilde b_{L,R}$) are exactly degenerate.

\subsection{Production at Tevatron Run II}

While the cross sections for the individual pair-production of the $\tilde
c_{L,R}$, $\tilde s_{L,R}$ and $\tilde b_{L,R}$ are essentially the same as
that for a top-squark of the same mass, the situation is more complicated
for squarks of the first generation. The latter depend sensitively on the
gluino mass because of the presence of $t$-channel diagrams. Moreover,
processes such as $\bar u u \to \tilde u_L \tilde u_R^*$ or $d d \to \tilde
d_{L, R} \tilde d_{L, R}$ become possible and relevant. Of course, in the
limit of very large gluino mass, the squark production processes are driven
essentially by QCD and dominated by the production of pairs of mass
eigenstates, analogous to the top squark production considered already. In
particular, at the leading order, the total production cross section for the
ten degenerate squarks of a given mass is simply ten times the corresponding
top squark production cross section.

\begin{figure}[h]
\centerline{
\resizebox{0.8\textwidth}{!}{
\rotatebox{270}{\includegraphics{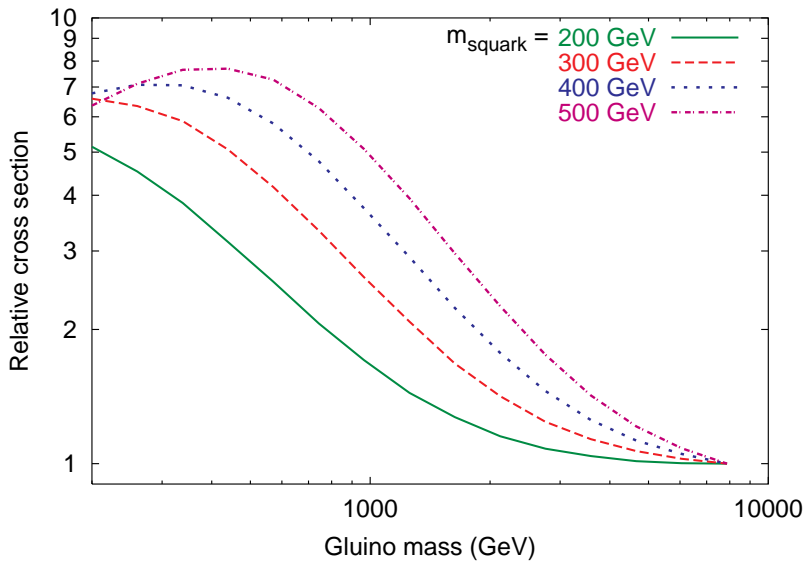}}
}}
\caption{Gluino mass dependence of the NLO cross section for production of
ten degenerate squarks in 2 TeV $p\bar{p}$ collisions, from PROSPINO~ 
\protect\cite{PROSPINO}. Shown are the cross sections for $\tilde{q}\tilde{q}%
^{\ast }$ production normalized to the value at large $m_{\tilde{g}}$, for
common squark masses of 200, 300, 400 and 500 GeV. Production of $\tilde{q}%
\tilde{q}$ is small at a $p\bar{p}$ collider and is neglected here; it
yields an additional 3-15\% increase in the total cross section at low $m_{%
\tilde{g}}$ for this range of squark masses. Cross sections are evaluated at
the scale $\protect\mu =m_{\tilde{q}}$. }
\label{fig:10sqgluino}
\end{figure}

Since a relatively light gluino only serves to increase the total cross
section (see Fig.~\ref{fig:10sqgluino}), it can be argued that the heavy
gluino limit is a \emph{conservative one}. To avoid considering an
additional free parameter, we shall perform our analysis in this limit. To a
first approximation, the signal cross sections presented below will scale%
\footnote{%
That the gluino exchange diagram has a different topology as compared to the
(dominant) quark-initiated QCD diagram indicates that corresponding angular
distributions would be somewhat different. Thus, the efficiency after cuts
is not expected to be strictly independent of the gluino mass. For the most
part, though, this is only a subleading effect.} with the gluino mass
approximately as shown in Fig.~\ref{fig:10sqgluino}. 

Like top squarks, the 10 degenerate squarks can also be produced via cascade
decays of heavier supersymmetric particles. To be conservative, we again
neglect this source of squark production by assuming that the masses of the
heavier supersymmetric particles are large enough that their production rate
at Tevatron energies can be neglected.

\begin{figure}[tbp]
\resizebox{0.75\textwidth}{!}{
\rotatebox{270}{\includegraphics{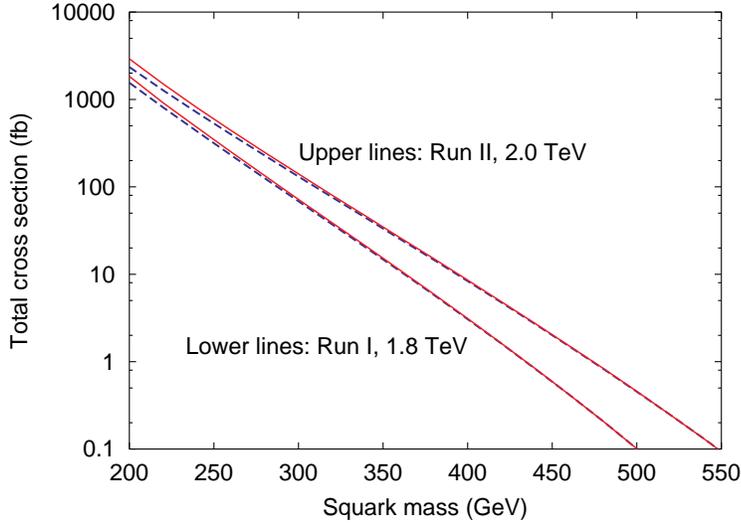}}
}
\caption{LO (dashed) and NLO (solid) cross sections for the production of
ten degenerate squarks in the heavy gluino limit in $p \bar p$ collisions at
Tevatron Run I (1.8 TeV) and Run II (2.0 TeV), from PROSPINO~\protect\cite
{PROSPINO}. Cross sections are evaluated at the scale $\protect\mu =
m_{\tilde q}$. }
\label{fig:10sqxsecmass}
\end{figure}

The NLO cross sections for production of ten degenerate squarks including
QCD and SUSY-QCD corrections have been implemented numerically in PROSPINO~ 
\cite{PROSPINO}. We generate squark production events using the LO cross
section evaluated at the scale $\mu = m_{\tilde q}$, improved by the NLO
K-factor obtained from PROSPINO~\cite{PROSPINO} (see Fig.~\ref
{fig:10sqxsecmass}), in the limit that the gluino is very heavy. The
K-factor varies between 1 and 1.25 for $m_{\tilde q}$ decreasing from 550 to
200 GeV. As in the case of the top squark analysis, we use the CTEQ5 parton
distribution functions~\cite{CTEQ5} and neglect the shift in the $p_T$
distribution of the squarks due to gluon radiation at NLO. 


\subsection{Signals in low-energy SUSY breaking}

The decays of the ten degenerate squarks are very simple. As long as they
are heavier than the lightest neutralino,\footnote{%
As before, we do not allow the possibility of cascade decays through other
neutralinos/charginos. Were we to allow these, the channel we are
considering would be somewhat suppressed, but additional, more spectacular,
channels would open up.} they decay via $\tilde q \to q \tilde \chi_1^0 \to
q \gamma \tilde G$. The signal and backgrounds are then identical to those
of the two-body FC stop decay discussed in Sec.~\ref{sec:2body}, and
consequently we use the same selection cuts. In fact, in view of the tenfold
increase in the signal strength, we could afford more stringent cuts so as
to eliminate virtually all backgrounds, but this is not quite necessary.

\begin{figure}[tbp]
\vspace*{-8ex} 
\resizebox{0.85\textwidth}{!}{
\includegraphics*[0,0][550,550]{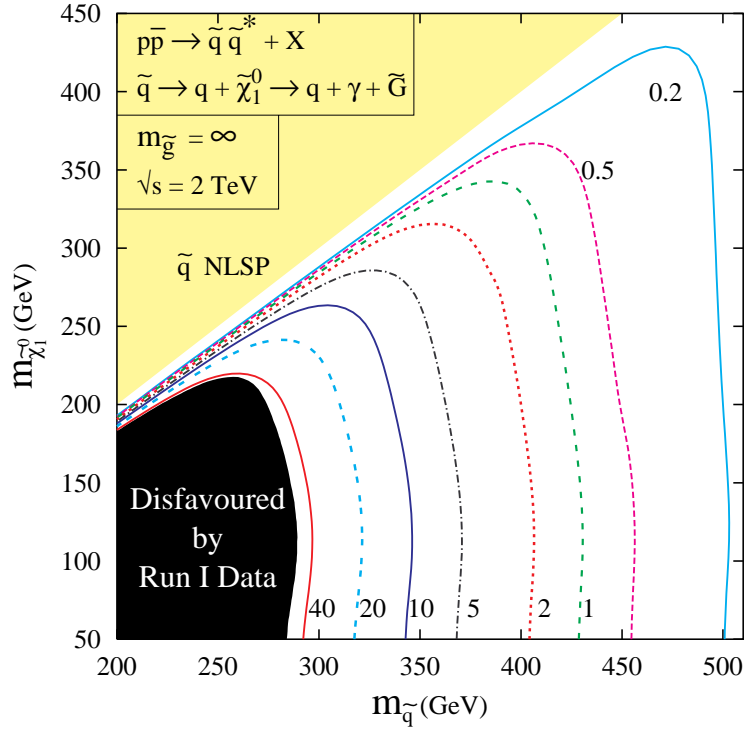}
}
\caption{Cross section in fb for production of 10 degenerate squarks in Run
II in the heavy gluino limit with $\tilde q \to q \protect\gamma \tilde G$,
after cuts. The black area is excluded by non-observation of $jj\protect%
\gamma\protect\gamma\not\!\!E_T$ events in Run I. }
\label{fig:10squarks}
\end{figure}

The signal cross section after cuts (but before efficiencies) for production
of ten degenerate squarks in the heavy gluino limit is shown in Fig.~\ref
{fig:10squarks} as contours in the $m_{\tilde q}$--$m_{\tilde \chi_1^0}$
plane. We assume that the squark $\tilde q$ decays predominantly into $q
\tilde \chi_1^0$. The branching ratio of $\tilde \chi_1^0 \to \gamma \tilde
G $ is taken from Fig.~\ref{fig:binoBR} assuming that $\tilde \chi_1^0$ is a
pure bino. Clearly, the effect of the kinematic cuts on the signal is very
similar to that in the case of the 2-body FC decay of the stop. The mass
reach, of course, is much larger due to the tenfold increase in the total
cross section; also, unlike in the case of the stop, the 2-body decay is
dominant throughout the entire parameter space. 

The non-observation of $jj\gamma\gamma \not\!\!E_T$ events at Run I of the
Tevatron excludes the region of parameter space shown in black in Fig.~\ref
{fig:10squarks}. As in our stop analysis, in this excluded region, at least
3 signal events would have been detected in the 100 pb$^{-1}$ of Run I data,
with negligible background. In particular, we estimate that Run I data
excludes the ten degenerate squarks up to a common mass of about 280 GeV. As
in the case of 2-body FC stop decays, when the mass splitting between the
squarks and the neutralino is too small (\textit{i.e.}, less than about 40
GeV), the jets become soft and the signal efficiency decreases dramatically,
leaving an unexcluded region of parameter space. The D{\O} search for
inclusive $p \bar p \to \tilde \chi_2^0 + X$ with $\tilde \chi_2^0 \to
\gamma \tilde \chi_1^0$ \cite{DZerosquark} yields a limit on the production
cross section of about 0.5 pb for parent squark masses above about 250 GeV.
Interpreted in terms of pair production of 10 degenerate squarks in the
large $m_{\tilde g}$ limit with $\tilde \chi_2^0 \to \gamma \tilde \chi_1^0$
reidentified as $\tilde \chi_1^0 \to \gamma \tilde G$ again increases the
signal efficiency by a factor of $\sim 2.7$ because every event contains two
photons \cite{DZerosquark}; the D{\O} analysis then yields a bound on the
common squark mass of about 275 GeV, again in rough agreement with our
result.\footnote{%
Ref.~\cite{DZerosquark} quotes a squark mass bound of 320 GeV for the case $%
m_{\tilde q} \ll m_{\tilde g}$ in the context of low-energy SUSY breaking;
we expect that this is due to the contribution of chargino and neutralino
production to the total SUSY cross section in their analysis.}

Taking the total background cross section to be 2 fb (see Table~\ref
{tab:2bodybg}), we show in Table~\ref{tab:10squarkevents} the expected
maximum discovery mass reach for ten degenerate squarks at Tevatron Run~II
for various amounts of integrated luminosity.\footnote{%
Again, if data from the CDF and D{\O} detectors are combined, the integrated
luminosity of the machine is effectively doubled.} In particular, with 4 fb$%
^{-1}$ a squark discovery can be expected in this channel if $m_{\tilde q} <
360$ GeV. Again, if the lightest neutralino is not a pure bino, the reach at
large neutralino masses will be reduced. However, this will have very little
effect on the maximum squark mass reach quoted here, because this maximum
reach occurs for $m_{\tilde \chi_1^0} \sim 100 - 150$ GeV, where all
neutralino decay modes other than $\gamma \tilde G$ suffer a large kinematic
suppression. From Fig.~\ref{fig:binoBR}, it is evident that the neutralino
branching fraction into $\gamma \tilde G$ will be reduced by no more than
10\% in this mass range as long as the neutralino is at least 50\% bino;
such a reduction in the neutralino branching fraction will lead to a
reduction of only a few GeV in the maximum squark mass reach.

\begin{table}[tbp]
\begin{center}
\begin{tabular}{|c|c|c|c|c|}
\hline
$\int\mathcal{L}$ & B & S for a 5$\sigma$ discovery & $\sigma_S \times
\epsilon_{\gamma}^2$ & Maximum squark mass reach \\ \hline
2 fb$^{-1}$ & 4 & 14 & 7.0 fb & 345 GeV \\ 
4 fb$^{-1}$ & 8 & 18 & 4.5 fb & 360 GeV \\ 
15 fb$^{-1}$ & 30 & 31 & 2.1 fb & 390 GeV \\ 
30 fb$^{-1}$ & 60 & 42 & 1.4 fb & 405 GeV \\ \hline
\end{tabular}
\end{center}
\caption{ Number of signal events (S) required for a 5$\protect\sigma$
squark discovery at Tevatron Run~II, assuming production of 10 degenerate
squarks in the limit that the gluino is very heavy, and the corresponding
signal cross section after cuts and efficiencies and maximum squark mass
reach. We take $\protect\epsilon_{\protect\gamma} = 0.80$. The number of
background events (B) is based on a background cross section of 2 fb from
Table~\ref{tab:2bodybg}. }
\label{tab:10squarkevents}
\end{table}


\chapter{Concluding remarks\label{sec:conclusions}}

We have examined the possibility of the existence of a non-minimal Higgs
sector. In particular, we consider the next to minimal Higgs sector
consisting of two Higgs doublets. There are several motivations to explore
those scenarios, among them we have: 1)\ The fact that the Higgs sector is
totally unknown so far, 2) the unnatural hierarchy predicted by SM between
the Yukawa couplings of the third family of quarks, 3) the possibility of
generating either spontaneous or explicit CP violation, 4) the generation of
flavor changing neutral currents (inspired in the increasing evidence on
neutrino oscillations) still compatible with the strong experimental limits
on them, and 5) the possibility that some models with larger symmetries end
up at low energies in a non minimal Higgs sector as in the case of SUSY
models. In particular if the SUSY particles are heavy enough, the MSSM Higgs
sector acquires the form of a constrained 2HDM type II.

The two Higgs doublet model predicts the existence of five Higgs particles:
One CP odd Higgs boson $\left( A^{0}\right) $,\ two CP even ones $\left(
h^{0},H^{0}\right) $,$\;$and two charged Higgs particles $\left( H^{\pm
}\right) $. If a charged Higgs were discovered first, it would be a clear
signature of the presence of a non-minimal Higgs sector. The detection of a
CP-odd Higgs boson would also yield evidence on physics beyond the SM
because of the significant deviations of its couplings to SM particles
respect to the ones of a CP even Higgs. In particular, it worths to mention
that couplings of $A^{0}VV$ should be absent, or at least highly suppressed.
Of course, another possible signal would be the discovery of more than one
Higgs. Finally, if only one CP even Higgs is discovered and the same
experiment is unable to detect other Higgs modes, deviations on its
couplings from the ones predicted by the SM can contain the information on
certain underlying Higgs spectrum, in the case of the 2HDM even in the
framework of the decoupling limit in which the couplings at tree level of
the lighter Higgs boson are SM-like, precision measurements of the couplings
could detect such deviations via radiative corrections. If a SM-like Higgs
is detected, one of the most promising couplings to look for possible
deviations from the SM behavior is the trilinear self coupling $%
h^{0}h^{0}h^{0}$.

The most important sources for production of Higgs bosons in Hadron
Colliders are the gluon-gluon fusion and $qq^{\prime }\;$production, while
in $e^{+}e^{-}\;$colliders the most important sources consists of $WW,ZZ\;$%
fusion\ and\ Higgstraghlung. Additionally, detection depends on the
subsequent decay of the Higgs bosons, the dominant decay channels are highly
model dependent and the estimated rate production should be calculated for
specific regions of the free parameters. Finally, other interesting
perspectives for Higgs production might be provided by $\gamma \gamma \;$%
fusion in $\gamma \gamma \;$colliders dominated by $\rho W^{-}\;$or $t%
\overline{t}\;$loop, and also by $\mu \mu \;$colliders if they run at the
Higgs resonance$\;\sqrt{s}=m_{H}$.

On the other hand, the presence of FCNC could be an indirect signature of
the existence of an extended Higgs sector, perhaps the strongest motivation
to look for these rare processes lies on the evidence of neutrino
oscillations. Despite processes with FCNC are strongly suppressed by some
underlying principle still unknown, there is an important region of
parameters in which those processes are still compatible with experimental
constraints. As for the quark sector, the FCNC involving the first family
are strongly constrained, while the bounds on the other mixings are
considerably softened. In this work we concentrate on FCNC in the lepton
sector. If LFV is expected in the neutral sector of leptons because of the
neutrino oscillations, it is reasonable to consider that those processes are
also present in the charged lepton sector as well. Useful constraints
involving the vertices $\xi _{\mu \tau },\xi _{e\tau },\xi _{\tau \tau },\xi
_{\mu \mu }\;$can be derived from the $\left( g-2\right) _{\mu }\;$factor,
and the decays $\mu \rightarrow e\gamma ,\;\tau \rightarrow \mu \gamma ,$\
and $\tau \rightarrow \mu \mu \mu $. In turn, based on the information
gotten for these vertices some upper limits on the decay widths $\Gamma
(\tau \rightarrow e\gamma )$, and $\Gamma (\tau \rightarrow eee)$ are
obtained. Those estimations were calculated by assuming $m_{h^{0}}\approx
115GeV\;$and $m_{A^{0}}\gtrsim m_{h^{0}}$. Since the most recent estimations
of $\Delta a_{\mu },$\ still provides an important window for new Physics,
we obtain from it very interesting lower and upper bounds on the mixing
vertex $\xi _{\mu \tau }\;$at 95\% C.L. Specifically, an allowed interval of 
$7.62\times 10^{-4}\lesssim \xi _{\mu \tau }^{2}\lesssim 4.44\times
10^{-2}\; $was found in a quite wide region of parameters. Of course, we
should realize that both SM test and experimental measurements of $a_{\mu
}\; $are still being scrutinized and current results are not definitive at
all. However, if not severe changes occur in forthcoming experiments and/or
SM estimations, these constraints could continue being valid at least at a
lower confidence level. Future improvements on both estimations should
elucidate this point.

Based on these constraints, and on the leptonic decays $\mu \rightarrow
e\gamma ,\;\tau \rightarrow \mu \gamma ,\;$and $\tau \rightarrow \mu \mu \mu
\;$we got the following conservative and quite general bounds of LFV
vertices 
\begin{eqnarray*}
\xi _{e\tau }^{2} &\lesssim &2.77\times 10^{-14}\;, \\
-1.8\times 10^{-2} &\lesssim &\xi _{\tau \tau }\lesssim 2.2\times
10^{-2}\;;\;\left| \xi _{\mu \mu }\right| \lesssim 0.13\;\;\;\;\text{for\ }%
m_{A^{0}}>>m_{h^{0}} \\
\left| \xi _{\tau \tau }\right| &\lesssim &1.0\times 10^{-2}\;;\;\left| \xi
_{\mu \mu }\right| \lesssim 0.15\;\;\;\;\;\;\;\;\text{for\ }m_{A^{0}}\approx
m_{h^{0}}
\end{eqnarray*}
It is remarkable the very strong hierarchy among the allowed values for $\xi
_{e\tau }^{2}\;$and $\xi _{\mu \tau }^{2}.$ Moreover, these bounds on $\xi
_{\tau \tau }\;$and $\xi _{\mu \mu }\;$are considerably relaxed for certain
specific values of $m_{A^{0}}$. In that case, the room available for them is
such that they permit either a strong enhancement or a strong suppression of
the couplings $H\tau \overline{\tau }\;$and/or $H\mu \overline{\mu }\;$for
any neutral Higgs boson.

Furthermore, we estimate upper limits on the decay widths $\Gamma (\tau
\rightarrow e\gamma )$, and$\;\Gamma (\tau \rightarrow eee)$, finding that
they are basically hopeless to look for, in near future experiments, at
least in the framework of the 2HDM type III with heavy Higgs bosons.

In addition, we see that processes with flavor changing charged currents are
intimately related to the processes with flavor changing neutral currents.
The same mixing elements that produce the FCNC correct the FCCC in the quark
sector (respect to the CKM elements) while producing the FCCC in the
leptonic sector. It worths to point out that the study of the FCCC at tree
level in the 2HDM (III) depend on a less number of free parameters since
only one scalar particle is exchanged, and the Higgs boson involved does not
couple through a mixing angle. For instance, based on the decay width $%
\Gamma \left( \mu ^{-}\rightarrow \nu _{e}e^{-}\overline{\nu }_{\mu }\right)
\;$we can get the upper limit $\left| \xi _{e\mu }/m_{H^{+}}\right| \leq
3.8\times 10^{-3}GeV^{-1}\;$without any further assumption on the free
parameters of the model.

It worths to say that the phenomenological constraints described so far,
were calculated in the fundamental parametrization of the 2HDM(III) with
only one VEV. Nevertheless, it is possible to use other parametrizations for
the 2HDM type III in which both VEV's are taken different from zero.\ In
that case, the part of the Lagrangian involving the Higgs sector is written
in terms of the spurious parameter $\tan \beta $, in such a way that the
Yukawa Lagrangian type III is expressed as the one in the model type I plus
some FC interactions, or as the model type II plus some FC interactions.
These particular parametrizations are equivalent to the fundamental one in
which one of the VEV is zero and $\tan \beta \;$vanishes. However, these non
trivial bases could be useful when we want to compare the model type III
with the models type I or II. Additionally, these parametrizations
facilitates the study of the decoupling limit of the 2HDM. By setting a
fixed value of $\tan \beta \;$we are using a particular basis to write the
model, in that sense, the parameter $\tan \beta \;$acts as a ``gauge
fixing'' term.

Now, based on non trivial parametrizations of model III in which models type
I and II become apparent, we get bounds on the Higgs boson masses, by making
reasonable assumptions on the FC vertices. Since the contribution of the
pseudoscalar Higgs boson to $\Delta a_{\mu }\;$is negative while the
estimated value is positive definite at 95\%\ CL, the estimated value of $%
\Delta a_{\mu }\;$impose lower bounds on the Pseudoscalar Higgs mass.
Specifically, we have taken for $\widetilde{\eta }\left( \widetilde{\xi }%
\right) _{\mu \tau }$ the geometric average of the Yukawa couplings, and we
also utilized values one order of magnitude larger and one order of
magnitude smaller. Taking these three values for the FC vertex we find that
the smaller value for $\widetilde{\eta }\left( \widetilde{\xi }\right) _{\mu
\tau }\;$the more stringent lower bounds for $m_{A^{0}}\;$. Additionally,
assuming $m_{H^{0}}=m_{h^{0}},\;$we show that in the limit of small (large) $%
\tan \beta \;$the lower bound of $m_{A^{0}}\;$becomes merely $%
m_{A^{0}}\approx m_{h^{0}}\;$for parametrization of type I (II). In the case
of different scalar masses, there is still a lower asymptotic limit for $%
m_{A^{0}}$. Notwithstanding, these lower constraints on $m_{A^{0}}\;$should
be considered carefully, since for $\widetilde{\eta }\left( \widetilde{\xi }%
\right) _{\mu \tau }\;$we can only make reasonable estimations but they are
unknown so far.

On the other hand, as a matter of addendum, the production and decays of
top-squarks at the Tevatron collider is examined in the framework of general
low energy supersymmetry breaking models. In models of low-energy SUSY
breaking, signatures of SUSY particle production generically contain two
hard photons plus missing energy due to the decays of the two neutralino
NLSPs produced in the decay chains. Standard Model backgrounds to such
signals are naturally small at Run II of the Tevatron. We studied the
production and decay of top squarks at the Tevatron in such models in the
case where the lightest Standard Model superpartner is a light neutralino
that predominantly decays into a photon and a light gravitino. We considered
2-body flavor-changing and 3-body decays of the top squarks. The reach of
the Tevatron in such models is larger than in the standard supergravity
models and than in models with low-energy SUSY breaking in which the stop is
the NLSP, rather than the neutralino. We estimate that top squarks with
masses below about 200 GeV can be excluded based on Run~I data, assuming
that 50 GeV $\mathrel{\raise.3ex\hbox{$<$\kern-.75em\lower1ex\hbox{$\sim$}}}%
m_{\tilde{\chi}_{1}^{0}}\mathrel{\raise.3ex\hbox{$<$\kern-.75em\lower1ex%
\hbox{$\sim$}}}m_{\tilde{t}}-10$ GeV. For a modest final Run II luminosity
of 4 fb$^{-1}$, stop masses up to 285 GeV are accessible in the 2-body decay
mode, and up to 320 GeV in the 3-body decay mode.

Top squarks can also be produced in top quark decays. We found that, within
the context of low-energy SUSY breaking with the stop as the
next-to-next-to-lightest SUSY particle, the region of parameter space in
which stop production in top quark decays is possible is almost entirely
excluded by Run I data if the lighter stop is predominantly right-handed;
however, an interesting region is still allowed if the lighter stop is
predominantly left-handed, due to the smaller branching ratio of $%
t\rightarrow \tilde{t}_{L}\tilde{\chi}_{1}^{0}$. Run II will cover the
entire parameter space in which top decays to stop are possible.

We also studied the production and decay of the ten squarks associated with
the five light quarks, assumed to be degenerate. In models of low-energy
SUSY breaking, the decays of the ten degenerate squarks lead to signals
identical to those for the 2-body flavor-changing stop decays. The cross
section for production of ten degenerate squarks at the Tevatron is
significantly larger than that of the top squark. We estimate that the 10
degenerate squarks with masses below about 280 GeV can be excluded based on
Run~I data, assuming that $m_{\tilde{\chi}_{1}^{0}}\mathrel{\raise.3ex%
\hbox{$<$\kern-.75em\lower1ex\hbox{$\sim$}}}m_{\tilde{q}}-40$ GeV. For a
final Run II luminosity of 4 fb$^{-1}$, squark masses as large as 360 GeV
are easily accessible in the limit that the gluino is very heavy. The
production cross section, and hence the discovery reach, increases further
with decreasing gluino mass.

\appendix
\setcounter{chapter}{0}

\chapter{Determination of the minima of the potential and the mass
eigenstates\label{minima of the pot}}

\section{Minimum condition for the potential}

The potential can be written in terms of the following hermitian gauge
invariant operators 
\begin{eqnarray*}
\widehat{A} &\equiv &\Phi _{1}^{\dagger }\Phi _{1}\;,\;\widehat{B}\equiv
\Phi _{2}^{\dagger }\Phi _{2},\;\widehat{C}\equiv \frac{1}{2}\left( \Phi
_{1}^{\dagger }\Phi _{2}+\Phi _{2}^{\dagger }\Phi _{1}\right) =\text{Re}%
\left( \Phi _{1}^{\dagger }\Phi _{2}\right) ,\; \\
\widehat{D} &\equiv &-\frac{i}{2}\left( \Phi _{1}^{\dagger }\Phi _{2}-\Phi
_{2}^{\dagger }\Phi _{1}\right) =\text{Im}\left( \Phi _{1}^{\dagger }\Phi
_{2}\right)
\end{eqnarray*}
The most general renormalizable and $SU\left( 2\right) \times U\left(
1\right) \;$invariant potential is given by 
\begin{eqnarray}
V_{g} &=&-\mu _{1}^{2}\widehat{A}-\mu _{2}^{2}\widehat{B}-\mu _{3}^{2}%
\widehat{C}-\mu _{4}^{2}\widehat{D}+\lambda _{1}\widehat{A}^{2}+\lambda _{2}%
\widehat{B}^{2}+\lambda _{3}\widehat{C}^{2}+\lambda _{4}\widehat{D}^{2} 
\notag \\
&&+\lambda _{5}\widehat{A}\widehat{B}+\lambda _{6}\widehat{A}\widehat{C}%
+\lambda _{8}\widehat{A}\widehat{D}+\lambda _{7}\widehat{B}\widehat{C}%
+\lambda _{9}\widehat{B}\widehat{D}+\lambda _{10}\widehat{C}\widehat{D}
\label{Vg}
\end{eqnarray}
For the sake of simplicity, we shall use in this appendix the following
parametrization for the doublets 
\begin{equation}
\Phi _{1}=\left( 
\begin{array}{c}
\phi _{1}+i\phi _{2} \\ 
\phi _{3}+i\phi _{4}
\end{array}
\right) \;\;\;;\;\;\;\Phi _{2}=\left( 
\begin{array}{c}
\phi _{5}+i\phi _{6} \\ 
\phi _{7}+i\phi _{8}
\end{array}
\right)  \label{8 fields}
\end{equation}
we will assume from now on, that both VEV can be taken real (i.e. there is
not spontaneous $CP-$ violation) such that$\;\langle \phi _{3}\rangle =v_{1}/%
\sqrt{2}\;;\;\langle \phi _{7}\rangle =v_{2}/\sqrt{2}$. Now, we examine the
minimum conditions (tadpoles).

\begin{equation*}
T_{i}=\left. \frac{\partial V}{\partial \phi _{i}}\right| _{\phi _{3}=v_{1}/%
\sqrt{2},\phi _{7}=v_{2}/\sqrt{2}}=0
\end{equation*}
where $\phi _{i}$, with $i=1,..,8\;$denotes each gauge eigenstate of the
scalars defined in (\ref{8 fields}). We obtain the following non trivial
equations 
\begin{eqnarray}
0 &=&T_{3}=\frac{1}{4}\lambda _{7}v_{2}^{3}+\lambda _{1}v_{1}^{3}+\frac{3}{4}%
\lambda _{6}v_{1}^{2}v_{2}-\mu _{1}^{2}\allowbreak v_{1}+\frac{1}{2}\lambda
_{3}v_{2}^{2}v_{1}+\frac{1}{2}\lambda _{5}v_{1}v_{2}^{2}-\frac{1}{2}\mu
_{3}^{2}\allowbreak v_{2}  \notag \\
0 &=&T_{4}=\left( -2\mu _{4}^{2}+\lambda _{9}v_{2}^{2}+\lambda
_{8}v_{1}^{2}+\allowbreak \lambda _{10}v_{2}v_{1}\right) v_{2}  \notag \\
0 &=&T_{7}=\frac{3}{4}\lambda _{7}v_{2}^{2}v_{1}-\mu _{2}^{2}v_{2}+\lambda
_{2}v_{2}^{3}-\frac{1}{2}\mu _{3}^{2}\allowbreak v_{1}+\frac{1}{4}\lambda
_{6}v_{1}^{3}+\frac{1}{2}\lambda _{3}v_{1}^{2}v_{2}+\frac{1}{2}\lambda
_{5}v_{2}\allowbreak v_{1}^{2}  \notag \\
0 &=&T_{8}=\left( 2\mu _{4}^{2}+\lambda _{9}v_{2}^{2}+\allowbreak \lambda
_{8}v_{1}^{2}+\lambda _{10}v_{2}v_{1}\right) v_{1}  \label{min conditions}
\end{eqnarray}

Now, the mass matrix elements are gotten from the terms quadratic in the
scalars 
\begin{equation}
M_{ij}^{2}=\frac{1}{2}\left. \frac{\partial ^{2}V}{\partial \phi
_{i}\partial \phi _{j}}\right| _{\phi _{3}=v_{1}/\sqrt{2},\phi _{7}=v_{2}/%
\sqrt{2}}  \label{second deriv}
\end{equation}

from (\ref{second deriv}, \ref{Vg}), and (\ref{8 fields}) the $8\times 8\;$%
mass matrix becomes

\begin{equation}
\left( 
\begin{array}{cccccccc}
M_{11}^{2} & 0 & 0 & 0 & M_{15}^{2} & M_{16}^{2} & 0 & 0 \\ 
0 & M_{22}^{2} & 0 & 0 & M_{25}^{2} & M_{26}^{2} & 0 & 0 \\ 
0 & 0 & M_{33}^{2} & M_{34}^{2} & 0 & 0 & M_{37}^{2} & M_{38}^{2} \\ 
0 & 0 & M_{34}^{2} & M_{44}^{2} & 0 & 0 & M_{47}^{2} & M_{48}^{2} \\ 
M_{15}^{2} & M_{25}^{2} & 0 & 0 & M_{55}^{2} & 0 & 0 & 0 \\ 
M_{16}^{2} & M_{26}^{2} & 0 & 0 & 0 & M_{66}^{2} & 0 & 0 \\ 
0 & 0 & M_{37}^{2} & M_{47}^{2} & 0 & 0 & M_{77}^{2} & M_{78}^{2} \\ 
0 & 0 & M_{38}^{2} & M_{48}^{2} & 0 & 0 & M_{78}^{2} & M_{88}^{2}
\end{array}
\right)  \label{complete mass matrix}
\end{equation}
where the matrix elements are given by

\begin{eqnarray*}
M_{11}^{2} &=&-\allowbreak \mu _{1}^{2}+\lambda _{1}v_{1}^{2}+\frac{1}{2}%
\lambda _{5}v_{2}^{2}+\frac{1}{2}\lambda _{6}v_{1}v_{2} \\
M_{15}^{2} &=&-\frac{1}{2}\mu _{3}^{2}+\frac{1}{2}\allowbreak \lambda
_{3}v_{1}v_{2}+\frac{1}{4}\lambda _{6}v_{1}^{2}+\frac{1}{4}\lambda
_{7}v_{2}^{2} \\
M_{16}^{2} &=&-\frac{1}{2}\mu _{4}^{2}\allowbreak +\frac{1}{4}\lambda
_{10}v_{1}v_{2}+\frac{1}{4}\lambda _{8}v_{1}^{2}+\frac{1}{4}\lambda
_{9}v_{2}^{2} \\
M_{22}^{2} &=&-\allowbreak \mu _{1}^{2}+\lambda _{1}v_{1}^{2}+\frac{1}{2}%
\lambda _{5}v_{2}^{2}+\frac{1}{2}\lambda _{6}v_{1}v_{2} \\
M_{25}^{2} &=&\allowbreak -\frac{1}{4}\lambda _{9}v_{2}^{2}-\frac{1}{4}%
\lambda _{10}v_{1}v_{2}+\frac{1}{2}\mu _{4}^{2}-\frac{1}{4}\lambda
_{8}v_{1}^{2} \\
M_{26}^{2} &=&-\allowbreak \frac{1}{2}\mu _{3}^{2}+\frac{1}{4}\lambda
_{7}v_{2}^{2}+\frac{1}{4}\lambda _{6}v_{1}^{2}+\frac{1}{2}\lambda
_{3}v_{1}v_{2} \\
M_{33}^{2} &=&-\allowbreak \mu _{1}^{2}+3\lambda _{1}v_{1}^{2}+\frac{1}{2}%
\lambda _{3}v_{2}^{2}+\frac{1}{2}\lambda _{5}v_{2}^{2}+\frac{3}{2}\lambda
_{6}v_{1}v_{2}\allowbreak \\
M_{34}^{2} &=&\allowbreak -\frac{1}{2}\lambda _{8}v_{1}v_{2}-\frac{1}{4}%
\lambda _{10}v_{2}^{2} \\
M_{37}^{2} &=&-\allowbreak \frac{1}{2}\mu _{3}^{2}+\frac{3}{4}\lambda
_{7}v_{2}^{2}+\frac{3}{4}\lambda _{6}v_{1}^{2}+\lambda
_{5}v_{1}v_{2}+\lambda _{3}\allowbreak v_{1}v_{2} \\
M_{38}^{2} &=&\allowbreak \frac{1}{4}\lambda _{9}v_{2}^{2}+\frac{1}{2}%
\lambda _{10}v_{1}v_{2}-\frac{1}{2}\mu _{4}^{2}+\frac{3}{4}\lambda
_{8}v_{1}^{2}
\end{eqnarray*}
\begin{eqnarray}
M_{44}^{2} &=&-\allowbreak \mu _{1}^{2}+\lambda _{1}v_{1}^{2}+\frac{1}{2}%
\lambda _{4}v_{2}^{2}+\frac{1}{2}\lambda _{5}v_{2}^{2}+\frac{1}{2}\lambda
_{6}v_{1}v_{2}\;  \notag \\
M_{47}^{2} &=&\allowbreak -\frac{3}{4}\lambda _{9}v_{2}^{2}-\frac{1}{2}%
\lambda _{10}v_{1}v_{2}+\frac{1}{2}\mu _{4}^{2}-\frac{1}{4}\lambda
_{8}v_{1}^{2}  \notag \\
M_{48}^{2} &=&-\allowbreak \frac{1}{2}\mu _{3}^{2}+\frac{1}{4}\lambda
_{7}v_{2}^{2}+\frac{1}{4}\lambda _{6}v_{1}^{2}+\frac{1}{2}\lambda
_{3}v_{1}v_{2}-\frac{1}{2}\lambda _{4}v_{1}\allowbreak v_{2}  \notag \\
M_{55}^{2} &=&\allowbreak -\mu _{2}^{2}+\lambda _{2}v_{2}^{2}+\frac{1}{2}%
\lambda _{5}v_{1}^{2}+\frac{1}{2}\lambda _{7}v_{1}v_{2}  \notag \\
M_{66}^{2} &=&-\allowbreak \mu _{2}^{2}+\lambda _{2}v_{2}^{2}+\frac{1}{2}%
\lambda _{5}v_{1}^{2}+\frac{1}{2}\lambda _{7}v_{1}v_{2}  \notag \\
M_{77}^{2} &=&-\allowbreak \mu _{2}^{2}+3\lambda _{2}v_{2}^{2}+\frac{1}{2}%
\lambda _{3}v_{1}^{2}+\frac{1}{2}\lambda _{5}v_{1}^{2}+\frac{3}{2}\lambda
_{7}v_{1}v_{2}\allowbreak  \notag \\
M_{78}^{2} &=&\allowbreak \frac{1}{2}\lambda _{9}v_{2}v_{1}+\frac{1}{4}%
\lambda _{10}v_{1}^{2}  \notag \\
M_{88}^{2} &=&-\allowbreak \mu _{2}^{2}+\lambda _{2}v_{2}^{2}+\frac{1}{2}%
\lambda _{4}v_{1}^{2}+\frac{1}{2}\lambda _{5}v_{1}^{2}+\frac{1}{2}\lambda
_{7}v_{1}v_{2}  \label{complete mass terms}
\end{eqnarray}
Observe that $v_{2}=0$,$\;$is a valid solution for Eqs. (\ref{min conditions}%
). Now, if we solve using $v_{2}=0,\;$and replace the other minimum
conditions into the mass terms (\ref{complete mass terms}) we get 
\begin{eqnarray*}
M_{11}^{2} &=&0\;;\;M_{15}^{2}=0\;;\;M_{16}^{2}=\frac{1}{2}\lambda
_{8}v_{1}^{2}\;;\;M_{22}^{2}=0\;;\;M_{25}^{2}=\allowbreak -\frac{1}{2}%
\lambda _{8}v_{1}^{2} \\
M_{26}^{2} &=&0;\;M_{33}^{2}=2\lambda _{1}v_{1}^{2}\allowbreak
\;;\;M_{34}^{2}=\allowbreak 0\;;\;M_{37}^{2}=\frac{1}{2}\lambda
_{6}v_{1}^{2}\;;\;M_{38}^{2}=\lambda _{8}v_{1}^{2}
\end{eqnarray*}
\begin{eqnarray*}
M_{44}^{2} &=&0\;;\;M_{47}^{2}=\allowbreak -\frac{1}{2}\lambda
_{8}v_{1}^{2}\;;\;M_{48}^{2}=0\;;\;M_{55}^{2}=\allowbreak -\mu _{2}^{2}+%
\frac{1}{2}\lambda _{5}v_{1}^{2} \\
M_{66}^{2} &=&-\allowbreak \mu _{2}^{2}+\frac{1}{2}\lambda
_{5}v_{1}^{2}\;;\;M_{77}^{2}=-\allowbreak \mu _{2}^{2}+\frac{1}{2}\lambda
_{3}v_{1}^{2}+\frac{1}{2}\lambda _{5}v_{1}^{2}\allowbreak \\
M_{78}^{2} &=&\frac{1}{4}\lambda _{10}v_{1}^{2}\;;\;M_{88}^{2}=-\allowbreak
\mu _{2}^{2}+\frac{1}{2}\lambda _{4}v_{1}^{2}+\frac{1}{2}\lambda
_{5}v_{1}^{2}
\end{eqnarray*}

We can see that the potential \ref{Vg} violates CP explicitly by means of
the terms $M_{16},\;M_{25},\;M_{34},\;M_{38},\;M_{47},\;M_{78}\;$in the mass
matrix (\ref{complete mass matrix}), since they mix real parts with
imaginary parts of the complex neutral fields $\phi _{1}^{0},\;\phi _{2}^{0}$%
, see Eqs. (\ref{8 fields}, \ref{first param}). As we shall see below, these
terms vanish in the potentials (\ref{Lag10},\ \ref{VAP}, \ref{VB}) avoiding
explicit CP violation. Notwithstanding, potentials (\ref{Lag10},\ \ref{VA})
can exhibit spontaneous CP violation while potentials (\ref{VAP}, \ref{VB})
do not allow spontaneous CP violation either.

Before continuing with some special cases of the potential, we shall
introduce some useful formulae. If we have a $2\times 2\;$real symmetric
matrix 
\begin{equation*}
A=\left( 
\begin{array}{cc}
a & c \\ 
c & b
\end{array}
\right)
\end{equation*}
its eigenvalues and orthonormal eigenvectors are given by 
\begin{eqnarray}
k_{1,2} &=&\frac{1}{2}\left[ a+b\pm \sqrt{\left( a-b\right) ^{2}+4c^{2}}%
\right]  \notag \\
\left( 
\begin{array}{c}
\cos \delta \\ 
\sin \delta
\end{array}
\right) &\equiv &\overrightarrow{u}_{1}\leftrightarrow k_{1}\;\;;\;\;\left( 
\begin{array}{c}
-\sin \delta \\ 
\cos \delta
\end{array}
\right) \equiv \overrightarrow{u}_{2}\leftrightarrow k_{2}  \notag \\
\sin 2\delta &=&\frac{2c}{\sqrt{\left( a-b\right) ^{2}+4c^{2}}}\;\;;\;\;\cos
2\delta =\frac{\left( a-b\right) }{\sqrt{\left( a-b\right) ^{2}+4c^{2}}}
\label{diagonalization}
\end{eqnarray}
from which we get 
\begin{eqnarray*}
UAU^{\dagger } &=&\left( 
\begin{array}{cc}
k_{1} & 0 \\ 
0 & k_{2}
\end{array}
\right) \\
U &\equiv &\left( 
\begin{array}{c}
\overrightarrow{u}_{1}^{T} \\ 
\overrightarrow{u}_{2}^{T}
\end{array}
\right) =\left( 
\begin{array}{cc}
\cos \delta & \sin \delta \\ 
-\sin \delta & \cos \delta
\end{array}
\right)
\end{eqnarray*}
further, if we start from a doublet of gauge eigenstates 
\begin{equation*}
\Omega \equiv \left( 
\begin{array}{c}
\Omega _{1} \\ 
\Omega _{2}
\end{array}
\right)
\end{equation*}
and settle the following definitions 
\begin{equation*}
H\equiv U\Omega \equiv \left( 
\begin{array}{c}
H_{1} \\ 
H_{2}
\end{array}
\right) \;,\;M\equiv UAU^{\dagger }=\left( 
\begin{array}{cc}
k_{1} & 0 \\ 
0 & k_{2}
\end{array}
\right)
\end{equation*}
we can check that, owing to the unitarity of the rotation matrix $U\;$the
following identity is held 
\begin{equation*}
\Omega ^{\dagger }A\Omega =H^{\dagger }MH
\end{equation*}
and since $M\,\;$is diagonal, we say that the doublet $H\;$contains the mass
eigenstates $H_{1},\;H_{2}$. They$\;$are obtained by rotating the original
gauge eigenstates through the unitary matrix $U$;$\;$and $k_{1},\;k_{2}\;$%
correspond to the squared masses of the mass eigenstates $H_{1},\;H_{2}\;$%
respectively. These results will be applied widely in foregoing issues.

From now on, we shall examine some special cases of the potential (\ref{Vg}).

\subsection{The potential with $C-$invariance\label{minima of Lag10}}

According to section (\ref{Higgspot 2HDM}), if we demand from the potential
to be $C-$invariant, we ought to settle $\mu _{4}^{2}=\lambda _{8}=\lambda
_{9}=\lambda _{10}=0.\;$Additionally, we can always make a rotation on the
two doublets in such a way that only one of them get a VEV (see appendix \ref
{rotation pot} for details). So we can take $\langle \phi _{3}\rangle =v_{1}/%
\sqrt{2}\;;\;\langle \phi _{7}\rangle =v_{2}/\sqrt{2}=0$, without any loss
of generality.

With the assumptions above, the minimum conditions are reduced to only two
equations

\begin{equation*}
\mu _{1}^{2}\allowbreak =\lambda _{1}v_{1}^{2}\;\;;\;\;\mu
_{3}^{2}\allowbreak =\frac{\lambda _{6}v_{1}^{2}}{2}
\end{equation*}
replacing them into the mass matrix elements we get

\begin{eqnarray*}
M_{15}^{2} &=&-\frac{1}{2}\mu _{3}^{2}+\frac{1}{4}\lambda
_{6}v_{1}^{2}\;;\;M_{33}^{2}=\allowbreak 2\lambda
_{1}v_{1}^{2}\;;\;M_{37}^{2}=\frac{1}{2}\lambda _{6}v_{1}^{2} \\
M_{55}^{2} &=&-\allowbreak \mu _{2}^{2}+\frac{1}{2}\lambda
_{5}v_{1}^{2}\;;\;M_{66}^{2}=-\allowbreak \mu _{2}^{2}+\frac{1}{2}\lambda
_{5}v_{1}^{2} \\
M_{77}^{2} &=&-\allowbreak \mu _{2}^{2}+\frac{1}{2}\lambda _{3}v_{1}^{2}+%
\frac{1}{2}\lambda _{5}v_{1}^{2}\allowbreak \;;\;M_{88}^{2}=-\allowbreak \mu
_{2}^{2}+\frac{1}{2}\lambda _{4}v_{1}^{2}+\frac{1}{2}\lambda _{5}v_{1}^{2}
\end{eqnarray*}
the other terms vanish. Now, we can relabel the matrix in the following way $%
1,2,3,4,5,6,7,8\rightarrow 1,2,5,6,3,7,4,8\;$the new positions of the above
elements in this new matrix are\footnote{%
It is equivalent to relabel the scalar fields $\phi _{i}\;$which clearly
does not affect the physical content of the mass matrix.} 
\begin{equation}
\left( 
\begin{array}{cccccccc}
0 & 0 & 0 & 0 & 0 & 0 & 0 & 0 \\ 
0 & 0 & 0 & 0 & 0 & 0 & 0 & 0 \\ 
0 & 0 & M_{55}^{2} & 0 & 0 & 0 & 0 & 0 \\ 
0 & 0 & 0 & M_{66}^{2} & 0 & 0 & 0 & 0 \\ 
0 & 0 & 0 & 0 & M_{33}^{2} & M_{37}^{2} & 0 & 0 \\ 
0 & 0 & 0 & 0 & M_{73}^{2} & M_{77}^{2} & 0 & 0 \\ 
0 & 0 & 0 & 0 & 0 & 0 & 0 & 0 \\ 
0 & 0 & 0 & 0 & 0 & 0 & 0 & M_{88}^{2}
\end{array}
\right)  \label{matrix lag10}
\end{equation}
which can be decomposed in the following way 
\begin{equation}
\left( 
\begin{array}{cccc}
0 & 0 & 0 & 0 \\ 
0 & 0 & 0 & 0 \\ 
0 & 0 & M_{55}^{2} & 0 \\ 
0 & 0 & 0 & M_{66}^{2}
\end{array}
\right) \oplus \left( 
\begin{array}{cc}
M_{33}^{2} & M_{37}^{2} \\ 
M_{73}^{2} & M_{77}^{2}
\end{array}
\right) \oplus \left( 
\begin{array}{cc}
0 & 0 \\ 
0 & M_{88}^{2}
\end{array}
\right)  \label{decompose lag10}
\end{equation}

Additionally, since $M_{55}^{2}=M_{66}^{2}\;$we get for the first matrix

\begin{equation}
\left( 
\begin{array}{cccc}
0 & 0 & 0 & 0 \\ 
0 & 0 & 0 & 0 \\ 
0 & 0 & M_{55}^{2} & 0 \\ 
0 & 0 & 0 & M_{66}^{2}
\end{array}
\right) =\left( 
\begin{array}{cc}
0 & 0 \\ 
0 & M_{55}^{2}
\end{array}
\right) \otimes I_{2\times 2}  \label{adddecomp lag10}
\end{equation}

With this decomposition it is easier to diagonalize the matrix (\ref{matrix
lag10}) since it is equivalent to diagonalize each of the submatrices given
above. Let us start by diagonalizing the matrix (\ref{adddecomp lag10})
corresponding to the indices $1,2,5,6.\;$Its orthonormalized eigenvectors
and eigenvalues are given by 
\begin{eqnarray}
\left\{ \left( 
\begin{array}{cccc}
1 & 0 & 0 & 0
\end{array}
\right) ^{T},\left( 
\begin{array}{cccc}
0 & 1 & 0 & 0
\end{array}
\right) ^{T}\right\} &\rightarrow &m_{G^{+}}^{2}=0  \notag \\
\left\{ \left( 
\begin{array}{cccc}
0 & 0 & 1 & 0
\end{array}
\right) ^{T},\left( 
\begin{array}{cccc}
0 & 0 & 0 & 1
\end{array}
\right) ^{T}\right\} &\rightarrow &m_{H^{+}}^{2}=-\mu _{2}^{2}+\frac{1}{2}%
\lambda _{5}v_{1}^{2}  \label{charged eigen}
\end{eqnarray}
thus each eigenvalue is doubly degenerate. Therefore, they correspond to
charged eigenstates. The eigenvalue $m_{G^{+}}^{2}=0\;$correspond to a would
be Goldstone boson (massless scalar) while $m_{H^{+}}^{2}=-\mu _{2}^{2}+%
\frac{1}{2}\lambda _{5}v_{1}^{2}\;$correspond to the mass of a charged Higgs
boson. According to the eigenvectors obtained in (\ref{charged eigen}) the
mass eigenstates are given by 
\begin{equation*}
\left( 
\begin{array}{c}
G_{1} \\ 
G_{2} \\ 
H_{1} \\ 
H_{2}
\end{array}
\right) =\left( 
\begin{array}{cccc}
1 & 0 & 0 & 0 \\ 
0 & 1 & 0 & 0 \\ 
0 & 0 & 1 & 0 \\ 
0 & 0 & 0 & 1
\end{array}
\right) \left( 
\begin{array}{c}
\phi _{1} \\ 
\phi _{2} \\ 
\phi _{5} \\ 
\phi _{6}
\end{array}
\right)
\end{equation*}

But taking into account that to define a scalar charged Higgs boson we
require two degrees of freedom, we realize that the first two scalar fields
define the charged would be Goldstone boson $G^{+}$,$\;$which is written as
a\ linear combination of the scalars $\phi _{1},\phi _{2}$. The assigment of
both degrees of freedom can be done as $\left( \phi _{1},\phi _{2}\right)
^{T}\rightarrow \phi _{1}+i\phi _{2}\equiv \phi _{1}^{+}$;$\;$in the same
way $\left( G_{1},G_{2}\right) ^{T}\rightarrow G_{1}+iG_{2}\equiv G^{+},\;$%
something similar happens to $\left( \phi _{5},\phi _{6}\right) ^{T}\equiv
\phi _{2}^{+}\;$and$\;\left( H_{1},H_{2}\right) ^{T}\equiv H^{+}$.
Therefore, the relation among the gauge and mass eigenstates of the charged
scalar fields can be simplified as\footnote{%
Noteworthy, this is the opposite procedure respect to the one developed in
Sec. (\ref{kin 2HDM}), in which we extend the dimension of the
representation, see for example Eq. (\ref{doubling}). In this case we are
``shrinking'' the dimension of the representation.} 
\begin{equation*}
\left( 
\begin{array}{c}
G^{+} \\ 
H^{+}
\end{array}
\right) =\left( 
\begin{array}{cc}
1 & 0 \\ 
0 & 1
\end{array}
\right) \left( 
\begin{array}{c}
\phi _{1}^{+} \\ 
\phi _{2}^{+}
\end{array}
\right)
\end{equation*}

The second matrix in (\ref{decompose lag10}) correspond to the indices 3,7. 
\begin{equation*}
\left( 
\begin{array}{cc}
M_{33}^{2} & M_{37}^{2} \\ 
M_{73}^{2} & M_{77}^{2}
\end{array}
\right) =\left( 
\begin{array}{cc}
\allowbreak 2\lambda _{1}v_{1}^{2} & \frac{1}{2}\lambda _{6}v_{1}^{2} \\ 
\frac{1}{2}\lambda _{6}v_{1}^{2} & -\allowbreak \mu _{2}^{2}+\frac{1}{2}%
\lambda _{3}v_{1}^{2}+\frac{1}{2}\lambda _{5}v_{1}^{2}
\end{array}
\right)
\end{equation*}
applying Eqs. (\ref{diagonalization}) the eigenvalues and orthonormalized
eigenvectors are: 
\begin{equation*}
m_{H^{0},h^{0}}^{2}=\left( \lambda _{1}+\frac{1}{2}\lambda _{+}\right)
v_{1}^{2}-\allowbreak \frac{1}{2}\mu _{2}^{2}\pm \sqrt{\allowbreak \left[
\left( \lambda _{1}-\frac{1}{2}\lambda _{+}\right) v_{1}^{2}+\allowbreak 
\frac{1}{2}\mu _{2}^{2}\right] ^{2}+\left( \frac{1}{2}\lambda
_{6}v_{1}^{2}\right) ^{2}}
\end{equation*}
\begin{eqnarray*}
\left( 
\begin{array}{c}
\cos \alpha \\ 
\sin \alpha
\end{array}
\right) &\leftrightarrow &m_{H^{0}}^{2}\;\;;\;\;\left( 
\begin{array}{c}
-\sin \alpha \\ 
\cos \alpha
\end{array}
\right) \leftrightarrow m_{h^{0}}^{2} \\
\tan 2\alpha &=&\frac{\lambda _{6}v_{1}^{2}}{\left( \allowbreak 2\lambda
_{1}-\lambda _{+}\right) v_{1}^{2}+\allowbreak \mu _{2}^{2}}\;\;;\;\;\lambda
_{+}\equiv \frac{1}{2}\left( \lambda _{3}+\lambda _{5}\right)
\end{eqnarray*}
therefore, the diagonalization process is performed by the following
rotation of the gauge eigenstates 
\begin{equation*}
\left( 
\begin{array}{cc}
\cos \alpha & \sin \alpha \\ 
-\sin \alpha & \cos \alpha
\end{array}
\right) \left( 
\begin{array}{c}
\sqrt{2}\phi _{3}-v_{1} \\ 
\sqrt{2}\phi _{7}-v_{2}
\end{array}
\right) =\left( 
\begin{array}{c}
H^{0} \\ 
h^{0}
\end{array}
\right)
\end{equation*}
but according to the parametrizations (\ref{8 fields}) and (\ref{doub param}%
) of the Higgs doublets

\begin{equation*}
\left( \sqrt{2}\phi _{3,7}-v_{1,2}\right) \rightarrow h_{1,2}
\end{equation*}
obtaining 
\begin{equation*}
\left( 
\begin{array}{cc}
\cos \alpha & \sin \alpha \\ 
-\sin \alpha & \cos \alpha
\end{array}
\right) \left( 
\begin{array}{c}
h_{1} \\ 
h_{2}
\end{array}
\right) =\left( 
\begin{array}{c}
H^{0} \\ 
h^{0}
\end{array}
\right)
\end{equation*}

Finally, we diagonalize the submatrix corresponding to the elements 4,8 in
Eq. (\ref{decompose lag10})

\begin{equation*}
\left( 
\begin{array}{cc}
M_{44}^{2} & M_{48}^{2} \\ 
M_{84}^{2} & M_{88}^{2}
\end{array}
\right) =\left( 
\begin{array}{cc}
0 & 0 \\ 
0 & \allowbreak -\mu _{2}^{2}+\frac{1}{2}\left( \lambda _{4}+\lambda
_{5}\right) v_{1}^{2}
\end{array}
\right)
\end{equation*}
whose eigenvectors and eigenvalues are

\begin{equation*}
\left\{ \left( 
\begin{array}{c}
1 \\ 
0
\end{array}
\right) \right\} \leftrightarrow 0,\left\{ \left( 
\begin{array}{c}
0 \\ 
1
\end{array}
\right) \right\} \leftrightarrow -\mu _{2}^{2}+\frac{1}{2}v_{1}^{2}\lambda
_{4}+\frac{1}{2}v_{1}^{2}\lambda _{5}
\end{equation*}
the first eigenvalue correspond to a neutral, massless would Goldstone boson,%
$\allowbreak \;$while the second one is associated to another neutral Higgs
boson. 
\begin{equation*}
m_{G^{0}}=0\;\text{;\ }m_{A^{0}}=-\allowbreak \mu _{2}^{2}+\frac{1}{2}\left(
\lambda _{4}+\lambda _{5}\right) v_{1}^{2}
\end{equation*}

The mass eigenstates are given by

\begin{equation*}
\left( 
\begin{array}{cc}
1 & 0 \\ 
0 & 1
\end{array}
\right) \left( 
\begin{array}{c}
\phi _{4} \\ 
\phi _{8}
\end{array}
\right) =\left( 
\begin{array}{c}
G^{0} \\ 
A^{0}
\end{array}
\right)
\end{equation*}

\subsection{The potential with a $Z_{2}\;$invariance}

As explained in section (\ref{Higgspot 2HDM}) a way to avoid spontaneous $%
CP\;$violation consists of demanding invariance under the $Z_{2}$ symmetry\ $%
\Phi _{1}\rightarrow \Phi _{1},\;\Phi _{2}\rightarrow -\Phi _{2}\;\;$(as
well as the $C-$invariance) and correspond to setting $\mu _{3}^{2}=\mu
_{4}^{2}=\lambda _{6}=\lambda _{7}=\lambda _{8}=\lambda _{9}=\lambda
_{10}=0\;$in (\ref{Vg}).$\;$The minimum conditions (\ref{min conditions})
are reduced to two equations:

\begin{eqnarray*}
v_{1}\left[ -\mu _{1}^{2}\allowbreak +\lambda _{1}v_{1}^{2}+\lambda
_{+}v_{2}^{2}\right] &=&0 \\
v_{2}\left[ -\mu _{2}^{2}+\lambda _{2}v_{2}^{2}+\lambda _{+}v_{1}^{2}\right]
&=&0
\end{eqnarray*}

where$\;\lambda _{+}=\frac{1}{2}\left( \lambda _{3}+\lambda _{5}\right) .\;$%
So we have two sets of independent solutions

a) 
\begin{equation*}
v_{1}^{2}=\frac{\lambda _{2}\mu _{1}^{2}-\lambda _{+}\mu _{2}^{2}}{\lambda
_{1}\lambda _{2}-\lambda _{+}^{2}}\;\;\;;\;\;\;v_{2}^{2}=\frac{\lambda
_{1}\mu _{2}^{2}-\lambda _{+}\mu _{1}^{2}}{\lambda _{1}\lambda _{2}-\lambda
_{+}^{2}}
\end{equation*}
or b)

\begin{equation*}
v_{2}^{2}=0\;\;\;;\;\;\;v_{1}^{2}=\frac{\mu _{1}^{2}}{\lambda _{1}}
\end{equation*}

\subsubsection{First set of solutions}

The decomposed mass matrix, after using the minimum conditions, relabeling
and taking into account that $M_{15}^{2}=M_{26}^{2},\;M_{11}^{2}=M_{22}^{2},%
\;$and $M_{55}^{2}=M_{66}^{2}\;$becomes

\begin{equation*}
M_{tot}^{2}=\left[ \left( 
\begin{array}{cc}
M_{11}^{2} & M_{15}^{2} \\ 
M_{15}^{2} & M_{55}^{2}
\end{array}
\right) \otimes I_{2\times 2}\right] \oplus \left( 
\begin{array}{cc}
M_{33}^{2} & M_{37}^{2} \\ 
M_{37}^{2} & M_{77}^{2}
\end{array}
\right) \oplus \left( 
\begin{array}{cc}
M_{44}^{2} & M_{48}^{2} \\ 
M_{48}^{2} & M_{88}^{2}
\end{array}
\right)
\end{equation*}
\ 
\begin{eqnarray*}
M_{11}^{2} &=&-\frac{1}{2}v_{2}^{2}\lambda _{3}\;;\;M_{15}^{2}=\frac{1}{2}%
\allowbreak \lambda _{3}v_{1}v_{2}\;;\;M_{22}^{2}=-\frac{1}{2}%
v_{2}^{2}\lambda _{3}\;;\;M_{26}^{2}=\frac{1}{2}\lambda _{3}v_{1}v_{2} \\
M_{33}^{2} &=&2\lambda _{1}v_{1}^{2}\;;\;M_{37}^{2}=2\lambda _{+}\allowbreak
v_{1}v_{2}\;;\;M_{44}^{2}=\lambda _{-}v_{2}^{2}\;;\;M_{48}^{2}=-\lambda
_{-}v_{1}\allowbreak v_{2} \\
M_{55}^{2} &=&\allowbreak -\frac{1}{2}v_{1}^{2}\lambda _{3}\;;\;M_{66}^{2}=-%
\frac{1}{2}v_{1}^{2}\lambda _{3}\;;\;M_{77}^{2}=2\lambda
_{2}v_{2}^{2}\allowbreak \;;\;M_{88}^{2}=\lambda _{-}v_{1}^{2}
\end{eqnarray*}

where $\lambda _{-}\equiv \frac{1}{2}\left( \lambda _{4}-\lambda _{3}\right) 
$.\ We first diagonalize the submatrix corresponding to $1,5$.\ The
eigenvalues and orthonormal eigenvectors are 
\begin{eqnarray*}
\left\{ \frac{1}{\sqrt{1+\left( \frac{v_{2}}{v_{1}}\right) ^{2}}}\left( 
\begin{array}{c}
1 \\ 
\frac{v_{2}}{v_{1}}
\end{array}
\right) \right\} &\leftrightarrow &0\;,\allowbreak \\
\left\{ \frac{1}{\sqrt{1+\left( \frac{v_{2}}{v_{1}}\right) ^{2}}}\left( 
\begin{array}{c}
-\frac{v_{2}}{v_{1}} \\ 
1
\end{array}
\right) \right\} &\leftrightarrow &\frac{1}{2}\frac{\left(
v_{1}^{2}+v_{2}^{2}\right) \left( \mu _{3}^{2}-\lambda _{3}v_{1}v_{2}\right) 
}{v_{1}v_{2}}
\end{eqnarray*}
which can be written as 
\begin{eqnarray*}
\left( 
\begin{array}{c}
\cos \beta \\ 
\sin \beta
\end{array}
\right) &\leftrightarrow &0\;\;;\;\;\left( 
\begin{array}{c}
-\sin \beta \\ 
\cos \beta
\end{array}
\right) \leftrightarrow -\frac{1}{2}v_{1}^{2}\lambda _{3}-\frac{1}{2}%
v_{2}^{2}\lambda _{3} \\
\cos \beta &=&\frac{v_{1}}{\sqrt{v_{1}^{2}+v_{2}^{2}}}\;\;;\;\allowbreak
\sin \beta =\frac{v_{2}}{\sqrt{v_{1}^{2}+v_{2}^{2}}}
\end{eqnarray*}

So the mass Higgs bosons, and mass eigenstates are 
\begin{eqnarray*}
\left( 
\begin{array}{c}
G^{+} \\ 
H^{+}
\end{array}
\right) &=&\left( 
\begin{array}{cc}
\cos \beta & \sin \beta \\ 
-\sin \beta & \cos \beta
\end{array}
\right) \left( 
\begin{array}{c}
\phi _{1}^{+} \\ 
\phi _{2}^{+}
\end{array}
\right) \\
m_{G^{+}}^{2} &=&0\;,\;m_{H^{+}}^{2}=-\frac{1}{2}\lambda _{3}\left(
v_{1}^{2}+v_{2}^{2}\right)
\end{eqnarray*}

As for the indices 3,7; the eigenvectors and eigenvalues are evaluated from (%
\ref{diagonalization}) 
\begin{eqnarray*}
m_{H^{0},h^{0}} &=&\lambda _{1}v_{1}^{2}+\lambda _{2}v_{2}^{2}\pm \sqrt{%
\left( \lambda _{1}v_{1}^{2}-\lambda _{2}v_{2}^{2}\right) ^{2}+4\lambda
_{+}^{2}v_{1}^{2}v_{2}^{2}} \\
\tan 2\alpha &=&\frac{2v_{1}v_{2}\lambda _{+}}{\left( \lambda
_{1}v_{1}^{2}-\lambda _{2}v_{2}^{2}\right) } \\
\left( 
\begin{array}{c}
H^{0} \\ 
h^{0}
\end{array}
\right) &=&\left( 
\begin{array}{cc}
\cos \alpha & \sin \alpha \\ 
-\sin \alpha & \cos \alpha
\end{array}
\right) \left( 
\begin{array}{c}
h_{1} \\ 
h_{2}
\end{array}
\right)
\end{eqnarray*}

Finally, for the indices 4,8; it is obtained

\begin{equation*}
\left( 
\begin{array}{cc}
\cos \beta & \sin \beta \\ 
-\sin \beta & \cos \beta
\end{array}
\right) \left( 
\begin{array}{c}
\phi _{4} \\ 
\phi _{8}
\end{array}
\right) =\left( 
\begin{array}{c}
G^{0} \\ 
A^{0}
\end{array}
\right)
\end{equation*}
\begin{equation*}
m_{G^{0}}^{2}=0,\;\;\;m_{A^{0}}^{2}=\frac{1}{2}\left( \lambda _{4}-\lambda
_{3}\right) \left( v_{1}^{2}+v_{2}^{2}\right)
\end{equation*}

\subsubsection{Second set of solutions}

The mass matrix combined with the minimal conditions is given by:

\begin{eqnarray*}
M_{ij}^{2} &=&\left[ \left( 
\begin{array}{cc}
M_{11}^{2} & M_{15}^{2} \\ 
M_{15}^{2} & M_{55}^{2}
\end{array}
\right) \otimes I_{2\times 2}\right] \oplus \left( 
\begin{array}{cc}
M_{33}^{2} & M_{37}^{2} \\ 
M_{37}^{2} & M_{77}^{2}
\end{array}
\right) \oplus \left( 
\begin{array}{cc}
M_{44}^{2} & M_{48}^{2} \\ 
M_{48}^{2} & M_{88}^{2}
\end{array}
\right) \\
&=&\left[ \left( 
\begin{array}{cc}
0 & 0 \\ 
0 & -\mu _{2}^{2}+\frac{1}{2}\lambda _{5}v_{1}^{2}
\end{array}
\right) \otimes I_{2\times 2}\right] \oplus \left( 
\begin{array}{cc}
2\lambda _{1}v_{1}^{2} & 0 \\ 
0 & -\allowbreak \mu _{2}^{2}+\frac{1}{2}\lambda _{3}v_{1}^{2}+\frac{1}{2}%
\lambda _{5}v_{1}^{2}
\end{array}
\right) \\
&&\oplus \left( 
\begin{array}{cc}
0 & 0 \\ 
0 & -\allowbreak \mu _{2}^{2}+\frac{1}{2}\lambda _{4}v_{1}^{2}+\frac{1}{2}%
\lambda _{5}v_{1}^{2}
\end{array}
\right)
\end{eqnarray*}

So in this case the matrix is already diagonal and consequently the mass
eigenstates coincide with the gauge eigenstates, the elements in the
diagonal are the eigenvalues i.e. the squared masses of the Higgs bosons.
Since $M_{55}^{2}=M_{66}^{2}$,$\;$they correspond to two particles
degenerate in mass (i.e. two charged particles), in addition, other two
degrees of freedom should be associated to their corresponding Goldstone
bosons. So we get

\begin{eqnarray*}
\phi _{1}^{+} &=&H^{+}\;;\;\phi _{2}^{+}=G^{+}\;;\;\phi _{3}=H^{0}\;;\;\phi
_{7}=h^{0}\;;\;\phi _{4}=G^{0}\;;\;\phi _{8}=A^{0}\;;\; \\
m_{G^{+}}^{2} &=&0\;;\;m_{H^{+}}^{2}=-\mu _{2}^{2}+\frac{1}{2}\lambda
_{5}v_{1}^{2}\;;\;m_{H^{0}}^{2}=2\lambda _{1}v_{1}^{2} \\
m_{h^{0}}^{2} &=&-\allowbreak \mu _{2}^{2}+\frac{1}{2}\left( \lambda
_{3}+\lambda _{5}\right)
v_{1}^{2}\;;\;m_{G^{0}}^{2}=0\;;m_{A^{0}}^{2}=-\allowbreak \mu _{2}^{2}+%
\frac{1}{2}\left( \lambda _{4}+\lambda _{5}\right) v_{1}^{2}\;
\end{eqnarray*}

\subsection{The potential with the global $U\left( 1\right) \;$symmetry}

As explained in section (\ref{Higgspot 2HDM}) another way to avoid
spontaneous $CP\;$violation consists of imposing invariance under the global 
$U\left( 1\right) $ symmetry\ $\Phi _{2}\rightarrow e^{i\varphi }\Phi _{2}\;$%
(as well as the $C-$invariance) and correspond to setting $\lambda
_{6}=\lambda _{7}=\lambda _{8}=\lambda _{9}=\lambda _{10}=\mu _{4}^{2}=0\;$%
and $\lambda _{3}=\lambda _{4}\;$in (\ref{Vg}).$\;$In this case, we should
assume that in general $v_{1},v_{2}\neq 0.\;$The minimum conditions (\ref
{min conditions}) are reduced to two equations:

\begin{eqnarray*}
\lambda _{1}v_{1}^{3}-\mu _{1}^{2}\allowbreak v_{1}+\frac{1}{2}\lambda
_{3}v_{2}^{2}v_{1}+\frac{1}{2}\lambda _{5}v_{1}v_{2}^{2}-\frac{1}{2}\mu
_{3}^{2}\allowbreak v_{2} &=&0 \\
-\mu _{2}^{2}v_{2}+\lambda _{2}v_{2}^{3}-\frac{1}{2}\mu _{3}^{2}\allowbreak
v_{1}+\frac{1}{2}\lambda _{3}v_{1}^{2}v_{2}+\frac{1}{2}\lambda
_{5}v_{2}\allowbreak v_{1}^{2} &=&0
\end{eqnarray*}
and after reorganizing the matrix elements as $1,2,3,4,5,6,7,8\rightarrow
1,2,5,6,3,7,4,8$;\ using the minimum conditions, and taking into account
that $M_{15}^{2}=M_{26}^{2},\;M_{11}^{2}=M_{22}^{2},\;M_{55}^{2}=M_{66}^{2}$
the decomposed matrix becomes

\begin{equation*}
M_{tot}^{2}=\left[ \left( 
\begin{array}{cc}
M_{11}^{2} & M_{15}^{2} \\ 
M_{15}^{2} & M_{55}^{2}
\end{array}
\right) \otimes I_{2\times 2}\right] \oplus \left( 
\begin{array}{cc}
M_{33}^{2} & M_{37}^{2} \\ 
M_{37}^{2} & M_{77}^{2}
\end{array}
\right) \oplus \left( 
\begin{array}{cc}
M_{44}^{2} & M_{48}^{2} \\ 
M_{48}^{2} & M_{88}^{2}
\end{array}
\right)
\end{equation*}
with 
\begin{eqnarray*}
M_{11}^{2} &=&\frac{1}{2}\mu _{3}^{2}\allowbreak \frac{v_{2}}{v_{1}}-\frac{1%
}{2}\lambda _{3}v_{2}^{2}\;\;;\;\;M_{15}^{2}=-\frac{1}{2}\mu _{3}^{2}+\frac{1%
}{2}\allowbreak \lambda _{3}v_{1}v_{2} \\
M_{33}^{2} &=&\allowbreak \frac{1}{2}\mu _{3}^{2}\allowbreak \frac{v_{2}}{%
v_{1}}+2\lambda _{1}v_{1}^{2}\allowbreak \;\;;\;\;M_{37}^{2}=-\allowbreak 
\frac{1}{2}\mu _{3}^{2}+\lambda _{5}v_{1}v_{2}+\lambda _{3}\allowbreak
v_{1}v_{2}
\end{eqnarray*}
\begin{eqnarray*}
M_{44}^{2} &=&\frac{1}{2}\mu _{3}^{2}\allowbreak \frac{v_{2}}{v_{1}}%
\;;\;M_{48}^{2}=-\allowbreak \frac{1}{2}\mu
_{3}^{2}\;;\;M_{55}^{2}=\allowbreak \frac{1}{2}\mu _{3}^{2}\allowbreak \frac{%
v_{1}}{v_{2}}-\frac{1}{2}\lambda _{3}v_{1}^{2}\;\;\; \\
M_{77}^{2} &=&\frac{1}{2}\mu _{3}^{2}\allowbreak \frac{v_{1}}{v_{2}}%
+2\lambda _{2}v_{2}^{2}\allowbreak \;;\;M_{88}^{2}=\frac{1}{2}\mu
_{3}^{2}\allowbreak \frac{v_{1}}{v_{2}}
\end{eqnarray*}
the eigenvalues and the matrix of rotation corresponding to the submatrix of
indices $1,5\;$are 
\begin{eqnarray*}
\left( 
\begin{array}{c}
G^{+} \\ 
H^{+}
\end{array}
\right) &=&\left( 
\begin{array}{cc}
\cos \beta & \sin \beta \\ 
-\sin \beta & \cos \beta
\end{array}
\right) \left( 
\begin{array}{c}
\phi _{1}^{+} \\ 
\phi _{2}^{+}
\end{array}
\right) \\
m_{G^{+}} &=&0\;,\;m_{H^{+}}=\frac{1}{2}\frac{\left(
v_{1}^{2}+v_{2}^{2}\right) \left( \mu _{3}^{2}-\lambda _{3}v_{1}v_{2}\right) 
}{v_{1}v_{2}}
\end{eqnarray*}
for the indices 3,7, the eigenvalues and the rotation angle read

\begin{eqnarray*}
m_{H^{0},h^{0}}^{2} &=&\lambda _{1}v_{1}^{2}+\lambda _{2}v_{2}^{2}+\frac{1}{4%
}\mu _{3}^{2}\left( \tan \beta +\cot \beta \right) \pm R_{\lambda } \\
\tan 2\alpha &=&\frac{2v_{1}v_{2}\lambda _{+}-\frac{1}{2}\mu _{3}^{2}}{%
\lambda _{1}v_{1}^{2}-\lambda _{2}v_{2}^{2}+\frac{1}{4}\mu _{3}^{2}\left(
\tan \beta -\cot \beta \right) }\;
\end{eqnarray*}
\begin{equation*}
R_{\lambda }\equiv \sqrt{\left[ \lambda _{1}v_{1}^{2}-\lambda _{2}v_{2}^{2}+%
\frac{1}{4}\mu _{3}^{2}\left( \tan \beta -\cot \beta \right) \right]
^{2}+\left( 2v_{1}v_{2}\lambda _{+}-\frac{1}{2}\mu _{3}^{2}\right) ^{2}}
\end{equation*}
finally for the matrix 4,8, the rotation angle is $\beta \;$once again 
\begin{equation*}
\left( 
\begin{array}{cc}
\cos \beta & \sin \beta \\ 
-\sin \beta & \cos \beta
\end{array}
\right) \left( 
\begin{array}{c}
\sqrt{2}\phi _{4}-v_{1} \\ 
\sqrt{2}\phi _{8}-v_{2}
\end{array}
\right) =\left( 
\begin{array}{c}
G^{0} \\ 
A^{0}
\end{array}
\right)
\end{equation*}
and the eigenvalues are 
\begin{equation*}
m_{G^{0}}=0,\;m_{A^{0}}=\frac{1}{2}\mu _{3}^{2}\frac{v_{1}^{2}+v_{2}^{2}}{%
v_{1}v_{2}}
\end{equation*}

Finally, it worths to say that the parametrization of the Higgs Hunter's
Guide (in the CP conserving case i.e. $\xi =0$, and $\lambda _{5}=\lambda
_{6}$ in Eq. (4.8) of Ref. \cite{Hunter}), can be gotten from the following
associations 
\begin{eqnarray*}
\mu _{1}^{2} &\rightarrow &\left[ 2\lambda _{1}v_{1}^{2}+2\lambda _{3}\left(
v_{1}^{2}+v_{2}^{2}\right) \right] \\
\mu _{2}^{2} &\rightarrow &\left[ 2\lambda _{2}v_{2}^{2}+2\lambda _{3}\left(
v_{1}^{2}+v_{2}^{2}\right) \right] \\
\mu _{3}^{2} &\rightarrow &2\lambda _{5}v_{1}v_{2}\;\;;\;\lambda
_{1}\rightarrow \lambda _{1}+\lambda _{3} \\
\lambda _{2} &\rightarrow &\lambda _{2}+\lambda _{3}\;\;;\;\;\lambda
_{3}\rightarrow \lambda _{5}-\lambda _{4} \\
\lambda _{4} &\rightarrow &\lambda _{6}-\lambda _{4}=\lambda _{5}-\lambda
_{4} \\
\lambda _{5} &\rightarrow &\lambda _{4}+2\lambda _{3}
\end{eqnarray*}
where the parameters on left are the ones in our parametrization Eqs. (\ref
{VB}, \ref{VBP}), and the parameters on right correspond to the ones in the
Higgs Hunter's Guide.

\appendix
\setcounter{chapter}{1}

\chapter{Rotation of the Yukawa Lagrangian in the 2HDM (III)\label%
{rotation Yuk}}

We start defining two Higgs doublets with VEV 
\begin{equation}
\Phi _{1,2}^{\prime }=\left( 
\begin{array}{c}
\left( \phi _{1,2}^{+}\right) ^{\prime } \\ 
\left( \phi _{1,2}^{0}\right) ^{\prime }
\end{array}
\right) \;\;\;\;\text{and\ \ \ \ \ }\langle \Phi _{1,2}^{\prime }\rangle
=v_{1,2}  \label{doubletnotri}
\end{equation}
and writing the Yukawa Lagrangian as 
\begin{eqnarray}
-\pounds _{Y} &=&\widetilde{\eta }_{ij}^{U,0}\overline{Q}_{iL}^{0}\widetilde{%
\Phi }_{1}^{\prime }U_{jR}^{0}+\widetilde{\eta }_{ij}^{D,0}\overline{Q}%
_{iL}^{0}\Phi _{1}^{\prime }D_{jR}^{0}+\widetilde{\xi }_{ij}^{U,0}\overline{Q%
}_{iL}^{0}\widetilde{\Phi }_{2}^{\prime }U_{jR}^{0}+\widetilde{\xi }%
_{ij}^{D,0}\overline{Q}_{iL}^{0}\Phi _{2}^{\prime }D_{jR}^{0}  \notag \\
&&+\text{lepton sector}+h.c.  \label{Yuknotri}
\end{eqnarray}
we can make a rotation between the doublets 
\begin{equation}
\left( 
\begin{array}{c}
\Phi _{1} \\ 
\Phi _{2}
\end{array}
\right) \equiv \left( 
\begin{array}{cc}
\cos \theta & \sin \theta \\ 
-\sin \theta & \cos \theta
\end{array}
\right) \left( 
\begin{array}{c}
\Phi _{1}^{\prime } \\ 
\Phi _{2}^{\prime }
\end{array}
\right)  \label{rotdoublet}
\end{equation}
We shall deal with the quark sector only henceforth, the results that we are
going to obtain are also valid for the lepton sector with the appropiate
replacements. In terms of $\Phi _{1},\;\Phi _{2}$,\ the Yukawa Lagrangian
could be rewritten as 
\begin{eqnarray*}
-\pounds _{Y} &=&\widetilde{\eta }_{ij}^{U,0}\overline{Q}_{iL}^{0}\left(
\cos \theta \widetilde{\Phi }_{1}-\sin \theta \widetilde{\Phi }_{2}\right)
U_{jR}^{0} \\
&&+\widetilde{\eta }_{ij}^{D,0}\overline{Q}_{iL}^{0}\left( \cos \theta \Phi
_{1}-\sin \theta \Phi _{2}\right) D_{jR}^{0} \\
&&+\widetilde{\xi }_{ij}^{U,0}\overline{Q}_{iL}^{0}\left( \sin \theta 
\widetilde{\Phi }_{1}+\cos \theta \widetilde{\Phi }_{2}\right) U_{jR}^{0} \\
&&+\widetilde{\xi }_{ij}^{D,0}\overline{Q}_{iL}^{0}\left( \sin \theta \Phi
_{1}+\cos \theta \Phi _{2}\right) D_{jR}^{0}+h.c.
\end{eqnarray*}
\begin{eqnarray*}
-\pounds _{Y} &=&\overline{Q}_{iL}^{0}\left( \cos \theta \widetilde{\eta }%
_{ij}^{U,0}+\sin \theta \widetilde{\xi }_{ij}^{U,0}\right) \widetilde{\Phi }%
_{1}U_{jR}^{0} \\
&&+\overline{Q}_{iL}^{0}\left( \cos \theta \widetilde{\eta }_{ij}^{D,0}+\sin
\theta \widetilde{\xi }_{ij}^{D,0}\right) \Phi _{1}D_{jR}^{0} \\
&&+\overline{Q}_{iL}^{0}\left( -\sin \theta \widetilde{\eta }%
_{ij}^{U,0}+\cos \theta \widetilde{\xi }_{ij}^{U,0}\right) \widetilde{\Phi }%
_{2}U_{jR}^{0} \\
&&+\overline{Q}_{iL}^{0}\left( -\sin \theta \widetilde{\eta }%
_{ij}^{D,0}+\cos \theta \widetilde{\xi }_{ij}^{D,0}\right) \Phi
_{2}D_{jR}^{0}+h.c.
\end{eqnarray*}
defining 
\begin{equation}
\left( 
\begin{array}{c}
\eta _{ij}^{\left( U,D\right) ,0} \\ 
\xi _{ij}^{\left( U,D\right) ,0}
\end{array}
\right) =\left( 
\begin{array}{cc}
\cos \theta & \sin \theta \\ 
-\sin \theta & \cos \theta
\end{array}
\right) \left( 
\begin{array}{c}
\widetilde{\eta }_{ij}^{\left( U,D\right) ,0} \\ 
\widetilde{\xi }_{ij}^{\left( U,D\right) ,0}
\end{array}
\right)  \label{rotmat}
\end{equation}
the Lagrangian could be written as 
\begin{eqnarray}
-\pounds _{Y} &=&\overline{Q}_{iL}^{0}\eta _{ij}^{U,0}\widetilde{\Phi }%
_{1}U_{jR}^{0}+\overline{Q}_{iL}^{0}\eta _{ij}^{D,0}\Phi _{1}D_{jR}^{0} 
\notag \\
&&+\overline{Q}_{iL}^{0}\xi _{ij}^{U,0}\widetilde{\Phi }_{2}U_{jR}^{0}+%
\overline{Q}_{iL}^{0}\xi _{ij}^{D,0}\Phi _{2}D_{jR}^{0}+h.c.  \label{Yukfund}
\end{eqnarray}
with the same form as the original Lagrangian if we forget the prime
notation. Consequently, the combined rotations (\ref{rotdoublet},\ref{rotmat}%
) do not have physical consequences since it is basically a change of basis.
In particular we can choose $\theta =\beta \;$such that

\begin{eqnarray}
\langle \Phi _{1}\rangle &=&\cos \beta \langle \Phi _{1}^{\prime }\rangle
+\sin \beta \langle \Phi _{2}^{\prime }\rangle =\frac{v_{1}^{2}+v_{2}^{2}}{%
\sqrt{v_{1}^{2}+v_{2}^{2}}}=\sqrt{v_{1}^{2}+v_{2}^{2}}\equiv v  \notag \\
\langle \Phi _{2}\rangle &=&-\sin \beta \langle \Phi _{1}^{\prime }\rangle
+\cos \beta \langle \Phi _{2}^{\prime }\rangle =0  \label{VEV zero}
\end{eqnarray}
In whose case we managed to get $\langle \Phi _{2}\rangle =0.\;$Since this
lagrangian contains exactly the same physical information as the first one,
we conclude that in model type III the parameter $\tan \beta \;$is totally
spurious and we can assume without any loss of generality that one of the
VEV is zero \footnote{%
Indeed the rotation should be made in all the Lagrangians involving Higgs
doublets, consequently, we have to ensure that the physical content of the
potential is not altered by the rotation, in appendix (\ref{rotation pot})
we see that the rotation can be performed in the potential\ $V\;$defined in
Eq. (\ref{Lag10}).}. We shall settle $\theta =\beta \;$henceforth.

On the other hand, it is possible to reverse the steps above and start from
the representation in which $\langle \Phi _{2}\rangle =0\;$(the
``fundamental representation'') and make a rotation of the Higgs doublets
from which the $\tan \beta \;$parameter arises. If we assume the fundamental
representation as our ``starting point'', we see that by making a rotation
of the doublets, the mixing matrices are rotated such that 
\begin{equation}
\left( 
\begin{array}{c}
\widetilde{\eta }_{ij}^{\left( U,D\right) ,0} \\ 
\widetilde{\xi }_{ij}^{\left( U,D\right) ,0}
\end{array}
\right) =\left( 
\begin{array}{cc}
\cos \beta & -\sin \beta \\ 
\sin \beta & \cos \beta
\end{array}
\right) \left( 
\begin{array}{c}
\eta _{ij}^{\left( U,D\right) ,0} \\ 
\xi _{ij}^{\left( U,D\right) ,0}
\end{array}
\right)  \label{rotmatinv}
\end{equation}
from this rotation we can see that $\widetilde{\eta }_{ij}^{\left(
U,D\right) ,0}\;$and $\widetilde{\xi }_{ij}^{\left( U,D\right) ,0}$ depend
on the $\tan \beta $ parameter (since $\eta _{ij}^{\left( U,D\right) ,0}\;$%
and $\xi _{ij}^{\left( U,D\right) ,0}\;$come from the fundamental
representation, they do not depend on $\tan \beta $). In the non trivial
parametrization the VEV's are $v_{1}\;$and $v_{2}\;$where $\tan \beta
=v_{2}/v_{1},\;\sin \beta =v_{2}/\sqrt{v_{1}^{2}+v_{2}^{2}},\;$and the
Yukawa Lagrangian in terms of the mass eigenstates can be written in the
following ways (see section (\ref{Yuk III})) 
\begin{eqnarray}
-\pounds _{Y}^{I} &=&-\pounds _{Y}\left( \text{Type\ I,}\;\tan \beta \right)
+\widetilde{\eta }^{U}\left( \eta ^{U},\xi ^{U},\tan \beta \right)  \notag \\
&&+\widetilde{\eta }^{D}\left( \eta ^{D},\xi ^{D},\tan \beta \right)
\label{Yuk1} \\
-\pounds _{Y}^{II} &=&-\pounds _{Y}\left( \text{Type\ II,}\;\tan \beta
\right) +\widetilde{\eta }^{U}\left( \eta ^{U},\xi ^{U},\tan \beta \right) 
\notag \\
&&+\widetilde{\xi }^{D}\left( \eta ^{D},\xi ^{D},\tan \beta \right)
\label{Yuk2}
\end{eqnarray}
the fundamental parametrization is clearly independent on $\tan \beta ,\;$%
and since all these parametrizations are physically equivalent, the
parametrizations defined by Eqs. (\ref{Yuk1},\ref{Yuk2}) cannot depend on $%
\tan \beta \;$either. I shall restrict the discussion to $-\pounds
_{Y}^{II}\;$but the same ideas are applicable to $-\pounds _{Y}^{I}.\;$In $%
-\pounds _{Y}^{II}$, it$\;$is very clear that $-\pounds _{Y}\left( \text{%
Type\ II,}\;\tan \beta \right) $ depends on $\tan \beta \;$explicitly.
However, as the whole Lagrangian must be independent of it; the mixing
matrices should depend on $\tan \beta \;$in such a way that they cancel the
dependence on this parameter from $-\pounds _{Y}\left( \text{Type\ I,}\;\tan
\beta \right) .\;$This emphasize the fact that the mixing matrices are basis
dependent (changing the basis means changing $\tan \beta )$.

Let us see explicitly the relation among the parameters in the fundamental
representation $(\xi ^{U,D},\eta ^{U,D},\Phi _{1},\Phi _{2},\alpha )\;$and
the ones in the non trivial representation $(\widetilde{\xi }^{U,D},%
\widetilde{\eta }^{U,D},\Phi _{1}^{\prime },\Phi _{2}^{\prime },\alpha
^{\prime },\beta )$. First of all, Eqs. (\ref{rotdoublet}) provides us with
the relation between $(\Phi _{1},\Phi _{2})\;$and $(\Phi _{1}^{\prime },\Phi
_{2}^{\prime })$.\ On the other hand, Eq. (\ref{rotmat}) provides the
relation among$\;(\xi ^{\left( U,D\right) ,0},\eta ^{\left( U,D\right)
,0})\; $and $(\widetilde{\xi }^{\left( U,D\right) ,0},\widetilde{\eta }%
^{\left( U,D\right) ,0})$. Notwithstanding, the most useful relations would
be the ones among$\;(\xi ^{U,D},\eta ^{U,D})\;$and$\;(\widetilde{\xi }^{U,D},%
\widetilde{\eta }^{U,D})$ i.e. the mixing matrices when the Lagrangian is
written in terms of mass eigenstates. In the non-trivial basis the relation
between gauge and mass Higgs eigenstates can be taken from (\ref{mass
eigenstates}) with the appropiate change of notation. 
\begin{eqnarray}
\left( 
\begin{array}{c}
\left( \phi _{1}^{+}\right) ^{\prime } \\ 
\left( \phi _{2}^{+}\right) ^{\prime }
\end{array}
\right) &=&\left( 
\begin{array}{cc}
\cos \beta & -\sin \beta \\ 
\sin \beta & \cos \beta
\end{array}
\right) \left( 
\begin{array}{c}
G^{+} \\ 
H^{+}
\end{array}
\right)  \notag \\
\left( 
\begin{array}{c}
h_{1}^{\prime } \\ 
h_{2}^{\prime }
\end{array}
\right) &=&\left( 
\begin{array}{cc}
\cos \alpha ^{\prime } & -\sin \alpha ^{\prime } \\ 
\sin \alpha ^{\prime } & \cos \alpha ^{\prime }
\end{array}
\right) \left( 
\begin{array}{c}
H^{0} \\ 
h^{0}
\end{array}
\right)  \notag \\
\left( 
\begin{array}{c}
g_{1}^{\prime } \\ 
g_{2}^{\prime }
\end{array}
\right) &=&\left( 
\begin{array}{cc}
\cos \beta & -\sin \beta \\ 
\sin \beta & \cos \beta
\end{array}
\right) \left( 
\begin{array}{c}
G^{0} \\ 
A^{0}
\end{array}
\right)  \label{Eigennotri}
\end{eqnarray}
performing the rotation (\ref{rotdoublet}) and using (\ref{doubletnotri}) we
get

\begin{eqnarray*}
\Phi _{1} &=&\cos \beta \Phi _{1}^{\prime }+\sin \beta \Phi _{2}^{\prime } \\
&=&\cos \beta \left( 
\begin{array}{c}
\left( \phi _{1}^{+}\right) ^{\prime } \\ 
\left( h_{1}^{\prime }+v_{1}+ig_{1}^{\prime }\right) /\sqrt{2}
\end{array}
\right) +\sin \beta \left( 
\begin{array}{c}
\left( \phi _{2}^{+}\right) ^{\prime } \\ 
\left( h_{2}^{\prime }+v_{2}+ig_{2}^{\prime }\right) /\sqrt{2}
\end{array}
\right) \\
&\equiv &\left( 
\begin{array}{c}
\phi _{1}^{+} \\ 
\left( h_{1}+v+ig_{1}\right) /\sqrt{2}
\end{array}
\right) \\
\Phi _{2} &=&-\sin \beta \Phi _{1}^{\prime }+\cos \beta \Phi _{2}^{\prime }
\\
&=&-\sin \beta \left( 
\begin{array}{c}
\left( \phi _{1}^{+}\right) ^{\prime } \\ 
\left( h_{1}^{\prime }+v_{1}+ig_{1}^{\prime }\right) /\sqrt{2}
\end{array}
\right) +\cos \beta \left( 
\begin{array}{c}
\left( \phi _{2}^{+}\right) ^{\prime } \\ 
\left( h_{2}^{\prime }+v_{2}+ig_{2}^{\prime }\right) /\sqrt{2}
\end{array}
\right) \\
&\equiv &\left( 
\begin{array}{c}
\phi _{2}^{+} \\ 
\left( h_{2}+ig_{2}\right) /\sqrt{2}
\end{array}
\right)
\end{eqnarray*}
so the conversion from the gauge eigenstates in the non trivial basis to the
gauge eigenstates in the fundamental representation is given by 
\begin{eqnarray}
\left( 
\begin{array}{c}
\phi _{1}^{+} \\ 
\phi _{2}^{+}
\end{array}
\right) &=&\left( 
\begin{array}{cc}
\cos \beta & \sin \beta \\ 
-\sin \beta & \cos \beta
\end{array}
\right) \left( 
\begin{array}{c}
\left( \phi _{1}^{+}\right) ^{\prime } \\ 
\left( \phi _{2}^{+}\right) ^{\prime }
\end{array}
\right) \;,\;  \notag \\
\left( 
\begin{array}{c}
h_{1} \\ 
h_{2}
\end{array}
\right) &=&\left( 
\begin{array}{cc}
\cos \beta & \sin \beta \\ 
-\sin \beta & \cos \beta
\end{array}
\right) \left( 
\begin{array}{c}
h_{1}^{\prime } \\ 
h_{2}^{\prime }
\end{array}
\right) \;,  \notag \\
\left( 
\begin{array}{c}
g_{1} \\ 
g_{2}
\end{array}
\right) &=&\left( 
\begin{array}{cc}
\cos \beta & \sin \beta \\ 
-\sin \beta & \cos \beta
\end{array}
\right) \left( 
\begin{array}{c}
g_{1}^{\prime } \\ 
g_{2}^{\prime }
\end{array}
\right) \;.  \label{gaueigconv}
\end{eqnarray}
And the relation between the gauge eigenstates and the mass eigenstates in
the fundamental parametrization, becomes particularly simple.

\begin{eqnarray}
\left( 
\begin{array}{c}
\phi _{1}^{+} \\ 
\phi _{2}^{+}
\end{array}
\right) &=&\left( 
\begin{array}{cc}
\cos \beta & \sin \beta \\ 
-\sin \beta & \cos \beta
\end{array}
\right) \left( 
\begin{array}{cc}
\cos \beta & -\sin \beta \\ 
\sin \beta & \cos \beta
\end{array}
\right) \left( 
\begin{array}{c}
G^{+} \\ 
H^{+}
\end{array}
\right) =\left( 
\begin{array}{c}
G^{+} \\ 
H^{+}
\end{array}
\right) \;,  \notag \\
\left( 
\begin{array}{c}
g_{1} \\ 
g_{2}
\end{array}
\right) &=&\left( 
\begin{array}{c}
G^{0} \\ 
A^{0}
\end{array}
\right) \;,  \notag \\
\left( 
\begin{array}{c}
h_{1} \\ 
h_{2}
\end{array}
\right) &=&\left( 
\begin{array}{cc}
\cos \beta & \sin \beta \\ 
-\sin \beta & \cos \beta
\end{array}
\right) \left( 
\begin{array}{c}
h_{1}^{\prime } \\ 
h_{2}^{\prime }
\end{array}
\right)  \notag \\
&=&\left( 
\begin{array}{cc}
\cos \beta & \sin \beta \\ 
-\sin \beta & \cos \beta
\end{array}
\right) \left( 
\begin{array}{cc}
\cos \alpha ^{\prime } & -\sin \alpha ^{\prime } \\ 
\sin \alpha ^{\prime } & \cos \alpha ^{\prime }
\end{array}
\right) \left( 
\begin{array}{c}
H^{0} \\ 
h^{0}
\end{array}
\right)  \notag \\
\left( 
\begin{array}{c}
h_{1} \\ 
h_{2}
\end{array}
\right) &=&\left( 
\begin{array}{cc}
\cos \alpha & -\sin \alpha \\ 
\sin \alpha & \cos \alpha
\end{array}
\right) \allowbreak \left( 
\begin{array}{c}
H^{0} \\ 
h^{0}
\end{array}
\right)  \label{gaumastri}
\end{eqnarray}
where we have defined 
\begin{equation}
\alpha \equiv \alpha ^{\prime }-\beta  \label{alfaconv}
\end{equation}
On the other hand, from (\ref{rotmat}) and (\ref{Rotation Iu}) we get

\begin{equation*}
\eta ^{U,0}=\cos \beta \widetilde{\eta }^{U,0}+\sin \beta \widetilde{\xi }%
^{U,0}=\cos \beta \widetilde{\eta }^{U,0}+\sin \beta \left( \frac{\sqrt{2}}{%
v_{2}}T_{L}^{\dagger }M_{U}T_{R}-\frac{v_{1}}{v_{2}}\widetilde{\eta }%
^{U,0}\right)
\end{equation*}
and after applying the unitarity transformations defined in (\ref{Up transf}%
) we obtain 
\begin{equation*}
T_{L}\eta ^{U,0}T_{R}^{\dagger }=\cos \beta \left( T_{L}\widetilde{\eta }%
^{U,0}T_{R}^{\dagger }\right) +\sin \beta \left[ \frac{\sqrt{2}}{v_{2}}M_{U}-%
\frac{v_{1}}{v_{2}}\left( T_{L}\widetilde{\eta }^{U,0}T_{R}^{\dagger
}\right) \right]
\end{equation*}
we should note that the unitary matrices $T_{L,R}\;$and $V_{L,R}\;$defined
in Eq. (\ref{Up transf}) are independent on the basis since they indicate
the rotations of the spinors from gauge to mass eigenstates, and rotations (%
\ref{rotdoublet},\ref{rotmat}) do not modify such spinors. Consequently the
transformation$\;T_{L}X^{0}T_{R}^{\dagger }$ defines the conversion of the
mixing matrices from the gauge eigenstates to the mass eigenstates,
therefore 
\begin{eqnarray*}
T_{L}\eta ^{U,0}T_{R}^{\dagger } &=&\cos \beta \widetilde{\eta }^{U}+\sin
\beta \left( \frac{\sqrt{2}}{v_{2}}M_{U}-\frac{v_{1}}{v_{2}}\widetilde{\eta }%
^{U}\right) \\
\eta ^{U} &=&\frac{v_{1}}{v}\widetilde{\eta }^{U}+\frac{\sqrt{2}}{v}M_{U}-%
\frac{v_{1}}{v}\widetilde{\eta }^{U}=\frac{\sqrt{2}}{v}M_{U}
\end{eqnarray*}
so we obtain finally 
\begin{equation*}
\eta ^{U}=\frac{\sqrt{2}}{v}M_{U}
\end{equation*}
i.e. the matrix$\;\eta ^{U}\;$is associated only to the mass of the fermion
as expected. Similarly from Eqs. (\ref{rotmat}, \ref{Rotation IIu}, \ref{Up
transf}) 
\begin{eqnarray*}
\xi ^{U,0} &=&-\sin \beta \widetilde{\eta }^{U,0}+\cos \beta \widetilde{\xi }%
^{U,0}=-\sin \beta \left[ \frac{\sqrt{2}}{v_{1}}T_{L}^{\dagger }M_{U}T_{R}-%
\frac{v_{2}}{v_{1}}\widetilde{\xi }^{U,0}\right] \\
&&+\cos \beta \widetilde{\xi }^{U,0} \\
T_{L}\xi ^{U,0}T_{R}^{\dagger } &=&-\tan \beta \frac{\sqrt{2}}{v}M_{U}+\sin
\beta \tan \beta \widetilde{\xi }^{U}+\cos \beta \widetilde{\xi }^{U} \\
\xi ^{U} &=&\sec \beta \widetilde{\xi }^{U}-\frac{\sqrt{2}\tan \beta }{v}%
M_{U}
\end{eqnarray*}
furthermore, Eq. (\ref{Rotation IIu}) is also valid for $\eta ^{\left(
U,D\right) ,0}$ by setting $v_{2}=0,\;v_{1}=v.\;$Using Eq. (\ref{Rotation
IIu}) applied to $\eta ^{U,0}\;$and Eq. (\ref{rotmat}) we get

\begin{eqnarray*}
\widetilde{\xi }^{U,0} &=&\sin \beta \left[ T_{L}^{\dagger }M_{U}T_{R}\frac{%
\sqrt{2}}{v}\right] +\cos \beta \xi ^{U,0} \\
\widetilde{\xi }^{U} &=&\left[ \frac{\sqrt{2}\sin \beta }{v}M_{U}\right]
+\cos \beta \xi ^{U}
\end{eqnarray*}
similarly 
\begin{equation*}
\widetilde{\eta }^{U}=M_{U}\frac{\sqrt{2}}{v}\cos \beta -\sin \beta \xi ^{U}
\end{equation*}
for the down sector is exactly the same. Summarizing, we get the following
links among the mixing matrices $\left( \eta ^{U,D},\xi ^{U,D}\right) \;$in
the fundamental parametrization with the ones in the non trivial basis $%
\left( \widetilde{\eta }^{U,D},\widetilde{\xi }^{U,D}\right) $: 
\begin{eqnarray}
\widetilde{\eta }^{U,D} &=&M_{U,D}\frac{\sqrt{2}}{v}\cos \beta -\sin \beta
\xi ^{U,D}\;\;;\;\;\;\widetilde{\xi }^{U,D}=\frac{\sqrt{2}\sin \beta }{v}%
M_{U,D}+\cos \beta \xi ^{U,D}  \notag \\
\eta ^{U,D} &=&\frac{\sqrt{2}}{v}M_{U,D}\;\;\;\;\;\;\;\;\;\;\;\;;\;\;\;\xi
^{U,D}=\sec \beta \widetilde{\xi }^{U,D}-\frac{\sqrt{2}\tan \beta }{v}M_{U,D}
\label{parametlink}
\end{eqnarray}
and the relations among the gauge Higgs eigenstates and mass Higgs
eigenstates in the fundamental parametrization, are given by

\begin{eqnarray*}
\left( 
\begin{array}{c}
\phi _{1}^{+} \\ 
\phi _{2}^{+}
\end{array}
\right) &=&\left( 
\begin{array}{c}
G^{+} \\ 
H^{+}
\end{array}
\right) \;\;\;;\;\;\left( 
\begin{array}{c}
g_{1} \\ 
g_{2}
\end{array}
\right) =\left( 
\begin{array}{c}
G^{0} \\ 
A^{0}
\end{array}
\right) \\
\left( 
\begin{array}{c}
h_{1} \\ 
h_{2}
\end{array}
\right) &=&\left( 
\begin{array}{cc}
\cos \alpha & -\sin \alpha \\ 
\sin \alpha & \cos \alpha
\end{array}
\right) \allowbreak \left( 
\begin{array}{c}
H^{0} \\ 
h^{0}
\end{array}
\right) \\
\alpha &\equiv &\alpha ^{\prime }-\beta
\end{eqnarray*}
while the analogous relations for the non-trivial bases, are given by Eqs. (%
\ref{mass eigenstates}).

As a proof of consistency, we shall use the equations \ref{parametlink} and
the Yukawa couplings in the non trivial parametrization Eqs. (\ref{Yukawa 1u}%
, \ref{Yukawa 1d}, \ref{Yukawa 2u}, \ref{Yukawa 2d}) in order to obtain the
Yukawa couplings in the fundamental parametrization Eq. (\ref{YukexpandIII}%
). Let us check first the couplings $\overline{D}DA^{0}\;$in the
parametrization in which the model type II becomes apparent, where $D\;$%
refers to the three down fermions. From Lagrangian (\ref{Yukawa 2d}) and
expressions (\ref{parametlink}) 
\begin{equation*}
\overline{D}DA^{0}:-\frac{ig\tan \beta }{2M_{W}}\overline{D}M_{D}\gamma
_{5}DA^{0}+\frac{i}{\sqrt{2}\cos \beta }\overline{D}\widetilde{\xi }%
^{D}\gamma _{5}DA^{0}
\end{equation*}
\begin{eqnarray*}
&=&-\frac{ig\tan \beta }{2M_{W}}\overline{D}M_{D}\gamma _{5}DA^{0}+\frac{i}{%
\sqrt{2}\cos \beta }\overline{D}\left[ \frac{\sqrt{2}\sin \beta }{v}%
M_{D}+\cos \beta \xi ^{D}\right] \gamma _{5}DA^{0} \\
\overline{D}DA^{0} &:&\frac{i}{\sqrt{2}}\overline{D}\xi ^{D}\gamma _{5}DA^{0}
\end{eqnarray*}
now let us examine $\overline{D}DH^{0}$%
\begin{eqnarray*}
\overline{D}DH^{0} &:&\frac{g}{2M_{W}\cos \beta }\overline{D}M_{D}D\cos
\alpha ^{\prime }H^{0}+\frac{1}{\sqrt{2}\cos \beta }\overline{D}\widetilde{%
\xi }^{D}D\sin \left( \alpha ^{\prime }-\beta \right) H^{0} \\
&=&\frac{g}{2M_{W}\cos \beta }\overline{D}M_{D}D\cos \alpha ^{\prime }H^{0}
\\
&&+\frac{1}{\sqrt{2}\cos \beta }\overline{D}\left[ \frac{\sqrt{2}\sin \beta 
}{v}M_{D}+\cos \beta \xi ^{U}\right] D\sin \left( \alpha ^{\prime }-\beta
\right) H^{0} \\
&=&\overline{D}\frac{M_{D}}{v}\cos \left( \alpha ^{\prime }-\beta \right)
DH^{0}+\frac{1}{\sqrt{2}}\overline{D}\xi ^{U}D\sin \left( \alpha ^{\prime
}-\beta \right) H^{0} \\
&=&\frac{g}{2M_{W}}\overline{D}M_{D}DH^{0}\cos \alpha +\frac{1}{\sqrt{2}}%
\overline{D}\xi ^{D}DH^{0}\sin \alpha
\end{eqnarray*}
now $\overline{D}Dh^{0}$%
\begin{eqnarray*}
\overline{D}Dh^{0} &:&-\frac{g}{2M_{W}\cos \beta }\overline{D}M_{D}D\sin
\alpha ^{\prime }h^{0}+\frac{1}{\sqrt{2}\cos \beta }\overline{D}\widetilde{%
\xi }^{D}D\cos \left( \alpha ^{\prime }-\beta \right) h^{0} \\
&=&-\frac{g}{2M_{W}\cos \beta }\overline{D}M_{D}D\sin \alpha ^{\prime }h^{0}
\\
&&+\frac{1}{\sqrt{2}\cos \beta }\overline{D}\left[ \frac{\sqrt{2}\sin \beta 
}{v}M_{D}+\cos \beta \xi ^{D}\right] D\cos \left( \alpha ^{\prime }-\beta
\right) h^{0} \\
&=&\overline{D}\left[ -\frac{gM_{D}}{2M_{W}\cos \beta }\left( \sin \alpha
^{\prime }-\sin \beta \cos \left( \alpha ^{\prime }-\beta \right) \right) %
\right] Dh^{0} \\
&&+\frac{1}{\sqrt{2}}\overline{D}\xi ^{D}D\cos \left( \alpha ^{\prime
}-\beta \right) h^{0} \\
&=&-\frac{g}{2M_{W}}\overline{D}M_{D}D\sin \left( \alpha ^{\prime }-\beta
\right) h^{0}+\frac{1}{\sqrt{2}}\overline{D}\xi ^{D}D\cos \left( \alpha
^{\prime }-\beta \right) h^{0} \\
&=&-\frac{g}{2M_{W}}\overline{D}M_{D}Dh^{0}\sin \alpha +\frac{1}{\sqrt{2}}%
\overline{D}\xi ^{D}Dh^{0}\cos \alpha
\end{eqnarray*}
finally for the charged Higgs, in the Lagrangian where model type II becomes
apparent i.e. (\ref{Yukawa 1u}) plus (\ref{Yukawa 2d}) the Yukawa coupling
is given by 
\begin{eqnarray*}
\overline{U}DH^{+} &:&-\frac{g\cot \beta }{\sqrt{2}M_{W}}\overline{U}%
M_{U}KP_{L}DH^{+}+\frac{1}{\sin \beta }\overline{U}\widetilde{\eta }%
^{U}KP_{L}DH^{+} \\
&&-\frac{g\tan \beta }{\sqrt{2}M_{W}}\overline{U}KM_{D}^{d}P_{R}DH^{+}+\frac{%
1}{\cos \beta }\overline{U}K\widetilde{\xi }^{D}P_{R}DH^{+} \\
&=&-\frac{g\cot \beta }{\sqrt{2}M_{W}}\overline{U}M_{U}KP_{L}DH^{+}-\frac{%
g\tan \beta }{\sqrt{2}M_{W}}\overline{U}KM_{D}^{d}P_{R}DH^{+} \\
&&+\frac{1}{\sin \beta }\overline{U}\left( \frac{\sqrt{2}g}{2M_{W}}M_{U}\cos
\beta -\sin \beta \xi ^{U}\right) KP_{L}DH^{+} \\
&&+\frac{1}{\cos \beta }\overline{U}K\left[ \frac{\sqrt{2}g\sin \beta }{%
2M_{W}}M_{D}^{d}+\cos \beta \xi ^{D}\right] P_{R}DH^{+} \\
&=&-\overline{U}\xi ^{U}KP_{L}DH^{+}+\overline{U}K\xi ^{D}P_{R}DH^{+} \\
&=&\overline{U}\left( K\xi ^{D}P_{R}-\xi ^{U}KP_{L}\right) DH^{+}
\end{eqnarray*}
all these couplings coincide with the ones describe in the Lagrangian for
the fundamental parametrization, Eq. (\ref{YukexpandIII}). Therefore, with
this procedure we can check that all Lagrangians generated from Eqs. (\ref
{Yukawa 1u}, \ref{Yukawa 1d}, \ref{Yukawa 2u}, \ref{Yukawa 2d}) i.e. $%
-\pounds _{Y\left( U\right) }^{I}-\pounds _{Y\left( D\right)
}^{I},\;-\pounds _{Y\left( U\right) }^{I}-\pounds _{Y\left( D\right)
}^{II},\;-\pounds _{Y\left( U\right) }^{II}-\pounds _{Y\left( D\right)
}^{II},\;-\pounds _{Y\left( U\right) }^{II}-\pounds _{Y\left( D\right)
}^{I},\;$coincides with the Lagrangian (\ref{YukexpandIII}) if we take into
account the expressions (\ref{parametlink}).

The results expressed by Eqs. (\ref{parametlink}), show us that the value of
the flavor changing vertices is basis dependent, though the value of the
couplings are basis independent as it must be. The mixing angles between
Higgs gauge eigenstates and Higgs mass eigenstates are also basis dependent
as expected. The transformations (\ref{rotdoublet}, \ref{rotmat}) reveals an 
$SO(2)\;$global symmetry of the model type III. This is like a ``\emph{%
global gauge invariance of the 2HDM type III}'' in which $\tan \beta $\
fixes the gauge.

In that sense, we can realize that the models type I and II have a
remarkable difference respect to the model type III, since it is well known
that the former two ones are highly dependent on the $\tan \beta \;$%
parameter while the latter is not. In writing the parametrization $\pounds
_{Y}^{II}\;$we can see easily the reason: $-\pounds _{Y}^{II}\left( \text{%
Type}\;\text{II}\right) \;$clearly depend on $\tan \beta \;$but when we add
the mixing parameters, they acquire the precise values to cancel such
dependence. In other words, the model type II does not have mixing
parameters at the tree level to absorb their $\tan \beta $\ dependence. We
can see the difference from the point of view of symmetries, the 2HDM is
constructed in such a way that we make an exact ``duplicate'' of the SM
Higgs doublet.

\begin{equation*}
\Phi _{1}=\left( 
\begin{array}{c}
\phi _{1}^{+} \\ 
\phi _{1}^{0}
\end{array}
\right) \;\;\;\;\Phi _{2}=\left( 
\begin{array}{c}
\phi _{2}^{+} \\ 
\phi _{2}^{0}
\end{array}
\right) \;\;\;Y_{1}=Y_{2}=1
\end{equation*}
These doublets have the same quantum numbers and are consequently
indistinguishable (at least at this step). Owing to this indistinguibility
we can perform the rotation

\begin{equation*}
\left( 
\begin{array}{c}
\Phi _{1}^{\prime } \\ 
\Phi _{2}^{\prime }
\end{array}
\right) =\left( 
\begin{array}{cc}
\cos \theta & \sin \theta \\ 
-\sin \theta & \cos \theta
\end{array}
\right) \left( 
\begin{array}{c}
\Phi _{1} \\ 
\Phi _{2}
\end{array}
\right)
\end{equation*}
over an arbitrary angle $\theta \;$without any physical consequences (it is
in fact a change of basis). It means that the model is invariant under a
global SO(2) transformation of the ``bidoublet'' $\left( 
\begin{array}{cc}
\Phi _{1} & \Phi _{2}
\end{array}
\right) ^{T}$. However, it is very common to impose a discrete symmetry on
the Higgs doublets $\left( \Phi _{1}\rightarrow \Phi _{1},\;\Phi
_{2}\rightarrow -\Phi _{2}\right) $ or a global $U\left( 1\right) $ symmetry 
$\left( \Phi _{1}\rightarrow \Phi _{1},\;\Phi _{2}\rightarrow e^{i\varphi
}\Phi _{2}\right) $ to prevent dangerous FCNC. In that case, we are
introducing a distinguibility between the doublets, because they acquire
very different couplings to the fermions (models type I and II). Of course,
we could have defined the symmetry in the opposite way $\left( \Phi
_{1}\rightarrow -\Phi _{1},\;\Phi _{2}\rightarrow \Phi _{2}\right) \;$but
once we have chosen one of them, we cannot interchange $\Phi
_{1}\leftrightarrow \Phi _{2}\;$anymore, without changing the physical
content. Such fact breaks explicity the SO(2) symmetry of the ``bidoublet''.
On the other hand, it is precisely this symmetry what allows us to absorb
the $\tan \beta \;$parameter, and since models type I and II do not have
that symmetry, we are not able to absorb it properly.

\appendix
\setcounter{chapter}{2}

\chapter{Rotation in the Higgs potential\label{rotation pot}}

\section{Transformation of the parameters in the potential}

Let us start from an arbitrary parametrization in which both VEV are in
general different from zero. The most general renormalizable and gauge
invariant potential read 
\begin{eqnarray}
V_{g} &=&-\widetilde{\mu }_{1}^{2}\widehat{A}^{\prime }-\widetilde{\mu }%
_{2}^{2}\widehat{B}^{\prime }-\widetilde{\mu }_{3}^{2}\widehat{C}^{\prime }-%
\widetilde{\mu }_{4}^{2}\widehat{D}^{\prime }+\widetilde{\lambda }_{1}%
\widehat{A}^{\prime 2}+\widetilde{\lambda }_{2}\widehat{B}^{\prime 2}+%
\widetilde{\lambda }_{3}\widehat{C}^{\prime 2}+\widetilde{\lambda }_{4}%
\widehat{D}^{\prime 2}  \notag \\
&&+\widetilde{\lambda }_{5}\widehat{A}^{\prime }\widehat{B}^{\prime }+%
\widetilde{\lambda }_{6}\widehat{A}^{\prime }\widehat{C}^{\prime }+%
\widetilde{\lambda }_{8}\widehat{A}^{\prime }\widehat{D}^{\prime }+%
\widetilde{\lambda }_{7}\widehat{B}^{\prime }\widehat{C}^{\prime }+%
\widetilde{\lambda }_{9}\widehat{B}^{\prime }\widehat{D}^{\prime }+%
\widetilde{\lambda }_{10}\widehat{C}^{\prime }\widehat{D}^{\prime }
\label{Vg2}
\end{eqnarray}
where we have defined four independent gauge invariant hermitian operators 
\begin{eqnarray*}
\widehat{A}^{\prime } &\equiv &\Phi _{1}^{\prime \dagger }\Phi _{1}^{\prime
}\;,\;\widehat{B}^{\prime }\equiv \Phi _{2}^{\prime \dagger }\Phi
_{2}^{\prime },\;\widehat{C}^{\prime }\equiv \frac{1}{2}\left( \Phi
_{1}^{\prime \dagger }\Phi _{2}^{\prime }+\Phi _{2}^{\prime \dagger }\Phi
_{1}^{\prime }\right) =\text{Re}\left( \Phi _{1}^{\prime \dagger }\Phi
_{2}^{\prime }\right) ,\; \\
\widehat{D}^{\prime } &\equiv &-\frac{i}{2}\left( \Phi _{1}^{\prime \dagger
}\Phi _{2}^{\prime }-\Phi _{2}^{\prime \dagger }\Phi _{1}^{\prime }\right) =%
\text{Im}\left( \Phi _{1}^{\prime \dagger }\Phi _{2}^{\prime }\right)
\end{eqnarray*}
the doublets and the VEV are denoted as 
\begin{equation}
\Phi _{1,2}^{\prime }=\left( 
\begin{array}{c}
\left( \phi _{1,2}^{+}\right) ^{\prime } \\ 
\left( \phi _{1,2}^{0}\right) ^{\prime }
\end{array}
\right) =\left( 
\begin{array}{c}
\left( \phi _{1,2}^{+}\right) ^{\prime } \\ 
\left( \frac{h_{1,2}+v_{1,2}+ig_{1,2}}{\sqrt{2}}\right) ^{\prime }
\end{array}
\right) \;\;\text{and\ \ \ }\langle \Phi _{1,2}^{\prime }\rangle
=v_{1,2}^{\prime }  \label{doubletnotri2}
\end{equation}
In appendix \ref{rotation Yuk}, we have seen that in the Yukawa Lagrangian
type III, we are able to make a rotation of the doublets as

\begin{equation}
\left( 
\begin{array}{c}
\Phi _{1} \\ 
\Phi _{2}
\end{array}
\right) \equiv \left( 
\begin{array}{cc}
\cos \theta & \sin \theta \\ 
-\sin \theta & \cos \theta
\end{array}
\right) \left( 
\begin{array}{c}
\Phi _{1}^{\prime } \\ 
\Phi _{2}^{\prime }
\end{array}
\right)  \label{rotdoublet2}
\end{equation}
without changing the physical content of the Lagrangian. However, we must
demonstrate that such rotation can be carried out in the potential without
changing its physical content either. In order to show it, we shall
calculate the way in which $\widetilde{\mu }_{i},\,\widetilde{\lambda }%
_{i}\; $parameters transform under this rotation. First, we calculate the
way in which the operators $\widehat{A}^{\prime },\;\widehat{B}^{\prime },\;%
\widehat{C}^{\prime },\;\widehat{D}^{\prime }$ transform. Taking into
account Eq. (\ref{rotdoublet2}) we get 
\begin{eqnarray*}
\widehat{A}^{\prime } &\equiv &\Phi _{1}^{\prime \dagger }\Phi _{1}^{\prime
}=\left( \Phi _{1}^{\dagger }\cos \theta -\Phi _{2}^{\dagger }\sin \theta
\right) \left( \Phi _{1}\cos \theta -\Phi _{2}\sin \theta \right) \\
&=&\Phi _{1}^{\dagger }\Phi _{1}\cos ^{2}\theta -\Phi _{1}^{\dagger }\Phi
_{2}\cos \theta \sin \theta -\Phi _{2}^{\dagger }\Phi _{1}\cos \theta \sin
\theta +\Phi _{2}^{\dagger }\Phi _{2}\sin ^{2}\theta \\
&=&\Phi _{1}^{\dagger }\Phi _{1}\cos ^{2}\theta -2\cos \theta \sin \theta
\left( \frac{\Phi _{1}^{\dagger }\Phi _{2}+\Phi _{2}^{\dagger }\Phi _{1}}{2}%
\right) +\Phi _{2}^{\dagger }\Phi _{2}\sin ^{2}\theta \\
&=&\widehat{A}\cos ^{2}\theta +\widehat{B}\sin ^{2}\theta -\sin 2\theta 
\widehat{C}
\end{eqnarray*}
similarly, we obtain the transformation for the other operators, the results
read 
\begin{eqnarray*}
\widehat{A}^{\prime } &=&\widehat{A}\cos ^{2}\theta +\widehat{B}\sin
^{2}\theta -\widehat{C}\sin 2\theta \\
\widehat{B}^{\prime } &=&\widehat{A}\sin ^{2}\theta +\widehat{B}\cos
^{2}\theta +\widehat{C}\sin 2\theta \\
\widehat{C}^{\prime } &=&\frac{1}{2}\widehat{A}\sin 2\theta -\frac{1}{2}%
\widehat{B}\sin 2\theta +\widehat{C}\cos 2\theta \\
\widehat{D}^{\prime } &=&\widehat{D}
\end{eqnarray*}
\begin{eqnarray*}
\widehat{A}^{\prime 2} &=&\widehat{A}^{2}\cos ^{4}\theta +\widehat{B}%
^{2}\sin ^{4}\theta +\widehat{C}^{2}\sin ^{2}2\theta +\frac{1}{2}\widehat{A}%
\widehat{B}\sin ^{2}2\theta \\
&&-2\widehat{A}\widehat{C}\sin 2\theta \cos ^{2}\theta -2\widehat{B}\widehat{%
C}\sin ^{2}\theta \sin 2\theta \\
\widehat{B}^{\prime 2} &=&\widehat{A}^{2}\sin ^{4}\theta +\widehat{B}%
^{2}\cos ^{4}\theta +\widehat{C}^{2}\sin ^{2}2\theta +\frac{1}{2}\widehat{A}%
\widehat{B}\sin ^{2}2\theta \\
&&+2\widehat{A}\widehat{C}\sin 2\theta \sin ^{2}\theta +2\widehat{B}\widehat{%
C}\sin 2\theta \cos ^{2}\theta \\
\widehat{C}^{\prime 2} &=&\frac{1}{4}\left( \widehat{A}^{2}+\widehat{B}%
^{2}\right) \sin ^{2}2\theta +\widehat{C}^{2}\cos ^{2}2\theta -\frac{1}{2}%
\widehat{A}\widehat{B}\sin ^{2}2\theta \\
&&+\frac{1}{2}\widehat{A}\widehat{C}\sin 4\theta -\frac{1}{2}\widehat{B}%
\widehat{C}\sin 4\theta \\
\widehat{D}^{\prime 2} &=&\widehat{D}^{2}
\end{eqnarray*}
\begin{eqnarray*}
\widehat{A}^{\prime }\widehat{B}^{\prime } &=&\left( \frac{1}{4}\widehat{A}%
^{2}+\frac{1}{4}\widehat{B}^{2}-\widehat{C}^{2}\right) \sin ^{2}2\theta +%
\widehat{A}\widehat{B}\left( \cos ^{4}\theta +\sin ^{4}\theta \right) \\
&&+\left( \widehat{A}\widehat{C}-\widehat{B}\widehat{C}\right) \sin 2\theta
\cos 2\theta \\
\widehat{A}^{\prime }\widehat{C}^{\prime } &=&\frac{1}{2}\widehat{A}^{2}\sin
2\theta \cos ^{2}\theta -\frac{1}{2}\widehat{B}^{2}\sin ^{2}\theta \sin
2\theta -\widehat{C}^{2}\sin 2\theta \cos 2\theta -\frac{1}{4}\widehat{A}%
\widehat{B}\sin 4\theta \\
&&+\widehat{A}\widehat{C}\left( 4\cos ^{2}\theta -3\right) \cos ^{2}\theta +%
\widehat{B}\widehat{C}\left( 4\cos ^{2}\theta -1\right) \sin ^{2}\theta \\
\widehat{A}^{\prime }\widehat{D}^{\prime } &=&\widehat{A}\widehat{D}\cos
^{2}\theta +\widehat{B}\widehat{D}\sin ^{2}\theta -\widehat{C}\widehat{D}%
\sin 2\theta
\end{eqnarray*}
\begin{eqnarray}
\widehat{B}^{\prime }\widehat{C}^{\prime } &=&\frac{1}{2}\widehat{A}^{2}\sin
2\theta \sin ^{2}\theta -\frac{1}{2}\widehat{B}^{2}\sin 2\theta \cos
^{2}\theta +\frac{1}{2}\widehat{C}^{2}\sin 4\theta +\frac{1}{4}\widehat{A}%
\widehat{B}\sin 4\theta  \notag \\
&&+\widehat{A}\widehat{C}\left( \cos 2\theta +2\cos ^{2}\theta \right) \sin
^{2}\theta +\widehat{B}\widehat{C}\left( \cos 2\theta -2\sin ^{2}\theta
\right) \cos ^{2}\theta  \notag \\
\widehat{B}^{\prime }\widehat{D}^{\prime } &=&\widehat{A}\widehat{D}\sin
^{2}\theta +\widehat{B}\widehat{D}\cos ^{2}\theta +\widehat{C}\widehat{D}%
\sin 2\theta  \notag \\
\widehat{C}^{\prime }\widehat{D}^{\prime } &=&\frac{1}{2}\left( \widehat{A}%
\widehat{D}-\widehat{B}\widehat{D}\right) \sin 2\theta +\widehat{C}\widehat{D%
}\cos 2\theta  \label{operator transf}
\end{eqnarray}

Now, we can build up a new parametrization of the potential such that 
\begin{eqnarray}
V_{g} &=&-\mu _{1}^{2}\widehat{A}-\mu _{2}^{2}\widehat{B}-\mu _{3}^{2}%
\widehat{C}-\mu _{4}^{2}\widehat{D}+\lambda _{1}\widehat{A}^{2}+\lambda _{2}%
\widehat{B}^{2}+\lambda _{3}\widehat{C}^{2}+\lambda _{4}\widehat{D}^{2} 
\notag \\
&&+\lambda _{5}\widehat{A}\widehat{B}+\lambda _{6}\widehat{A}\widehat{C}%
+\lambda _{8}\widehat{A}\widehat{D}+\lambda _{7}\widehat{B}\widehat{C}%
+\lambda _{9}\widehat{B}\widehat{D}+\lambda _{10}\widehat{C}\widehat{D}
\label{Vg rotated}
\end{eqnarray}
in order to find the values of $\mu _{i},\;\lambda _{i}\;$in terms of $%
\widetilde{\mu }_{i},\;\widetilde{\lambda }_{i}$, we use the Eqs. (\ref{Vg2}%
), and (\ref{operator transf}) to write e.g. the coefficient proportional to
the operator $\widehat{A}$, and these terms are compared with the term
proportional to the operator $\widehat{A}\;$in Eq. (\ref{Vg rotated})
obtaining 
\begin{equation*}
-\mu _{1}^{2}\widehat{A}=\left( -\widetilde{\mu }_{1}^{2}\cos ^{2}\theta -%
\widetilde{\mu }_{2}^{2}\sin ^{2}\theta -\widetilde{\mu }_{3}^{2}\sin \theta
\cos \theta \right) \widehat{A}
\end{equation*}
therefore, the coefficient $\mu _{1}^{2}\;$is related to the parameters $%
\widetilde{\mu }_{i},\;\widetilde{\lambda }_{i}\;$in the following way 
\begin{equation*}
\mu _{1}^{2}=\left( \widetilde{\mu }_{1}^{2}\cos ^{2}\theta +\widetilde{\mu }%
_{2}^{2}\sin ^{2}\theta +\frac{1}{2}\widetilde{\mu }_{3}^{2}\sin 2\theta
\right)
\end{equation*}
by the same token, the other sets$\;$of $\mu _{i},\;\lambda _{i}$ parameters
are related to the $\widetilde{\mu }_{i},\;\widetilde{\lambda }_{i}\;$%
parameters in the following way 
\begin{eqnarray*}
\mu _{1}^{2} &=&\left( \widetilde{\mu }_{1}^{2}\cos ^{2}\theta +\widetilde{%
\mu }_{2}^{2}\sin ^{2}\theta +\frac{1}{2}\widetilde{\mu }_{3}^{2}\sin
2\theta \right) \\
\mu _{2}^{2} &=&\left( \widetilde{\mu }_{1}^{2}\sin ^{2}\theta +\widetilde{%
\mu }_{2}^{2}\cos ^{2}\theta -\frac{1}{2}\widetilde{\mu }_{3}^{2}\sin
2\theta \right) \\
\mu _{3}^{2} &=&\left( -\widetilde{\mu }_{1}^{2}\sin 2\theta +\widetilde{\mu 
}_{2}^{2}\sin 2\theta +\widetilde{\mu }_{3}^{2}\cos 2\theta \right) \\
\mu _{4}^{2} &=&\widetilde{\mu }_{4}^{2}
\end{eqnarray*}

\begin{eqnarray*}
\lambda _{1} &=&\left( \widetilde{\lambda }_{1}\cos ^{4}\theta +\widetilde{%
\lambda }_{2}\sin ^{4}\theta +\frac{1}{4}\left( \widetilde{\lambda }_{3}+%
\widetilde{\lambda }_{5}\right) \sin ^{2}2\theta \right. \\
&&\left. +\frac{1}{2}\left( \widetilde{\lambda }_{6}\cos ^{2}\theta +%
\widetilde{\lambda }_{7}\sin ^{2}\theta \right) \sin 2\theta \right) \\
\lambda _{2} &=&\left( \widetilde{\lambda }_{1}\sin ^{4}\theta +\widetilde{%
\lambda }_{2}\cos ^{4}\theta +\frac{1}{4}\left( \widetilde{\lambda }_{3}+%
\widetilde{\lambda }_{5}\right) \sin ^{2}2\theta \right. \\
&&\left. -\frac{1}{2}\left( \widetilde{\lambda }_{6}\sin ^{2}\theta +%
\widetilde{\lambda }_{7}\cos ^{2}\theta \right) \sin 2\theta \right) \\
\lambda _{3} &=&\left( \left( \widetilde{\lambda }_{1}+\widetilde{\lambda }%
_{2}-\widetilde{\lambda }_{5}\right) \sin ^{2}2\theta +\widetilde{\lambda }%
_{3}\cos ^{2}2\theta +\frac{1}{2}\left( \widetilde{\lambda }_{7}-\widetilde{%
\lambda }_{6}\right) \sin 4\theta \right) \\
\lambda _{4} &=&\widetilde{\lambda }_{4}
\end{eqnarray*}

\begin{eqnarray*}
\lambda _{5} &=&\left( \frac{1}{2}\left( \widetilde{\lambda }_{1}+\widetilde{%
\lambda }_{2}-\widetilde{\lambda }_{3}\right) \sin ^{2}2\theta +\widetilde{%
\lambda }_{5}\left( \cos ^{4}\theta +\sin ^{4}\theta \right) \right. \\
&&\left. +\frac{1}{4}\left( \widetilde{\lambda }_{7}-\widetilde{\lambda }%
_{6}\right) \sin 4\theta \right) \\
\lambda _{6} &=&2\left( \widetilde{\lambda }_{2}\sin ^{2}\theta -\widetilde{%
\lambda }_{1}\cos ^{2}\theta \right) \sin 2\theta +\frac{1}{2}\left( 
\widetilde{\lambda }_{3}+\widetilde{\lambda }_{5}\right) \sin 4\theta \\
&&+\widetilde{\lambda }_{6}\left( 4\cos ^{2}\theta -3\right) \cos ^{2}\theta
+\widetilde{\lambda }_{7}\left( \cos 2\theta +2\cos ^{2}\theta \right) \sin
^{2}\theta \\
\lambda _{8} &=&\left( \widetilde{\lambda }_{8}\cos ^{2}\theta +\widetilde{%
\lambda }_{9}\sin ^{2}\theta +\frac{1}{2}\widetilde{\lambda }_{10}\sin
2\theta \right)
\end{eqnarray*}

\begin{eqnarray}
\lambda _{7} &=&2\left( \widetilde{\lambda }_{2}\cos ^{2}\theta -\widetilde{%
\lambda }_{1}\sin ^{2}\theta \right) \sin 2\theta -\frac{1}{2}\left( 
\widetilde{\lambda }_{3}+\widetilde{\lambda }_{5}\right) \sin 4\theta  \notag
\\
&&+\widetilde{\lambda }_{6}\left( 4\cos ^{2}\theta -1\right) \sin ^{2}\theta
+\widetilde{\lambda }_{7}\left( \cos 2\theta -2\sin ^{2}\theta \right) \cos
^{2}\theta  \notag \\
\lambda _{9} &=&\left( \widetilde{\lambda }_{8}\sin ^{2}\theta +\widetilde{%
\lambda }_{9}\cos ^{2}\theta -\frac{1}{2}\widetilde{\lambda }_{10}\sin
2\theta \right)  \notag \\
\lambda _{10} &=&\left( \left( \widetilde{\lambda }_{9}-\widetilde{\lambda }%
_{8}\right) \sin 2\theta +\widetilde{\lambda }_{10}\cos 2\theta \right)
\label{paramet rotated}
\end{eqnarray}

\subsection{Tadpoles}

From now on, we shall consider the potential with invariance under charge
conjugation, i.e. $\mu _{4}=\lambda _{8}=\lambda _{9}=\lambda _{10}=0$ (see
section \ref{Higgspot 2HDM}). In that case the tadpoles are given by 
\begin{eqnarray*}
T &=&\left( -\mu _{1}^{2}v_{1}-\frac{1}{2}\mu _{3}^{2}v_{2}+\lambda
_{1}v_{1}^{3}+\frac{1}{2}\lambda _{3}v_{1}v_{2}^{2}+\frac{1}{2}\lambda
_{5}v_{1}v_{2}^{2}+\frac{3}{4}\lambda _{6}v_{1}^{2}v_{2}\right. \\
&&\left. +\frac{1}{4}\lambda _{7}v_{2}^{3}\right) h_{1}+\left( -\mu
_{2}^{2}v_{2}-\frac{1}{2}\mu _{3}^{2}v_{1}+\lambda _{2}v_{2}^{3}+\frac{1}{2}%
\lambda _{3}v_{1}^{2}v_{2}\right. \\
&&\left. +\frac{1}{2}\lambda _{5}v_{1}^{2}v_{2}+\frac{1}{4}\lambda
_{6}v_{1}^{3}+\frac{3}{4}\lambda _{7}v_{2}^{2}v_{1}\right) h_{2}
\end{eqnarray*}
\begin{eqnarray}
T_{3} &=&\left( -\mu _{1}^{2}v_{1}-\frac{1}{2}\mu _{3}^{2}v_{2}+\lambda
_{1}v_{1}^{3}+\frac{1}{2}\lambda _{3}v_{1}v_{2}^{2}\right.  \notag \\
&&\left. +\frac{1}{2}\lambda _{5}v_{1}v_{2}^{2}+\frac{3}{4}\lambda
_{6}v_{1}^{2}v_{2}+\frac{1}{4}\lambda _{7}v_{2}^{3}\right)  \notag \\
T_{7} &=&\left( -\mu _{2}^{2}v_{2}-\frac{1}{2}\mu _{3}^{2}v_{1}+\lambda
_{2}v_{2}^{3}+\frac{1}{2}\lambda _{3}v_{1}^{2}v_{2}\right.  \notag \\
&&\left. +\frac{1}{2}\lambda _{5}v_{1}^{2}v_{2}+\frac{1}{4}\lambda
_{6}v_{1}^{3}+\frac{3}{4}\lambda _{7}v_{2}^{2}v_{1}\right)
\label{tadp no rot}
\end{eqnarray}
these tadpoles coincide with the minimum conditions as we see from Eq. (\ref
{min conditions}) and applying $\mu _{4}=\lambda _{8}=\lambda _{9}=\lambda
_{10}=0$. Now, we find the relation among the tadpoles in both
parametrizations by using Eq.(\ref{tadp no rot}), and the third of Eqs. (\ref
{gaumastri}).

\begin{eqnarray*}
T_{3}h_{1}+T_{7}h_{2} &=&T_{3}\left( h_{1}^{\prime }\cos \theta
+h_{2}^{\prime }\sin \theta \right) +T_{7}\left( -h_{1}^{\prime }\sin \theta
+\cos \theta h_{2}^{\prime }\right) \\
&=&\left( T_{3}\cos \theta -T_{7}\sin \theta \right) h_{1}^{\prime }+\left(
T_{3}\sin \theta +T_{7}\cos \theta \right) h_{2}^{\prime } \\
&=&T_{3}^{\prime }h_{1}^{\prime }+T_{7}^{\prime }h_{2}^{\prime }
\end{eqnarray*}
from \ which we see that the tadpoles in both parametrizations are related
through the rotation 
\begin{equation}
\left( 
\begin{array}{c}
T_{3}^{\prime } \\ 
T_{7}^{\prime }
\end{array}
\right) \equiv \left( 
\begin{array}{cc}
\cos \theta & -\sin \theta \\ 
\sin \theta & \cos \theta
\end{array}
\right) \left( 
\begin{array}{c}
T_{3} \\ 
T_{7}
\end{array}
\right)  \label{tadp relation}
\end{equation}
As a proof of consistency, we can check that from (\ref{tadp no rot}), and
from (\ref{paramet rotated}) the following relation is gotten after a bit of
cumbersome algebra 
\begin{eqnarray}
T_{3}\cos \theta -T_{7}\sin \theta &=&\left( -\widetilde{\mu }%
_{1}^{2}v_{1}^{\prime }-\frac{1}{2}\widetilde{\mu }_{3}^{2}v_{2}^{\prime }+%
\widetilde{\lambda }_{1}v_{1}^{\prime 3}+\frac{1}{2}\widetilde{\lambda }%
_{3}v_{1}^{\prime }v_{2}^{\prime 2}\right.  \notag \\
&&\left. +\frac{1}{2}\widetilde{\lambda }_{5}v_{1}^{\prime }v_{2}^{\prime 2}+%
\frac{3}{4}\widetilde{\lambda }_{6}v_{1}^{\prime 2}v_{2}^{\prime }+\frac{1}{4%
}\widetilde{\lambda }_{7}v_{2}^{\prime 3}\right)  \notag \\
T_{3}\sin \theta +T_{7}\cos \theta &=&\left( -\widetilde{\mu }%
_{2}^{2}v_{2}^{\prime }-\frac{1}{2}\widetilde{\mu }_{3}^{2}v_{1}^{\prime }+%
\widetilde{\lambda }_{2}v_{2}^{\prime 3}+\frac{1}{2}\widetilde{\lambda }%
_{3}v_{1}^{\prime 2}v_{2}^{\prime }\right.  \notag \\
&&\left. +\frac{1}{2}\widetilde{\lambda }_{5}v_{1}^{\prime 2}v_{2}^{\prime }+%
\frac{1}{4}\widetilde{\lambda }_{6}v_{1}^{\prime 3}+\frac{3}{4}\widetilde{%
\lambda }_{7}v_{2}^{\prime 2}v_{1}^{\prime }\right)  \label{tadp relat}
\end{eqnarray}
where 
\begin{equation*}
\left( 
\begin{array}{c}
v_{1}^{\prime } \\ 
v_{2}^{\prime }
\end{array}
\right) \equiv \left( 
\begin{array}{cc}
\cos \theta & -\sin \theta \\ 
\sin \theta & \cos \theta
\end{array}
\right) \left( 
\begin{array}{c}
v_{1} \\ 
v_{2}
\end{array}
\right)
\end{equation*}
relates the VEV between both parametrizations\footnote{%
Observe that, such rotation keeps invariant the quantity $%
v_{1}^{2}+v_{2}^{2}=v_{1}^{\prime 2}+v_{2}^{\prime 2}=\frac{2m_{w}^{2}}{g^{2}%
}$, as it should be.}. The terms on right of Eqs. (\ref{tadp relat}) are
precisely the tadpoles in the ``prime parametrization'', then the relations
given by Eqs. (\ref{tadp relation}) are held as expected. Therefore,
tadpoles are preserved by the rotation.

\subsection{Higgs boson masses}

Another important proof of consistency is to verify that both
parametrizations predict the same masses for the Higgs bosons. We shall use
once again, the potential with $C-$invariance Eq. (\ref{Lag10}). In a
general parametrization, the minimal conditions can be taken from \ref{min
conditions} by using $\mu _{4}=\lambda _{8}$ $=\lambda _{9}=\lambda _{10}=0$%
, as it corresponds to the Lagrangian in Eq. (\ref{Lag10}). Those conditions
are reduced to 
\begin{eqnarray*}
\mu _{1}v_{1} &=&\left( -\frac{1}{2}\mu _{3}v_{2}+\lambda _{1}v_{1}^{3}+%
\frac{1}{2}\lambda _{3}v_{2}^{2}v_{1}+\frac{1}{2}\lambda _{5}v_{2}^{2}v_{1}+%
\frac{3}{4}\lambda _{6}v_{1}^{2}v_{2}+\frac{1}{4}\lambda _{7}v_{2}^{3}\right)
\\
\mu _{2}v_{2} &=&\left( -\frac{1}{2}\mu _{3}v_{1}+\lambda _{2}v_{2}^{3}+%
\frac{1}{2}\lambda _{3}v_{1}^{2}v_{2}+\frac{1}{2}\lambda _{5}v_{1}^{2}v_{2}+%
\frac{1}{4}\lambda _{6}v_{1}^{3}+\frac{3}{4}\lambda _{7}v_{1}v_{2}^{2}\right)
\end{eqnarray*}
the mass matrix is obtained from (\ref{complete mass matrix}), and (\ref
{complete mass terms}) by using once again $\mu _{4}=\lambda _{8}$ $=\lambda
_{9}=\lambda _{10}=0$.\ Let us start with the matrix elements corresponding
to $m_{H^{0}},m_{h^{0}}$. If we assume that both VEV are different from zero
and utilize the minimum conditions, we obtain the following mass matrix 
\begin{equation}
\left( 
\begin{array}{cc}
M_{33}^{2} & M_{37}^{2} \\ 
M_{37}^{2} & M_{77}^{2}
\end{array}
\right)  \label{mass37param}
\end{equation}
with

\begin{eqnarray}
M_{33}^{2} &=&\frac{1}{4v_{1}}\left( 2\mu _{3}^{2}v_{2}+8\lambda
_{1}v_{1}^{3}+3\lambda _{6}v_{1}^{2}v_{2}-\lambda _{7}v_{2}^{3}\allowbreak
\right)  \notag \\
M_{37}^{2} &=&-\allowbreak \frac{1}{2}\mu _{3}^{2}+\frac{3}{4}\lambda
_{7}v_{2}^{2}+\frac{3}{4}\lambda _{6}v_{1}^{2}+\lambda _{3}\allowbreak
v_{1}v_{2}+\lambda _{5}v_{1}v_{2}  \notag \\
M_{77}^{2} &=&\frac{1}{4v_{2}}\left( 2\mu _{3}^{2}v_{1}+8\lambda
_{2}v_{2}^{3}-\lambda _{6}v_{1}^{3}+3\lambda _{7}v_{2}^{2}v_{1}\right)
\label{terms37param}
\end{eqnarray}
For the sake of simplicity, we just show that the determinant of this matrix
(i.e. the product of the squared masses), coincides for two parametrizations
connected by a transformation like (\ref{rotdoublet2}). The mass matrix in
any other parametrization with both VEV different from zero, have the same
form as (\ref{mass37param}, \ref{terms37param}) but replacing $\mu
_{i}^{2}\rightarrow \widetilde{\mu }_{i}^{2},\;\lambda _{i}\rightarrow 
\widetilde{\lambda }_{i}.\;$It is a fact of cumbersome algebra to demostrate
that 
\begin{equation*}
M_{33}^{2}M_{77}^{2}-\left( M_{37}^{2}\right) ^{2}=\widetilde{M}_{33}^{2}%
\widetilde{M}_{77}^{2}-\left( \widetilde{M}_{37}^{2}\right) ^{2}
\end{equation*}
this demostration is carried out by taking into account the relations (\ref
{paramet rotated}) among the parameters in both bases. In a similar fashion,
we can show that the eigenvalues coincide in both bases. Therefore, the mass
Higgs bosons are equal in both parametrizations as it must be. Finally, if
the angle of rotation is chosen such that one of the VEV is zero, (e.g. $%
v_{2}=0$) in one of the bases, then the minimum conditions and mass matrix
elements become much simpler (see Sec. \ref{minima of Lag10}), and the
equality is easier to demonstrate.

By the same token, we can check that for the other Higgs mass matrices the
determinants and eigenvalues are invariant under the transformation (\ref
{rotdoublet2}). Showing that the observables are not altered by this change
of basis.

\end{document}